%% file: main.tex
\documentclass[superscriptaddress,aps,prx,eqsecnum,nofootinbib,notitlepage,10pt,longbibliography]{revtex4-2}

\makeatletter
\def\switch@array{\let\@array@sw\@array@sw@LaTeX}
\makeatother

\usepackage{amsmath, amssymb, amstext, amsthm}
\usepackage{siunitx}
\usepackage{graphics}
\usepackage[dvipsnames, table]{xcolor}
\usepackage{enumitem}
\setenumerate[0]{label = (\roman*)}
\usepackage{tikz}

\usepackage{booktabs,makecell,siunitx}
\usetikzlibrary{quantikz2}
\usepackage{tabularx}
\usepackage{subcaption}
\usepackage{ragged2e}
\usepackage{subcaption}
\DeclareCaptionJustification{fulljust}{\justifying}
\usepackage{array,multirow}

\usepackage{zref-clever}

\usepackage[normalem]{ulem}

\newcommand\myshade{90}
\usepackage[pagebackref=false]{hyperref}
\hypersetup{
	plainpages = false,
	unicode = false,
	pdftoolbar = true, 
	pdfmenubar = true, 
	pdfnewwindow = true,
	colorlinks = true,
	linkcolor = NavyBlue!\myshade!black,
	citecolor = PineGreen!\myshade!black,
	filecolor = NavyBlue!\myshade!black,
	urlcolor = NavyBlue!\myshade!black
}

\input{commands}

\makeatletter
\renewcommand\tableofcontents{%
  \twocolumngrid
  \begingroup
    \hypersetup{linkcolor=black}%
    \let\appendix\relax 
    \@starttoc{toc}%
  \endgroup
  \onecolumngrid
}
\def\l@subsubsection#1#2{}
\makeatother

\begin{document}

\title{Ground-state energy estimation compiled on cat qubits: 7\,164 qubits for FeMoco-54 in 23.5 hours}

\author{Larissa Kroell}
\affiliation{Alice \& Bob, 49 Bd du Général Martial Valin, 75015 Paris, France}
\author{Hugo Jacinto}
\affiliation{Alice \& Bob, 49 Bd du Général Martial Valin, 75015 Paris, France}
\affiliation{Université Paris--Saclay, CEA, CNRS, Institut de Physique Théorique, \num[detect-all]{91191} Gif-sur-Yvette, France}
\author{Victor Hertz}
\affiliation{Alice \& Bob, 49 Bd du Général Martial Valin, 75015 Paris, France}
\author{Axel Pappalardo}
\affiliation{Alice \& Bob, 49 Bd du Général Martial Valin, 75015 Paris, France}
\author{Adel Ben Moussa}
\affiliation{Alice \& Bob, 49 Bd du Général Martial Valin, 75015 Paris, France}
\author{Manuel Rispler}
\affiliation{Alice \& Bob, 49 Bd du Général Martial Valin, 75015 Paris, France}
\author{Peter Shanahan}
\affiliation{Alice \& Bob, 49 Bd du Général Martial Valin, 75015 Paris, France}
\affiliation{School of Informatics, University of Edinburgh, United Kingdom}
\author{\'Elie Gouzien}
\affiliation{Alice \& Bob, 49 Bd du Général Martial Valin, 75015 Paris, France}
\author{Christophe Vuillot}
\affiliation{Alice \& Bob, 49 Bd du Général Martial Valin, 75015 Paris, France}
\author{Vivien Londe}
\affiliation{Alice \& Bob, 49 Bd du Général Martial Valin, 75015 Paris, France}

\begin{abstract}
We provide a detailed resource estimate of the state-of-the-art sum-of-squares block encoding algorithm for electronic structure Hamiltonians of Low et al. on superconducting cat qubits.
To achieve this, we present compilations with closed-form costs of the dominant subroutines such as skew-tree \textsc{Qrom}, a \textsc{Swap} network and a \textsc{Phaseup} for both a repetition-cat and an LDPC-cat architecture.
Our compilation represents algorithmic qubits in the $X$ basis, taking advantage of the asymmetric cost of $X$ and $Z$ access that the concatenation of cat qubits with phase-flip classical code brings.
Additionally, we provide a protocol for $\vert i \rangle$-state distillation and a two-level Toffoli-like $\vert \text{C}_X\text{C}_X X \rangle$ magic state distillation protocol, including three factory layouts.
To minimize the cost of Clifford corrections induced by magic state injections we tailor injection gadgets to each subroutine.
Combining our compilation work with the adder compilation from the companion work Pappalardo et al., we provide global layouts for the block encoding in both architectures which we use to obtain resource estimates for ground-state energy estimation for the benchmarks of FeMoco-54 and FeMoco-76.
Assuming a cycle time of \qty{1}{\micro\second}, a reaction time of \qty{10}{\micro\second}, a phase estimation precision of \qty{1}{\milli\hartree} and circuit-level phase-flip noise with probability $p_Z = 5 \times 10^{-4}$ with negligible bit-flip error, FeMoco-54 requires fewer than $7\,500$ cat qubits with an expected runtime of under one day for a single phase estimation run using an LDPC-cat architecture.
Relaxing the bit-flip requirement to $p_X = 10^{-11}$ with an additional layer of elevator codes raises this to an estimated $13\,000$ cat qubits.
\end{abstract}
\maketitle

\tableofcontents

\input{sections/introduction.tex}
\input{sections/high_level.tex}
\input{sections/compilation}
\input{sections/physical_resources.tex}
\input{sections/ldpc_circuit_noise}
\input{sections/acknowledgments}

\appendix
\counterwithin*{equation}{section}
\renewcommand\theequation{\thesection\arabic{equation}}
\input{sections/appendix}

\input{sections/app_i_state_dist.tex}

\input{sections/app_Clifford_gates}

\newpage
\bibliographystyle{alpha-revtex}
\bibliography{references_femoco}

\end{document}

%% file: commands.tex
\zcsetup{cap}

\newcommand{\bfr}{\mathbf{r}}
\newcommand{\bfu}{\mathbf{u}}
\newcommand{\bfx}{\mathbf{x}}
\newcommand{\bfv}{\mathbf{v}}

\newcommand{\cO}{\mathcal{O}}
\newcommand{\cS}{\mathcal{S}}
\newcommand{\cH}{\mathcal{H}}

\newcommand{\R}{\mathbb{R}}
\newcommand{\I}{\mathbb{I}}
\newcommand{\bC}{\mathbb{C}}

\newcommand{\ii}{\mathrm{i}}  

\newcommand{\DFTHC}{\mathrm{DFTHC}}

\newcommand{\srt}{\mathrm{sqrt}}
\newcommand{\sos}{\mathrm{SOS}}
\newcommand{\fro}{\mathrm{fro}}
\newcommand{\reg}{\mathrm{reg}}
\DeclareMathOperator{\argmin}{arg\hspace{1pt}min}
\newcommand{\Relu}{\mathrm{Relu}}
\newcommand{\gap}{\mathrm{gap}}
\newcommand{\SF}{\mathrm{SF}}
\newcommand{\nega}{\mathrm{neg}}
\newcommand{\pos}{\mathrm{pos}}
\newcommand{\transpose}{\mathrm{T}}

\newcommand{\coef}{\mathrm{coef}}
\newcommand{\rot}{\mathrm{rot}}
\newcommand{\BE}[1]{\mathrm{BE}[#1]}

\newcommand{\LDPC}{\mathrm{LDPC}}
\newcommand{\rep}{\mathrm{rep}}

\newcommand{\xor}{\oplus}

\newcommand{\C}{\mathord{\mathrm{C}}}

\newcommand{\CZ}{\mathord{\C Z}}

\newcommand{\CCZ}{\mathord{\C\C Z}}

\newcommand{\CXCXX}{\mathord{\C_X \C_X X}}
\newcommand{\CXCXZ}{\mathord{\C_X \C_X Z}}
\newcommand{\CXX}{\C_X X }

\newcommand{\ketXmag}{\ket{X^{1/4}}}
\newcommand{\keti}{\ket{\ii}}

\DeclareSIUnit{\hartree}{Ha}

\newcommand{\outcome}[1]{\mathord{:}#1}

%% file: sections/introduction.tex
\section{Introduction}
\label{sec:intro}

The iron-molybdenum cofactor (FeMoco) of nitrogenase has been a standard benchmark for fault-tolerant ground-state energy estimation for electronic structure Hamiltonians since \cite{Reiher+Elucidating17}.
Usually two active space models are discussed, FeMoco-54  \cite{Reiher+Elucidating17} and FeMoco-76 \cite{Li+FeMocoComplexity19}. While the ground-state energy of FeMoco-54 was shown to be accessible to classical simulation in \cite{Li+FeMocoComplexity19}, the ground-state energy calculations for FeMoco-76 remained out of reach until \cite{Zhai+ClassicalFeMoco26}. In this recent work, the authors estimate the ground-state energy of FeMoco-76 to about chemical accuracy by extrapolating coupled-cluster and density matrix renormalization group calculations at an idealized cost of under a minute on the Frontier supercomputer. These calculations come without a formal error bound. A variational calculation reaching the same accuracy without extrapolation would take an estimated 22.9 hours. We therefore do not treat FeMoco as a demonstration of quantum advantage but as a benchmark: its Hamiltonians are fully specified, so differences between resource estimates can be attributed to the algorithm, the error-correcting code, and the compilation. 

The cost of ground-state energy estimation for these benchmarks has decreased with each new representation of the Hamiltonian, from sparse and low-rank qubitization \cite{Berry+QubitizationSparse19} to double factorization \cite{vBurg+QCCompCat21}, tensor hypercontraction \cite{Lee+EvenMore21}, and block-invariant symmetry shifts \cite{LoaizaIzmaylovBLISS23,Rocca+DFBLISS24,Caesura+THCBLISS25,Patel+Min1-normBliss25}. In this work, we use the state-of-the-art double-factorized tensor hypercontraction (DFTHC) ansatz of \cite{Low+FastQuantumSA2025}, which writes the Hamiltonian as a sum of squares (SOS) and combines its block encoding with spectral amplification \cite{LowChuangUniformSA17,ZlokapaSomma24,King+SOSSA26}.  
In this work, we use the same Hamiltonians, DFTHC parameters, and precisions as in \cite[Table~XII]{low2026denser} and follow the block-encoding structure of \cite{Low+FastQuantumSA2025,low2026denser}.

\paragraph{Cat-qubit architectures and hardware-tailored compilation.}
Architectures based on bosonic cat qubits reduce the overhead of error correction from the hardware side.
In a dissipatively stabilized cat qubit, bit-flip errors are suppressed exponentially with the mean photon number, while phase-flip errors grow only linearly \cite{Mirrahimi+CatQubits14,Guillaud_2019}.
The resulting biased qubit only needs to be protected against phase flips, which can be done with a classical code such as a repetition code \cite{Guillaud_2019} or a classical LDPC code that remains implementable in 2D \cite{Ruiz_2025}. For the repetition code, the number of physical qubits per logical qubit then grows linearly with the code distance $d$ rather than quadratically as in the surface code, and LDPC codes reduce it further through a higher encoding rate. 

\begin{figure}[!htb]
  \centering
  \includegraphics[width=0.9\linewidth]{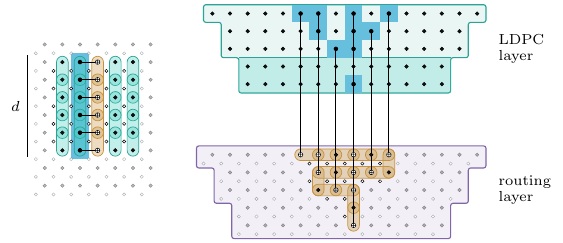}
  \caption{Access to the logical operators of algorithmic qubits, shown for the step common to all operations requiring $Z$ access: the support of $\bar Z$ (blue) is coupled by transversal \textsc{Cnot}s to an auxiliary repetition code (amber) prepared in $\ket{0}_L$. (Left) Repetition-cat architecture: $Z_L$ spans all $d$ qubits of the code, so $Z$ access is paid in space: a corridor of $d$ data qubits alongside the code, which is shared by a pair of logical qubits. (Right) Hybrid LDPC-cat architecture, with the LDPC layer drawn above the routing layer on which it is stacked: $Z$ access is provided by a snake-shaped repetition code prepared in the routing layer underneath the support of $Z_L$. No corridors are needed, but all logical qubits share the routing layer, so $Z$ access is paid in time: operations that would congest the routing layer are performed sequentially.}
  \label{fig:access}
\end{figure}

The price of this efficiency is a restricted set of logical operations \cite{Guillaud_2019,barizien2025accessiblequantumgatesclassical}. As the codes are built from $X$-type stabilizers only, lattice surgery is restricted to $X$-type Pauli measurements \cite{Gouzien_2023,Ruiz_2025,Simultaneous2026prep}, and $Z$-type operations rely on the bias-preserving \textsc{Cnot}, which is transversal between repetition codes. 
The $X$ logical operator in both architectures can be accessed requiring interaction with only one physical data qubit. 
On the contrary, accessing the $Z$ logical operator requires interaction with several qubits and comes at different costs for the different architectures, as illustrated in \zcref{fig:access}.
In the repetition-cat architecture, $Z$ access requires an access corridor alongside the code and is thus paid in space, whereas in the hybrid LDPC-cat architecture \cite{Ruiz_2025} it is provided by a routing layer shared by all logical qubits. In this shared routing layer, access to $Z$ logical operators can conflict, and the access is thus paid in time.
Additionally, Clifford resource states such as $\ket{i}$ have to be distilled alongside the non-Clifford magic states \cite{Ruiz_2026}.
This together with the asymmetric access considerations detailed above distinguishes compilation on a cat-qubit architecture from its surface code counterpart.
Compiling an algorithm for a cat-qubit architecture thus requires for each subroutine to be laid out and scheduled anew, and since $Z$ access is paid in space or in time, the best choice of subroutines depends on the architecture.

Our main contribution is given by closed form compilation results on cat-qubits for both a repetition-cat and an LDPC-cat architecture. To compile the block-encoding efficiently with respect to the asymmetric access requirement, we view algorithmic qubits as represented in the $X$ basis, so that a Toffoli gate becomes a $\CXCXZ$ gate and a multi-target \textsc{Cnot} becomes a parity-controlled \textsc{Not}. In this color-changed convention, the address and system registers only require $X$ access (unless they are used as dirty ancillae), so they can be stored densely in the repetition-cat architecture and are more easily accessible in the LDPC-cat architecture. Using this framework, we give lattice-surgery compilations of the skew-tree \textsc{Qrom} \cite{KhattarGidneyCondClean25,low2026denser}. We present the first compilation of \textsc{Swap} networks \cite{low2024trading} on cat-qubit architectures, in which some of the $X^{1/2}$ corrections arising from state injection are merged into one per address qubit. Furthermore, we provide a compilation of a \textsc{Phaseup} together with a specific strategy for its practical compilation that minimizes the number of phase gates to apply, thus minimizing the number of Clifford resource states consumed.
For the resource states, we propose a distillation protocol for $\ket{i}$ states in repetition codes, based on \cite[Section~V]{Ruiz_2026}, as well as two-level factories for $\ket{\CXCXX}$ states with three pipelined layouts, reaching output error rates of about $10^{-11}$ at $p_Z = 10^{-3}$ and $2\times10^{-13}$ at $p_Z = 5\times10^{-4}$. 
Finally, we assemble global layouts and end-to-end resource counts for FeMoco-54 and FeMoco-76 on both architectures. The latter relies on the compilation of an adder, which is done in our companion work \cite{Shor2026prep} that shares our compilation framework and uses the results obtained here.

We use the standard assumptions of a \qty{1}{\micro\second} cycle time and a \qty{10}{\micro\second}  reaction time.
We consider circuit-level noise with phase-flip errors only, with an input phase-flip error rate of $p_Z = 5\times10^{-4}$ and, as a second operating point, $p_Z = 10^{-3}$.
We assume bit-flip errors to be negligible and discuss an additional layer of bit-flip protection in \zcref{sec:elevator-codes}.
We also provide a more precise logical error model of the LDPC-cat codes~\cite{Ruiz_2025} obtained from new circuit-level simulations in \zcref{sec:ldpc_circuit_noise}.
Throughout our compilation, Toffoli gates are executed serially using a single two-level factory in the simultaneous layout of \zcref{subsec: CxCxX_factories}, the algorithmic region takes the footprint admitted by its largest subroutine, and within a given footprint all remaining choices are made to minimize the runtime (see \zcref{sec: Resource Count BE}).
The distance $d$ is also chosen by minimizing the expected runtime. 

We compare both architectures on a reference configuration given by the smallest possible compute zone, which loads $\lambda = 2$ rotation angles at a time.
This ensures optimal space usage by all compiled primitives without adding additional qubits solely as ancillae.
The results of this comparison are given in \zcref{tab:intro-comparison}.
For an input phase-flip error rate of $p_Z = 5\times10^{-4}$, we find that for both FeMoco-54 and FeMoco-76 the LDPC-cat architecture leads to space footprint reduction of 36\% to 38\% for comparable runtime.

\begin{table}[!htb]
  \centering
  \begin{tabular}{llrrrrrr}
    \toprule
    & & \multicolumn{3}{c}{FeMoco-54} & \multicolumn{3}{c}{FeMoco-76} \\
    \cmidrule(lr){3-5}\cmidrule(lr){6-8}
    Architecture & Dirty ancillae & Qubits & Toffolis & $\langle t\rangle$ & Qubits & Toffolis & $\langle t\rangle$ \\
    \midrule
    Repetition-cat & no  & $\mathbf{11\,192}$ & $\mathbf{2.89\times 10^{9}}$ & \textbf{0.95 d} & $\mathbf{12\,568}$ & $\mathbf{1.36\times 10^{10}}$ & \textbf{4.49 d} \\
    Repetition-cat & yes & $13\,028$ & $1.99\times 10^{9}$ & 0.68 d & $15\,152$ & $6.70\times 10^{9}$ & 2.20 d \\
    LDPC-cat & no  & $7\,164$ & $2.91\times 10^{9}$ & 1.10 d & $7\,818$ & $1.38\times 10^{10}$ & 5.09 d \\
    LDPC-cat & yes & $\mathbf{7\,164}$ & $\mathbf{2.01\times 10^{9}}$ & \textbf{0.98 d} & $\mathbf{7\,818}$ & $\mathbf{6.94\times 10^{9}}$ & \textbf{3.31 d} \\
    \bottomrule
  \end{tabular}
    \caption{Physical qubits, Toffoli count, and expected duration of a single phase-estimation run at $p_Z = 5\times10^{-4}$ for respectively $d_\text{LDPC} = 16$ and $d_\text{rep} = 15$, with and without using the system qubits as dirty ancillae for data loading.
    Bold entries are the compilations of \zcref{tab: Rep code error accounting,tab: LDPC code error accounting}. All entries use the reference configuration, i.e. the smallest possible compute zone ($\lambda = 2$).
    For FeMoco-54, at the same footprint, the repetition-cat architecture admits a faster configuration with $\lambda = 3, d = 13$ (see \zcref{fig:tradeoff lambda sweep intro}).}
  \label{tab:intro-comparison}
\end{table}

\begin{figure}[h]
    \centering
    \includegraphics[width=\linewidth]{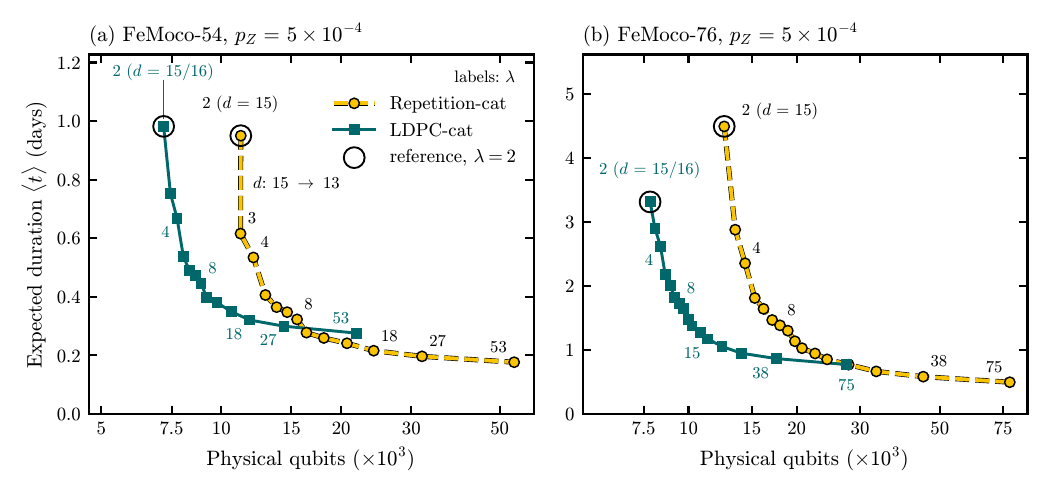}
    \caption{The space--time trade--off curve showing physical qubits against expected runtime for both FeMoco-54 and FeMoco-76 as a function of an increased compute zone.
    The size of the compute zone is parametrized by the number of rotation angles loaded, $\lambda$.
    Each resource estimate is obtained using the fastest algorithm for each fixed layout and the distance is chosen to minimize expected runtime. Note that for the LDPC-cat architecture this process leads to $d_\rep = 15$ and $d_\LDPC = 16$ throughout, and for the repetition-cat architecture it leads to $d_\rep = 15$ for $\lambda = 2$ for FeMoco-54 and for all $\lambda$ for FeMoco-76, and $d_\rep = 13$ for $\lambda \geq 3$ for FeMoco-54.  The circled data point for each architecture marks the reference configuration.}
    \label{fig:tradeoff lambda sweep intro}
\end{figure}

\zcref{fig:tradeoff lambda sweep intro} shows what happens when we increase the compute zone to allow for more rotation angles to be loaded at the same time for a phase-flip error rate of $p_Z = 5\times10^{-4}$.
Increasing the compute zone reduces the number of lookup table applications necessary to load the rotation angles and thus decreases the expected runtime.
This compute zone sweep also reveals a smaller footprint configuration for the repetition-cat architecture and FeMoco-54.
Loading $\lambda = 3$ rotation angles at a time leads to enough of a runtime reduction, so that the distance can be dropped to $d=13$ as opposed to $d=15$ for $\lambda = 2$.
The distance reduction offsets the increase from the compute zone in physical qubits, leading to our smallest repetition-cat footprint compilation for FeMoco-54 on $11\,176$ physical qubits and an expected runtime of 14.8 hours.
\zcref{fig:tradeoff lambda sweep intro} also reveals a smallest spacetime compilation for each molecule for this specific trade-off. More precisely, for FeMoco-54 the smallest spacetime compilation has a spacetime volume of 0.088 megaqubit-hours and is given by the LDPC-cat architecture loading $\lambda = 9$ rotation angles at a time.
For FeMoco-76, the smallest spacetime compilation is also given by the LDPC-cat architecture with a spacetime volume of  0.314 megaqubit-hours at $\lambda = 19$. 

\paragraph{Architecture dependence of compiled costs.}\label{sec: intro architecture dependence}

The two architectures pay for $Z$ access differently (\zcref{fig:access}), which changes the optimal choice of data loading.
To separate the effect of the architecture from the effect of the data loading, \zcref{tab:intro-comparison} also reports the cost of running the algorithm optimized for one architecture on the other. For the repetition-cat architecture this means adding access corridors for the system qubits to use them as dirty ancillae. 
One interesting learning from the comparison of both architectures' spacetime tradeoff behavior is that, for a given circuit, repetition-cat is typically faster as it allows for easier parallelization.
However, speed-up from algorithmic spacetime tradeoff tends to favor LDPC-cat, as increasing the number of logical qubits is proportionally cheaper than in repetition-cat architecture.
In both architectures, loading and applying the Givens rotation angles dominates the runtime, with around $65\%$ and $70\%$ of the cycles in the repetition-cat and the LDPC-cat architecture, for the reference configuration respectively.
The second-most costly routine is given by state preparation for the inner block encoding necessary for the two-electron terms in the SOS ansatz, which we compile using pure state preparation.
In terms of primitives, \textsc{Qrom} constitutes nearly 90\% of cycles for the repetition cat architecture.
For the LDPC-cat architecture the split is given by 45\% \textsc{Qrom} and 32\% \textsc{Swap}, since the system qubits can be used as dirty ancillae.

\paragraph{Comparison, limitations and scope.}

Comparing our results with the state-of-the-art surface code compilation in \cite{low2026denser} with regular non-biased noise qubits, we see that for a given expected runtime, across all compute-zone sizes and both molecules, the surface-code estimate needs $8$--$22\times$ more physical qubits than the repetition-cat architecture and $15$--$37\times$ more than the LDPC-cat architecture to be equally fast.
As for high-rate superconducting qLDPC architectures~\cite{Bravyi_2024} this overhead reduction comes at the price of increased hardware complexity.
However, developing architecture-tailored compilation offers a useful comparison to showcase the relevance of noise-biased hardware development.
In general, each hardware platforms comes with its own pros and cons. 
Neutral atoms~\cite{cain_shors_2026,quera_qldpc_VHER, webster2026pinnaclearchitecturereducingcost} have flexible long-range connectivity but low clock rate.
Transmons~\cite{gidney2025factor} offer a fast clock rate but are hard to connect beyond nearest neighbor planar connectivity, thus limiting the logical encoding rate.
Now, our estimates indicate that dissipative cat qubits~\cite{Chamberland_2022,reglade2024quantum,rousseau2025enhancing,putterman2025hardware,paradina2026engineering} offer a competitive route to low-overhead fault-tolerant quantum computation, at the cost of more complex qubit designs but with only local connectivity.

The next natural step would be to better explore how resistant the high overhead reduction is to lower noise bias. 
In particular, we leave it for upcoming work to build a proper logical gate-set and lead a compilation effort on a concatenated architecture correcting for a small amount of bit flips as well.
Indeed, neglecting bit flips as we have done in the resource estimates above requires a physical bit-flip error rate of $p_X \approx 10^{-18}$.  With an additional layer of elevator codes \cite{shanahan2026elevator}, a physical bit-flip error rate of $p_X = 10^{-11}$ suffices, and FeMoco-54 at $p_Z = 5\times10^{-4}$ in the reference configuration is then extrapolated to cost $19\,830$ cat qubits in the repetition-cat architecture and an estimated $12\,694$ in the LDPC-cat architecture as explained in \zcref{sec:elevator-codes}.
As resource state factories account for a substantial amount of the total footprint, it would be valuable to develop more efficient strategies by for example leveraging multiple-output protocols~\cite{Haah_2018,singh2026borrowed} or finding ways to eliminate the second level distillation stage by using larger unfolded distillation protocols~\cite{londe2026} while still mitigating hardware constraints.
Having in mind space efficiency, our current compilation work relies on the serial application of Toffoli gates with a fixed single $\ket{\CXCXX}$ factory.
Regarding the resource estimates, we do not parallelize phase gradient addition to apply several Givens rotations at the same time and the method of state preparation is also fixed to pure state preparation.
Leveraging the availability of several magic state per logical cycle and allowing for changing other subroutines, could thus further reduce the expected runtime. 

The paper is organized as follows. \zcref{sec:factorization} reviews the DFTHC ansatz with spectral amplification, and \zcref{sec:block-encoding} describes the resulting block encoding, following \cite{Low+FastQuantumSA2025,low2026denser}. \zcref{sec:subroutines} summarizes the data-loading subroutines from the literature that we compile. \zcref{sec:compilation} introduces the compilation conventions and the resource state factories, and compiles the subroutines on both architectures. \zcref{sec: Resource Count BE} assembles the global layouts and resource counts for the reference configuration as well as an increased compute zone, compares the two architectures, and discusses bit-flip protection. \zcref{sec:ldpc_circuit_noise} presents the logical error rates of the LDPC-cat codes. \zcref{app:factorization} collects details on the Hamiltonian factorizations, \zcref{app:istate_distillation} on the distillation of $\ket{i}$ states, and \zcref{app: Clifford} on the circuit identity used of Clifford gates injections.

%% file: sections/high_level.tex
\section{Block encoding overview}\label{sec:alg description}
\subsection{Factorization techniques for electronic structure Hamiltonians}
\label{sec:factorization}

We consider an electronic structure Hamiltonian in the spin-free formalism \cite{Low+FastQuantumSA2025}, i.e., 

\begin{equation}\label{eq: H definition}
H = \sum_{i,j} h_{ij}^{(1)} E_{ij} + \frac12 \sum_{i,j,k,\ell} h_{ijk\ell}^{(2)} E_{ij} E_{k\ell}, \quad E_{ij} = \sum_{\sigma \in \{0,1\}} a_{i\sigma}^\dagger a_{j \sigma},
\end{equation}
where $a_{i\sigma}^\dagger, a_{j\sigma}$ denote fermionic creation and annihilation operators respectively. Here, the one-body operator $ h_{ij}^{(1)}$ and the two-body Coulomb operator $h_{ijk\ell}^{(2)}$ correspond to the standard one- and two-electron integrals
\begin{align*}
h_{ij}^{(1)} &= \int \phi_i ( \bfr)\bigg( -\frac12 \nabla^2 - U(\bfr)\bigg) \phi_j(\bfr) \mathrm{d}\bfr ,\\
h_{ijk\ell}^{(2)} &= \int \frac{\phi_i(\bfr_1) \phi_j(\bfr_1) \phi_k(\bfr_2) \phi_\ell(\bfr_2)}{\Vert \bfr_1 - \bfr_2 \Vert} \mathrm{d}\bfr_1 \mathrm{d} \bfr_2
\end{align*}
for real molecular orbitals $\{ \phi_i(\bfr)\}$ and nuclear potential $U(\bfr)$. Observe that both operators are real, the one-body operator is symmetric and the Coulomb operator satisfies the following symmetries $h_{i j k \ell}^{(2)} = h_{j i k \ell}^{(2)} = h_{j i \ell k}^{(2)} = h_{\ell k j i}^{(2)} = h_{i j \ell k}^{(2)} = h_{\ell k i j}^{(2)} = h_{k \ell i j}^{(2)} = h_{k \ell j i}^{(2)}$ for all $i,j,k,\ell$. The one- and two-body integrals can be classically computed and are given as input to the electronic structure problem. In order to efficiently encode the Hamiltonian, it is necessary to reduce the amount of coefficients and thus, we are interested in compressions of the two-body Coulomb operator $h_{ijk\ell}^{(2)}$. In what follows, we write $[n] = \{0,\ldots, n-1\}$ for a positive integer $n$, with the convention $[0] = \varnothing$.

\subsubsection{Double Factorization}\label{sec: DF}
As the name suggests, double factorization is based on applying spectral decomposition twice. In the context of quantum computing, this was applied to electronic structure Hamiltonians in \cite{vBurg+QCCompCat21, Motta2021LowRank}. First, observe that  we can view $h_{ijk\ell}^{(2)} = h_{(ij),(k\ell)}^{(2)}$ as a real symmetric $N^2 \times N^2$ matrix. Thus, consider its spectral decomposition $h_{ijk\ell}^{(2)} = \sum_{r\in [N^2]} \lambda_r \Tilde{L}_{ij}^{(r)} \Tilde{L}_{k\ell}^{(r)} = \sum_{r\in [N^2]} L_{ij}^{(r)} L_{k\ell}^{(r)}$ for real matrices $\Tilde{L}_{ij}^{(r)} $, where in the last equality we set  $L_{ij}^{(r)} = \sqrt{\lambda_r}\Tilde{L}_{ij}^{(r)}$. This is possible, since the Coulomb operator is positive semidefinite (see \zcref{app: Positivity Coulomb op}), hence $\lambda_r \geq 0$. In practice, this decomposition can be computed efficiently using for example a Cholesky decomposition. Next, note that utilizing the symmetries of $h_{ijk\ell}^{(2)}$, we can assume that $L_{ij}^{(r)}$ is also symmetric by simply considering $\tfrac12 \big(L_{ij}^{(r)} + L_{ji}^{(r)}\big)$ instead. More precisely, we have
\begin{equation*}
\sum_{r\in [N^2]} \frac14 \bigg(L_{ij}^{(r)} + L_{ji}^{(r)}\bigg)\bigg(L_{k\ell}^{(r)} + L_{\ell k}^{(r)}\bigg) = \frac14 \bigg(h_{ijk\ell}^{(2)} + h_{jik\ell}^{(2)} + h_{ij\ell k}^{(2)} + h_{ji\ell k}^{(2)}\bigg) = h_{ijk\ell}^{(2)}.
\end{equation*}
In sum, we can write $h_{ijk\ell}^{(2)} = \sum_{r\in [N^2]} L_{ij}^{(r)} L_{k\ell}^{(r)}$, for real symmetric matrices $L_{ij}^{(r)}$. Using this, we can again consider its spectral decomposition $L_{ij}^{(r)} = \sum_{b\in [N]} w_b^{(r)} u_{bi}^{(r)} u_{bj}^{(r)}$, for $w_b^{(r)} \in \R, \bfu_b \in \R^N$ and  $\bfu_b \cdot \bfu_{b^\prime} = \delta_{bb^\prime}$. At the end, we obtain the following equivalent expression for the Coulomb operator
\begin{equation*}
h_{ijk\ell}^{(2)} = \sum_{r \in [N^2]} \sum_{b, b^\prime \in [N]} w_b^{(r)} w_{b^\prime}^{(r)} u_{b^\prime i}^{(r)} u_{b^\prime j}^{(r)} u_{bk}^{(r)} u_{b\ell}^{(r)}.
\end{equation*}

As described, this factorization does not necessarily lead to a reduction in coefficients that need to be loaded, since we decompose $h_{ijk\ell}^{(2)}$ as $N^2$, $N\times N$ matrices giving $\cO(N^4)$ coefficients. However, the key idea is to reduce the range $r\in [N^2]$ to $r\in [R]$ for some choice of rank $R \ll N^2$ leading to $\cO(RN^2)$ coefficients. This can also be pushed further by restricting the range of $b\in [N]$  as done in \cite{vBurg+QCCompCat21}. Double factorization then results in the following Hamiltonian
\begin{equation}\label{eq: H DF}
    H_\mathrm{DF} = \sum_{i,j} h_{ij}^{(1)} E_{ij} + \frac12 \sum_{r \in [R]} \left( \sum_{b \in [B]} \sum_{i,j} w_b^{(r)} u_{bi}^{(r)} u_{bj}^{(r)}E_{ij} \right)^2,
\end{equation}
where $R, B$ are each chosen by appropriate truncation taking the approximation error into account. 

\subsubsection{Tensor Hypercontraction}\label{sec: THC}
The goal of Tensor Hypercontraction is to decompose $h^{(2)}_{ijk\ell}$ as a product of order-2 tensors. In particular, we have the following Hamiltonian
\begin{equation}\label{eq: H THC}
    H_\mathrm{THC} = \sum_{i,j} h_{ij}^{(1)} E_{ij} + \frac12 \sum_{i,j,k,\ell} \sum_{b,c \in [M]} \chi_{bi} \chi_{bj} \zeta_{bc} \chi_{ck} \chi_{c\ell} E_{ij} E_{k\ell},
\end{equation}
where $\zeta_{bc} = \zeta_{cb}$, and $M$ is a rank parameter. To obtain the coefficients \cite{Lee+EvenMore21} minimize the two-norm error:
\begin{equation*}
    \Vert h^{(2)} - G \Vert_2^2 = \sum_{i,j,k,\ell} \vert h^{(2)}_{ijk\ell} - G_{ijk\ell} \vert^2,\, \text{ for } \, G_{ijk\ell} = \sum_{b,c \in [M]} \chi_{bi} \chi_{bj} \zeta_{bc} \chi_{ck} \chi_{c\ell}.
\end{equation*}
The difficulty lies in the non-linear nature and subsequent dependency on the initial value for this minimization problem. If the real molecular orbitals $\phi_i(\bfr)$ are explicitly available, a robust initial guess for $\chi$ and $\zeta$ can be constructed by an interpolative decomposition of $\phi_i(\bfr) \phi_j(\bfr)$. However, if only $h^{(2)}$ is known, random initial guesses have to be performed. In \cite{Lee+EvenMore21} this was the case for FeMoco.

\subsubsection{Double Factorized-Tensor Hypercontraction Sum of Squares}
Inspired by both compression techniques mentioned above, \cite{Low+FastQuantumSA2025}  combine key aspects of the respective factorizations to obtain a mixed ansatz. We now give an overview of the ansatz and the corresponding optimization problem as described in \cite{Low+FastQuantumSA2025}, following their approach closely and adding some details when deemed necessary.  The DFTHC ansatz is given as follows depending on a choice of rank $R$, and two copy parameters $C$ and $B$:
\begin{equation}\label{eq: HDFTC full}
\begin{aligned}
    H_\mathrm{DFTHC} &= \sum_{i,j} \Tilde{h}_{ij}^{(1)} E_{ij} + \frac12 \sum_{r\in [R]} \sum_{c \in [C]} \left( W^{(rc)} \I + \sum_{b\in [B]} w_b^{(rc)} \sum_{i,j} u_{bi}^{(r)} u_{bj}^{(r)} E_{ij}\right)^2 - \frac12 \sum_{r \in [R]} \sum_{c\in [C]}\big( W^{(rc)}\big)^2 \I,\\
    \Tilde{h}_{ij}^{(1)} & = h_{ij}^{(1)}  - \sum_{r\in [R]} \sum_{c\in [C]} W^{(rc)} \sum_{b\in [B]} w_b^{(rc)} u_{bi}^{(r)} u_{bj}^{(r)}.
\end{aligned}
\end{equation}
Here, $w_b^{(rc)} \in \R, \bfu_b^{(r)} \in \R^N$ with $\Vert \bfu_b^{(r)} \Vert = 1$, but in contrast to double factorization, we do not require orthogonality. However, with an additional orthogonality constraint and setting $C=1, W^{(r)}=0$, we recover $H_\mathrm{DF}$ as in \eqref{eq: H DF}. Similarly, we observe a connection to factorized THC by setting $R=1, B = C = M$. To see the resemblance more clearly and to recognize DFTHC combines adding multiple directions via a copy parameter $C$ from THC as well as additional factorization from DF, note that the Hamiltonian in \eqref{eq: HDFTC full} simplifies to  
\begin{align}\label{eq: HDFTC simple}
    H_\mathrm{DFTHC} &= \sum_{i,j} h_{ij}^{(1)} E_{ij} + \frac12 \sum_{r\in [R]} \sum_{c \in [C]} \left(\sum_{b\in [B]} w_b^{(rc)} \sum_{i,j} u_{bi}^{(r)} u_{bj}^{(r)} E_{ij}\right)^2.
\end{align}
While this expression makes clear that only the two-body term is approximated, the inclusion of the free scalar parameter $W^{(rc)}$ is key to the fundamental feature of DFTHC: using different metrics to obtain favorable parameters and coefficients. As mentioned above, previously a combination of accuracy in the Hamiltonian approximation and the 1-norm of the resulting Hamiltonian was used to evaluate the quality of an ansatz. For $H_\DFTHC$ it is crucial to take into account an additional metric coming from spectral amplification. This leads to a sum-of-squares (SOS) decomposition. 

The main idea behind spectral amplification for Hamiltonian simulation is to encode a ``square-root'' of the Hamiltonian instead of the Hamiltonian itself. This amplifies the low-energy sector of the Hamiltonian leading to efficiency gains when one is interested in for example ground state energy estimation or simulating dynamics. The idea has been employed in previous works \cite{LowChuangUniformSA17, ZlokapaSomma24}. While it is not always possible to decompose $H$ into a square-root, it is always possible to find $\beta<0$ such that $H - \beta \I = H_\srt^\dagger H_\srt$, since $H$ is Hermitian. While taking the square root leads to a potential efficiency gain, considering a large shift $\vert \beta \vert$ can result in worse performance. To see this, assume first that $H$ is positive semi-definite and observe that by \cite[Appendix A.2]{Low+FastQuantumSA2025}, given a block encoding of $H_\mathrm{sqrt}$ with block encoding constant $\lambda_\mathrm{sqrt}$ we can obtain a block encoding of $2\tfrac{H}{\lambda_\mathrm{sqrt}^2} - \I$ for $H = H_\mathrm{sqrt}^\dagger H_\mathrm{sqrt}$. Using this block encoding and phase estimation, we can compute the lowest eigenvalue $E_\mathrm{gap}$ of $H$ by estimating $\arccos\big(2\tfrac{E_\mathrm{gap}}{\lambda_\mathrm{sqrt}^2} -1\big)$. Next, we determine the number of calls to the block encoding required to obtain a given accuracy. A first order approximation shows that the number of calls is proportional to $\lambda_\mathrm{eff} = \sqrt{E_\mathrm{gap} (\lambda_\mathrm{sqrt}^2- E_\mathrm{gap})}$. Note that it is therefore beneficial to aim for a small $E_\gap$ while avoiding a large discrepancy between $\lambda_\srt^2$ and $E_\gap$. In the case of spectral amplification, $H-\beta \I$ takes the role of $H$ and $E_\mathrm{gap} = E_\mathrm{gs} - \beta$, where $E_\mathrm{gs}$ is the ground state energy of $H$. Thus, it is important to find $\beta$ such that $\vert E_\mathrm{gs} - \beta \vert$ is small.  Note that by definition, if $H$ is not positive semi-definite to begin with, the largest $\beta$ that satisfies the above is $ E_\mathrm{gs}$. 

Crucially, the previous explanation does not take into account the existence of an implementable $H_\srt$, which can be efficiently simulated. To overcome this and to obtain near optimal shift $-\beta$ , \cite{Low+FastQuantumSA2025} introduce a semi-definite programming (SDP) formulation of this problem
\begin{equation}\label{eq: SDP}
    \max \beta \text{ s.t. } H - \beta \I = \sum_{\ell \in [L]} O_\ell^\dagger O_\ell \text{ for } O_\ell \in \mathrm{span}\{\cS\},
\end{equation}
where $\cS$ is a set of appropriate generators for the sum-of-square terms. For example, a canonical choice for $\cS$ are the fermionic ladder operators and their products, i.e., $\I, a_i, a_i^\dagger$, $a_i a_j^\dagger$, et cetera. For more information on the SDP and the choice of SOS algebra see \cite{King+SOSSA26, Rubin+Frustration26}. For the DFTHC ansatz, we consider a slightly simpler SOS algebra generated by $\cS_{\DFTHC} = \{\I, a_{i\sigma}, a_{i\sigma}^\dagger, \sum_{\sigma \in \{0,1\}} a_{i\sigma}^\dagger a_{j\sigma} \} =\{\I, a_{i\sigma}, a_{i\sigma}^\dagger, E_{ij} \} $. To see that $H_\DFTHC$ arises as a sum of squares from $\mathrm{span}\{\cS_{\DFTHC}\}$ observe that elements in this set arise as follows
\begin{equation}\label{eq: General SOS summand}
    O_\ell = \alpha^\ell \I + \sum_{i, \sigma} \bigg( \beta_{i\sigma}^\ell a_{i\sigma} + \delta_{i\sigma}^\ell a_{i \sigma}^\dagger\bigg) + \sum_{j, k} \xi_{jk}^\ell E_{jk}.
\end{equation}
Note that in the original expression \eqref{eq: HDFTC full} we can already read off SOS summands for the two-electron term. For the block encoding it will be beneficial to express this in terms of Majorana operators. We denote Majorana operators using the Greek letter $\gamma$, i.e., $a_{i\sigma} = \tfrac12 (\gamma_{i\sigma, 0}+ i \gamma_{i\sigma, 1})$. Utilizing the spectral decomposition of $\Tilde{h}^{(1)}_{ij}$ to obtain a SOS decomposition for the one electron terms, we obtain the following 
{\small
\begin{align}
    H_\DFTHC &= \sum_{s\in [N]} \sum_{\sigma \in \{0,1\}} O_{\pos,\sigma, s}^\dagger O_{\pos,\sigma, s} + \sum_{s\in [N]} \sum_{\sigma \in \{0,1\}} O_{\nega,\sigma ,s}^\dagger O_{\nega,\sigma, s} + \sum_{r\in [R]} \sum_{c\in [C]} O_{\SF, rc}^\dagger O_{\SF, rc} + E_\sos\I, \label{eq: DFTHC SOS}\\ 
    O_{\pos,\sigma, s} &= \frac{\sqrt{\mu_s^+}}{2} \big(\gamma_{\bfv^+_s, \sigma, 0} + i\gamma_{\bfv^+_s, \sigma, 1}\big), \hspace{10em} O_{\nega,\sigma, s} = \frac{\sqrt{\mu_s^-} }{2} \big(\gamma_{\bfv^-_s, \sigma, 0} - i \gamma_{\bfv^-_s,\sigma, 1}\big), \label{eq: 1 elec O} \\
    O_{\SF, rc} &= \frac{1}{\sqrt2} \left( \Big(W^{(rc)} + \sum_{b\in [B]} w_b^{(rc)}\Big)\I + \frac{i}{2} \sum_{b\in [B]} \sum_{\sigma \in \{0,1\}} w_b^{(rc)} \gamma_{\bfu_b^{(r)}, \sigma, 0 } \gamma_{\bfu_b^{(r)}, \sigma, 1 } \right), \nonumber \\
    &: = \frac{1}{\sqrt2} \Big( w_B^{(rc)}\I + \frac{i}{2} \sum_{b\in [B]} \sum_{\sigma \in \{0,1\}} w_b^{(rc)} \gamma_{\bfu_b^{(r)}, \sigma, 0 } \gamma_{\bfu_b^{(r)}, \sigma, 1 } \Big),\label{eq:2-elec O} \\
    E_\sos & = - 2 \sum_{s} \mu_s^- - \frac12 \sum_{r\in [R]} \sum_{c\in [C]} \big(W^{(rc)}\big)^2 \label{eq: E_sos}.
\end{align}}

Here, $\mu^\pm$ denote the non-negative and negative eigenvalues of $\Tilde{h}^{(1)}_{ij}$ respectively, $\bfv^\pm$ denote the corresponding eigenvectors,  and we define $\gamma_{\bfx, \sigma, 0}\ = \sum_{j\in [N]} x_j \gamma_{j\sigma, 0}$ for $\bfx \in \R^N$. For more details on this derivation, see \zcref{app: DFTHC Details}. Hence, one can view the DFTHC ansatz as a special case of $H = \sum_{\ell} O_\ell^\dagger O_\ell$ over the simplified SOS algebra generated by $\cS_\DFTHC$ with the additional constraint of encoding the one-body term exactly and restricting the rank of the two-body terms to $B$ for each summand $O_{\SF,rc}$. The copy parameter $C$ and rank parameter $R$ restrict the number of second-order summands we consider, and $R$ serves as an additional measure for how many decompositions of rank $B$ we allow for. The description of the shift $E_\sos$ in \eqref{eq: E_sos} also makes clear the importance of the introduced free identity parameter in the definition of $H_\DFTHC$ given in \eqref{eq: HDFTC full}. Without the inclusion of $W^{(rc)}$ the only possible way of optimizing $E_\sos$ is through optimizing the negative eigenvalues of the one-body term. Even though omitting the free identity parameter would lead to a simplified encoding of the Hamiltonian, the resulting $E_\sos$ lacks the needed flexibility to achieve an overall efficiency gain. This can be verified by considering the SDP in \eqref{eq: SDP} with $\cS = \{ a_{i\sigma}, a_{i\sigma}^\dagger, E_{ij} \}$. Observe also that we introduced the notation $w_B^{(rc)} := W^{(rc)} + \sum_{b\in [B]} w_b^{(rc)}$. This is helpful later for the block encoding, as the free identity parameter is only relevant for the 2-electron terms $O_\SF$ and thus it makes sense to load it as part of the $B$ register.

In the remainder of the section, we review how the coefficients are determined via optimization \cite{Low+FastQuantumSA2025}. For our resource estimate, we take the resulting parameters from \cite[Table XII]{low2026denser} and do not recompute them, so readers interested in compilation of the block encoding can skip forward to \zcref{sec:block-encoding}. To determine the coefficients, we consider three metrics: the approximation error for the two-body term $\Vert h^{(2)}_{ijk\ell}- h^{(2)\prime}_{ijk\ell} \Vert_\fro$, the block encoding constant $\Lambda$, and the shift $E_\sos$, weighted with respective hyperparameters $\varepsilon_\reg$, $\lambda_\reg$ and $E_\reg$. In sum, we want to solve
\begin{equation}\label{eq: easy optimization}
    \underset{\substack{u \in \R^{R\times B\times N},\\
    w \in \R^{R \times (B+1) \times C}}}{\argmin} \left[ \frac{\Vert h^{(2)} - h^{(2)\prime} \Vert_\fro}{\varepsilon_\reg} + \frac{\Lambda}{\lambda_\reg} + \Relu \left( \frac{E_\gap - E_\reg}{E_\reg}\right) \right],
\end{equation}
where $\Relu(x) : = \max (x, 0)$ for a given choice of $(R,B,C)$. We discuss the three quantities in detail below.
\begin{enumerate}
    \item As seen in the simplification \eqref{eq: HDFTC simple}, the only approximation to $H$ comes from the two-body term. Hence, if we are interested in good approximations to $H$, we need to minimize the error in the two-body term.  From the simplification we also see that in the $H_\DFTHC$ ansatz the corresponding two-body tensor is given by
    \begin{equation*}
        h_{ijk\ell}^{(2)\prime} = \sum_{r\in [R], c\in [C]} \left(\sum_{b\in [B]} w_b^{(rc)} u_{b,i}^{(r)} u_{b,j}^{(r)} \right) \left(\sum_{b^\prime\in [B]} w_{b^\prime}^{(rc)} u_{b^\prime,k}^{(r)} u_{b^\prime,\ell}^{(r)} \right).
    \end{equation*}
    \item The goal is to block encode $H_\srt$. As shown in \cite{Low+FastQuantumSA2025, King+SOSSA26}, we can obtain a block encoding of $\tfrac{H_\srt^\dagger H_\srt}{\Lambda} -\I$ from block encodings of the corresponding SOS summands $O_\ell$. In particular, given a block encoding $\mathrm{BE}\Big[ \frac{O_\ell}{\lambda_\ell}\Big]$ of $O_\ell$ with constant $\lambda_\ell$, then, $\Lambda = \tfrac12 \sum_{\ell} \lambda_\ell^2$. For the specific ansatz in \eqref{eq: DFTHC SOS}, the block encoding constants for $O_{\pos,\sigma, s}, O_{\nega,\sigma, s}, O_{\SF,rc}$ can be obtained using ordinary linear combination of unitaries (LCU) techniques leading to 
    \begin{equation}
        \Lambda = \Vert \Tilde{h}^{(1)}   \Vert_1 + \frac{1}{4} \sum_{r\in [R]}\sum_{c\in [C]} \Bigg( \Big\vert W^{(rc)} + \sum_{b\in [B]} w_{b}^{(rc)}\Big\vert + \sum_{b\in [B]} \Big\vert w_{b}^{(rc)} \Big\vert \Bigg)^2,
    \end{equation}
    where $\Vert \cdot \Vert_1$ denotes the Schatten 1-norm. In particular, $\Vert \Tilde{h}^{(1)}   \Vert_1 = \sum_{s \in [N]} \mu^\pm_s$.
    \item Since the optimization problem in \eqref{eq: easy optimization} is not equivalent to the original SDP formulation in \eqref{eq: SDP} with $\cS_\DFTHC$, optimality of the shift $E_\sos$ is not guaranteed. Thus, we use the last summand in the cost function as a penalty for high discrepancy from the optimal shift obtained from the SDP with $\cS_\DFTHC$. The $\Relu(\cdot)$ function is used to prevent minimizing for unrealistically small energy gaps. A sensible choice of hyperparameter is  choosing $E_\reg$ as coming from a solution of the SDP. Instead of simply defining $E_\reg$ as equal to the shift $E_\sos$,  \cite{Low+FastQuantumSA2025} define it as equal to $E_\gap^\mathrm{SDP} = E_\mathrm{gs} - E_\sos^\mathrm{SDP}$, where $E_\mathrm{gs}$ denotes the ground state energy of the original system. Equivalently, $E_\gap^\mathrm{SDP}$ is the aforementioned ground state energy of the SOS in \eqref{eq: DFTHC SOS} and $E_\reg$ can be chosen as the ground state energy of the solution to the SDP. Since $E_\mathrm{gs}$ and the ground state of the SOS are not readily available as quantities, one generally uses classical approximations to these quantities. We remark that if $E_\reg$ is chosen in this manner, we obtain the following simplification in the numerator of the $\Relu(\cdot)$ function, $E_\gap - E_\reg = E_\sos^\mathrm{SDP} - E_\sos$, where the latter denotes the optimal shift obtained by solving the corresponding SDP. However, if a solution to the SDP is not available, the hyperparameter needs to be chosen differently. For example, \cite{Low+FastQuantumSA2025} sweep over $E_\reg$ in a pre-determined interval. 
\end{enumerate}

\paragraph{Addition of Block-Invariant Symmetry Shifting (BLISS).}
Besides factoring the Hamiltonian, we can also look for shifted operators that are closely related to the original Hamiltonian and share desired behavior. For example if $\hat S$ is a symmetry of $H$, i.e., $[H, \hat S] = 0$, then $H, \hat S$ are simultaneously diagonalizable. In particular, if we are interested in the eigenvector $\ket{\psi}$ of $H$ and $S \ket{\psi} = s\ket{\psi}$, then $H-(\hat S -s\I)\ket{\psi} = H \ket{\psi}$, thus the spectrum of $H-(\hat S -s\I)$ with respect to $\ket{\psi}$ is unchanged but $H-(\hat S -s\I)$ might be easier to work with. In \cite{LoaizaIzmaylovBLISS23} this was generalized to operators of the form $H-\hat B(\hat S -s\I)$, where $\hat B$ is an arbitrary operator and coined Block-Invariant Symmetry Shifting (BLISS). BLISS can be used to search over appropriate operators to obtain a 1-norm reduction \cite{Patel+Min1-normBliss25} and has been used in combination with both double factorization \cite{Rocca+DFBLISS24} and THC \cite{Caesura+THCBLISS25} leading to improvements.

To combine BLISS with DFTHC, \cite{Low+FastQuantumSA2025} consider a shift with respect to the total particle number operator $\hat N = \sum_{i\in [N]} E_{ii}$. More specifically, fixing an eigenvalue $\eta$ of $\hat N$, we are interested in operators of the form 
\begin{equation}
    H_\mathrm{BLISS} = H - \frac12 \hat B (\hat N - \eta\I) - \alpha ( \hat N - \eta\I ), \text{ with } \hat B = \sum_{i,j \in [N]} h^{(sym)}_{ij} E_{ij}
\end{equation}
for a symmetric order-2 tensor $h^{(sym)}_{ij}$. Note that since $[\hat B, \hat N] = 0$, we can write $\hat B \hat N = \tfrac{1}{2} (\hat B \hat N + \hat N \hat B) = \frac12 \sum_{i, j, k, \ell} \big(h^{(sym)}_{ij} \delta_{k\ell} + \delta_{ij} h^{(sym)}_{k\ell}\big) E_{ij} E_{k\ell}$. Hence, we obtain
\begin{align*}
    H_\mathrm{BLISS} = \alpha \eta \I  + \sum_{ij} \Big( h^{(1)}_{ij} + \frac{\eta}{2} h^{(sym)}_{ij} - \alpha \I\Big) E_{ij} + \frac12 \sum_{i,j,k,\ell} \Big( h^{(2)}_{ijk\ell} + \frac{h^{(sym)}_{ij} \delta_{k\ell} + \delta_{ij} h^{(sym)}_{k\ell}}{2} \Big) E_{ij} E_{k\ell}.
\end{align*}
Thus, we can enlarge the set of Hamiltonians we compare the $H_\DFTHC$ ansatz against as follows
\begin{align*}
    \mathcal{H}_\eta = \Bigg\{ &h_S^{(0)} + \sum_{i,j} h_{S, ij}^{(1)} E_{ij} + \sum_{i,j, k , \ell} h_{S, ijk\ell}^{(2)} E_{ij} E_{k\ell} \, \Big\vert \, h_S^{(0)} = \alpha \eta\I, h_{S}^{(1)} = h^{(1)} - \frac{\eta}{2} h^{(sym)} + \alpha \I,\\
    & h_{S}^{(2)} = h^{(2)} + \frac{h^{(sym)} \otimes \I + \I \otimes h^{(sym)}}{2}, \alpha \in \R, h^{(sym)} = h^{(sym)\dagger} \in M_N(\R) \Bigg\} .
\end{align*}
Lastly, combining BLISS with the DFTHC ansatz leads to the following optimization problem
\begin{equation}\label{eq: full optimization}
    \underset{\substack{u \in \R^{R\times B\times N},\\
    w \in \R^{R \times (B+1) \times C}, \\ H_S \in \mathcal{H}_\eta}}{\argmin} \left[ \frac{\Vert h^{(2)}_S - h^{(2)\prime} \Vert_\fro}{\varepsilon_\reg} + \frac{\Lambda}{\lambda_\reg} + \Relu \left( \frac{E_\gap - E_\reg}{E_\reg}\right) \right].
\end{equation}
Note that the scalar shift $ \alpha \eta\I$ leads to an addition to $E_\sos$ of $\alpha\eta$ and thus $E_\gap$ will inherit this change.

\subsection{Description of the block encoding}
\label{sec:block-encoding}

We briefly recall a general method to block encode sum-of-squares Hamiltonians following the description in \cite{Low+FastQuantumSA2025, King+SOSSA26} culminating in a high-level description of the block encoding of the DFTHC ansatz based on \cite{Low+FastQuantumSA2025, low2026denser}. In this section, we avoid specifications on how subroutines are implemented as well as register optimization, but instead focus on the conceptual task of subroutines and registers. Specifications depending on our compilation target can be found in \zcref{sec: Resource Count BE}.

Given a Hamiltonian $H$ with a sum-of-squares decomposition $H = \sum_{\ell \in [L]} O_\ell^\dagger O_\ell := H_\srt^\dagger H_\srt$ for $O_\ell \in \bC^{k \times m}$, we are interested in obtaining a block encoding of $2 \frac{H}{\lambda_\srt^2}-\I$ from block encodings of  $O_\ell$, where $\lambda_\srt = \sqrt{\sum_{\ell \in [L]} \lambda_\ell^2}$ and $\lambda_\ell$ denotes the block encoding constant for $O_\ell$. To do so, we review the construction in \cite{Low+FastQuantumSA2025}. Suppose we have access to a controlled block encoding $\BE{O_\ell/\lambda_\ell}$ of $O_\ell$ with block encoding constant $\lambda_\ell \geq \Vert O_\ell \Vert_2$, i.e., $\BE{O_\ell/\lambda_\ell} \in M_n(\bC)$ is a unitary (with an efficient quantum circuit description) such that there exists projectors $\Pi_k, \Pi_m$ with $\Pi_k \BE{O_\ell/\lambda_\ell} \Pi_m = O_\ell/\lambda_\ell$, for all $\ell \in [L]$, where we can assume that $\Pi_m$ is the projector onto $\ket{0}$. Then, we obtain $\BE{H_\srt/\lambda_\srt}$ from utilizing a linear combination of unitaries from a \textsc{Select} subroutine and a state preparation subroutine \textsc{Prep}:
\begin{align*}
    \textsc{Sel} := \sum_{\ell \in [L]} \ket{\ell}\hspace{-3pt}\bra{\ell} \otimes \mathrm{BE}\left[\frac{O_\ell}{\lambda_\ell}\right], \qquad \textsc{Prep} \ket{0} :=  \frac{1}{\lambda_\srt} \sum_{\ell \in [L]} \lambda_\ell \ket{\ell}.
\end{align*}
Recall that we omit any specific implementation details for \textsc{Sel} and \textsc{Prep} such as extra registers. Observe that $\Pi_k \textsc{Sel}\textsc{Prep} \ket{0} = 1/\lambda_\srt \sum_{\ell \in [L]} \ket{\ell} \otimes O_\ell$ and thus we can use \textsc{Sel} and \textsc{Prep} to obtain the desired block encoding of $H$. Note that since $(\sum_{\ell \in [L]} \ket{\ell} \otimes O_\ell)^\dagger (\sum_{\ell \in [L]} \ket{\ell} \otimes O_\ell) = \sum_{\ell \in [L]} O_\ell^\dagger O_\ell = H$, we define $H_\srt := \sum_{\ell \in [L]} \ket{\ell} \otimes O_\ell$ and can use a typical reflection to obtain a block encoding of the Chebyshev polynomial $T_2(x) = 2x^2 -1$ resulting in a block encoding of $2 H/\lambda_\srt^2 - \I$ as desired. More precisely, letting $\textsc{Ref}_\mathrm{BE}$ denote the reflection over the subspace spanned by the columns of $H_\srt$, i.e., the reflection around $\oplus_{\ell \in [L]} \big(\vert \ell \rangle \langle \ell \vert \otimes \Pi_k\big) = I \otimes \Pi_k$, we obtain the following description
\begin{equation*}
    \mathrm{BE}\left[ 2\frac{H}{\lambda_\srt^2} - \I\right] = \frac{1}{\lambda_\srt^2} \left(\sum_{\ell \in [L]} \bra{\ell} \otimes O_\ell^\dagger\right) \textsc{Ref}_\mathrm{BE} \left(\sum_{\ell^\prime \in [L]} \ket{\ell^\prime} \otimes O_{\ell^\prime}\right) =  \textsc{Prep}^\dagger \textsc{Sel}^\dagger \textsc{Ref}_\mathrm{BE} \textsc{Sel} \textsc{Prep}.
\end{equation*}

\begin{figure}
    \centering
    \includegraphics[width = \textwidth]{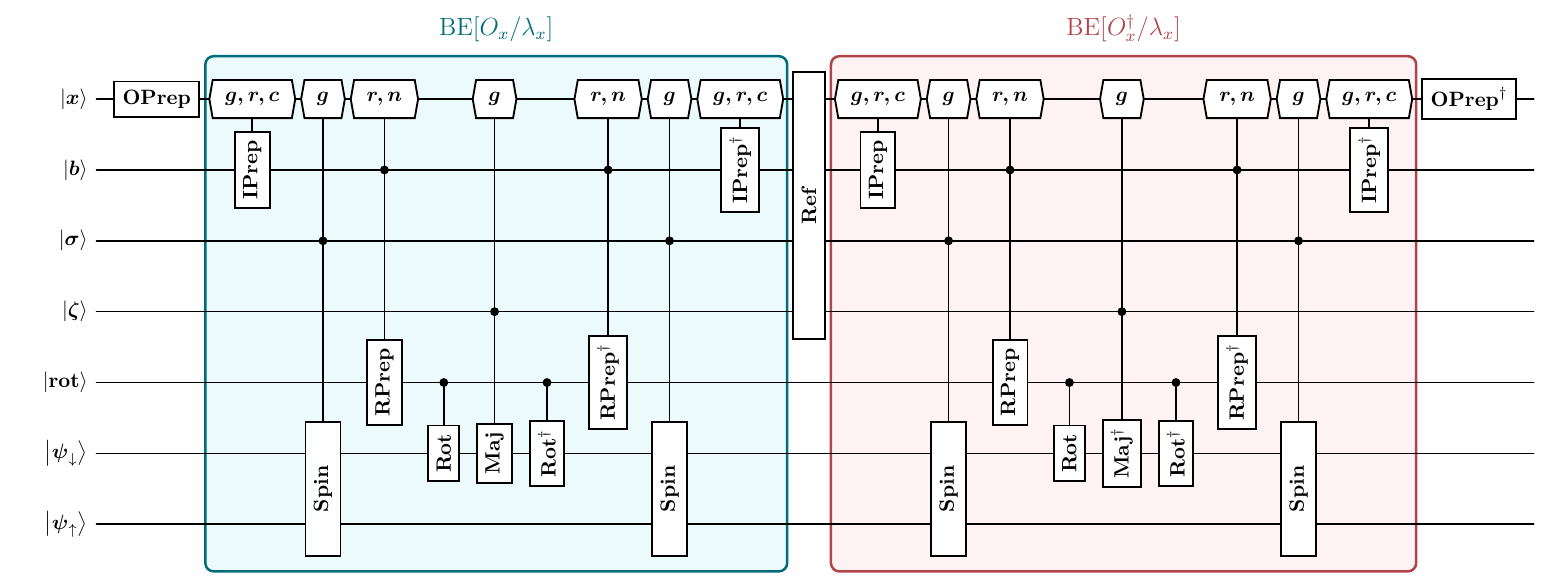}
    \caption{A high-level quantum circuit description of the sum-of-squares block encoding of the DFTHC ansatz for a quantum chemistry Hamiltonian. Here, \textsc{Oprep} and \textsc{Iprep} refer to \textsc{OuterPrep} and \textsc{InnerPrep} respectively. The circuit produces a block encoding of $H_\DFTHC/\Lambda -\I$ for $\Lambda = 1/2 \sum_{x\in [X]} \lambda_x^2$, where $\lambda_x$ is the block encoding constant of the SOS factor $O_x$. Here, $x$ denotes a flattened index containing information for $g,r,c$ and $n$. Hexagonal gate boxes indicate which subindex of $x$ is used as control. A black dot on a register indicates that the corresponding index is used as a control. Note that \textsc{RPrep} and \textsc{Rot} are merged for resource optimization.}
    \label{fig: high-level BE circ}
\end{figure}

Applying this method to the DFTHC ansatz, results in a block encoding circuit split into several parts as seen in \zcref{fig: high-level BE circ}. We give a high-level description below. First, the state carrying the amplitudes $\lambda_\ell$ needs to be prepared. To unify notation, write 
\begin{equation*}
    H_\DFTHC = \sum_{g \in \{\pos, \nega, \SF\}, r \in [R_g], c\in[C_g]} O_{g,rc}^\dagger O_{g,rc} = \sum_{x \in [X]} O_{x}^\dagger O_{x},
\end{equation*}
with $R_\pos = N_+, R_\nega = N-N_+, R_\SF = R, C_\pos = C_\nega = 0, C_\SF = C$, where $x$ denotes a flattened index
\begin{equation*}
    x(g,n, r,c) = \begin{cases}
        n, \qquad &g = \pos,\\
        N_+ + n, \qquad &g = \nega,\\
        N+ rC + c, \qquad &g =\SF,
    \end{cases} \qquad \text{ for } X = N + RC,
\end{equation*}
with $N_+$ denoting the number of positive eigenvalues of the 1-electron term. For this qualitative description of the block encoding, we omit details on how $x$ is unpacked and indices $N,R,C$ are accessed. Discussions on the exact index flattening scheme used and how to unpack it can be found in \zcref{sec: Resource Count BE} following \cite{low2026denser}. The corresponding $\lambda_x$ arise from the scalar factors for the descriptions $O_{\pos, \sigma, n}, O_{\nega,\sigma, n}, O_{\SF, rc}$ given in \eqref{eq: 1 elec O} and \eqref{eq:2-elec O}. More precisely $\lambda_x = \sqrt{\mu_n^\pm}$ for 1-electron terms, and $\lambda_x = 1/\sqrt{2} ( \vert W^{(rc)} + \sum_{b\in [B]} w_b^{(rc)} \vert + \sum_{b\in [B]} \vert w_b^{(rc)}\vert)$ for 2-electron terms. The first state-preparation \textsc{Prep} thus is used to prepare the state $\frac{1}{\lambda_\srt} \sum_{x\in [X]} \lambda(g_x, n_x, r_x, c_x) \ket{x}$, with 
\begin{equation*}
    \lambda(g, n, r, c) = \begin{cases}
        2\sqrt{\mu_n^+}, \qquad &g= \pos, \\
        2\sqrt{\mu_n^-}, \qquad &g= \nega, \\
        \frac{1}{\sqrt2} \left( \Big\vert W^{(rc)} + \sum_{b\in [B]} w_b^{(rc)} \Big\vert + \sum_{b\in [B]} \vert w_b^{(rc)}\vert\right), \qquad &g = \SF.
    \end{cases}
\end{equation*}
For this state preparation, the coefficients are loaded with precision $b_\coef$. We refer to this state preparation as \textsc{OuterPrep} following the convention in \cite{Low+FastQuantumSA2025, low2026denser}.

Next, for \textsc{Sel} we need to implement $\BE{O_{x}/\lambda_x}$. As $O_{x}/\lambda_x$ is obtained as an LCU, we first need to perform an inner state preparation, referred to as \textsc{InnerPrep}, which loads the coefficients in the descriptions of $O_x$ again with precision $b_\coef$. Since for the 1-electron terms, the Majorana operators in $O_{\pos, \sigma, n}$ share the same coefficient $\sqrt{\mu^+_n}/2$ (similarly $\sqrt{\mu^-_n}/2$ for  $O_{\nega, \sigma, n}$), and the linear combination of the form $\gamma_0 \pm i\gamma_1$ is obtained through the Majorana operator application, no additional inner state prep is required. Thus, \textsc{InnerPrep} concerns only 2-electron terms and depending on the index pair $(r,c)$ an inner state preparation loading the coefficients $\vert W^{(rc)} + \sum_{b\in [B]} w_b^{(rc)} \vert$, and $\vert w_b^{(rc)}\vert$ for $b\in [B]$ needs to be performed. 

Second, we need to take the spin as well as the different Majorana operators $\gamma_{\bfv^\pm_n, \sigma, 0}$ and  $\gamma_{\bfv^\pm_n, \sigma, 1}$ into account. Note that while for the operators stemming from the 1-electron terms, the consideration of the spin occurs in the outer sum, as $O_{\pos, \sigma, n}, O_{\nega, \sigma, n}$ explicitly depend on $\sigma$, the spin for operators $O_{\SF, rc}$ stemming from the 2-electron terms needs to be accounted for in the block encoding of $O_{\SF, rc}$, i.e., the inner sum. To account for spin-selection in the inner sum, a controlled \textsc{Swap} between spin-up and spin-down system qubits controlled on a spin register is used. This method is analogous to the description given in \cite{vBurg+QCCompCat21} arising from double factorization. In \zcref{fig: high-level BE circ}, this is denoted as \textsc{Spin} controlled on $g$, as it requires selecting the $O_{\SF, rc}$ terms. Even though $O_{\pos, \sigma, n}$ and $O_{\nega, \sigma, n}$ do not require spin selection in the inner sum, to account for $\gamma_{\bfv^\pm_n, \sigma, 0} \pm i \gamma_{\bfv^\pm_n, \sigma, 1}$ in the application of the Majorana operators, another single flag qubit register ($\ket{\zeta}$ in \zcref{fig: high-level BE circ}) is used, which will be relevant in the Majorana operator application. 

Third, the Majorana operators need to be rotated using Givens rotations \cite{vBurg+QCCompCat21}. This operation is denoted as \textsc{Rot} in \zcref{fig: high-level BE circ}. Note that the rotation angles for $O_{\SF, rc}$ given by $ \bfu^{(r)}_b$ depend on $R, B$, but not on $C$. Moreover, the rotation angles for 1-electron terms $O_{\pos, \sigma, n}, O_{\nega, \sigma, n}$ given by $\bfv_n^\pm$ only depend on $n \in [N]$. Thus, the rotation angles are loaded controlled on $r,n$ in $x$ and $b$ using a data loading procedure \textsc{RPrep}. The rotation angles are loaded with a precision $b_\rot$ and thus, $(N-1)b_\rot$ rotation angles need to be loaded. After the Givens rotations have been performed, the Majorana operators need to be applied to a single qubit based on the Jordan-Wigner transform. To discern between $O_{\SF, rc}$ and $O_{\pos, \sigma, n}, O_{\nega,\sigma,n}$, this operation is controlled on $g$. Furthermore, to obtain the sum $\gamma_{\bfv^\pm_n, \sigma, 0} \pm i \gamma_{\bfv^\pm_n, \sigma, 1}$ for the 1-electron SOS factors, the additional control qubit $\ket{\zeta}$ is used. We denote this operation by \textsc{Maj}. In sum, after performing the necessary uncompuations we arrive at a block encoding of $O_x/\lambda_x$ as a combination of $\textsc{InnerPrep}^\dagger  \textsc{Spin}^\dagger \textsc{RPrep}^\dagger \textsc{Rot}^\dagger \textsc{Maj} \textsc{Rot} \textsc{RPrep}\textsc{Spin} \allowbreak \textsc{InnerPrep}.$ 

Lastly, having now applied the block encoding of $O_x/\lambda_x$, we perform a reflection around the block encoding as mention in the beginning of this section. Next, block encodings for $O_x^\dagger$ are performed by reversing the previous inner circuits. This only leads to a change from \textsc{Maj} to $\textsc{Maj}^\dagger$ and finishes the high-level discussion of \textsc{Sel}. To obtain the overall block encoding $\textsc{OuterPrep}^\dagger$ is applied. \zcref{fig: high-level BE circ} gives a quantum circuit description of the block encoding, including explicit registers for the spin $\sigma$ and for the 1-electron term Majorana selection $\zeta$. We refrain from stating size requirements on the registers as this depends on the specific subroutines chosen and is discussed in \zcref{sec: Resource Count BE} in more detail. Similarly, we explicitly mention a register for storing the rotation angles to allow for a distinction between the rotation angle loading \textsc{RPrep} and the application of the rotation \textsc{Rot}, but other ancilla registers have been omitted. 
 
\section{Detailed subroutine descriptions}\label{sec:subroutines}

Most subroutines used for the block encoding of the DFTHC ansatz are described in \cite{Lee+EvenMore21, vBurg+QCCompCat21, Low+FastQuantumSA2025, low2026denser}. We review the subroutines for which we optimize the physical layout of cat qubits. We don't optimize the physical layout for the subroutines with small spacetime volume. This includes the implementation of the Majorana operator (see \cite[Appendix B6]{Low+FastQuantumSA2025} and \cite[Appendix E.2]{low2026denser}), reflections for $T_2$ \cite{Low+FastQuantumSA2025} and for the walk operator and Fourier resource state preparation. 

\subsection{Lookup table}
Given classical data in form of a table $T$ of $M$ addresses, a lookup table describes the implementation of the following unitary operator $U_T\colon \cH^{\otimes m}\otimes \cH^{\otimes b}  \to  \cH^{\otimes m} \otimes \cH^{\otimes b} \colon \ket{k} \ket{z} \mapsto \ket{k} \ket{z \oplus T(k)}$, where $m = \lceil \log M \rceil$ and $\cH^{\otimes b}$ corresponds to a $b$-bit output register. This is also often referred to as quantum read-only memory (\textsc{Qrom}) in its simplest form, while some specific subroutines come with their own naming convention. We can view this implementation in two parts, first we need to range over all possible $k$ values, second $T(k)$ needs to be applied. In case of data loading, $T(k)$ is relatively easy to implement, since it is just a multi-target \textsc{Cnot} gate. Ranging over possible $k$ values, often also referred to as a \textsc{Select} subroutine, is more involved as it entails toggling bits on and off using Toffoli gates. Since loading classical data is a core task in quantum algorithms, many improvements and optimizations can be found in the literature. For the physical layout optimization, we focus on a specific subroutine for the \textsc{Select} as well as a method of reducing Toffoli count utilizing clean and potentially dirty ancilla.

\subsubsection{Skew tree QROM}\label{sec: Skew-tree QROM description}
\begin{figure}
    \centering
    \includegraphics[width=\textwidth]{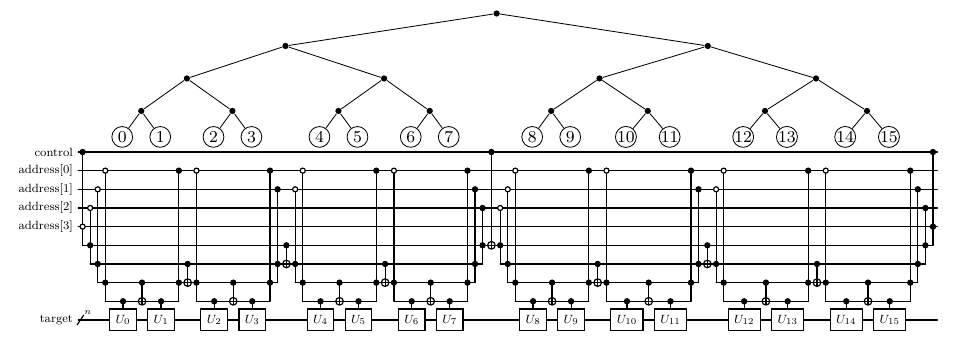}
    \caption{Quantum Circuit for unary iteration as described in \cite{babbush2018encoding} for four address qubits. Superimposed is a binary tree to illustrate that this circuit can be obtained by traversing a balanced tree, where numbered nodes correspond to the application of the unitary $U_i$.}
    \label{fig:balanced-tree controlled}
\end{figure}
Ranging over all possible addresses $0, \ldots, M-1$ can be achieved naively by implementing $M$ multi-controlled \textsc{Not} gates each of which can be decomposed into $\cO(\log M)$ ordinary Toffoli gates leading to an approximate cost of $\cO(M\log M )$. This was improved to linear Toffoli scaling using unary iteration introduced in \cite{babbush2018encoding}. Instead of having a separate multi-controlled \textsc{Not} gate per possible index, they key idea of unary iteration is to reuse qubits encoding indices by utilizing a sequence of \textsc{And} gate ladders, $\lceil\log M\rceil$ clean ancilla, and measurement based uncomputation (MBU) leading to $M-1$ Toffoli gates in case of a controlled \textsc{Select} subroutine. A similar improvement but without MBU resulting in $1.5 M-1$ Toffoli gates was given in \cite{Childs+TowardsQSpeedup18}, where it was also pointed out that the structure of this subroutine follows a binary tree traversal as pictured in \zcref{fig:balanced-tree controlled}. In the case of using clean ancilla and MBU, edges correspond to \textsc{And} gates or \textsc{And}$^\dagger$ uncomputations depending on whether they are traversed down or upward. Furthermore, switching direction at a node leads to a \textsc{Cnot} gate. Only numbered nodes correspond to data loading.

\begin{figure}
\centering

\begin{subfigure}[b]{\textwidth}
  \includegraphics[width=\linewidth]{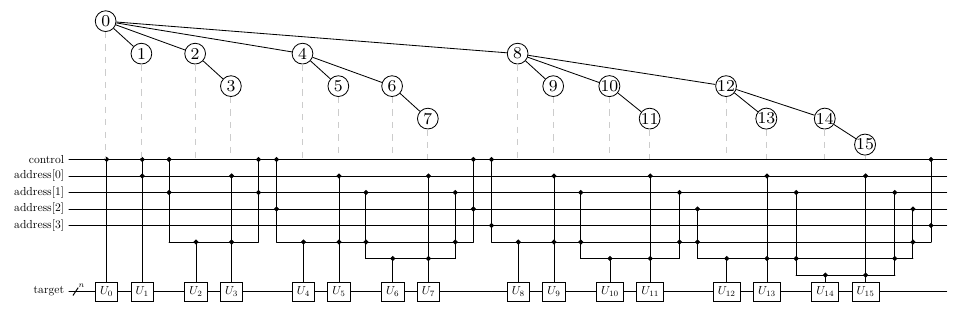}
  \caption{Quantum Circuit of a controlled skew tree \textsc{Qrom} for 4 address qubits. Here $U_i$ loads the Möbius transform of the data. Note that data loading for odd nodes $U_{2k+1}$ for $k=0,\ldots,7$ denotes a short hand for implementation using an additional \textsc{And}/\textsc{And}$^\dagger$ gate pair.}
  \label{fig:skew-tree controlled}
\end{subfigure}

\medskip
\begin{subfigure}[b]{\textwidth}
  \includegraphics[width=\linewidth]{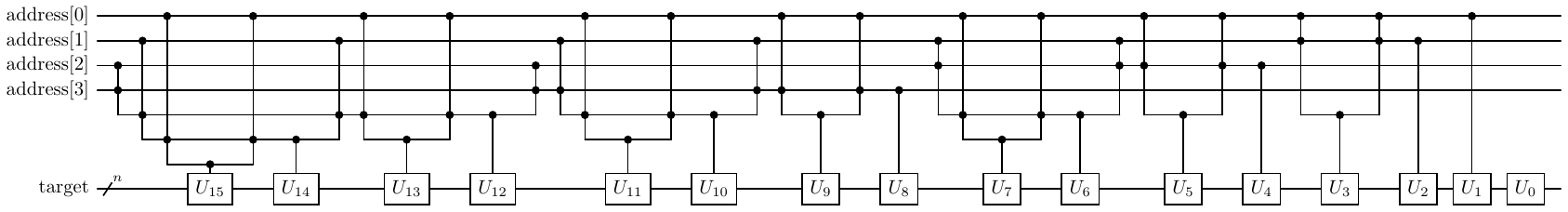}
  \caption{Quantum circuit of an uncontrolled skew tree \textsc{Qrom} for 4 address qubits, where the tree is traversed in reverse depth first search. Again $U_i$ loads the Möbius transform of the data.}
  \label{fig:skew tree uncont}
\end{subfigure}
\end{figure}
While the aforementioned binary tree traversal does take advantage of structural properties of successive bits, it still loads exactly $T(k)$ when $k$ is on by utilizing laddered Toffoli gates to select the corresponding state. This requirement was softened by \cite{KhattarGidneyCondClean25}, where they considered changing the Toffoli order by using a skew tree traversal resulting in operations being applied not only when the specific bit is on but also to other bits, which needs to be compensated by classically transforming the data. The change in order has several benefits: As focused on in \cite{KhattarGidneyCondClean25} it has better Toffoli count when not relying on MBU in comparison with the ordinary binary tree version ($1.25 M-1$ Toffoli gates as opposed to $1.5 M-1$) and saves one ancilla qubit for data amounts that are not a power of 2. Additionally, as shown in \cite{low2026denser}[Appendix C] using the skew tree traversal in the reversed order also allows for classical absorption of Clifford corrections when injecting a $\CCZ$ magic state. An example circuit together with the corresponding skew tree is depicted in \zcref{fig:skew-tree controlled} and \zcref{fig:skew tree uncont} shows the uncontrolled counterpart, where we traverse the tree starting from 15. 

The idea behind the skew tree \textsc{Qrom} is the following: instead of utilizing a Toffoli ladder before data is loaded to ensure the toggling on of the correct quantum state, after each Toffoli gate we perform a data load leading to multiple quantum states receiving a given data load. This manifests in the tree as each node contains a corresponding data element as opposed to the binary tree where only leaves were used for data loading. Specifically, loading corresponding to the quantum state $\ket{p}$ happens as soon as the corresponding control is on, regardless of whether this leads to data loading on other quantum states. Thus, anything that is performed when the control $p$ is on affects all quantum states $\ket{q}$ such that $p \preceq_2 q$, where $p \preceq_2 q$ if and only if whenever a bit is 1 in $p$ it must also be $1$ in $q$. Now, focusing on a fixed $\ket{q}$, it will accumulate everything that is loaded on $\ket{p}$ with $p \preceq_2 q$. In other words, what is loaded on $\ket{q}$ is given by $f(q) = \oplus_{p\preceq_2 q} T(p)$, which is exactly the Möbius transform of the data. Since the Möbius transform is an involution, if our goal is to load exactly $T(q)$ on $\ket{q}$, instead of loading the data directly, we need to load the Möbius transform of the data. In sum, $U_q$ in \zcref{fig:skew-tree controlled}/\zcref{fig:skew tree uncont} corresponds to loading $f(q) =  \oplus_{p\preceq_2 q} T(p)$. While this discussion only concerns data loading using Pauli $X$ gates, the skew tree \textsc{Select} can in principle be used for any function $T$ with the property that  $\oplus_{p\preceq_2 q} T(p)$ has an efficient out-of-place implementation for all $p \in \{0,1\}^m$. 

\paragraph{Classically absorbed corrections.} As evident when comparing \zcref{fig:balanced-tree controlled} and \zcref{fig:skew-tree controlled}, the skew tree \textsc{Qrom} lacks \textsc{And} gate ladders, easing the scheduling requirements from a compilation standpoint. In particular, each Toffoli application is followed by a multi-target \textsc{Cnot} and the \textsc{And} gates are applied periodically. While this is indeed a benefit of the skew tree \textsc{Qrom}, its structure allows to absorb corrections coming from the injection of the Toffoli-like magic state. Specifically, when injecting a $\CCZ$ state, we require two \textsc{Cnot} corrections between the controls and the target of the \textsc{And} gate and a $\CZ$ phase correction between the controls. The phase correction commutes with the other \textsc{And} gate controls in the address register and accumulates phases when interacting with \textsc{And}$^\dagger$, leading to multi-controlled $\CZ$ gates, which are later combined with the \textsc{Phaseup} for the \textsc{Qrom} uncomputation (see \zref{sec: phase fixup}) . Hence, these corrections do not need to be applied immediately. Similarly, \cite{low2026denser}[Theorem 3] shows that the \textsc{Cnot} corrections do not need to be performed at all, but instead the classical data can be updated. 

The main idea behind \cite[Theorem 3]{low2026denser} is the notion of a ``wandering \textsc{And}'' gate as introduced in \cite[Appendix A.3]{gidney2025factor}: instead of computing $a \wedge b$ for controls $a,b$, a wandering \textsc{And} computes $(a\oplus m_1) \wedge (b \oplus m_2)$, where $m_1,m_2$ are the measurement outcomes relevant for the \textsc{Cnot} corrections. Now, this updated value can be propagated through all operations controlled on this qubit, which affects exactly all data loads lying on the subbranch in the tree. The affected nodes can then be corrected by updating the classical data, which is only possible if the skew tree is traversed using reversed depth first search (DFS), so that any necessary updates only occur on future data loads. Reversed DFS means the $M-1$th node is traversed first, followed by $M-2$ etc., and thus, manifests in the circuit as reading \zcref{fig:skew-tree controlled} from right to left. \zcref{fig:skew tree uncont} shows an uncontrolled skew tree \textsc{Qrom} in reversed DFS. 

One important assumption of classical correction absorption is that this only holds for the uncontrolled \textsc{Qrom} as shown in \cite[Remark 2]{low2026denser}, otherwise any power of two node still needs to be corrected. This still gives the benefit of foregoing corrections for most data loads though, which allows us to have a smaller magic state injection footprint as opposed to the auto-corrected injection, while not being reaction time limited. This is true as long as the classical data update can be performed in time. From \cite{low2026denser}[Theorem 3] it follows that the shortest time between a faulty injection and a necessary classical update is given by the time of applying a multi-target \textsc{Cnot} and an \textsc{And}$^\dagger$ gate. Since the latter corresponds to a measurement whose associated Clifford correction is performed after the unary iteration, it has no time footprint. The former is applied through lattice surgery. Thus, as long as the time for one round of lattice surgery is larger than the time necessary for the classical update, no wait occurs. 

\paragraph{Iterating over an arbitrary range.} In some instances, it might be beneficial to consider a unary iteration over the interval $[M-K, M-1]$ for $0<K<M-1$ as opposed to taking the full interval $[0,M-1]$ into account. Naively, this can be achieved by simply pruning leaves that do not require data loads and by instead starting with an \textsc{And} gate ladder of $\lceil\log(M-K)\rceil$. Afterwards, the cost is given by the usual skew tree \textsc{Qrom} cost over $K$ addresses. To see that this is compatible with the Möbius transformed data loading of the skew tree \textsc{Qrom}, observe that this corresponds to loading $\Tilde{T}$ instead of $T$ with $\tilde{T}(p) = T(p)$ for $p \geq M-K$ and $\tilde{T}(p) = 0$ otherwise. Then, the Möbius transform of the data is given by $\tilde{f}(p) = \oplus_{q \preceq_2 p} \tilde{T}(q)$ and in particular, since $q \preceq_2 p$ implies $q \leq p$, we have $\tilde{f}(p) = 0$ for all $p < M-K$. In sum, no data loading is required on entries $p$ not traversed by the pruned tree. However, this is not compatible with absorbing classical corrections as described above, since some corrections would need to be performed at nodes $p \leq M-K-1$. As such nodes are not traversed, \textsc{Cnot} corrections would need to be accounted for directly, when considering a skew tree \textsc{Qrom} for loading only high values. Since we are interested in a small qubit footprint with few latency constraints, we focus on the skew tree \textsc{Qrom} that allows for classical absorption of corrections.

\subsubsection{SelectSwap (aka QROAM) clean and dirty}
For large address registers, it can be beneficial to reduce the linear Toffoli scaling with the number of address states. To do so, in 2018 \cite{low2024trading} introduced   a trade-off between adding ancilla qubits and a reduction in Toffoli count.  An idea that is based on a well-known back-and-forth in quantum algorithm design: the possibility to trade propagation in time for expansion in space and vice versa. Instead of unraveling the \textsc{Select} solely in time using unary iteration, Low et al. split the address registers into two parts, the first being encoded in space using additional ancilla and the second still being resolved in time as a unary iteration, but now reduced in size. 

More precisely, suppose as before that we have an address register of size $M$ and want to load $b$-bit data. We then split the address register into $M/\alpha$ and $\alpha$ states, where $\alpha \in [M]$ is a tuning parameter and instead of a single $b$-bit register, we duplicate it $\alpha$ times. For simplicity of explanation assume $\alpha = 2^a$, otherwise we first compute $\ket{k} \mapsto \ket{x}\ket{y}$ with $x = \lfloor k/\alpha \rfloor$ and $y = k \mod \alpha$ leading to an extra $\cO(\log M \log \alpha)$ Toffoli gates in addition to the costs from the \textsc{SelectSwap} subroutine. In the power of two case, the address register is split into $m-a$ and $a$ qubits. Utilizing the ($m-a$)-qubit register, we preform a \textsc{Select} subroutine using a skew tree \textsc{Qrom} as described in \zcref{sec: Skew-tree QROM description}, the output of which loads $\alpha$ $b$-bit numbers. In the end, the ($b\alpha$)-qubit output register holds the data corresponding to the lookup table $T$ taking the space multiplicity into account,i.e., 
$$\ket{k}\ket{0} \mapsto \ket{k}\ket{T\big(k\tfrac{M}{\alpha}\big), T\big( k\tfrac{M}{\alpha}+1\big), \ldots, T\big(k\tfrac{M}{\alpha}+ \alpha-1\big)}.$$
Now, the remaining address register of size $a$ is used to perform a multiplexed permutation implemented using a network of controlled swap gates to bring the correct table entry forward. We refer to this subroutine simply as \textsc{Swap}. This leads to a combined Toffoli cost of $\cO(M/\alpha + \alpha b)$ with the former coming from the \textsc{Select} and the latter from the \textsc{Swap} network. If the goal is to minimize the Toffoli count, choosing $\alpha = \cO(\sqrt{M/b})$ achieves the desired. However, in practice the trade-off between available ancilla and Toffoli count reduction needs to be taken into account. Furthermore, for large $b$ the necessity of setting aside a ($b\alpha$)-qubit output register can act as a strong deterrent of this method.

\begin{figure}
    \centering
    \includegraphics[width=0.6\linewidth]{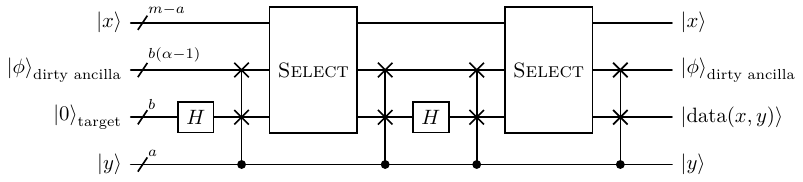}
    \caption{\textsc{SelectSwap} circuit loading $b$-bit data utilizing dirty ancilla. The address register of $\lceil \log M \rceil = m$ qubits is split into two registers $\ket{x}\ket{y}$ with the first holding the $m-a$ most significative qubits, and the second the $a$ least significative qubits, where $\alpha = 2^a$.}
    \label{fig: dirty SelectSwap}
\end{figure}

Instead of using clean ancilla as described in the previous paragraph, the \textsc{SelectSwap} circuit can also be performed using $b(\alpha-1)$ dirty ancilla. This was already observed in \cite{low2024trading} and slightly improved by \cite{Berry+QubitizationSparse19}. We briefly present the latter construction. The idea here is that one application \textsc{Select} \textsc{Swap} leads to the dirty ancilla being modified with loaded data and a reshuffling of qubits. However, by shuffling first with an additional application of the \textsc{Swap} network before applying \textsc{Select} \textsc{Swap$^\dagger$}, only the modification with loaded data remains. Another application of \textsc{Swap} \textsc{Select} \textsc{Swap$^\dagger$} thus restores the state of the dirty ancilla. To ensure that data is still loaded, the first \textsc{Swap} \textsc{Select} \textsc{Swap$^\dagger$} subroutine is conjugated by a layer of Hadamard gates as seen in \zcref{fig: dirty SelectSwap} which makes the state invariant under data loads. We refer to this construction as \emph{dirty} \textsc{SelectSwap}. In total, we observe four \textsc{Swap} networks and two applications of the skew tree \textsc{Qrom} leading to a Toffoli cost of $2M/\alpha + 4b(\alpha-1)$ utilizing $b(\alpha -1)$ dirty ancilla, $m-a$ clean ancilla for the skew tree \textsc{Qrom} and $b$ clean qubits to store the output register.

Recently, \cite{motlagh2026halving} introduced a similar method by replacing the  \textsc{Swap} network with a controlled \textsc{Copy} operation leading to further improvements in Toffoli cost. However, controlled \textsc{Copy} is implemented using an additional \textsc{Select} subroutine coupled with Toffoli gates instead of multi-target \textsc{Cnot} gates.  This additional \textsc{Select} subroutine  leads to a higher ancilla cost. As we are primarily interested in a small footprint compilation, we thus refrain from utilizing the version relying on \textsc{Copy}. Observe also, that the application of Toffoli gates instead of multi-target \textsc{Cnot} gates in the \textsc{Select} makes the usage of skew tree unary iteration non feasible, since the accumulation of Toffoli gates is inefficient.

\subsubsection{Lookup Table batching}\label{sec: Lookup table batching}

Regardless of the table lookup method, loading a large amount of potentially high-precision data requires a large output register. If the number of qubits is limited, it can thus be advantageous to divide the data into smaller chunks, perform sequential table lookups on a smaller register, use the data, and then discard the data by either overwriting or uncomputing the output. We refer to this process as \emph{lookup table batching}, the idea of which has been observed before in the literature (see, for example, \cite{motlagh2026halving, low2026denser}). Batching is possible as long as the data can be processed and safely overwritten or  discarded in between table lookups. While reduction of output register size is one advantage, depending on the lookup table method, one can also achieve potential Toffoli cost benefits. We observe the advantage for the dirty \textsc{SelectSwap} subroutine below.

Suppose that we want to load $bJ$-bit data by performing $J$ sequential data loads with $b$-bit lookup tables $T_1, \ldots, T_J$ for a fixed output register utilizing \textsc{SelectSwap} with dirty ancilla.  Naively, we could perform $J$ dirty \textsc{SelectSwaps} (leading to $2J$ (\textsc{Swap} \textsc{Select} \textsc{Swap$^\dagger$}) applications per lookup table application), where each application returns the dirty ancilla to their original state. This leads to a Toffoli cost of  $2J[M/\alpha + 2b(\alpha-1)]$.  Notably, since the dirty ancilla are returned to their original state after each dirty \textsc{SelectSwap} application, naive batching allows to change dirty ancilla between lookup table applications.  This can be useful if we want to use the output of the table lookup to interact with qubits previously used as dirty ancilla. Observe also, that to achieve the correct data load for each of the $J$ lookup tables, there are two options. First, the output register could be uncomputed using MBU resulting in phase corrections on the address register which need to be corrected using a \textsc{Phaseup} (see \zcref{sec: phase fixup}). These corrections can in principle be pushed through to the end of the $J$ lookup table procedures and combined into one joint operation as long as operations in between lookup tables do not interfere, since during each (dirty) \textsc{SelectSwap} the address register is only used as control. A benefit of uncomputing part or all of the output register is the ability to free the output register to be used as clean ancilla for operations performed between lookup tables. Second, if we do not want to keep track of additional phase corrections and to save the time to measure the output register, instead of loading $T_j$ for the $j$th table lookup, we load the difference $T_j \oplus T_{j-1}$.

If we can afford for dirty ancilla to be in an altered state during the full batching procedure, we can eliminate the second application of \textsc{Swap} \textsc{Select} \textsc{Swap$^\dagger$} in between data loads. We refer to this method as \emph{recycling batching} as opposed to \emph{naive batching} above. This elimination is due to the observation that each \textsc{Swap} application is based on the same permutation implementation. More precisely, we can track the accumulation on the dirty ancilla through subsequent \textsc{SelectSwap} applications, since they share the same pattern. Since \textsc{Swap} and \textsc{Swap$^\dagger$} define inverse permutations of registers for every $j \in [\alpha]$, each \textsc{Swap} \textsc{Select} \textsc{Swap$^\dagger$} combination returns the dirty ancilla to their original place, with the \textsc{Select} application adding the data load of the same index to the same dirty ancilla as the previous application. Thus, performing $J$ subsequent \textsc{Swap} \textsc{Select} \textsc{Swap$^\dagger$} combinations leads to the following data accumulation on the dirty ancilla: If we use the second option mentioned above and load only $T_j \oplus T_{j-1}$ at the $j$th step, the dirty ancilla hold an additional $T_J$. If we instead uncompute the output register between lookup table applications and load only $T_j$, the dirty ancilla hold $\bigoplus_{j\in[J]} T_j$. Hence, after $J$ \textsc{Swap} \textsc{Select} \textsc{Swap$^\dagger$} applications, we can return the dirty ancilla to their original state by using the same method as in \zcref{fig: dirty SelectSwap}, i.e., the output register is conjugated by a Hadamard layer and then we apply \textsc{SwapSelectSwap$^\dagger$} with a \textsc{Qrom} loading either $T_J$ or $\bigoplus_{j\in[J]} T_j$ depending on which uncomputing strategy is used. This dirty ancilla state recovery can be performed before or after the $J$ (\textsc{Swap} \textsc{Select} \textsc{Swap$^\dagger$}) applications. In sum, we obtain a Toffoli cost of $(J+1)[M/\alpha + 2b(\alpha-1)]$ for recycling batching. For comparison, the Toffoli cost of \textsc{SelectSwap} with dirty ancilla without batching is $(2[M/\alpha + 2bJ(\alpha-1)]$, assuming the availability of an output register of size $b'=bJ$.  This approach requires $Jb(\alpha-1)$ dirty ancilla.

\subsubsection{Phaseup}\label{sec: phase fixup}

Efficient uncomputation of a lookup table is performed using measurement based uncomputation. The output register is measured in the $X$ basis and depending on the outcome of these measurements, certain quantum states need to be phase corrected. Additionally, phase corrections in form of potentially multi-controlled $\CZ$ gates accumulate throughout the block encoding algorithm, for example due to corrections from magic state injections or MBU of \textsc{AND}$^\dagger$ gates. In most cases, the application of these phase corrections can be combined into one joint \textsc{Phaseup}. As with many other quantum algorithms, this can either be implemented by encoding the address register in space and time  \cite{Berry+QubitizationSparse19}[Figure 6] or by encoding it fully in space \cite{gidney2025factor}[Appendix A.3]. Since the \textsc{Phaseup} often follows the uncomputation of a lookup table, freshly uncomputed output register qubits can be used as clean ancillae, and therefore the fully space encoded version can be beneficial. If there are not enough ancillae available, we opt for the space-time split. We detail both methods in the following. 

\paragraph{Space-space split.} For a \textsc{Phaseup}, we are given an address register of size $m$ and a function $P \colon [M] \to \{0,1\}$ indicating which of $M = 2^m$ addresses requires a negative phase correction. We often think of $P$ as given in form of a table with $M$ rows indexed by $M$ addresses and a one entry if the address requires a phase-flip and a zero entry otherwise. First, split the address register in half and let $\ell$ denote the  $L=\lfloor m/2\rfloor$ least significant qubits register and $k$  the  $K=m-\lfloor m/2\rfloor$ most significant qubits register. The goal is to apply phase corrections as $\CZ$ gates between objects encoded in these split registers. For the fully space encoded version, we encode both registers in the monomial basis by applying a sequence of \textsc{And} gates to each register producing all possible products $\ket{\prod_{s\in S} \ell_s}, \ket{\prod_{s\in S^\prime} k_s}$ for all non-empty $S\subseteq [L], S^\prime \subseteq [K]$. This construction is referred to as a \emph{power-product} in \cite{gidney2025factor} and offers a high possibility of parallelization. To determine which power-product states share a phase correction with a $\CZ$ gate, we compute $P$ in its monomial basis expansion, i.e., $P(x) = \bigoplus_{S\subseteq [m]} c_S \prod_{s\in S} x_s$, where $x_{m-1}\ldots x_0$ is the binary decomposition of the integer $x$.
The coefficients $c_S \in \{0,1\}$ for $S\subseteq [m]$ then dictate necessary $\CZ$ phase corrections.
To compute the coefficients, observe that for a given address $x_{S'}$, whose binary decomposition is defined by $(x_{S^\prime})_i=1$ if and only if $i\in S'$, we have 
\begin{equation*}
    P(x_{S^\prime}) = \bigoplus_{S\in [M]} c_S \prod_{j\in S} (x_{S^\prime})_j = \bigoplus_{S\subseteq S^\prime} c_S,
\end{equation*}
since $\prod_{s\in S} X_s$ evaluated at $x_{S^\prime}$ is one if and only if $S$ is supported on $S^\prime$. Thus, we obtain $c_S$ by again applying the Möbius transform, i.e., $c_S = \bigoplus_{S^\prime \subseteq S} P(x_{S^\prime})$.

\begin{figure}
\begin{subfigure}[b]{\textwidth}
\centering
  \includegraphics[width=0.75\linewidth]{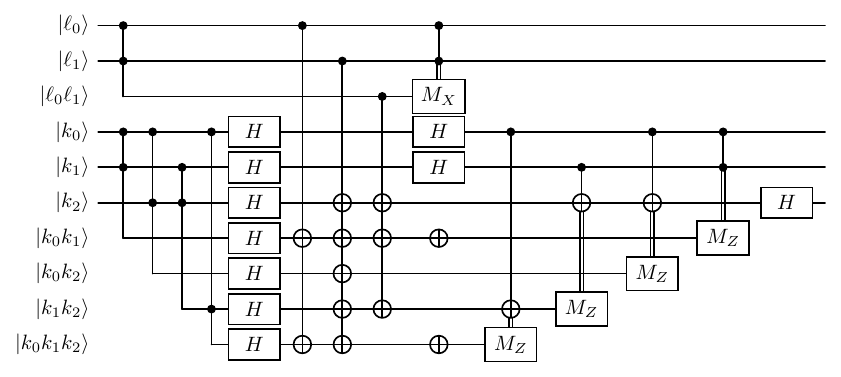}
    \caption{Example of a \textsc{Phaseup} circuit applying phase corrections on the addresses 10111, 01100, 10010 and 11110 in the space/space split. The address register is split into least-significant qubits $\ell_0, \ell_1$ and most-significant qubits $k_0, k_1, k_2$. The phases are corrected in the monomial basis using power-products and multi-target \textsc{Cnot} gates after a Hadamard conjugation. Power-products are then uncomputed using MBU and by performing the necessary corrections taking possible simplifications from Hadamard gates into account. }
  \label{fig:phase fixup circ}
\end{subfigure}

\medskip 
\begin{subfigure}[b]{\textwidth}
\centering
  \includegraphics[width=0.75\linewidth]{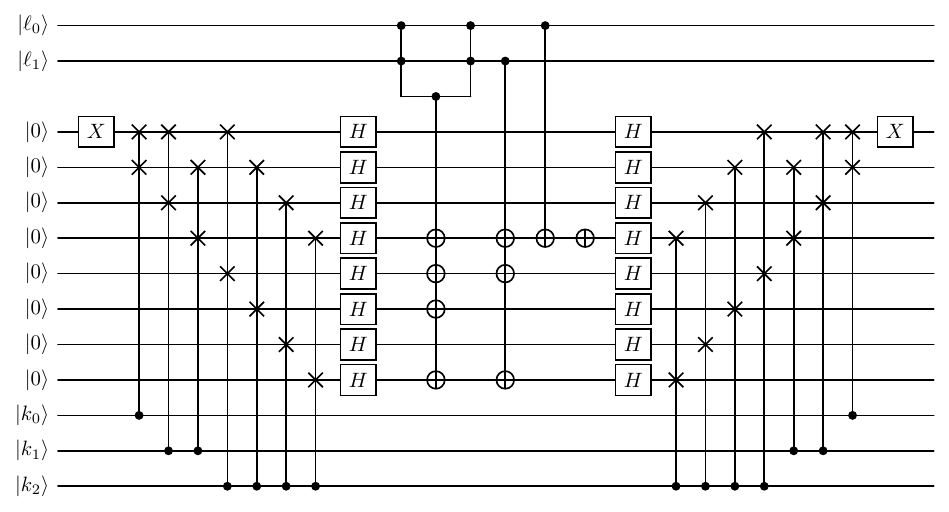}
  \caption{Quantum circuit of the above \textsc{Phaseup} using the space/time split. The phases are corrected using a skew-tree \textsc{Qrom} to encode half the address register in time, while a network of controlled \textsc{Swap} gates produces a unary-one-hot encoding in space. A conjugation by a Hadamard layer allows for the phases to be corrected using multi-target \textsc{Cnot} gates, which are determined using the Möbius transform necessary for the skew-tree \textsc{Qrom}.}
  \label{fig:phase fixup space/time}
\end{subfigure}
\end{figure}

Next, having obtained a list of coefficients $c_S, S\subseteq [m]$, we split the set of binary decomposition indices $S$ into two set of indices $S_\ell$, $S_k$ according to least and most-significant bits. Indeed, an address $x=x_{m-1}\ldots x_0$ also reads as $x=k_{K-1}\ldots k_0 \ell_L\ldots \ell_0$, so an address $x_S$ defined by its set of non-zero bit indices $S$ can equivalently be defined as $k_{S_k} \ell_{S_\ell}$. Since each $S_\ell, S_k $ has a single representative in the power-product representation, namely $ \prod_{s\in S_\ell}\ell_{s}$ and $\prod_{s\in S_k}k_{s}$, we can implement $c_S$ as a $\CZ$ gate between the representative of $S_\ell$ and $S_k$ if both are non-empty. If one of $S_\ell, S_k$ is empty, the phase correction simplifies to a Pauli $Z$ gate on the corresponding non-empty representative. The case of $S_\ell = S_k = \varnothing$ leads to a global phase.  Instead of applying $\CZ$ gates, we could also conjugate one of the registers by a layer of Hadamard gates, so that the phase corrections are implemented using \textsc{Cnot} gates. Note that the Hadamard layer can be applied to either of the registers. Usually this choice is made with an ease of compilation in mind. After applying the phase corrections, the power products are uncomputed using MBU. See \zcref{fig:phase fixup circ} for an example circuit, where we conjugate the most significant qubits by a layer of Hadamard gates to implement the phase corrections with multi-target \textsc{Cnot} gates. Overall, this construction leads to a Toffoli cost of $2^L + 2^K - m$ utilizing $2^L + 2^K - m$ clean ancilla and $2^K + K$ Hadamard gates and $2^L$ multi-target \textsc{Cnot} gates. 

We end by discussing the necessary corrections. First, observe that any phase corrections resulting from the magic state injection in the power product can be deferred to the \textsc{Phaseup} by updating the phase correction table $P$ before applying the \textsc{Cnot} gates, as originally noticed in \cite[Appendix A.3]{gidney2025factor}. This needs to be done during the application of the first Hadamard layer. Furthermore, note that similar to the skew-tree \textsc{Qrom}, the \textsc{And} gates for the power-products can be implemented using  wandering \textsc{And} gates as any errors also just lead to a classical update of $P$ to be performed during the application of the first Hadamard layer. Thus, the power-products can be implemented using non-autocorrected magic states. Only the phase corrections from the \textsc{And}$^\dagger$ gates in the power-product uncomputation need to be performed, some of which can be reduced to \textsc{Cnot} gates utilizing the layer of Hadamard gates (see \zcref{fig:phase fixup circ} for an example). Note that regardless of the correction taking the form of a $\CZ$ or \textsc{Cnot} gate, the reaction time must be taken into account when compiling the \textsc{Phaseup}. However, this does not lead to linear scaling in reaction time because uncomputations of a single product layer can be resolved simultaneously. Thus, the procedure scales logarithmically in the number of addresses in terms of reaction time and similar to the preparation of the power products, both can be uncomputed in parallel. 

\paragraph{Space-time split.} Following the approach of \cite{Berry+QubitizationSparse19} instead of resolving both address register halves simultaneously in space, one can be resolved in time using a \textsc{Qrom}, while the other half remains encoded in space using a unary-one-hot encoding. Due to the temporal encoding, the space-time split can be chosen to use less ancilla than the space-space split in most cases. Assuming again a split in terms of $L = \lfloor m/2\rfloor$ least significant qubits and $K = m- L$ most significant qubits, we need $L-1$ ancilla for the \textsc{Qrom} contribution and $2^K$  for the unary-one-hot encoding. If less ancilla are available, we can reduce the ancilla demand by increasing the number of least significant bits we are considering while decreasing the ancilla demand of the spatial unary-one-hot encoding. The unary-one-hot encoding is obtained using a network of controlled \textsc{Swap} gates controlled on $\ket{k}$ using $2^K$ clean ancilla. \zcref{fig:phase fixup space/time} shows an example circuit. The mechanism of the space-time split follows the space-space split closely: multi-controlled phase gates can now be performed as $\CZ$ gates between unary-one-hot encoded qubits and the flag qubit of the \textsc{Qrom} when the appropriate controls are on. Utilizing the same conjugation trick by a Hadamard layer, the phase gates can be converted and grouped to obtain multi-target \textsc{Cnot} gates. Since we are not utilizing power products, the corresponding multi-target \textsc{Cnot} gates are determined using $P$ without moving to its monomial basis expansion. However, we again use the skew-tree \textsc{Qrom} and thus, the final multi-target \textsc{Cnot} gates are computed using the Möbius transform. 

In terms of cost, the space-time split requires roughly $2^k$ more Toffoli gates and it requires slightly more Hadamard gates due to the uncomputation of the unary-one-hot encoding. Additionally, the nature of the \textsc{Swap} network does not allow for the implementation of a wandering \textsc{And} gate leading to a larger footprint for the magic state injection. However, the lack of measurement-based uncomputation also means that the space encoding of the space-time split does not incur reaction time wait. The larger footprint for the magic state injection is also not necessarily detrimental if a \textsc{Swap} network is used somewhere else in the circuit. In contrast, the skew-tree \textsc{Qrom} can be implemented using wandering \textsc{And} gates. For the phase corrections in the \textsc{Qrom}, one in principle needs to account for the phase correction for each \textsc{And}$^\dagger$ on the spot incurring reaction time wait. Alternatively, the phase corrections can still be pushed through and combined into a single \textsc{Phaseup} on $2^\ell$ addresses, leading to the same \textsc{Phaseup} procedure on a drastically smaller address register. We repeat this method until all phases have been resolved allowing us to forgo any reaction time wait. Note that this can also be done for the space-space split, if the Hadamard gates are not absorbed into the uncomputation. However, since both address register halves incur new phase corrections, two smaller \textsc{Phaseup}s need to be performed in the next step as opposed to one.  

\subsection{Rotation from a Fourier resource state}

To implement rotations, we use a Fourier resource state and phase gradient addition \cite{gidney2018halving}. Given a bit precision $b$, let $\ket{\varphi} = \frac{1}{2^{b/2}}\sum_{k =0}^{2^b-1} e^{-2\pi i k/ 2^b} \ket{k}$ denote the corresponding Fourier resource state. This is a catalytic state with respect to addition, since given a $b$-bit adder $\textsc{Adder}_b \colon \mathcal{H}^b \otimes \mathcal{H}^b \to \mathcal{H}^b \otimes \mathcal{H}^b \colon: \ket{x}\ket{y} \mapsto \ket{x} \ket{x+y \bmod 2^b}$ and a rotation angle $\ket{\theta}$ we obtain
\begin{equation*}
    \textsc{Adder}_b \ket{\theta} \ket{\varphi} = \ket{\theta} \ket{\varphi + \theta \bmod 2^b} = \frac{1}{2^{b/2}}\sum_{k =0}^{2^b-1} e^{-2\pi i k/ 2^b} \ket{\theta}\ket{k + \theta}  = e^{2\pi i \theta/2^b} \ket{\theta}\ket{\varphi}.
\end{equation*}
This can be used to implement a multiplexed rotation $R_Z$ of a state $\ket{c}$, only using one addition with the Fourier resource state as proven in \cite[Lemma 5]{low2026denser}. More precisely, the authors show the operation can be implemented using two multi-target \textsc{Cnot} gates controlled on the qubit that should be rotated $\ket{c}$ and targeting the Fourier resource state, i.e., 
\begin{equation}\label{eq:phase-gradient addition}
    \left( \bigotimes_{k=0}^{b-1} \textsc{Cnot}_{c, \varphi_k} \right) \textsc{Adder}_b  \left( \bigotimes_{k=0}^{b-1} \textsc{Cnot}_{c, \varphi_k} \right) \ket{\theta}\ket{\varphi} \ket{c} = e^{(-1)^c 2\pi i \theta / 2^b} \ket{\theta} \ket{\varphi} \ket{c},
\end{equation}
where $\textsc{Adder}_b$ acts on the first two registers.  Note that the Fourier resource state is a product state, as $\ket{\varphi} = \otimes_{k=0}^{b-1} 1/\sqrt{2} (\ket{0} + e^{-2\pi i/2^k} \ket{1})$. Thus, using a $b$-bit Fourier resource state allows for rotations of precision $b^\prime \leq b$. 

For the resource estimation in this paper, we use the adder of \cite{gidney2025factor} as compiled in \cite{Shor2026prep}, to which we refer for details. It acts out of place, $\textsc{Adder}_b\ket{x}\ket{y}\ket{0} = \ket{x}\ket{0}\ket{x+y \bmod 2^b}$, which realizes the map above up to relabeling the output register. We set $x$ to the rotation angle loaded as lookup table output, which is left intact and later cleared by measurement-based uncomputation, and $y$ to the Fourier resource state, which is then moved to the third register. Thus, the second multi-target \textsc{Cnot} in \eqref{eq:phase-gradient addition} targets this register.

%% file: sections/compilation.tex
\section{Compilation}
\label{sec:compilation}

In this section, we will detail the compilation of the main subroutines of the FeMoco algorithm onto a cat qubit architecture.
We will use the following notations 
\begin{equation} \label{eq: X14 X12 and S defs}
    X^{1/4} \coloneqq \ket{+}\bra{+} + e^{ \ii \pi /4} \ket{-}\bra{-} \, , \qquad 
    X^{1/2} \coloneqq \ket{+}\bra{+} + \ii \ket{-}\bra{-} = HSH \, , \qquad 
    \CXCXX \coloneqq H^{\otimes3}\CCZ H^{\otimes3} \,  .
\end{equation}
Working with highly biased-noise qubits concatenated with a classical code protecting against phase-flip imposes constraints on the implementation of a universal set of logical operations~\cite{barizien2025accessiblequantumgatesclassical,Guillaud_2019}. 
In particular, while resource states techniques, such as magic states factories, are most commonly reserved for operations outside of the Clifford group, we also use such techniques for $\keti=S\ket{+} = X^{1/2}\ket{1}$  resource states to implement Hadamard and $S$ gates.
Besides, similarly to surface code architectures~\cite{gidney2025factor,low2026denser}, we also use magic state distillation to prepare non-Clifford resource states as well~\cite{Ruiz_2026}.
As stabilizer codes against phase-flip are built out of $X$-stabilizers~\cite{Ruiz_2025}, we implement $X$-type lattice surgery operations between codes blocks~\cite{Horsman_2012, Gouzien_2023, Simultaneous2026prep}.
We also leverage the use of the biased-preserving \textsc{Cnot}~\cite{Guillaud_2019} to implement logical \textsc{Cnot}s transversally between repetition codes, compensating for the absence of $Z$-type lattice surgery operations.
In the end, our set of elementary logical operations is $\mathcal{E}=\{\mathcal{P}_{\ket+},\mathcal{P}_{\ket{0}},M_{Z},M_{X},Z,X,\textsc{Cnot}_L, M_{X\ldots X}\}$ where the $L$ subscript here emphasizes the locality of the transversal \textsc{Cnot}.
When completed with distilled resource states $\mathcal{E}\cup\{\mathcal{P}^*_{\ket {\CXCXX}},\mathcal{P}^*_{\ket{i}}\}$ is now universal, where $*$ denotes that those state preparations requires distillation.

We are interested in two main architectures: a repetition-cat architecture utilizing a repetition code with $X$-type stabilizers~\cite{Guillaud_2019,Gouzien_2023} and an LDPC-cat architecture as described in \cite{Ruiz_2025}. In the repetition-cat architecture, we assume a checkerboard layout, i.e., data qubits lie on a square lattice and ancillae are placed in the center of four cornering data qubits.
The ancilla is coupled to the four cornering data qubits leading to every qubit having connectivity of degree four.
To perform a transversal \textsc{Cnot} gate between adjacent repetition code columns, we would require \textsc{Cnot} gates between horizontally neighboring data qubits, which the above connectivity assumption does not permit.
However, we can mediate such \textsc{Cnot} gate by an ancilla adjacent to both.
More precisely, the ancilla is prepared in $\ket{0}$, afterwards it is the target of a \textsc{Cnot} gate from the control, later acts as the control of a \textsc{Cnot} gate onto the target, and lastly is measured in the $X$ basis.
The outcome of the $X$ basis measurement is then absorbed in the Pauli frame of the control data qubit.
We assume that all physical qubits are identical, such that any qubit of this lattice can serve as a data qubit or an ancilla.
The LDPC-cat architecture~\cite{Ruiz_2025} is based on a flip chip of two facing layers, with the bottom, \emph{repetition layer}, following the description above, and the top,  \emph{LDPC layer}, holding the data qubits of an LDPC-cat code on the same square lattice.
Each stabilizer check is assigned one ancilla, which is coupled to the at most four data qubits of its check.
Each data qubit of the LDPC layer is coupled to at most four ancillae, and to the facing qubit of the repetition layer below.

\begin{figure}
    \centering
\includegraphics[width=\linewidth]{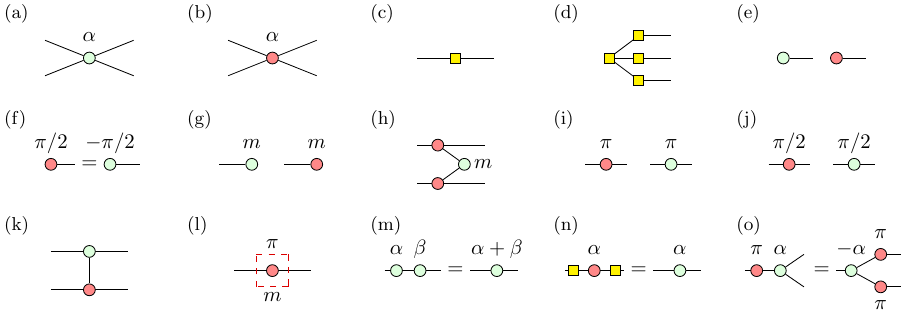}
    \caption{ZX-calculus reminders.
    (a),(b) Z resp.\@ red node with phase $\alpha$, $\ket{0\cdots0}\!\bra{0\cdots0}+e^{i\alpha}\ket{1\cdots1}\!\bra{1\cdots1}$ and resp.\@ $\ket{+\cdots+}\!\bra{+\cdots+}+e^{i\alpha}\ket{-\cdots-}\!\bra{-\cdots-}$. A node without a label has phase $0$.
    (c) Hadamard gate.
    (d) The $\ket{\CXCXX}=H^{\otimes 3}\ket{CCZ}$ state: the three-legged box is an H-box~\cite{Backens_2023}, $\sum_x(-1)^{x_1x_2x_3}\ket{x}$, with a Hadamard on each leg.
    (e) Preparations of $\ket{+}$ (left) and $\ket{0}$ (right). (f) Two equivalent writing of the $\ket{-i}$ state.
    (g) Measurement in the $X$ basis (left) and in the $Z$ basis (right) with outcome $m\in\{0,1\}$, integer phases implicitly include a factor of $\pi$.
    (h) Pauli-product measurement $M_{XX}$ with outcome $m$, $M_{X\cdots X}$ has one red node per qubit all linked to a single green node.
    (i) Pauli $X$ (left) and $Z$ (right).
    (j) $\sqrt{X}$ (left) and $S$ (right).
    (k) \textsc{Cnot}, with the green node on the control and the red node on the target.
    (l) An operation in a red dashed box is conditioned on the outcome $m$ written below it being $1$.
    (m) Fusion rule, works for either color.
    (n) Color change: Hadamards on every leg swap the color of a node.
    (o) $\pi$-commutation: a $\pi$ spider of one color passes through a spider of the other color, negating its phase and copying onto its other legs. 
    }
    \label{fig:ZX-reminders}
\end{figure}

When compiling the subroutines as well as the full FeMoco algorithm using the cat-qubit architecture, we view algorithmic qubits as represented in the $X$ basis as opposed to the usual computational basis, to take advantage of the asymmetric access requirements for $X$- and $Z$-logical operators.
At the logical level, this change of basis can be implemented by conjugating each qubit by a layer of Hadamards.
Subsequently, each gate can also be thought of as being conjugated by a layer of Hadamards.
More concretely, a Toffoli gate $\C\C X$ corresponds to $\CXCXZ$ in the color-changed convention and a phase gate $\C Z$ to a $\CXX$ gate.
Similarly, a multi-target $\textsc{Cnot}$ corresponds to a parity-controlled $\textsc{Not}$ and vice versa.
Using the language of ZX-calculus~\cite{vandewetering2020zxcalculusworkingquantumcomputer}, a diagrammatic language for quantum computation, of which a subset of rules used in this work is summarized in \zcref{fig:ZX-reminders}, this basis change corresponds exactly to switching red and green nodes.
Thus, we refer to this formalism as color-changed.
This viewpoint also makes clear that it leads to a role reversal of targets and controls. 
The resource estimates reported in this section are counted in terms of data qubits unless otherwise specified, i.e.\@, they do not explicitly account for ancilla qubits for stabilizer measurements.
For the physical qubit count, as we consider a square grid layout with face centered ancillae, we simply double the data qubit count.
Furthermore, when considering the temporal footprint of a subroutine, we do not explicitly account for operations whose durations are not directly proportional to the syndrome extraction cycle time $t_c$.
More precisely, for qubits stored in the repetition code, reported duration count formulas do not include measurements, initializations and transversal \textsc{Cnot}s, but instead report an explicit formula for the logical cycles.
Subroutine counts also do not include buffer staging time.

\subsection{Clifford resource state production}\label{subsec:i state factory}

\paragraph{$\keti$-state preparation protocol.}
In this section, we present a protocol to prepare $\ket{i}$ resource states in a cat-qubit repetition code. It is based on a generic distillation protocol built from two parts, a standard quantum circuit and an output selection procedure.
The circuit begins with the preparation of a Bell state between the repetition code and an external logical qubit, encoded in a so-called distillation code.
It ends with a logical operation that destructively measures the distillation code and, in the absence of noise, deterministically prepares the repetition code in a resource state.
This standard circuit has three main degrees of freedom: the logical operation, the distillation code and the Bell pair preparation protocol. The choice of logical operation at the second step fully controls which resource state is returned by the circuit, 
whereas the choice of distillation code and of Bell pair preparation can be tailored to meet physical requirements and fault tolerance of the implementation.
What makes the protocol particularly robust against noise is its selection strategy: at the level of the distillation code, the circuit is not protected by error correction but rather by error detection since the procedure is fully reset upon detection of particular syndromes. 
We summarize this high-level protocol in the case of $\keti$ state distillation in \zcref{fig:istate_distillation_high_lvl} in \zcref{app:istate_distillation}.

A fully compiled version of this scheme, called ``unfolded distillation'' \cite{Ruiz_2026}, uses a so-called ``unfolded'' implementation of a quantum Reed-Müller code as distillation code to produce $\ketXmag$ magic states.
Simulation results indicate that the protocol provides a favorable trade off in this case between logical error rate and space-time cost, especially since the distillation code works directly with physical qubits for both data and ancilla qubits.
This is mainly due to its error-protection strategy, which has been simplified by the intrinsic bit-flip protection of cat qubits and specialized to the targeted resource state.
While that paper focuses on $\ketXmag$ states, it was suggested in \cite[\textsection V]{Ruiz_2026} that adapting the protocol to distill $\keti$ states on cat qubit repetition codes should yield similarly interesting results for the same reasons.
However, despite the compiled circuit specially proposed there, no simulation results have been reported.
Here, we develop the idea further and propose a new compiled variant of the $\keti$ state distillation protocol, backed up with simulation results.
The full details can be found in \zcref{app:istate_distillation} ; we here only highlight a few key points where our compilation choices depart from the model of \cite{Ruiz_2026}.  

As already noted in \cite[\textsection V]{Ruiz_2026}, a reasonable choice of an $\keti$ state distillation code is the distance-$3$ triangular color code, which is the [[7,1,3]] Steane code, because its logical $X^{1/2}$ gate has a transversal implementation, see \zcref{app:istate_distillation}.
For our purposes, it is however desirable to use a triangular code with a distance larger than 3 in order to reach better logical error rates than the announced lower bound of $7(2 p_z)^3$ \cite[\textsection V]{Ruiz_2026}.
Yet, the circuit proposed there does not directly extend to higher distances.
Indeed, it uses a joint $X_LX_L$ measurement for Bell pair preparation, i.e.,\ a merge operation between the triangular code and the repetition code. This has weight 4 for a code of distance 3 but more qubits are involved at higher distances. Such an operation becomes not only difficult to make fault tolerant, it is also hard to reconcile with the connectivity-4 constraint of our layout. 
We thus depart from the original proposal of \cite[\textsection V]{Ruiz_2026} by preparing the Bell pair differently: the distillation code is initialized in a logical $\ket{+}_L$ state, the repetition code in a logical $\ket{0}_L$ state, and both codes are entangled by a logical \textsc{Cnot}. Since the latter is conveniently transversal in our setting, see \zcref{fig:istate_logical_cnot} in \zcref{app:istate_distillation}, it is fault tolerant.

Errors behave differently in our setting compared to \cite{Ruiz_2026}, so we have adapted its selection procedure accordingly, in order to retain the good results obtained there for $\ketXmag$ distillation.
First, because the circuit we are considering now contains only Clifford gates, syndrome decoding can be safely postponed to the end of the circuit, meaning that preselection in the sense of unfolded distillation \cite{Ruiz_2026} is no longer needed.
Moreover, hook errors can now become a major source of logical errors at the Bell pair preparation step if they are not handled properly.
Indeed, unfolded $\ketXmag$  distillation depends only on $X$ stabilizer measurements, including the merge operation, to prepare the logical Bell pair. Although phase flips make $X$-stabilizer measurements less reliable, they cannot propagate as 
hook errors to data qubits.
For these reasons, the authors of \cite{Ruiz_2026} simply protected the Bell pair by measuring $X$ stabilizers for as many ``rounds'' as required for the decoded measurement results to become reliable.
This makes the number of rounds a key parameter in their trade off between error rate and execution time.
Our situation is very different in this respect as, instead of $X$ stabilizers, we measure $Z$ stabilizers on the distillation code side: they potentially propagate phase-flip errors, so we have limited the number of rounds, see \zcref{app:istate_distillation}.

\input{graphics/istate_distillation/distillation_results.tex}

Our simulation results were obtained via usual Monte Carlo simulations on \texttt{stim}~\cite{gidney2021stim} using the decoder Tesseract~\cite{beni2025tesseractsearchbaseddecoderquantum}.
They demonstrate the conjecture from \cite[\textsection V]{Ruiz_2026} that a proper selection strategy can make the logical error rate  proportional to $p_z^{d_c}$, where $d_c$ is the distillation code distance, see \zcref{app:istate_distillation} and \zcref{fig:i_distillation_ler_vs_pz_plot}.
However, our prefactors are much higher than those predicted in that reference. 
Although reaching a $d_c$ circuit-level distance for our protocol requires taking it long enough, see \zcref{fig:i_distillation_rep_saturating_params}, we make the choice here to also consider shorter protocols because time is a key resource for our purposes.
The space-time parameters we consider are described in \zcref{tab:istate_distillation_results}. Our results are detailed in \zcref{tab:istate_ler} and visualized in \zcref{fig:istate_protocol_plots}.

\paragraph{Resource estimation.}
Anticipating the next sections, we use our results to quantify the space-time resource cost of $\keti$-state usage in our implementation of the FeMoco algorithm. 

Note first that, provided layout constraints are not an issue, the geometry of our protocol in principle allows preparing repetition codes from any of the three sides of the triangular code.
An $\ket{i}$-state factory therefore has three output ports, and it can prepare a new $\keti$ state on one port while a previously prepared state still occupies another.
We use this property to limit the number of factories required to cover a given demand as follows.
The amount of time it takes to fully produce a state is the sum of two terms: the duration $t_i$ of the quantum circuit and the reaction time $t_r$ that is in practice required for the decoder to process the measurements acquired over the protocol. We think in terms of repetition-code cycles and normalize both terms by the protocol's acceptance rate $\alpha$ in order to account for the number of times an output is discarded before it gets validated by selection.
An output port is then on average occupied for $W=\frac{t_i+t_r}{\alpha}+d$ cycles where $d$, the repetition-code distance, corresponds to the number of cycles it takes to fault-tolerantly route a produced $\keti$ state into the buffer area. Note that while $d$ and $d_r$ both designate the same physical quantity, we keep the latter notation for the distinct setting of our $\keti$-state preparation simulations, where $d_r$ is a swept parameter. 
Since a factory runs at most one $\keti$-state preparation every $t_i/\alpha$ cycles and each of its three ports is occupied for $W$ cycles per state, it delivers on average $3\frac{r}{\max(3t_i/\alpha,W)}$ states to the buffer every $r$ cycles. 

We apply this general reasoning to the concrete numbers of \zcref{sec: Resource Count BE}, taking $\ket{\CXCXX}$ production cadence $r=21$ cycles, reaction time $t_r=10$~\cite{Ruiz_2026, Gouzien_2023} and distance $d \leq 19$. Assuming that the latter parameter does not affect acceptance rate $\alpha$ once the saturation regime of \zcref{fig:i_distillation_rep_saturating_params} is attained, both $\alpha$ and $t_i/\alpha$ can then be read from our \zcref{tab:istate_ler}. We then get $W \leq 39$ for all the short and medium protocols in the table, leading to an average of $3\frac{21}{39} \simeq 1.6$ states delivered every $r=21$ cycles.
Since our most demanding subroutine, the \textsc{Swap} network, consumes 4 states every $r=21$ cycles, see \zcref{sec: Swap comp}, our demand in $\keti$ states is then covered by 3 factories. The global layout in \zcref{sec: Resource Count BE} features these three factories and make their nine ports available.

\subsection{Non-Clifford Magic state production}\label{subsec: CxCxX_factories}

\begin{figure}[!htb]
    \centering
    \includegraphics[width=0.9\linewidth]{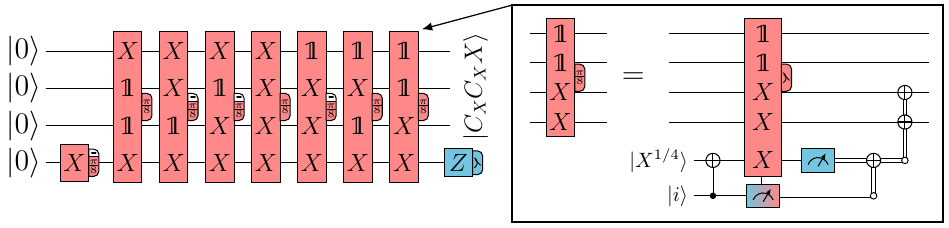}
    \caption{Circuit of the $8X^{1/4}\to1\CXCXX$ distillation protocol (left) as well as gate injection circuit (right) used to implement the protocol. 
    The two native operations used are local transversal \textsc{Cnot} and $X$-Pauli QND measurements via lattice surgery.
    The measurement basis of the $\ket{i}$ state is conditioned on the outcome of the lattice surgery operation.
    Note that the fourth qubit of the circuit is the check qubit, the protocol post-select on the $+1$ outcome of its final measurement.
    }
    \label{fig:second_level_distillation_circuit}
\end{figure}

Preparing high-fidelity non-Clifford magic states is a widely studied subject in which two main types of techniques have emerged: distillation protocols~\cite{Bravyi_2005,Bravyi_2012,Haah_2018} that consume many low quality states to produce a few high-quality ones, and cultivation protocols that rely on the ability to perform Clifford basis measurements~\cite{gidney2024_cultivation,sahay2025foldtransversalsurfacecodecultivation}.
Previous cat-qubit architectures proposed Toffoli-state preparation methods~\cite{Chamberland_2022,Gouzien_2023} using physical Toffoli gates, the implementation of which remains challenging.
In our architecture, very low error rate ($<10^{-10}$) non-Clifford magic states are obtained using a two-level distillation protocol. 
While two-level protocols are commonly used in many architectures~\cite{xu2026_bicycle_distillation,Litinski_Notascostly,gidney2025factor}, 
most of them concatenate a first stage of magic state cultivation with a $8T\to1\CCZ$ distillation protocol~\cite{jones_low-overhead_2013}.
For biased noise qubits, a physical level Hadamarized $15T\to1T$ protocol appears to be more efficient than standard surface code cultivation~\cite{Ruiz_2026}.
Such protocols do distill $\ket{X^{1/4}}$ rather than $T$ states.
Given an input phase-flip error rate of $p_Z=10^{-3}$  it achieves a first level output error rate of $p_{1}=3\times10^{-7}$.
We further concatenate this protocol with a $8X^{1/4}\to1\CXCXX$ distillation protocol, shown in~\zcref{fig:second_level_distillation_circuit}, in order to lower this error rate to a value as low as $p_{2}\geq 28 p_{1}^2=2.5\times10^{-12}$.
Note that in our case $\ket{i}$ states used for injection in the protocol are distilled as well, see~\zcref{subsec:i state factory}.
While in principle we could use distance $5$ triangular color-code factories to distill those states directly at the final error rate $p_2$, it is worth noting that an error on a $\ket{i}$ state will be detected by the protocol.
As a consequence, we've made the choice to use the first level distance $3$ $\ket{i}$ state distillation protocol, which yields states of error rate $p_{i}=6\times10^{-7}$ at $p_Z=10^{-3}$, becoming the limiting factor of the protocol, capping it at $p_{2}\gtrsim 1\times10^{-11}$. 
More precisely, in practice, half the time an error on $\ket{i}$ has no effect, while the other half of the time it does, depending on the chosen measurement basis.
So that it really adds $5 \times 10^{-12}$ to the $2.5 \times 10^{-12}$.
This is now sufficiently low to run Shor or FeMoco algorithms at relevant sizes.
At the second noise value used in this paper $p=5\times 10^{-4}$, simulations of level-1 factories yield $p_{1}=4\times 10^{-8}$ and $p_{i}=8\times10^{-8}$, following the modified short protocol results documented in  \zcref{tab:istate_ler_5e-4}.

\begin{figure}
    \centering
    \includegraphics[width=0.8\linewidth]{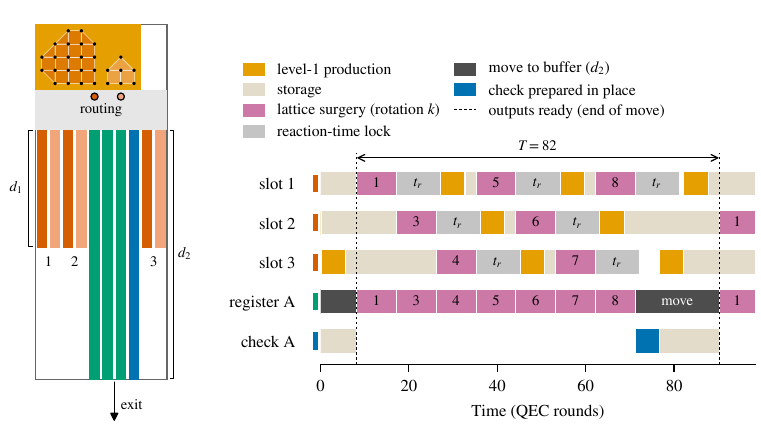}
    \caption{Naive implementation of a two-level factory, at $d_1=9$, $d_2=19$ and $d_m=9$.
Left: physical layout of the factory.
Two level-1 factories, one for $\ket{X^{1/4}}$ and one for $\ket{i}$ states, are represented as a $8\times5$ data qubits orange box.
It feeds respectively the dark orange and light orange repetition code slots through a three-row routing area (gray).
The output $\ket{\CXCXX}$ stored in green repetition codes and the check qubit in blue have length $d_2$ and move to the buffer from the bottom.
Right: steady-state schedule of the protocol.
The seven rotations are performed one at a time by lattice surgery taking $d_m$ rounds each.
They are labeled by their index $k$.
The first rotation of the circuit of \zcref{fig:second_level_distillation_circuit} is replaced by preparing the check qubit directly in $\ket{X^{1/4}}$.
After a lattice surgery step, the magic state slot is locked for the reaction time $t_r=10$ waiting to chose the basis of the $\ket{i}$ qubit measurement.
Then it is refilled by the level-1 factory in $t_1=5.5$ rounds.
The output codes are moved to the buffer in $d_2$ rounds before the next period starts.
The period $T=7d_m+d_2=82$ rounds is measured between two consecutive ends of moves when the output states become usable from the buffer.}
\label{fig:factories_layout_1}
\end{figure}

\begin{figure}
    \centering
    \includegraphics[width=0.8\linewidth]{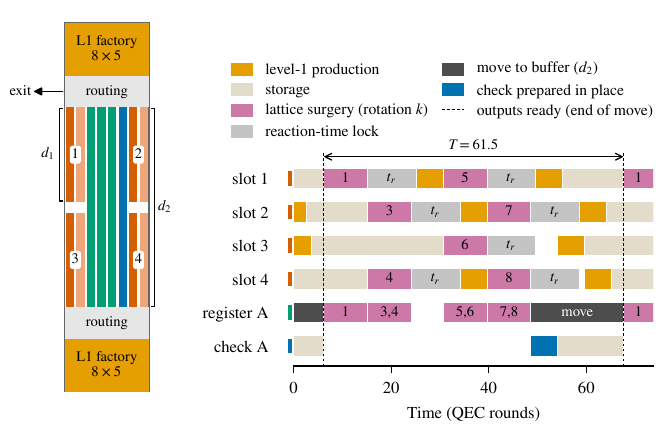}
    \caption{Second implementation of a two-level factory, at $d_1=9$, $d_2=19$ and $d_m=9$ with two routing areas available enabling parallelization.
    Same color conventions as \zcref{fig:factories_layout_1}.
    The duration of the protocol is reduced while the space footprint is slightly increased.
    The register waits between the second and third steps because slot recycling takes longer than one lattice surgery operation ($t_1+t_r>d_m$).
Hence, $T=3d_m+d_2+\max(d_m,\,t_1+t_r)=61.5$ rounds.}
    \label{fig:factories_layout_2}
\end{figure}

\begin{figure}[!htb]
    \centering
    \begin{subfigure}{\linewidth}
        \centering
        \includegraphics[width=\linewidth]{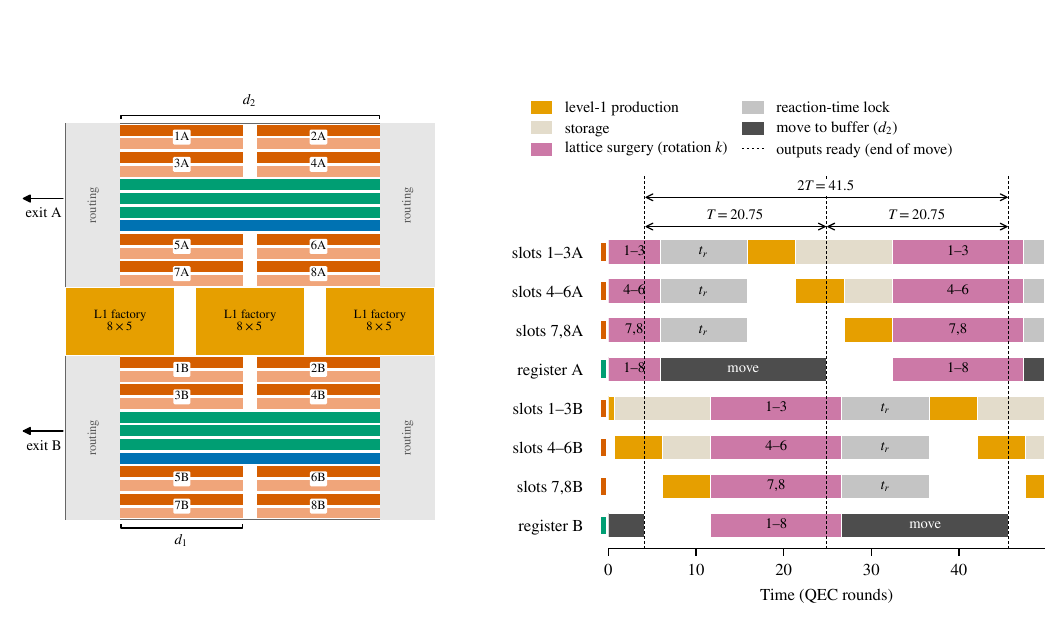}
        \caption{}
        \label{fig:factories_layout_3_layout}
    \end{subfigure}
    \par\medskip
    \begin{subfigure}{\linewidth}
        \centering
        \includegraphics[width=0.9\linewidth]{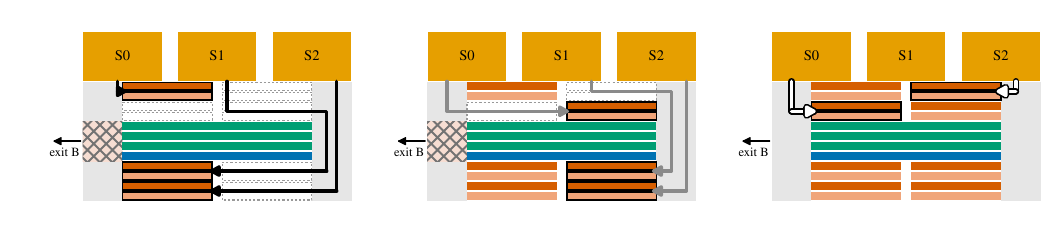}
        \caption{}
        \label{fig:factories_layout_3_fill_order}
    \end{subfigure}
    \caption{Simultaneous layout with three shared level-1 factories in one middle row, at $d_1=9$, $d_2=19$ and $d_{ms}=15$.
    \subref{fig:factories_layout_3_layout}~Layout and schedule, with the same color conventions as \zcref{fig:factories_layout_1}.
    Two registers, one above (A) and one below (B) the factory row, each sit between two four-line routing zones (gray), which carry both the routing of the simultaneous surgery and the transport of the level-1 outputs.
    All eight rotations of a register are performed in one simultaneous lattice surgery of $d_{ms}$ rounds using both routing zones.
    The two registers alternate half a period apart, so the factories refill one register while the other is in surgery, and the $d_2$-round move of the outputs overlaps the refill.
    Hence $T=\tfrac12\bigl[d_{ms}+\max(d_2,\,t_r+3t_1)\bigr]=20.75$ rounds between consecutive outputs, and $2T=41.5$ rounds per register.
    \subref{fig:factories_layout_3_fill_order}~Order of the three-step refill of the level-1 magic-state slots. This order does not conflict with the previously prepared magic state still moving to the buffer.}
    \label{fig:factories_layout_3}
\end{figure}

More concretely, we propose three possible compilations and layouts of two-level factories.
Each layout comes with its respective time schedule that aims at pipelining operations. 
A first minimalist version, mostly pedagogical, with only one first level magic state factory of each type, namely one $\ket{X^{1/4}}$ unfolded Reed-Muller factory and one $\ket{i}$ Steane code factory, both due to~\cite{Ruiz_2026}.
Together they do produce a pair of resource states in $t_1$ extraction cycles, namely the maximum of the duration of $\ket{X^{1/4}}$ and $\ket{i}$ distillation procedures.
In this first version, the gates of the protocol are applied serially, see~\zcref{fig:factories_layout_1}.
The 4 qubits in the circuit of \zcref{fig:second_level_distillation_circuit} are stored in repetition codes of distance $d_2$ while the gates of the circuit are performed by injecting magic states coming out of unfolded Reed-Muller factory and stored in repetition codes of distance $d_1$.
A third parameter, $d_m$, called timelike distance determines the number of rounds performed in each lattice surgery operation measuring a multiqubit $X$ operator.
Additionally, the reaction time $t_r$ quantifies the amount of time between a quantum measurement and a quantum operation -- a gate or measurement/preparation -- controlled on its output.
In particular, in our injection circuit it determines the time between the end of a lattice surgery operation and the measurement of the $\ket{i}$ magic state.
A second version leverages the presence of one bottom and one top routing area, see~\zcref{fig:factories_layout_2}, to perform two $X$-Pauli measurements at a time, thereby dividing the duration of the protocol by two.
A third version presented in \zcref{fig:factories_layout_3} uses simultaneous lattice surgery techniques, described in~\cite{Simultaneous2026prep}, to perform every multiqubit $X$-Pauli measurement of the circuit within a single lattice surgery operation, drastically reducing the duration of the protocol.
The timelike distance is increased to $d_{m,s}>d_m$ in this protocol as simultaneous lattice surgery induces correlated logical errors~\cite{Simultaneous2026prep}.
Typically $d_{m,s}$ is closer to $d_2$ because the protocol can detect and post-select any single lattice surgery errors but can't handle correlated wrong logical outcomes.
However, this speed-up comes at the price that the 8 magic states of each kind need to be ready at the start of the protocol rather than 1 or 2 for the first and second versions, respectively.
This induces a space overhead due to additional first-level factories and first-level magic state storage areas.
The total spacetime volume eventually remains smaller for this version. 

As explicitly shown on~\zcref{fig:second_level_distillation_circuit}, the distillation protocol consumes an equal number of $\ket{i}$ and $\ket{X^{1/4}}$ states.
To simplify the compilation, we model a pair of such factories as a $5\times 8$ sized hashed rectangle of data qubits.
This pair of factories produces magic states in two repetition codes attached at a specific position on the largest side of the rectangle.

From the different layouts it is straightforward to compute the sizes, $S_i$, of the factories --namely the size of the outer rectangles-- as a function of $d_1,d_2$ respectively the length of dark orange and green/blue repetition codes in \zcref{fig:factories_layout_1, fig:factories_layout_2, fig:factories_layout_3}.
\begin{align}
    S_1&= 10\times (d_2+3+5)\\
    S_2&=8\times(d_2+2\times(5+3))\\
    S_3&=(4+d_2+4)\times (2\times 12+5)
\end{align}
Note that we assumed $d_2\geq 2d_1$ for the second and third layout designs and $d_2+8\geq 2\times  8$ to compute $S_3$.
These two distance choices criteria are motivated by density-matrix simulations of the second-level protocol using similar techniques as~\cite{Litinski_Notascostly}.
In the simulation, we use different logical error rate models for memory, regular lattice surgery, and simultaneous lattice surgery, each fitted on simulations.
Here, we only count data qubits for clarity, while the full number, including ancillae, is given below in \zcref{tab:factory_summary}.

In each caption we also compute the duration of the two-level procedure using the time schedule given in the figures.
Note that this duration has in principle to be renormalized by the success probability of the protocol.
However, the rejection probability being around $10^{-6}$ makes this effect negligible.
Magic states are then moved to a buffer area before being used, which, by the way, is an opportunity to interleave them with $\ket{i}$ states to ease the later injection; see~\zcref{sec: Swap rep-code comp}.
While moving magic states to a buffer may seem time-inefficient, it is actually concurrent with the reaction time due to the measurement of the second-level factory check qubit, before which the success of the procedure is not established.
Indeed, one shouldn't inject the state into the computation without knowing whether the state is usable. 

Density-matrix simulations suggest that working with $d_1=9, d_2=19, d_m=9, d_{m,s}=15$ maintains an output error rate  close to $1\times 10^{-11}$, namely what a noiseless second level would yield. 
We also fix $t_1=5.5$ and $t_r=10$ using assumptions from ~\cite{Ruiz_2026} and ~\cite{Gouzien_2023}.
Each factory's volume and output error rate are gathered in \zcref{tab:factory_summary}.
Note that while $d_1$ is fixed to maximize the first-level factory performances, $d_2$ could be reduced to $17$ or $15$, particularly in the case $p=5\times 10^{-4}$. However, for simplicity of layout design, we decided to work with $d_2=19$ for both noise models. 
The duration and volume in this table are given for a move-to-buffer displacement that takes $d_2$ rounds.

\begin{table*}[!htb]
\centering
\small
\sisetup{exponent-product=\times, round-mode=none}
\setlength{\tabcolsep}{5pt}
\renewcommand{\arraystretch}{1.15}
\begin{tabular}{@{}l r l r S[table-format=1.2e1] S[table-format=1.2e-2] S[table-format=1.2e-2]@{}}
\toprule
 & Footprint $A$
 & \multicolumn{2}{c}{Time per output $T$ (cycles)}
 & {Volume $V$}
 & \multicolumn{2}{c}{Output error rate $p_2$} \\
\cmidrule(lr){3-4}\cmidrule(lr){6-7}
Layout & (qubits) & distance & value
 & {(qubits$\times$cycles)} & {$p=10^{-3}$} & {$p=5\times10^{-4}$} \\
\midrule
\makecell[tl]{first\\[-2pt] layout}
 & 540 & $d_m=9$ & 82
 & 4.43e4 & 1.32e-11 & 1.84e-13 \\
\addlinespace
\makecell[tl]{second\\[-2pt] layout}
 & 560 & $d_m=9$ & 61.5
 & 3.44e4 & 1.32e-11 & 1.81e-13 \\
\addlinespace
\makecell[tl]{third\\[-2pt] layout}
 & 1566 & $d_{ms}=15$ & 20.75
 & 3.25e4 & 1.48e-11 & 2.02e-13 \\
\bottomrule
\end{tabular}
\caption{Resources of the three two-level factory layouts, with $d_1=9$, $d_2=19$ and a noise bias of $\eta=10^{14}$
deduced from logical level density matrix simulation similarly to~\cite{Litinski_Notascostly}. 
Note that the space footprints here do include ancillae.}
\label{tab:factory_summary}
\end{table*}
\subsection{\textsc{Qrom} compilation}
\label{subsec:qrom compil}
In this section, we describe on a concrete layout the detailed schedule of the logical operations to implement the skew-tree \textsc{Qrom} described in~\zcref{sec: Skew-tree QROM description}.
We first do this compilation work on a repetition-cat architecture before doing it for a denser LDPC-cat architecture~\cite{Ruiz_2025}.

\subsubsection{Compilation on the repetition-cat architecture}\label{sec: QROM comp rep cat-code}

\begin{figure}
    \centering
    \begin{subfigure}[t]{0.5\textwidth}
        \centering
    \includegraphics[width=0.6\linewidth]{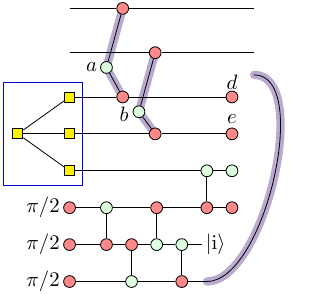}
    \caption{}
    \label{subfig:AND_ZX}
    \end{subfigure}%
    ~
    \begin{subfigure}[t]{0.5\textwidth}
    \centering
        \includegraphics[width=\linewidth]{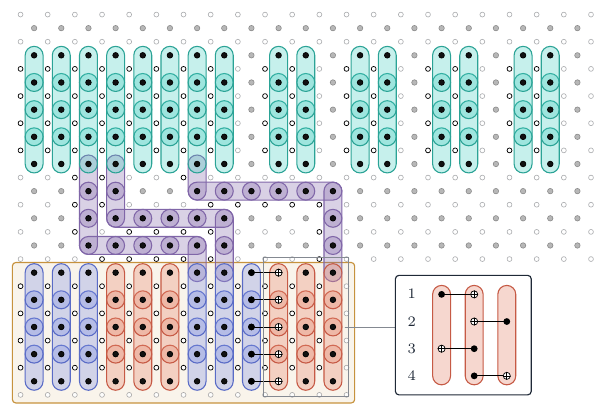}
        \caption{}
        \label{subfig:rep_code_layout_compiled}
    \end{subfigure}%
    \caption{a) is a ZX diagram that represents the \textsc{And} gate without Clifford and Pauli corrections in the color-changed convention. 
    This diagram consumes one $\ket{\CXCXX}$ state (framed in blue) and two $\ket{i}$ states with one additional catalytic $\ket{i}$ state.
    b) is the repetition code layout for the skew-tree \textsc{Qrom} with the lattice surgery operation highlighted in the color corresponding to the highlighted ZX edges of the same color.
    Algorithmic qubits are stored in teal repetition code while $\ket\CXCXX$ and $\ket{i}$ are respectively stored in the deep blue and orange repetition code.
    Address register and auxiliary qubits are packed while target register qubits are spaced with routing areas to access their $Z$ logical operators.
    The inset describes the catalytic Hadamard injection \textsc{Cnot} schedule also represented on the 3 last rows of the ZX-diagram.
    The circuit version of the Hadamard injection is shown in~\zcref{app: Clifford}.
    The labels 1,2,3,4 in the inset fixes the order in which the transversal \textsc{Cnot}s are performed.}
    \label{fig: compiled_and}
\end{figure}

For the repetition-cat code architecture~\cite{Guillaud_2019,Gouzien_2023}, we use a square grid layout of data qubits with face-centered ancilla qubits, see~\zcref{fig: compiled_and}.
Data and ancilla qubits are distinguished for the readability of the figures, although both are physical cat qubits.
The layout used for compiling the \textsc{Qrom} is essentially made up of three different areas: logical algorithmic qubits represented in teal, a routing area, and a resource state buffer area outlined in orange and containing buffer states represented in blue and orange, respectively, for $\ket{\CXCXX}$ and $\ket{i}$ magic states.
The logical qubits materialized by lines in the quantum circuit representation of the subroutines are called algorithmic qubits to differentiate them from logical qubits storing resource states or temporary auxiliary logical qubits used in specific gate implementations as described in circuit identities of \zcref{app: Clifford}.
In the layout, algorithmic qubits are ordered from left to right, with address registers in ascending order (leftmost repetition code is the least significant bit) then clean ancilla qubits resulting from an \textsc{And} gate, ordered by instantiation time and finally the target register stored with $Z$-access routing corridors.
We focus in this subsection on compiling the uncontrolled version in reversed DFS order of~\zcref{fig:skew tree uncont}, the motivations of this choice are described above in \zcref{sec: Skew-tree QROM description}.
This version of the skew-tree \textsc{Qrom} is made of \textsc{And} gates, multi-target \textsc{Cnot} gates, and measurement-based \textsc{And}$^\dagger$.
In this section we keep using the terminology of an \textsc{And} gate even though we work with its color-changed version.

The \textsc{And} injection scheme of~\zcref{subfig:AND_ZX} uses two $M_{XX}$ operations, 5 transversal local \textsc{Cnot}s and one logical qubit displacement from the resource state buffer area to an auxiliary qubit location.
Indeed, the output qubit of the \textsc{And} gate becomes an auxiliary qubit stored in the dedicated area.
We use $\ket{\CXCXX}$ state to perform an \textsc{And} gate, consequently a $H$ gate is applied on the third magic state qubit to make it a $\ket{\CXCXZ}$ magic state.
This Hadamard gate is performed using the catalytic injection circuit of \zcref{fig:H_w_cnot} in \zcref{app: Clifford}.
\zcref{subfig:rep_code_layout_compiled} shows that the injection step can be performed in a single $d$ rounds logical cycle without prohibitive routing costs.
This is possible because resource states are arranged in a convenient order in the buffer area.
Note that this parallelization comes at the price of having 3 routing alleys in-between the buffer area and the algorithmic qubits.
Adding horizontal routing alleys in the layout has a cost that is proportional only to the number of logical algorithmic qubits.
In particular, it is independent of the distance of the repetition code which is not the case for regular surface code architectures~\cite{Litinski_2019_GoS}. 
The algorithmic qubit ordering in the cyan area and the magic state qubits ordering in the buffer area are important for this compilation to work without resorting to crossing of lattice surgery repetition codes.
This compiled \textsc{And} gate takes $d$ rounds and consumes one $\ket{\CXCXX}$ state and three $\ket{i}$ states. 

\begin{figure}
    \centering
    \begin{subfigure}[t]{0.5\textwidth}
        \centering
    \includegraphics[width=0.6\linewidth]{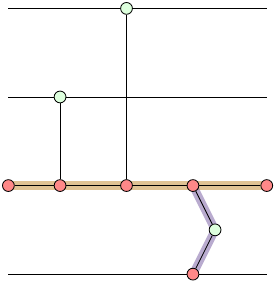}
    \caption{}
    \label{subfig:parity_NOT_ZX}
    \end{subfigure}%
    ~
    \begin{subfigure}[t]{0.5\textwidth}
    \centering
        \includegraphics[width=\linewidth]{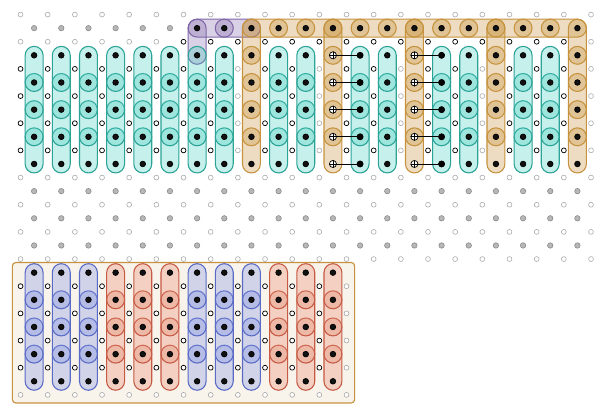}
        \caption{}
        \label{subfig:parity_NOT_layout}
    \end{subfigure}%
    
    \caption{a) is the ZX diagram of the parity-\textsc{Cnot} gate, whose circuit is shown in~\zcref{app: Clifford}, without Pauli corrections.
    b) is the repetition code layout for the skew-tree \textsc{Qrom}, drawn as in~\zcref{subfig:rep_code_layout_compiled}.
    Each operation is highlighted with the same color as in a): the auxiliary repetition code prepared in $\ket{0}_L$ is amber and the $M_{XX}$ lattice surgery is violet.
    The transversal \textsc{Cnot} represented in b) illustrates the case of a 2-qubit-parity-controlled-\textsc{Not} operation.}
    \label{fig: compiled_parityNOT}
\end{figure}

The second building-block to compile is the multi-target \textsc{Cnot} gate, or more accurately, the parity-controlled-\textsc{Not} gate (referred to as parity-\textsc{Cnot} gate in the sequel) in our color-changed implementation. Similarly to the lattice surgery implementation of the long-range \textsc{Cnot} introduced in~\cite{Gouzien_2023}, the circuit identity in~\zcref{fig:paritynot-gadget} of \zcref{app: Clifford} implements a long-range parity-\textsc{Cnot} gate using only local transversal gates and $M_{XX}$ lattice surgery operations.
In~\zcref{fig: compiled_parityNOT}, we compile this circuit onto the repetition code layout.
While no resource state are being used in this gate, an auxiliary logical qubit is prepared in $\ket{0}_L$ and stored in the tree-like repetition code shown in amber in~\zcref{subfig:parity_NOT_layout}.
Note that we arbitrarily distinguish the amber and violet regions, respectively for the $\ket{0}_L$ and the routing of the $M_{XX}$ operations, while in practice, data qubits of both regions are prepared in $\ket{0}$ and undergo $d$ rounds of syndrome extraction before being measured in the $Z$-basis.
The transversal CNOTs are performed right after data qubit initialization, as they commute with stabilizer measurements.
The duration of a parity-\textsc{Cnot} gate is therefore taken as $d$ rounds.
Finally, the \textsc{And}$^\dagger$ is performed using the measurement-based approach~\cite{gidney2018halving}.
The induced $\CXX$ phase correction can be pushed and merged with the \textsc{Phaseup} subroutine.

Note that whenever a parity-\textsc{Cnot} gate occurs right after an \textsc{And} gate, it can be performed simultaneously with the preceding \textsc{And} gate, as they use different routing areas.
This merge removes the $d$-cycle gap assumed in~\zcref{sec: Skew-tree QROM description}: the data of the following parity-\textsc{Cnot} $U_{i-1}$ depends on the injection outcomes of the merged \textsc{And}, which are only available at the end of the merged step.
To avoid waiting one reaction time, we prepare the auxiliary code of $U_{i-1}$ with $Z$ access to every output qubit, which does not depend on the data.
We then apply the transversal \textsc{Cnot}s selecting the relevant output qubits $t_r$ cycles after the end of the merged step, whenever the classical data is updated.
This is possible because these \textsc{Cnot}s commute with the stabilizer and $M_{XX}$ measurements of the auxiliary code; it only requires $d \geq t_r$.
Therefore, assuming sufficient magic state supply, the time and space costs of an $x$-bit address register (with $M\leq 2^x$ addresses) \textsc{Qrom} targeting a $N_{\mathrm{target}}$ bit output register in this repetition code layout are,
\begin{align}
    T_{\textsc{Qrom}}^{(\text{rep})}&=\underbrace{(2^x-1)}_{\#\text{ of parity-\textsc{Cnot}}}d +\underbrace{(2^{x-1}-x)}_{\# \text{ of \textsc{And} not followed by a parity-\textsc{Cnot}}}d, \\
    \Gamma^{(\text{rep})}_{\textsc{Qrom}}&= \Bigg(x + 3 \bigg\lceil \frac{N_\textrm{target}}{2} \bigg\rceil \Bigg)\left(d+3+1\right)+ N_{\text{buffer}}(3+3)d,
\end{align}
where we denote by $T_A^{(B)}$ the duration of subroutine $A$ in architecture $B$ and $\Gamma_A^{(B)}$ the space footprint of compiling $A$ in $B$ ignoring resource state factories footprints.
Note that the parity-\textsc{Cnot} directly preceded by an \textsc{And} gate are exactly the $U_i$ writing operations of~\zcref{fig:skew tree uncont} at odd $i$ different from 1.
More explicitly, in the uncontrolled skew-tree QROM compiled here, there are $2^x-1$ parity-\textsc{Cnot}s and $2^x-x-1$ \textsc{And}s among which $2^{x-1}-1$ are directly followed by a parity-controlled-\textsc{Not}.
In \zcref{subfig:rep_code_layout_compiled} of this section, we chose a setting with two buffer states --one buffer state being a $\ket{\CXCXX}$ and 3 $\ket{i}$-- in the buffer area, written $N_{\text{buffer}}=2$.
This is the minimal amount of buffer required to not slow down the computation in a cascade of \textsc{And} gates when $ t_{\CXCXX}\leq d$.
If magic states are produced at a lower rate than one magic state every $d$ rounds, namely one every $d\leq t_{\CXCXX}$, the computation is slowed down to at most
\begin{equation}\label{eq: QROM rep-code duration factory very limited}
\text{duration}=\underbrace{2^{x-1}}_{\#\text{ of Parity-\textsc{Cnot} not preceded by\textsc{And}}}d +\underbrace{(2^{x}-1-x)}_{\# \text{ of \textsc{And} }}\max (d,t_{\CXCXX}).
\end{equation}

\begin{figure}
    \centering
    \includegraphics[width=0.9\linewidth]{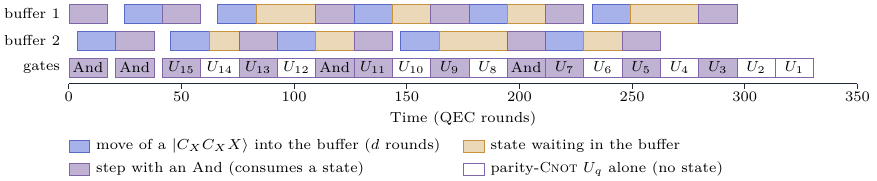}
    \caption{Non-naive schedule for the reverse DFS order of a skew-tree QROM whose circuit is shown in \zcref{fig:skew tree uncont}.
    Gates coloured in purple do consume magic state while white gates are parity-\textsc{Cnot} not preceded by an \textsc{And} gate.
    Here a minimal duration of $20.75-d$ is applied between the end of a magic state move in the buffer and the beginning of the next move-in as in the case of the third layout factory of \zcref{subsec: CxCxX_factories}.}
    \label{fig:qrom_schedule}
\end{figure}

Note that this is a pessimistic assumption, in practice parity-\textsc{Cnot} gates occur in between \textsc{And} gates, giving some time slack to the factory for producing the next magic state.
If additionally  $t_{\CXCXX}\leq 2d$, then we can have a tighter duration bound
\begin{equation}\label{eq: QROM rep-code duration factory limited}
\text{duration}=(2^{x}-1) d +\underbrace{(2^{x-1}-x)}_{\# \text{ of \textsc{And} not preceded by a parity-\textsc{Cnot}}}t_{\CXCXX}.
\end{equation}
We use this bound as we typically will work in the regime $d\leq t_{\CXCXX}\leq 2d$.
Finally, for an address register of $M$ addresses, this formula becomes 
\begin{equation}
     T_{\textsc{Qrom}}^{(\text{rep})} = (M-1)d + \bigg(\Big\lceil \frac{M}{2} \Big\rceil -\big\lceil \log M\big\rceil\bigg)t_{\CXCXX}
\end{equation}
An example time-schedule taking this feature into account for $x=4, d=17$ is shown in \zcref{fig:qrom_schedule}. 

\subsubsection{Compilation on the LDPC-cat architecture}\label{sec: QROM comp LDPC}

The main drawback of the compilation described above, when applied to the repetition-cat architecture, is that the repetition code encodes only one logical qubit using $d$ physical qubits resulting in a low encoding rate. In this section, we compile the same subroutine, the skew-tree \textsc{Qrom} in reverse DFS order of~\zcref{fig:skew tree uncont}, using the cat-LDPC architecture of~\cite{Ruiz_2025}.

\begin{figure}[!htb]
    \centering
    \begin{subfigure}[t]{0.5\textwidth}
        \centering
    \includegraphics[width=\linewidth]{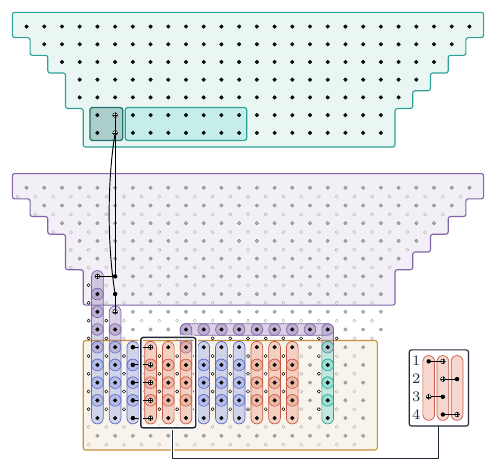}
    \caption{}
     \label{subfig: hybrid LDPC AND}
    \end{subfigure}%
    ~
    \begin{subfigure}[t]{0.5\textwidth}
    \centering
        \includegraphics[width=\linewidth]{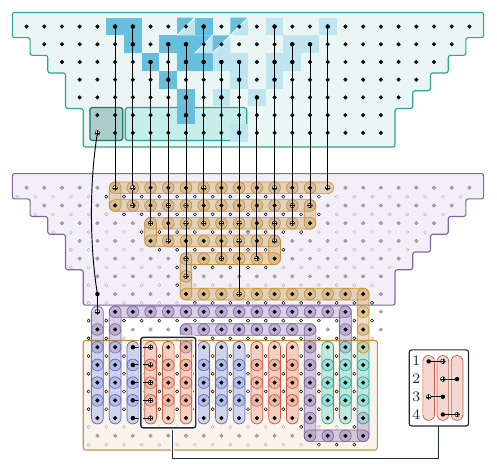}
        \caption{}
        \label{subfig: hybrid LDPC Parity + AND}
    \end{subfigure}%
    
    \caption{Compilation of \textsc{Qrom} primitives on a hybrid LDPC-repetition-cat architecture, with the LDPC layer (top) above the routing layer (bottom).
    The LDPC layer is the $d=16$ code of~\cite{Ruiz_2025} at width $L=17$. Only its data qubits are drawn, and the shaded band marks its first two rows, where the $X$ logical operators are accessed.
    Address qubits and output qubits are stored in the LDPC layer, in the registers marked by the brackets under its first two rows: address (left) and output (right).
    Buffer and ancilla qubits are stored in repetition codes at the bottom of the routing layer (lower box), below two extra routing rows. Blue denotes $\ket{\CXCXX}$ states, orange $\ket{i}$ states and teal ancilla qubits.
    Violet bands are lattice surgery operations.
    (a) shows the implementation of the \textsc{And} injection following the scheme in \zcref{subfig:AND_ZX}.
    Here, an $M_{XX}$ measurement is performed by initializing the routing qubit directly underneath the $X$ logical operator of the address qubit in $\ket{+}$ and performing the \textsc{Cnot} gates as depicted.
    The inset is the same as in~\zcref{fig: compiled_and}
    (b) shows the implementation of an \textsc{And} gate together with a parity-\textsc{Cnot} between the output qubits and an ancilla in the repetition code. To obtain the parity of the target qubits, the amber repetition code is initialized in $\ket{0}$ underneath the support of the XOR of the $Z$ logical operators of the involved target qubits (blue cells).
    The \textsc{Cnot}s are performed between qubits of the XOR of the $Z$ logicals and their corresponding data qubits in the computational layer.}

    \label{fig: hybrid LDPC AND, hybrid LDPC Parity + AND}
\end{figure}

In this setting, algorithmic qubits are stored in a classical LDPC code with a second routing layer underneath forming a flip-chip as described in~\cite{Ruiz_2025}.  Buffer qubits are allocated at the bottom of the routing layer in a dedicated zone between factory output and routing layer.
Ancilla qubits can either be allocated to this buffer zone, or instead be stored in the LDPC layer.
Since other subroutines often require a larger buffer zone, allocating the ancillae to the buffer area often does not increase the overall spatial footprint and it is thus our focus. Compiling the \textsc{And} gate is a clear extension of \zcref{subfig:rep_code_layout_compiled}, since it only requires accessing the $X$ operators of the logical qubits stored in the LDPC code and single qubit $X$ operators can be accessed on the first two data qubit rows of the chip.
More precisely, the $M_{XX}$ measurement between a $\ket{\CXCXX}$ buffer state and the address qubit in LDPC is implemented by initializing the qubit in the routing layer directly underneath the address qubit in $\ket{+}$, routing the buffer state towards this ancilla and computing the $X$-parity of the address qubit with the relevant buffer state data qubit using \textsc{Cnot} gates through the ancilla.  Note that our operation requires simultaneous access to at most two single qubit $X$ logical operators in the LDPC code.
Therefore, the situation where a data qubit of the routing layer underneath an $X$ logical operator is surrounded by other active data qubits does not happen, and we can ensure the two $M_{XX}$ measurements can be performed simultaneously. To avoid possible collisions from the buffer to the address qubits, we add two additional routing rows between the buffer and the bottom of the routing layer underneath the LDPC layer.
The qubit displacement between the buffer and the logical ancilla as well as the Hadamard injection are fully implemented in repetition code. The process is shown in \zcref{subfig: hybrid LDPC AND}. 

\begin{figure}[!htb]
    \centering
    \includegraphics[width=0.6\linewidth]{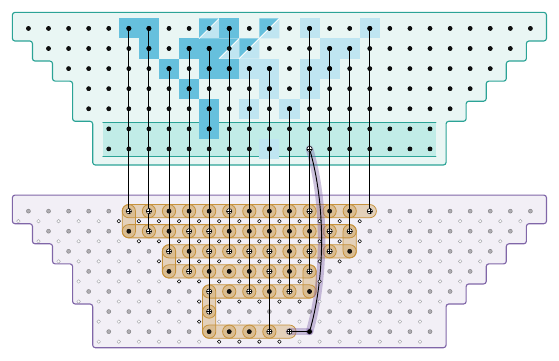}
    \caption{Parity-\textsc{Cnot} gate compiled on an LDPC-cat architecture, with the LDPC layer (top) above the computational layer (bottom).
    Colors and LDPC code correspond to the ones used in~\zcref{fig: compiled_parityNOT}.
    Here we represent a \textsc{Not} operation controlled by the parity of a pair of qubits, its target being also stored in the LDPC layer.
    The amber snake repetition code is prepared in $\ket{0}_L$ while the two shades of blue are the $Z$-logical operators of the two control qubits.
    The violet operation represents the $M_{XX}$ operation of~\zcref{subfig:parity_NOT_ZX}: a physical routing qubit prepared in $\ket{+}$, directly underneath the $X$ logical operator of the target, controls a \textsc{Cnot} on this logical and one on the end of the snake.
    The \textsc{Cnot}s are performed between qubits of the XOR of the support of the two logicals (black dots on plain blue cells) and their corresponding data qubits in the computational layer.
   }
\label{fig:LDPC_parity_controlled_not}
\end{figure}

To compile the parity-controlled \textsc{Not}, there are two cases which are compiled similarly.
Either, the target is an address qubit in the LDPC layer or the target is an ancilla qubit in the repetition code layer.
In both cases, the qubits determining the parity are stored in the LDPC layer.
Thus, to obtain the parity, a snake-like repetition code spanning the XOR of the support of the $Z$-logical operators of the qubits in the controlled parity is initialized in $\ket{0}_L$ in the routing layer.
The sequence of \textsc{Cnot} gates between the two chips is also determined by the XOR of the support: Any qubit in the XOR of the supports is the control of a \textsc{Cnot} with its corresponding data qubit in the routing layer.
Now, in the case that the target qubit is an ancilla qubit in repetition code, the necessary $M_{XX}$ is implemented as described in the \textsc{And} compilation.
The corresponding compilation is illustrated in \zcref{subfig: hybrid LDPC Parity + AND}, where the parity-\textsc{Cnot} is implemented at the same time as the previous \textsc{And} gate.
The case where the target is also in LDPC is illustrated in \zcref{fig:LDPC_parity_controlled_not}.
The only difference is that the $M_{XX}$ is now performed between the amber snake repetition code and the qubit in the LDPC layer.
Note that depending on the position of the logical $X$ of the target qubit, the snake may be made tree-shaped to make the snake discontinuous on the first row.
This is useful if the target logical $X$ stands in between $X$ logicals of qubits in the parity. 

Exactly as for the full repetition code setting, an \textsc{And} gate and a parity-\textsc{Cnot} can be compiled simultaneously (see \zcref{subfig: hybrid LDPC Parity + AND}).
As in~\zcref{sec: QROM comp rep cat-code}, we absorb the reaction time between the merged \textsc{And} and the next parity-\textsc{Cnot} by preparing the snake repetition code under the union of the $Z$-logical supports of all output qubits, and applying the inter-layer \textsc{Cnot}s on the support of the selected parity $t_r$ cycles after the end of the merged step.
Thus, the duration of all necessary primitives in the hybrid LDPC-repetition code matches the repetition code setting.
We allow for different distances in both layers: specifically let $d_\LDPC$ denote the distance of the LDPC layer and $d_\rep$ denote the distance for the repetition layer relevant for buffer storage.
Assuming that magic state factories produce a $\ket{C_XC_X X}$ state every $d = \max(d_\LDPC, d_\rep)$ cycles, we thus obtain the following spatial and temporal footprint for loading over $M$ addresses
\begin{align}
        \text{duration} &= (M-1)d + \bigg(\Big\lceil \frac{M}{2} \Big\rceil -x\bigg)d, \label{eq: QROM LDPC time non-slow}\\
        \Gamma^{(\text{LDPC hybrid})}_{\textsc{Qrom}} &= 2(HL + \Delta_H) + 2\max(L, 6N_\mathrm{buffer} + x) + \max(6N_\mathrm{buffer} + x - L, 0) + (6N_\mathrm{buffer} + x)d_\rep.\label{eq: QROM LDPC space buffer anc}
\end{align}

Here, $x = \lceil\log M \rceil$ denotes the number of address qubits, $H$ is the height of the LDPC layer, $L =\lceil x/2\rceil + \lceil N_\mathrm{target}/2 \rceil$ is the width of the LDPC layer ($N_\mathrm{target}$ being the number of output qubits), $\Delta_H$ denotes the number of extra qubits necessary for the triangular overhang in the LDPC layer and $N_\mathrm{buffer}$ is the number of buffers. Similar to the repetition code version, $N_\mathrm{buffer} = 2$ suffices. Note that the space formula takes into account the 2 routing rows above the buffer as well as the additional routing corridor next to the $x-1$ ancilla qubits. If $L$ is smaller than the buffer and ancilla zone, we need to provide an extra routing corridor for the parity-\textsc{Cnot}, which is reflected in the summand $\max(6N_\mathrm{buffer} + x - L, 0)$. For $d_\LDPC=16$, we have $H=7$ and $\Delta_H = 26$, whereas for $d_\LDPC = 22$, $H=8$ and $\Delta_H = 29$. If magic states are instead produced every $t_{\CXCXX}$ cycles with $d < t_{\CXCXX} \leq 2d$, the timing slows down to
\begin{equation}
     T_{\textsc{Qrom}}^{(\text{LDPC hybrid})} = (M-1)d + \bigg(\Big\lceil \frac{M}{2} \Big\rceil -x\bigg)t_{\CXCXX}
\end{equation}
similar to \eqref{eq: QROM rep-code duration factory limited}.
This is due to the fact that only \textsc{And} gates that are followed by another \textsc{And} gate need to be slowed down and any \textsc{And} gate followed by a parity-\textsc{Cnot} (the odd data loads) is followed by an \textsc{And}$^\dagger$ and another parity-\textsc{Cnot} giving enough time for the buffer to refill.

Instead of storing the ancilla in a repetition code next to the buffer area, they could be stored in the LDPC layer as well. This can be beneficial when considering a large number of addresses or if buffer space is limited.
This only leads to minor changes in the operations.
First, we assign the bottom row of the LPDC layer to address qubits and the row on top to ancillae.
The target qubits are still stored to the right of the address qubits. We now have to access at most three $X$ parities in the LDPC layer at the same time, which is still possible without congestion.
Second, the qubit displacement between the buffer $\ket{i}$ state and the ancilla has to be considered with care. In order for the displacement to work using an $M_{XX}$ measurement, ancillae need to be stored in $\ket{0}_L$.
This can be achieved by preparing all of them at the beginning of the computation.
Since the physical qubits making up the ancilla are also utilized in other logical qubits, this needs to be done using $M_Z$ measurements.
This requires accessing the $Z$-logical operator of the ancilla and thus we perform it serially taking $d$ cycles per preparation.
Once all ancillae have been set, they will be freed using an $M_Z$ measurement coming from the \textsc{And}$^\dagger$ in the color-changed convention and can be assumed to be in $\ket{0}$ before their next use.
Third, as already mentioned uncomputing an ancilla in the LDPC layer using MBU now relies on quantum non-demolition measurements and requires $d$ cycles. In our color-changed setting, since it requires access to the $Z$-logical operator, this can not necessarily happen at the same time as other operations taking up space in the routing layer.
Thus, these instances need to be actively accounted for.
However, any time an ancilla is initialized and then immediately uncomputed once the parity-\textsc{Cnot} is performed, we can simply forgo moving the ancilla into the LDPC layer and instead just utilize the corresponding buffer state in the repetition code. This occurs exactly for leaf data loads and does not add additional uncomputation time.
Hence, the extra time from uncomputation amounts to 
\begin{equation*}
   \underbrace{ (M-1-x)d}_{\text{\# total \textsc{And}}} \qquad - \hspace{-5pt}\underbrace{(\lceil M/2 \rceil -1)d}_{\text{\# immediately uncomputed \textsc{And}}} = \qquad (\lfloor M/2 \rfloor - x)d. 
\end{equation*}
Lastly, observe that an \textsc{And} together with a parity-\textsc{Cnot} only occurs at the leaf loads which are exactly the ones that do not require displacing the ancilla qubit into the LDPC layer.
Thus, these operations can be performed simultaneously with a similar reasoning to the ancilla in the buffer setting (see \zcref{subfig: hybrid LDPC Parity + AND}). In sum, we obtain the following formulas
\begin{align*}
   T_{\textsc{Qrom}}^{(\text{LDPC})}&= (M-1)d + \bigg(\Big\lceil \frac{M}{2} \Big\rceil -x\bigg)\max(d, t_{\CXCXX}) + \bigg(\Big\lfloor \frac{M}{2} \Big\rfloor -x\bigg)d + (x-2)d,\\
    \Gamma^{(\text{LDPC})}_{\textsc{Qrom}} &= 2(HL^\prime + \Delta_H) + 2\max(L^\prime, 6N_\mathrm{buffer}) + \max(6N_\mathrm{buffer} - L^\prime, 0) + 6N_\mathrm{buffer}d_\rep.
\end{align*}
Here, $(x-2)d$ is the time necessary to prepare the ancilla in $\ket{0}$ before the computation, $L^\prime = x + \lceil N_\mathrm{target}/2 \rceil$, and the rest of the notation follows \eqref{eq: QROM LDPC time non-slow} and \eqref{eq: QROM LDPC time non-slow}. Analogously to \eqref{eq: QROM rep-code duration factory limited}, we assume $t_{\CXCXX}\leq 2d$.

\subsection{\textsc{Swap} network}\label{sec: Swap comp}

In order to implement the \textsc{Select-Swap} circuit displayed in~\zcref{fig: dirty SelectSwap}, we compile the \textsc{Swap} network acting on multiple-qubit registers.
This operation is implemented as a network of \textsc{Ctrl-Swap} gates. The network size depends on the register sizes.
The network shown in \zcref{fig:swap-network} relies on the implementation of \textsc{Ctrl-Swap} logical gates which can be implemented as two \textsc{Cnot}s and one Toffoli gate.

\begin{figure*}[!htb]
  \centering
  \begin{tabular}{c@{\hskip 1.4cm}c}
    \begin{quantikz}[row sep={0.42cm,between origins}, column sep=0.42cm]
      \lstick{$\mathrm{address}[0]$} &                      & \ctrl{3} & \ctrl{5} & \ctrl{7} & \ctrl{9} &          &          &          \\
      \lstick{$\mathrm{address}[1]$} &                      &          &          &          &          & \ctrl{2} & \ctrl{6} &          \\
      \lstick{$\mathrm{address}[2]$} &                      &          &          &          &          &          &          & \ctrl{1} \\
      \lstick{$\ket{\phi_0}$}        & \qwbundle{b } & \swap{1} &          &          &          & \swap{2} &          & \swap{4} \\
      \lstick{$\ket{\phi_1}$}        & \qwbundle{b } & \targX{} &          &          &          &          &          &          \\
      \lstick{$\ket{\phi_2}$}        & \qwbundle{b } &          & \swap{1} &          &          & \targX{} &          &          \\
      \lstick{$\ket{\phi_3}$}        & \qwbundle{b } &          & \targX{} &          &          &          &          &          \\
      \lstick{$\ket{\phi_4}$}        & \qwbundle{b } &          &          & \swap{1} &          &          & \swap{2} & \targX{} \\
      \lstick{$\ket{\phi_5}$}        & \qwbundle{b } &          &          & \targX{} &          &          &          &          \\
      \lstick{$\ket{\phi_6}$}        & \qwbundle{b } &          &          &          & \swap{1} &          & \targX{} &          \\
      \lstick{$\ket{\phi_7}$}        & \qwbundle{b } &          &          &          & \targX{} &          &          &
    \end{quantikz}
    &
    \begin{tabular}{c@{\hskip 0.35cm}c@{\hskip 0.35cm}c}
      \begin{quantikz}[row sep={0.55cm,between origins}, column sep=0.4cm]
                                &              & \ctrl{1} & \\
        \lstick{$\ket{\phi_i}$} & \qwbundle{3} & \swap{1} & \\
        \lstick{$\ket{\phi_j}$} & \qwbundle{3} & \targX{} &
      \end{quantikz}
      & $=$ &
      \begin{quantikz}[row sep={0.42cm,between origins}, column sep=0.4cm]
                                   & \ctrl{1} & \ctrl{2} & \ctrl{3} & \\
        \lstick[3]{$\ket{\phi_i}$} & \swap{3} &          &          & \\
                                   &          & \swap{3} &          & \\
                                   &          &          & \swap{3} & \\
        \lstick[3]{$\ket{\phi_j}$} & \targX{} &          &          & \\
                                   &          & \targX{} &          & \\
                                   &          &          & \targX{} &
      \end{quantikz}
    \end{tabular}
    \\[0.5em]
    (a) & (b)
  \end{tabular}
  \caption{Figure from~\cite{low2024trading} showing a \textsc{Ctrl-Swap} network moving one of the $N=8$ data
    registers of $b $ qubits to the output, namely the top data register.
    (a) After the $\log_2 N=3$ layers the register initially storing
    $\ket{\phi_0}$ now holds $\ket{\phi_a}$ with
    $a$ the binary integer stored in the address register.
    (b) Each register-level \textsc{Ctrl-Swap} decomposes into $b $
    ordinary \textsc{Ctrl-Swap} on pairs of qubits sharing the same control
    (drawn for $b =3$), so the network costs $(N-1)b $
    \textsc{Ctrl-Swap} in total.}
  \label{fig:swap-network}
\end{figure*}

Note that as the Toffoli gate is performed through injection, one can rely on active injection \cite{Shor2026prep} to execute those two \textsc{Cnot} gates within the injection using a multi-target \textsc{Cnot} and a Parity-controlled \textsc{Not}, see~\zcref{eq:cswap-inject}.

\input{graphics/compilation_graphics/tikz/eq_CSWAP}

In the first two equalities, we consider the case where every measurement returns 0.
When this is not the case, there are additional Clifford corrections to apply.
Additionally, note that the last equality is true up to Pauli corrections only: when at least two of the measurement outcomes $m_i$ are $1$, there are extra Paulis that appear when the $m_i$ spider is commuted with the $m_{i'}$ correction.
While the skew-tree unary iteration from \cite{low2026denser} eliminates the complication of Clifford corrections in the \textsc{Qrom} compilation, here we have to deal with it.
Generally speaking, there are two ways to deal with such corrections.
The first possibility is to use an auto-corrected state~\cite{gidney2019flexiblelayoutsurfacecode}.
This way the Clifford correction is determined by choosing the measurement basis of the additional qubits in the auto-corrected magic state.
The second possibility is to apply the Clifford correction on the relevant algorithmic qubits before these qubits participates in a subsequent gate of the circuit.
This second possibility can slow down the computation significantly if the corrections take long to perform and if one of the qubits affected by the Clifford correction is involved in another (non-Clifford) gate shortly afterwards.

Note that for our cat architectures, different Clifford corrections have different costs: those requiring $\ket{i}$ states are particularly demanding as $\ket{i}$ states are distilled on our cat architectures.
Saving or spreading over time the consumption of $\ket{i}$ state is therefore particularly useful in this context.

The way these corrections are handled depends on the degree of parallelism being targeted. 
Concretely, the circuit of~\zcref{fig:swap-network} is made of $\log_2N$ layers, where $N$ is the number of addresses.
We say gates belong to the same layer if they have the same control qubit and commute with each other.
Therefore, the processing of each layer would take one logical cycle if enough magic states were supplied.
Beyond this obvious parallelization, time optimal quantum computation techniques~\cite{fowler2013timeoptimalquantumcomputation} allow to perform several layers of computation at the same time or, more precisely, to inject the Toffoli of several layers simultaneously. 
Indeed, performing a \textsc{Ctrl-Swap} gate is achieved by lattice surgery with a magic state; nothing prevents us from doing so simultaneously.
The time ordering of the circuit is then recovered by the magic-state measurement dependencies~\cite{fowler2013timeoptimalquantumcomputation}, and the result of the computation will not be available before these dependencies are resolved.
This is exactly the regime of reaction-limited computation.
In this work, since we are primarily interested in space minimization, we use a factory that produces one non-Clifford magic state every $\approx d$ cycles. Therefore, we do not rely on these two parallelization methods.

That said, examination of the \textsc{Ctrl-Swap} network circuit (\zcref{fig:swap-network}) reveals that, when the \textsc{Ctrl-Swap} gates are performed serially, qubits remain idle for at least one logical cycle following their involvement in a gate as no qubit outside of the address register is ever involved in two successive gates.
Note that this is true as long as $b  >1$ otherwise the two last gates of~\zcref{fig:swap-network} share a swapped qubit.
This indicates that some time is available for performing Clifford corrections.

As a consequence, we postpone the two \textsc{Cnot} corrections controlled by $m_1$ and $m_2$ outcomes (see~\zcref{eq:cswap-inject}).
They yield a \textsc{Swap} and a parity-\textsc{Cnot} correction, respectively.
Note that the \textsc{Swap} can be done in software by relabeling the indices of logical qubits.
More importantly, the $\CXX$ correction requires $3$ $\ket{i}$ states (although one of them could be used in a  catalytic way).
Looking closely at one injection scheme of this gate shows that the $\CXX$ correction is not atomic and can be decomposed through a 2-qubit gate injection and two $\sqrt{X}$ byproduct corrections, as explicit in~\zcref{eq:cxx-correction}

\begin{figure}[!htb]
    \centering
    \includegraphics[width=0.45\linewidth]{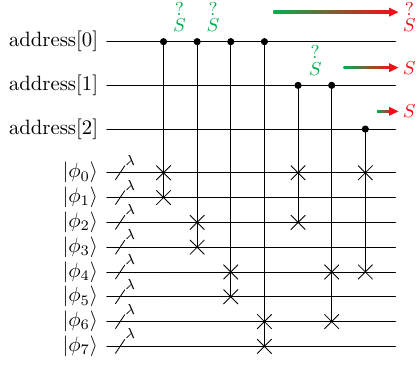}
    \caption{Circuit of a \textsc{Swap} network where the green $S$ correction arising from state injection can be deferred and merged into the fewest red $S$ corrections}
    \label{fig:postpone S}
\end{figure}

\input{graphics/compilation_graphics/tikz/eq_C_XX}
Note that we represented the first magic state as a $-\pi/2$ green dot, which, in the case of a single leg node, is the same as a $\pi/2$ red dot, namely a $\ket{i}$ state.
These 3 $\ket{i}$ state injections can either all be performed on the magic state to make it auto-corrected, or some of them may be injected later on the algorithmic qubits.
Remarkably, for one of those two single qubit corrections, it is particularly interesting to do it later on the control qubit of the \textsc{Ctrl-Swap} gate rather than on the magic state.
This is an interesting feature, as this correction commutes with all upcoming gates, contrary to every other Clifford correction here, which would create Clifford corrections with the magic states injected in the upcoming \textsc{Ctrl-Swap} layers.
This is illustrated in regular color convention on~\zcref{fig:postpone S}.
This way, instead of having one $\sqrt{X}$ (or $S$ in regular conventions) correction per injected \textsc{Ctrl-Swap} gate, we have one such correction per qubit in the address register.
As there are $\log_2N $ layers each composed of $ N/2^{k}$ register-\textsc{Ctrl-Swap}, themselves made of $b $ gates, with $k$ the index of the layer, this trick saves $b (N-1)- \log_2N$ $\ket{i}$ states as illustrated by \zcref{fig:postpone S}.
In fact, in the dirty-ancilla \textsc{Select-Swap} subroutine shown in~\zcref{fig: dirty SelectSwap}, there are 4 \textsc{Ctrl-Swap} networks sharing the same control register, which reduces the amount of correction by a factor of 4 as they can be postponed to the end of the \textsc{Select-Swap} circuit. This has even more impact in lookup table batching, where we perform $J+1$ \textsc{SelectSwap}s in a row.

For the two other injected $\ket{i}$ states, it is important that they are injected before the next \textsc{Ctrl-Swap} gate with overlapping support on the swapped qubits; however, doing this in an auto-corrected manner or as a classically controlled gate on the algorithmic qubits makes little difference.
Indeed, in any case, the buffer area would have to store simultaneously a $\ket{\CXCXX}$ state, 3 $\ket{i}$ states to do the Hadamard gate on this state and 2 $\ket{i}$ states either to make it partially auto-corrected or to perform the Clifford corrections of the previous injected gate if we don't use auto-correction.
We choose to use partial auto-correction in order to leverage the use of local transversal \textsc{Cnot} in the state preparation rather than long range multi-target \textsc{Cnot} as the correction on algorithmic qubits would require. 
This choice has the small advantage of decreasing the tension over the routing area by reducing the amount of lattice surgery at each cycle.

\begin{figure}[!hbt]
\centering
\includegraphics[width=0.7\linewidth]{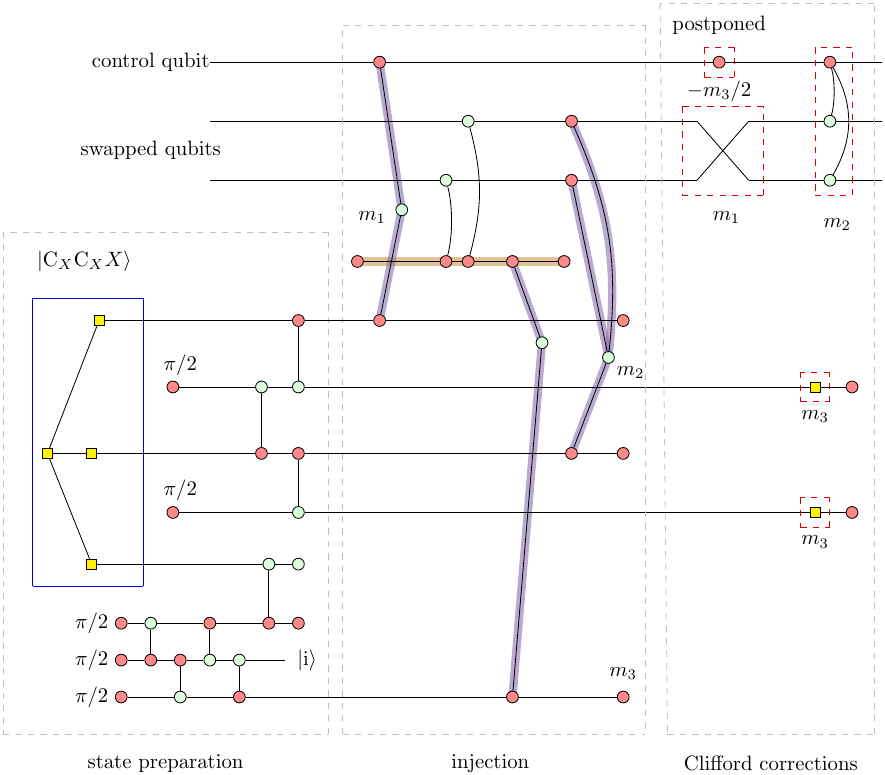}
    \caption{Partially auto-corrected injection of a \textsc{Ctrl-Swap} gate.
    Red-dashed outlined gates are conditioned on a measurement outcome. 
    Note that the red-dashed outlined Hadamard gates are not applied, they induce a change of basis of the final measurement.
    The colored edges refers to \zcref{fig:compiled_layout_CSWAP_No_correction} colored codes.
    Paulis corrections are omitted in this ZX-diagram, see \zcref{fig:zx_ctrl_Swap_w_corr} for the version containing all corrections.
    }
    \label{fig:compiled-swap-network-zx}
\end{figure}

\subsubsection{Compilation on the repetition-cat architecture}\label{sec: Swap rep-code comp}
\begin{figure}[!htb]
    \centering
    \includegraphics[width=0.6\linewidth]{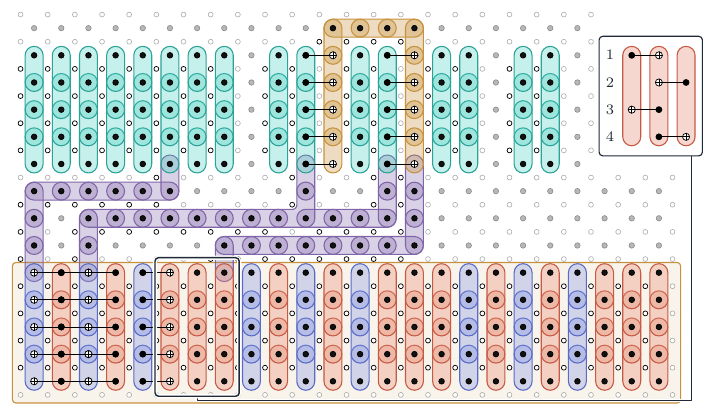}
    \caption{%
    Partially auto-corrected injection of a \textsc{Ctrl-Swap} gate compiled onto the physical
Layout, postponed parity-\textsc{not} Clifford correction is not represented. 
Lattice and color conventions are the same as in \zcref{subfig:rep_code_layout_compiled}.
Control qubits of \textsc{Ctrl-Swap} gates are on the top left packed blue rails while swapped qubits are on top right with routing corridors interleaved.
Inset is the Hadamard state preparation in \zcref{fig: compiled_and}.
Violet corridors implement surgery measurements of injection steps while
the amber repetition code is the logical auxiliary qubit used in the parity-controlled-\textsc{Not}, as highlighted in \zcref{fig:compiled-swap-network-zx}.}
\label{fig:compiled_layout_CSWAP_No_correction}
\end{figure}

An important feature of this compiled subroutine shown in~\zcref{fig:compiled-swap-network-zx} is that no access to the $Z$ logical operator of the control qubit is required.
This is a redundant feature with the fact that we've been able to remove the routing around the address register qubits in the \textsc{Qrom} compilation~\zcref{subfig:rep_code_layout_compiled}.
Indeed, in the circuit of~\zcref{fig: dirty SelectSwap} the complete address register made of $\ket{x}$ and $\ket{y}$ can be stored without routing alleys, while the dirty ancilla and the target register do require vertical routing alleys, providing access to their $Z$ logical operators.
Note, however, that the magic state used in the \textsc{Ctrl-Swap} network is slightly larger, which requires modifying the buffer area of \zcref{subfig:rep_code_layout_compiled}.
Moreover due to the classically controlled basis choice, $\ket{i}$ states are not instantaneously measured out after the lattice surgery injection step, they remain in the slot for $t_r$ before being measured.
As a consequence, following the operation schedule of \zcref{fig:ctrl_swap_schedule}, a third buffer state is necessary in the buffer area.
We compile the state preparation step, as well as the injection step on this repetition code layout in \zcref{fig:compiled_layout_CSWAP_No_correction}.
We further detail how we deal with the parity-\textsc{Not} and \textsc{Swap} corrections in \zcref{subsubsec:clifford corrections ctrl-swap}.

\subsubsection{Compiling the Clifford corrections}\label{subsubsec:clifford corrections ctrl-swap}

The swap correction is easily implemented by relabeling the qubits in the compiler.
This will not slow down the computation, as the reaction time is shorter than the injection time, and there are no two successive \textsc{Ctrl-Swap} gates that share one or two of their swapped qubits.

We propose three methods to deal with the $m_2$ controlled Clifford correction.
The first option is to make the magic state slightly more auto-corrected by adding one Bell pair in the state preparation, as is done for the \textsc{Adder} operation in~\cite{Shor2026prep}.
The downside of this option is that each buffer state is two logical qubits bigger.
As there are three magic states in the buffer area, this option leads to six additional logical qubits in total.

\begin{figure}[!htb]
    \centering
    \begin{subfigure}[t]{65pt}
        \centering
    \includegraphics{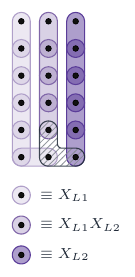}
        \caption{}
        \label{fig:Fork code}
    \end{subfigure}%
    \hspace{14pt}%
    \begin{subfigure}[t]{0.7\linewidth}
        \centering
        \includegraphics[width=0.9\linewidth]{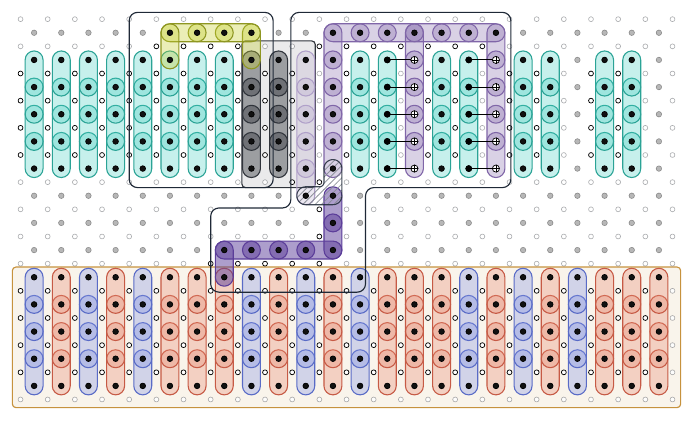}
        \caption{}
        \label{fig:delayed parity-not compiled}
    \end{subfigure}
    \caption{The fork code and the compilation of the third option to deal with the parity-\textsc{Cnot} correction.
    (a) The fork code, drawn with $n=6$ qubits per branch. The violet bands cover the supports of the weight-2 $X$-type stabilizers, one shade per branch, and the hatched L at the junction is the single weight-3 stabilizer. The key gives, for each branch, the logical operator implemented by an $X$ on any single qubit of that branch.
    (b) Color conventions are the same as in \zcref{fig:compiled_layout_CSWAP_No_correction}.
    The grey area holds the three parity-storage slots. Their repetition codes, in dark grey, store the parities of swapped qubits while they wait one reaction time to decide whether the parity-\textsc{Cnot} correction is applied.
    Right frame: the auxiliary fork code, with one shade of violet per branch as in \zcref{fig:Fork code}. Two transversal \textsc{Cnot} gates have their respective control on the two swapped logical qubits, and their target on the same branch of the fork code (medium violet). This branch is interpreted as holding the product of the two $X$ logical operators of the fork code.
    On a second branch (light violet), one of the two logical qubits is stored in the right-most slot of the gray area. On the third branch (dark violet), the other logical qubit is involved in an $M_{XX}$ measurement, which measures its joint $X$-parity with the output state of the Hadamard gadget in the buffer, as part of the injection step.
    Left frame: the parity-\textsc{Cnot} correction of the \textsc{Ctrl-Swap} gate before the previous one. The parity of the two qubits it swapped is stored in the left-most slot. Assuming its measurement outcome is $m_2=-1$, this parity is now injected through an $M_{XX}$ measurement with the control qubit, along the olive routing code.
    The middle slot holds the parity of the previous \textsc{Ctrl-Swap} gate, which waits one reaction time for its own $m_2$ outcome.
    For readability, we do not represent the state preparation and the $m_1$ and $m_2$ lattice surgery operations of~\zcref{fig:compiled-swap-network-zx}.}
\end{figure}

In the second and third options, the correction is applied before a subsequent~\textsc{Ctrl-Swap} gate acts on the same swapped qubit.
Sharing the same control qubit is not a problem, as the $m_2$ controlled correction commutes with such a subsequent \textsc{Ctrl-Swap} gate.
The challenge with these options is that without any information on qubit positions in the top right algorithmic qubit area, it is possible that the parity-\textsc{Cnot} correction of the previous gate requires the use of the same routing corridor as the one used for the current injection, namely the amber repetition code in~\zcref{fig:compiled_layout_CSWAP_No_correction}.
Such congestion is rather unlikely and could be dealt with on a case by case basis by a compiler aware of qubit positions.
However, for completeness, we offer two solutions to this problem.
The second option is to slow down the computation whenever this congestion problem appears and apply the $i$-th correction and the $(i+1)$-th injection steps sequentially.
In the worst case, the computation is slowed down by a factor of 2.
In practice, there is only a 50\% chance that the Clifford correction is required, and for large registers, it is very unlikely that such a congestion occurs.
Therefore, the time overhead of this second option is likely to be negligible in practice.
The third option, more involved, is to use a fork code that encodes two logical qubits in three branches to perform two logical \textsc{Cnot} gates through a single set of physical operations.
Indeed, if a fork code is used to store two auxiliary logical qubits $1,2$, then each physical qubit of one of the three branches supports the logical $X_{L,1}X_{L,2}$ operator, see \zcref{fig:Fork code}.
This feature is due to the following choice of logical basis: $Z_{L1}=\prod_j Z_{a,j}Z_{b,j}$ and $Z_{L2}=\prod_j Z_{c,j}\prod_{j\geq2} Z_{b,j}$, with the columns labeled $a,b,c$ from left to right in \zcref{fig:Fork code} and the rows $j=1,\dots,n$ starting from the junction.
As the code distance is given by the length of the smallest $Z$ logical operator, it is determined by the shortest path from one arm end to another.
As a consequence, if this branch is the target of a transversal \textsc{Cnot} between a repetition code and the fork code, then effectively, two logical \textsc{Cnot} gates are applied.
The injection step is subsequently performed by an $M_{XX}$ measurement between one of the two other branches of the fork code.
Then, two branches of the fork code are measured transversally in the $Z$ basis, effectively performing a $Z$ measurement of one of the logical qubits.
At this point, the parity-\textsc{Cnot} gate of the injection has been performed and the second logical qubit of the fork code is now held by a repetition code stored in a dedicated region.
This repetition code waits one reaction time for the $m_2$ measurement outcome. If $m_2=-1$, an $M_{XX}$ with the control qubit followed by a destructive $Z$ measurement of the repetition code is performed.
This applies the delayed $m_2$ Clifford correction.
If $m_2=+1$, the repetition code is measured in the $X$ basis.
This does not apply the $m_2$ Clifford correction.
In sum, the third option requires some space to store the resulting repetition code in the time between the $i^{th}$ injection and the end of the $i^{th}$ Clifford correction, namely for $d+t_r$ rounds.
A compiler is likely to find an available routing corridor if the swapped register is large enough. Otherwise, an area with three repetition code slots can be added between the control register and the swap register. This is the area represented in gray in~\zcref{fig:delayed parity-not compiled}.
The ZX-compilation exhibiting this way of managing the Clifford corrections is displayed in.
\begin{figure}
    \centering
    \includegraphics[width=0.8\linewidth]{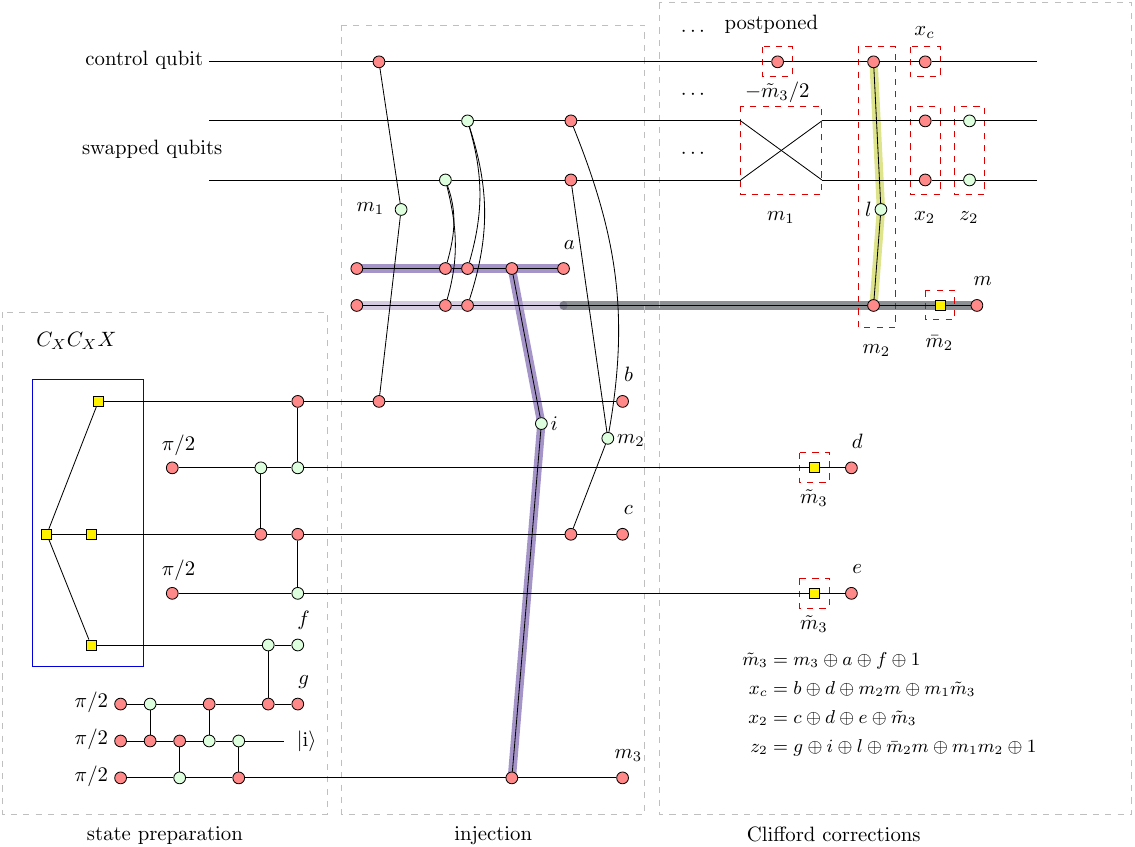}
    \caption{ZX-diagram of the compiled \textsc{Ctrl-swap} injection with the parity-\textsc{Cnot} correction using the transversal \textsc{cnot} duplication technique to store two parities in a Fork code.
    Every corrections including Paulis are represented on this diagram.
    The coloring of edges refers to \zcref{fig:delayed parity-not compiled} colored elements.
    In particular, the swapped qubits parity is stored in a Fork code represented by the lines of its two qubits turns into a dark grey repetition code storage when the first parity is measured.}
    \label{fig:zx_ctrl_Swap_w_corr}
\end{figure}
Note that the latter option is effectively free when compiling a \textsc{SelectSwap} routine. If the \textsc{Select} is implemented using the skew tree \textsc{Qrom} compilation outlined in \zcref{sec: QROM comp rep cat-code}, the clean ancilla necessary for the \textsc{Qrom} are situated between the address and target registers and are unused once the \textsc{Select} has concluded. Thus, these ancilla can be used as the aforementioned storage zone.

In \zcref{fig:delayed parity-not compiled}, the fork code auxiliary logical qubit method is compiled on the physical layout.
Note that the $m_1$ and $m_2$ operations are not represented in this figure for readability.
They would induce a crossing of the routing repetition codes, which is dealt with using three weight-three stabilizers at the intersection to do lattice surgery crossing as introduced in \cite{Simultaneous2026prep}.

\begin{figure}
    \centering
    \includegraphics[width=\linewidth]{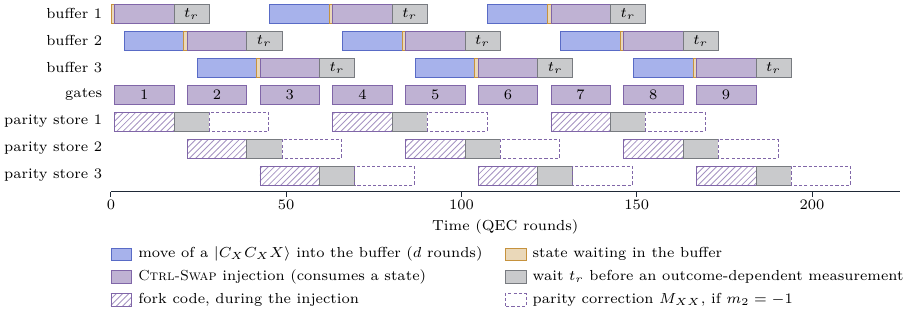}
    \caption{Schedule of a \textsc{Ctrl-Swap} network fed by the factory, for $N=4$ and $b =3$ (nine \textsc{Ctrl-Swap} gates), $d=17$ and $t_r=10$, starting from an empty buffer.
    A $\ket{C_XC_XX}$ state arrives every $t_{\CXCXX}=20.75$ rounds. Each injection keeps its buffer slot $t_r$ more rounds for the outcome-dependent measurements of its $\ket{i}$ codes, and each fork code holds a parity-storage slot for $2d+t_r$ rounds.
    With three buffer states and three storage slots, the gates follow the factory, one every $t_{\CXCXX}$ rounds.}

    \label{fig:ctrl_swap_schedule}
\end{figure}

To determine the spatial footprint of the routine, the same considerations as for the skew tree \textsc{Qrom} apply: address qubits can be stored in a packed manner whereas to be swapped qubits require $Z$ access corridors. Taking the larger buffer into account and considering the option where we store the additional parity-\textsc{Cnot} ancilla between the control and swap register, we obtain
\begin{equation}\label{eq: SWAP rep code space}
    \Gamma_{\textsc{Swap}\text{ network}}^{(\text{rep})} = (\lceil \log N \rceil + \lceil 3Nb/2 \rceil + 3)(d+4) + 8N_\mathrm{buffer}d,
\end{equation}
where $N$ denotes the number of addresses, $b$ the number of qubits in to be swapped patches, and $N_\mathrm{buffer}$ the number of buffer states. 
Contrary to the \textsc{Qrom} where no auto-correction were used, we set $N_\mathrm{buffer} =3$ to better pipeline operations, see \zcref{fig:ctrl_swap_schedule}.
Since all operations for the \textsc{Ctrl-Swap} injection can be performed simultaneously (see \zcref{fig:compiled_layout_CSWAP_No_correction, fig:delayed parity-not compiled}), a single \textsc{Ctrl-Swap} can be rounded to $d$ cycles.
Calculating the duration of the \textsc{Swap} network is then straightforward as we simply need the total number of \textsc{Ctrl-Swap}s performed and the time necessary to resolve the last Clifford corrections:
\begin{equation}\label{eq: SWAP rep code time}
    T_{\textsc{Swap}\text{ network}}^{(\text{rep})} = \underbrace{(N-1) b \max(d, t_{\CXCXX})}_{\textsc{Ctrl-Swap}} + \underbrace{t_r + (\lceil \log N \rceil +1)d}_\text{last Clifford corrections}.
\end{equation}
Here, $t_r$ denotes the reaction time and $t_{\CXCXX}$ the production time of a $\ket{\CXCXX}$ state. If $d<t_{\CXCXX}$, the computation is governed by the time it takes to refill the buffer. The correction term takes the reaction time of the last parity-\textsc{Cnot} into account as well as the time it takes to resolve it. 
We assume the worst-case scenario that all $X^{1/2}$ corrections need to be performed and that they are applied serially after the rest of the \textsc{Swap} network has concluded. 

\subsubsection{Compilation on the LDPC-cat architecture}\label{sec: Swap LDPC comp}

To compile the \textsc{Swap} network on an LDPC-cat architecture we take a similar approach to the compilation of the skew tree \textsc{Qrom}. The control register followed by the swap register are stored in the LDPC layer while buffer qubits are again stored in repetition code in a dedicated zone between the routing layer and the factories. We focus on the compilation of option 3 and store leftover ancillae holding the $Z$-parity of previously swapped qubits in three extra columns to the left of the buffer. If the \textsc{Swap} network is only a subpart of the algorithm with extra space available, the parity qubits could also be stored in an unused routing layer. However, this saves at most 3 qubits in repetition code and the space in the buffer zone is most likely already available from other subroutines such as the adder that uses a larger buffer area~\cite{Shor2026prep}. 

\begin{figure}[!htb]
    \centering
    \begin{subfigure}[t]{0.5\textwidth}
        \centering
    \includegraphics[width=\linewidth]{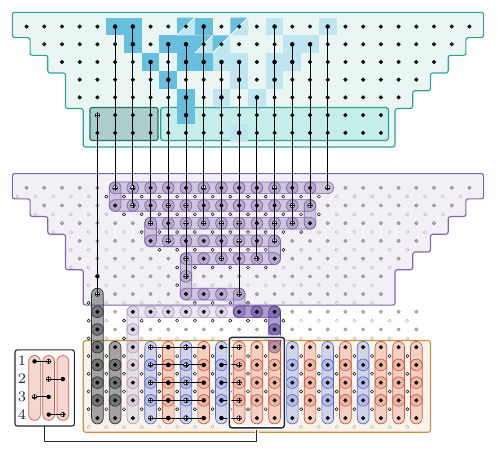}
    \caption{}
     \label{subfig: hybrid LDPC swap fork}
    \end{subfigure}%
    ~
    \begin{subfigure}[t]{0.5\textwidth}
    \centering
        \includegraphics[width=\linewidth]{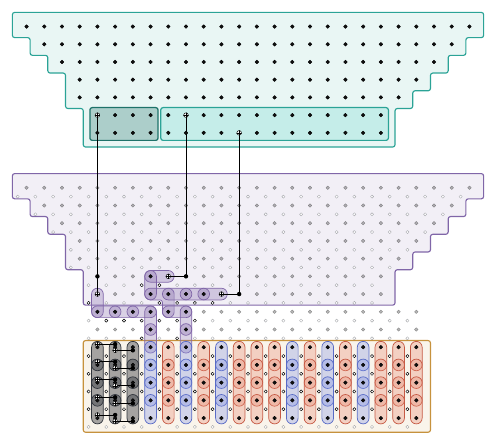}
        \caption{}
        \label{subfig: hybrid swap bus}
    \end{subfigure}%
    \caption{Compilation of \textsc{Ctrl-Swap} on an LDPC-repetition-cat architecture with additional storage for leftover parity qubits for Clifford corrections.
    Control qubits and swapped qubits are stored in the LDPC layer (teal box). On its first two rows, which give access to their $X$ logical operators, the control register is shaded in dark teal (left) and the swapped register in teal (right). The routing layer underneath it (violet box) is followed by two rows of routing and the buffer zone (orange box).
    The buffer holds two buffer states as it is enough for this slowed-down version.
    The three dark-gray codes on its left store the parities of previously swapped qubits.
    (a) shows the first phase: the parity-\textsc{Cnot}, together with buffer operations and the resolution of previous Clifford corrections.
    All buffer preparation steps are performed analogously to the repetition code setting in \zcref{fig:compiled_layout_CSWAP_No_correction} including the inset that gives the four transversal-\textsc{Cnot} steps of the catalytic Hadamard gate.
    The auxiliary fork code, with one shade of violet per branch as in \zcref{fig:Fork code}, is prepared in $\ket{00}_L$. Its medium-violet tine receives the parity of the swapped qubits and lies underneath the XOR of the supports of their $Z$ logical operators, drawn in two blues as in \zcref{fig:LDPC_parity_controlled_not}. 
    he dark tine performs the $M_{XX}$ with the output code of the Hadamard gate of the injection, while the light tine is routed into the right-most storage slot. Simultaneously, the left-most stored parity from a previous \textsc{Ctrl-Swap} injection is resolved, freeing its storage slot.
    (b) shows the $M_{XXX}$ measurement between the swapped qubits and one $\ket{C_XC_X X}$ qubit, as well as the remaining $M_{XX}$ between the control qubit and another $\ket{C_XC_X X}$ qubit, both along violet routing corridors. At the same time, the two stored parities shift over using transversal \textsc{Cnot} gates to free up the right-most storage slot.}
    \label{fig: hybrid LDPC ctrl-SWAP}
\end{figure}

Recall from \zcref{fig:compiled-swap-network-zx} that the compilation requires an $M_{XX}$ between the control qubit and a magic state qubit, an $M_{XXX}$ between the swapped qubits and a magic state qubit as well as a parity-\textsc{Cnot} between the swapped qubits and another $\ket{i}$ resource state qubit. While the parity-\textsc{Cnot} is limited to the $Z$-parity of only 2 qubits as opposed to potentially several in the setting of the \textsc{Qrom}, performing the $M_{XXX}$ and parity-\textsc{Cnot} simultaneously can lead to congestion in the routing layer, since the support of the routing qubits required for the XOR of two $Z$-logical operators can still have a large footprint.
Although this could be overcome by adding additional routing on top and on the side of the chip, we choose instead to perform these two operations sequentially.
This also allows to resolve any necessary parity-\textsc{Cnot} correction without any risk of congestion.
More precisely, in a first step, all the magic state preparation operations as well as the parity-\textsc{Cnot} using a fork code in the routing layer are performed. One tine of the fork is stored in the dedicated storage zone next to the buffer, another lies underneath the XOR of the support of the $Z$-logical operators of the swapped qubits to receive their $Z$-parity, and the remaining tine connects to the corresponding magic state.  Simultaneously, the parity-\textsc{Cnot} correction from an earlier \textsc{Ctrl-Swap} is performed on a control qubit using the left most stored parity ancilla. This is illustrated in \zcref{subfig: hybrid LDPC swap fork}.
Next, we perform the $M_{XX}$ with the control qubit and the $M_{XXX}$ for the multi-target $\textsc{Cnot}$ using a fork code reaching the tip of each target swap qubit to access their $X$ logical operator.
During this operation, the stored parity ancillae are moved using transversal \textsc{Cnot} gates to prepare for the next \textsc{Ctrl-Swap} as illustrated in \zcref{subfig: hybrid swap bus}. 

We again allow for separate distances for the LDPC and repetition layer, $d_\LDPC, d_\rep$ and set $d = \max(d_\LDPC, d_\rep)$. Calculating the temporal and spatial footprint of this subroutine is similar to the repetition-cat setting, except that now each \textsc{Ctrl-Swap} application takes $2d$ cycles. This holds under the assumption that the time $t_{\CXCXX}$ it takes to produce a $\ket{\CXCXX}$ state satisfies $t_{\CXCXX} \leq 2d$, since then we can guarantee timely buffer refilling. Thus, we obtain the following quantities
\begin{align}
    T_{\textsc{Swap}\text{ network}}^{(\text{LDPC})} &= \underbrace{2 (N-1) b d}_{\textsc{Ctrl-Swap}} + \underbrace{t_r + (\lceil \log N \rceil +1)d}_\text{last correction}, \\
    \Gamma_{\textsc{Swap}\text{ network}}^{(\text{LDPC})} &= 2(HL + \Delta_H) + 2\max(L, 8N_\mathrm{buffer} +3) +  (8N_\mathrm{buffer} +3)d_\rep,
\end{align}
where $N$ is the number of registers, i.e., $\lceil \log N \rceil$ address qubits control the \textsc{Ctrl-Swaps}, $b>1$ is the number of qubits per register, $L = \lceil (\log N)/2 \rceil + \lceil Nb/2 \rceil$ is the width of LDPC layer, $H$ its height, $\Delta_H$ denotes the number of qubits necessary for the triangular overhead, and $N_\mathrm{buffer}$ denotes the number of buffers.
Contrary to the repetition code setting, $N_\mathrm{buffer} =2$ suffices as the serialization of injection offer enough extra time for magic state production provided $t_{\CXCXX}+t_r \leq 2d$.
While \zcref{fig: hybrid LDPC ctrl-SWAP} depicts one row of routing between the buffer zone and the area underneath the LDPC layer, another row of routing are necessary if the buffer is placed farther to the right.
Thus, we by default account $2\max(L, 8N_\mathrm{buffer} +3)$ qubits for such routing.
The timing formula accounts two steps per \textsc{Ctrl-Swap} as well as the time necessary to resolve the remaining Clifford corrections.
This is costed analogously to the repetition code setting, consisting of the reaction time necessary to resolve the last parity-\textsc{Cnot} and accounting for the worst-case possible number of $X^{1/2}$ gates that are  performed serially. 

\subsection{\textsc{Phaseup} compilation}\label{sec: PhaseFixup comp}

As described in \zcref{sec: phase fixup}, the choice between space-time and space-space splits of a \textsc{Phaseup} depends on the number of available clean ancillae and the size of the register to fix.
In our final implementation, both are used in different places in the algorithm, depending on the subroutine and its size.
For the space-time split, 
a unary-one-hot encoding can be compiled using a \textsc{Swap} network as in \zcref{sec: Swap rep-code comp}.
In the color-changed convention, as outlined in \zcref{sec: phase fixup} the compilation of the space-time split of a \textsc{Phaseup} then follows directly from the compilation of the skew-tree \textsc{Qrom}, of the \textsc{Swap} network, and a layer of $H$ gates.
Similarly, the compilation of the space-space split of a \textsc{Phaseup} follows that of the \textsc{And} gate cascade, already compiled for the \textsc{Qrom}, and multi-target $\CXX$ gates.
We start by compiling the layer of $H$ gates and the multi-target $\CXX$, currently standing as the missing pieces, both on the repetition-cat and the LDPC-cat architecture, thus fixing the cost of \textsc{Phaseup}.
In a later point, we propose a technical improvement to minimize the number of $\CXX$ gates needed to perform a given fixup.
Although this improvement could reduce the cost of a concrete implementation, its effect is not taken into account in our resource estimation as we simply bound the number of multi-target $\CXX$ gates by the number of power products from the $\ell$ register corresponding to a worst-case scenario assumption. 

\subsubsection{Multi-target \texorpdfstring{$\CXX$ and $H$}{CXX and H} on the repetition-cat and LDPC-cat architecture.}\label{sec: Multi-target C_XX and H}
The $H$ gate layer appearing in the space-time split can be compiled assuming $Z$-access to the algorithmic qubits is available.
Indeed, in a space-time split \textsc{Phaseup} the layer of $H$ gate is applied to qubits that are then target qubits of a skew-tree \textsc{Qrom} for which they need $Z$-access anyway.

\begin{figure}[!htb]
    \centering
    \begin{subfigure}[t]{0.38\textwidth}\centering
        \includegraphics[width=\linewidth]{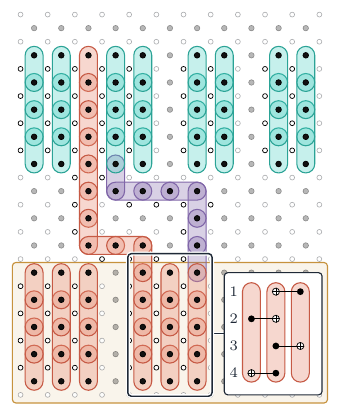}
        \caption{}\label{subfig:hzacc_rep_a}
    \end{subfigure}\hspace{0.05\textwidth}
    \begin{subfigure}[t]{0.38\textwidth}\centering
        \includegraphics[width=\linewidth]{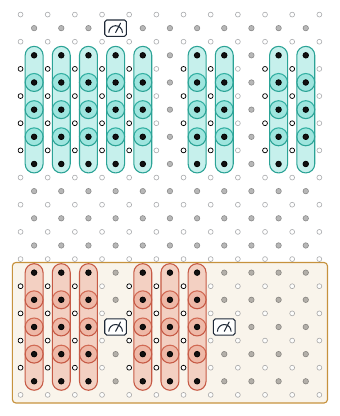}
        \caption{}\label{subfig:hzacc_rep_b}
    \end{subfigure}
    \caption{In-place catalyzed Hadamard injection from \zcref{fig:H_inplace} on a logical qubit $\ket{\psi}$ stored in a $Z$-access register of repetition codes.
    The buffer originally holds three repetition code $\ket{i}$ states.
    The inset displays a sequence of transversal \textsc{Cnot}s to turn them into the $\ket{CZ}$ pair and the catalyst $\ket{i}$, see \zcref{app: hadamard}.
    a) A first $M_{XX}$ with the target qubit does the interaction with the catalyzed state while a second $M_{XX}$ already teleports the output back to the algorithmic qubit area.
    b) a transversal measurement $M_Z$ on initial target qubit, left-most and right-most resource $\ket{i}$ state.}
    \label{fig:hadamard_inplace_rep}
\end{figure}

Each $H$ gate is applied following the circuit of \zcref{fig:H_inplace} of \zcref{app: Clifford}, describing Clifford state injection circuits. \zcref{fig:hadamard_inplace_rep} shows the compilation in the repetition-cat architecture, which takes $d$ rounds. 
Indeed, the duration of transversal measurements, transversal resets and transversal \textsc{Cnot}s are negligible compared to that of lattice surgery operations.
The cost of a layer of $H$ gates directly depends on the number of catalytic states available in the buffer simultaneously.
Given the fact that the $\ket{i}$-state factories are sized to provide four new $\ket{i}$ states per $\max(d,t_{\CXCXX})$ for the \textsc{Ctrl-Swap} network and the fact that the buffer area is bigger than 16 single qubits resource states (16 representing 2 catalytic states being used and 2 being refilled) for other subroutines, this means that $H$ gates can be performed two by two without any further assumptions.  For an $H$ layer on $N_H$ qubits stored in repetition codes, we have the following cost:
$$T_{\text{H layer}}^{(\text{rep})}= \big\lceil\frac{N_H}{2}\big\rceil d.$$

\begin{figure}[!htb]
    \centering
    \begin{subfigure}[t]{0.32\textwidth}\centering
        \includegraphics[width=\linewidth]{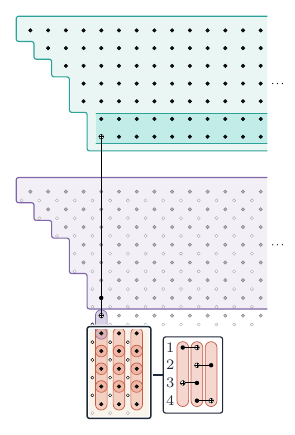}
        \caption{}\label{subfig:hinplace_ldpc_a}
    \end{subfigure}\hfill
    \begin{subfigure}[t]{0.32\textwidth}\centering
        \includegraphics[width=\linewidth]{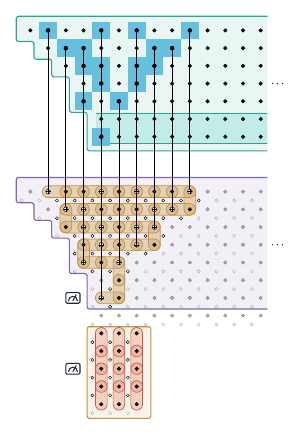}
        \caption{}\label{subfig:hinplace_ldpc_b}
    \end{subfigure}\hfill
    \begin{subfigure}[t]{0.32\textwidth}\centering
        \includegraphics[width=\linewidth]{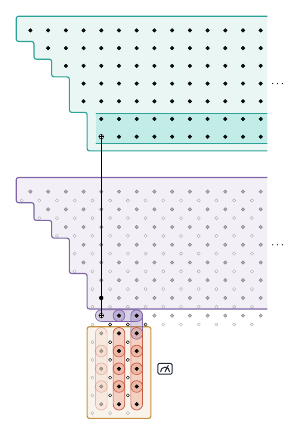}
        \caption{}\label{subfig:hinplace_ldpc_c}
    \end{subfigure}
    \caption{In-place catalyzed Hadamard injection, circuit from~\zcref{fig:H_inplace}, on a logical qubit $\ket{\psi}$ of the LDPC-cat code.
    The buffer and the inset are the same as in \zcref{fig:hadamard_inplace_rep}.
    a) First $M_{XX}$ using a routing qubit prepared in $\ket{+}$ under $X_L$ of $\ket{\psi}$ and a routing corridor merged with the resource state.
    b) QND $M_Z$ of the qubit in the LDPC layer using a $Z$ parity copy that lies on a repetition code (amber) prepared in $\ket{0}_L$, which is then measured, and the operation is repeated $d$ times~\cite{Ruiz_2025}. The left-most qubit of the resource state is transversally measured in the $Z$-basis once.
    c) Second $M_{XX}$ to teleport the qubit back into the LDPC layer. A transversal $M_Z$ of the right-most qubit of the resource state.}
    \label{fig:hadamard_inplace_ldpc}
\end{figure}

If, however, the qubits are stored in an LDPC code, in-place $H$ gates now require $3d$ cycles as the QND $M_Z$ measurement has to be repeated $d$ times, see~\zcref{fig:hadamard_inplace_ldpc}. 
Additionally, due to the QND $M_Z$ measurement that uses the routing layer, the parallelization of injection is not trivial anymore.
We hence do not assume that there are no conflicts and perform the Hadamard gates serially. 
This results in a cost of 
$$T_{\text{H layer}}^{(\text{LDPC})}= 3N_H d.$$
This, together with \zcref{subsec:qrom compil} and \zcref{sec: Swap comp} allows one to deduce the cost of a space-time split \textsc{Phaseup} for both architectures.

\begin{figure}[!htb]
    \centering
    \begin{subfigure}[t]{0.35\textwidth}\centering
        \includegraphics[width=\linewidth]{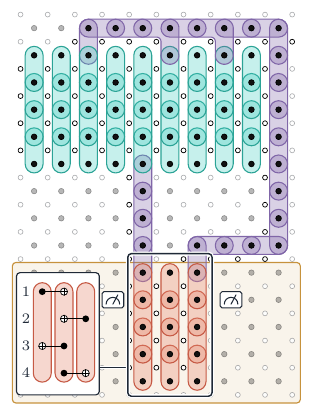}
        \caption{}\label{fig: CxXXX rep code}
    \end{subfigure}\hspace{0.05\textwidth}
    \begin{subfigure}[t]{0.38\textwidth}\centering
        \includegraphics[width=\linewidth]{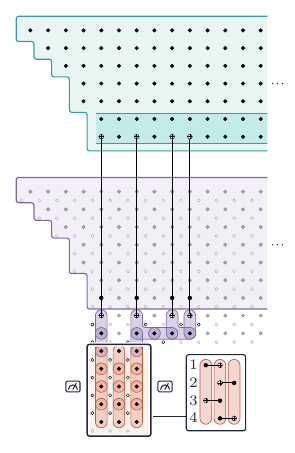}
        \caption{}\label{fig: CxXXX ldpc code}
    \end{subfigure}
    \caption{Compiled $C_X XXX$ injection of \zcref{fig: CxxXX} on a) the repetition-cat architecture, b) the LDPC-cat architecture.
    a) control qubit and target qubits are stored in an $X$-access only register of repetition codes drawn in teal.
    The buffer holds a catalytic state prepared via the four transversal \textsc{Cnot}s ordered from 1 to 4 in the inset, see \zcref{fig: catalytic state}.
    One $M_{XX}$ operation via the routing corridor in violet is applied between the control qubit and the catalytic state while a second $M_{XXXX}$ is applied on the three target qubits and the catalytic state.
    Then the two non-catalyzed legs of the resource state are measured in the $Z$ basis.
    b) same operations excepts that the $M_{X\ldots X}$ are performed using the violet routing code and a few straddling $M_{XX}$ between the routing and the LDPC layers.
    Indeed, each $X$ logical is reached through a routing qubit prepared in $\ket{+}$ under it, which controls the $X$ logical and the routing qubit below. }
    \label{fig: CxXXX compiled}
\end{figure}

On the other hand, the space-space split \textsc{Phaseup} is typically used after a skew-tree \textsc{Qrom} uncomputation as in this case many recently uncomputed qubits are available to be used as clean ancillae.
In the repetition-cat architecture, multi-target $\CXX$ gates thus have to be compiled using $X$-access only to the algorithmic qubits.
Indeed, the address register of a skew-tree \textsc{Qrom} are tightly packed with $X$-access only.
Fortunately, the injection circuit of the multi-target $\CXX$ gate from \zcref{fig: CxxXX} already uses joint $X$ measurement interactions only and is acting in-place.
There is no need for an additional teleportation step.
The injection is made of one catalytic state preparation and two $M_{X\ldots X}$, namely, one between the control qubit and the first leg of the catalytic state, and one between the target qubits and the third leg of the catalytic state.
Concretely, one of the $M_{X\ldots X}$ is performed using the bottom routing area, while the other one uses the top routing area, as shown on \zcref{fig: CxXXX rep code}.
While two catalytic states could, in principle, be filled every $d$ rounds, guaranteeing that the two injections don't induce congestion issues in the routing area requires a proper compiler schedule or surgery crossing techniques that are beyond the scope of this work.
Consequently, we limit ourselves to one $\CXX$ gate every $d$ rounds for the repetition-cat architecture.
For the LDPC-cat architecture, a true compiler would, for most configurations, find a routing pattern to measure both operations simultaneously as on \zcref{fig: CxXXX ldpc code}.
However, as a safe assumption, and because the \textsc{Phaseup} represents only a small fraction of the cost of the algorithm, we assume the measurements are done one at a time, bringing the duration of each multi-target $\CXX$ gate to $2d$.

As a summary, the space-space split \textsc{Phaseup} has a time-cost that is
\begin{multline}
    T_{\text{s-s Phaseup}}^{(\text{rep})}=\underbrace{\left(2^K+2^L-(K+L+2)\right)}_{\# \text{ of AND gates}}\max(d,t_{\CXCXX}) +\underbrace{\left(2^L-1\right)}_{\# \text{ of multi-target }\CXX}d + \underbrace{(K-1)}_{\# \text{ cascade depth in MBU}} t_r\\+ \frac{1}{2}\underbrace{\left(2^K-(K+1)\right)}_{\# \text{ of MBU Clifford corrections}} d,\\
    T_{\text{s-s Phaseup}}^{(\text{LDPC})}=\left(2^K+2^L-(K+L+2)\right)\max(d,t_{\CXCXX}) +\left(2^L-1\right)2d  +(K-1) t_r \\+\left(2^K-(K+1)\right) (d+\frac{1}{2}d),\\
\end{multline}
where $K$ is the size of the half most-significant bit register and $L$ is the size of the half least-significant bit register, as introduced in \zcref{sec: phase fixup}.
In the repetition-cat architecture, $M_Z$ are destructive and single shot, so the reaction time dependency slow-down between layers of MBU is taken into account separately from the eventual duration of Clifford corrections.
In the LDPC-cat architecture, $M_Z$ operations are not single-shot and can't be parallelized easily, so the measurement duration is added separately from the reaction time slow-down.
Note that although multi-target $\CXX$ have been assumed to take $2d$ rounds, here MBU $\CXX$ corrections are single control single target so that both $M_{XX}$ of the injection circuit can be routed at the same time without further assumptions.
Such $\CXX$ corrections only take $d$ rounds as in the repetition-cat architecture.
The $\textsc{And}$ gate compilation is the same as in the \textsc{Qrom} of \zcref{subsec:qrom compil}.
Indeed, Clifford corrections from $\textsc{And}$ injections are dealt with by classically updating the target of the upcoming $\CXX$ and the classical decisions in $\textsc{And}^\dagger$ MBU.
This feature was introduced as ``wandering'' $\textsc{And}$ gates in \cite[Appendix A.3]{gidney2025factor}.

\subsubsection{Minimizing the number of \texorpdfstring{$\CXX$}{CXX} gates }

In this subsection, we discuss a practical improvement that would be applied by the compiler to minimize the resource state cost and the duration of the space-space split \textsc{Phaseup} implementation.
Considering the space-space split circuit of \zcref{fig:phase fixup circ}, the high level idea is as follows: instead of performing one single control but multi-target $\CXX$ per possible monomial built out of the $\ell$ register, one can consider them as a whole and build a circuit of fewer parity-controlled multi-target $\CXX$ gates.
As each parity-controlled multi-target $\CXX$ gate costs one catalytic state just like one single control multi-target $\CXX$, this technique reduces the number of consumed resources as well as the duration. 

After the M\"obius transform of the phase table, every term supported on a single half of the split register is a Pauli $X$ operator on the corresponding power-product representative and is absorbed into the Pauli frame.
The remaining cross terms are collected in the \textsc{Phaseup} matrix $F$, whose rows and columns are indexed by the representatives of the two halves.
Writing $\mathbf{k}$ and $\boldsymbol{\ell}$ for the bit vectors of their $X$ eigenvalues, these terms implement the phase $(-1)^{\boldsymbol{\ell}^\top F \mathbf{k}}$.
We write $F = U V^\top$ over $\mathbb{F}_2$ with $\rho = \operatorname{rank}(F)$ columns in $U$.
Usually, the sequence of multi-target $\CXX$ ($\CZ$ in literature's conventions) is simply read out of the rows of $F$.
Each row corresponds to one gate, there are at most $2^L-1$ such gates with the notations of \zcref{sec: phase fixup}. 
However, a circuit of parity-controlled multi-target $\CXX$ where the $i$-th gate is controlled on the parity of the representatives in the support of the $i$-th column of $U$ and targets the representatives fixed by the $i$-th row of $V^\top$ yields the exact same phase correction.
Indeed, a multi-target $\mathrm{C}_X X$ controlled by the parity $\mathbf{u}^\top\boldsymbol{\ell}$ and targeting the representatives in the support of $\mathbf{v}$ applies the phase $(-1)^{(\mathbf{u}^\top\boldsymbol{\ell})(\mathbf{v}^\top\mathbf{k})}$, so the $\rho$ gates built from the columns $\mathbf{u}_i$ of $U$ and $\mathbf{v}_i$ of $V$ jointly apply $(-1)^{\sum_{i=1}^{\rho}(\mathbf{u}_i^\top\boldsymbol{\ell})(\mathbf{v}_i^\top\mathbf{k})} = (-1)^{\boldsymbol{\ell}^\top UV^\top\mathbf{k}} = (-1)^{\boldsymbol{\ell}^\top F\mathbf{k}}$. The naive circuit is the special case $U=\mathbb{I}$ and $V^\top=F$.

\begin{figure}[htbp]
  \centering
  \includegraphics[width=0.8\linewidth]{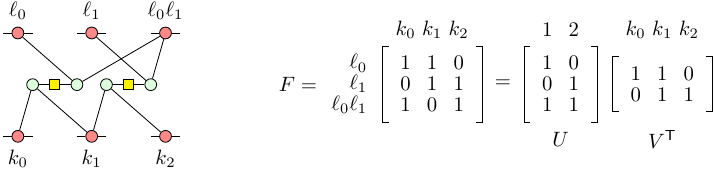}
  \caption{Example of a \textsc{Phaseup} operation which corresponds to a rank two matrix $F$.
  It is implemented with two parity-controlled multi-target $\CXX$ gates instead of the naive implementation with three multi-target $\CXX$ gates.
  The weight 2 and 3 products of $k_i$ addresses are not shown in this Figure.
  The $\CXX$ gate implementation is detailed in \zcref{app: Clifford}.
  }
\label{fig:phase_fixup_rank_factorisation}
\end{figure}

\zcref{fig:phase_fixup_rank_factorisation} illustrates the aforementioned mechanism with ZX diagrams.
The construction uses $\operatorname{rank}(F) \le \min(2^L, 2^K) - 1$ gates, that is \ $2\operatorname{rank}(F)$ $\ket{i}$ states when a catalyst is available.

The number of gates can be reduced further by letting both parities (control and target) of each $\mathrm{C}_X X$ gate span the two halves of the register.
Since $\ell_i\in\mathbb{F}_2$, $\ell_i^2=\ell_i$ and the product of two representatives of the same half is again a representative of that half (e.g., $\ell_0\cdot\ell_0\ell_1=\ell_0\ell_1$).
So the product of two parities supported on a single half is itself a parity of representatives of that half, i.e., a product of Pauli $X$ operators absorbed into the Pauli frame.
Consider a $\mathrm{C}_X X$ gate between the parities $p=\mathbf{a}^\top\boldsymbol{\ell}+\mathbf{b}^\top\mathbf{k}$ and $q=\mathbf{c}^\top\boldsymbol{\ell}+\mathbf{d}^\top\mathbf{k}$, which are sums of existing representatives, so that no additional power product is computed.
It applies the phase $(-1)^{pq}$, with
\begin{equation}
  pq=(\mathbf{a}^\top\boldsymbol{\ell})(\mathbf{c}^\top\boldsymbol{\ell})+(\mathbf{b}^\top\mathbf{k})(\mathbf{d}^\top\mathbf{k})+\boldsymbol{\ell}^\top\bigl(\mathbf{a}\mathbf{d}^\top+\mathbf{c}\mathbf{b}^\top\bigr)\mathbf{k}.
\end{equation}
The first two terms are absorbed into the Pauli frame by adding additional $X$ corrections on specific representatives, so that now a single gate implements the rank-two contribution $\mathbf{a}\mathbf{d}^\top+\mathbf{c}\mathbf{b}^\top$ to $F$.
Denoting again by $\mathbf{u}_i$ the $i$-th column of $U$ and by $\mathbf{v}_i^\top$ the $i$-th row of $V^\top$, the choice $(\mathbf{a},\mathbf{b},\mathbf{c},\mathbf{d})=(\mathbf{u}_{2j-1},\mathbf{v}_{2j},\mathbf{u}_{2j},\mathbf{v}_{2j-1})$ for the $j$-th gate therefore implements $(-1)^{\boldsymbol{\ell}^\top F\mathbf{k}}$ with $\lceil\rho/2\rceil$ gates.
This halves the number of consumed $\ket{i}$ states, from $2\rho$ to $2\lceil\rho/2\rceil$. 
The duration is halved as well as there are now half the number of gates to perform.
In the example of \zcref{fig:phase_fixup_rank_factorisation}, the single gate between $p=\ell_0+\ell_0\ell_1+k_1+k_2$ and $q=\ell_1+\ell_0\ell_1+k_0+k_1$ replaces the two gates shown: the $\ell$-half term $(\ell_0+\ell_0\ell_1)(\ell_1+\ell_0\ell_1)$ vanishes identically, and the $k$-half term $(k_1+k_2)(k_0+k_1)=k_0k_1+k_1+k_0k_2+k_1k_2$ is a Pauli-frame update.

This construction is optimal among circuits of $\mathrm{C}_X X$ gates between parities of representatives.
Indeed, the products of an $\ell$- and a $k$-representative are linearly independent functions of the address, so the contributions $\mathbf{a}\mathbf{d}^\top+\mathbf{c}\mathbf{b}^\top$ of the gates must sum to $F$, which requires at least $\lceil\rho/2\rceil$ of them.

%% file: graphics/istate_distillation/distillation_results.tex
\begin{table}[!htb]
\centering
\begin{subtable}[t]{0.48\textwidth}
\centering
\caption{Protocols by triangular code distance. }
\label{tab:istate_qubit_count}
    \input{graphics/istate_distillation/tables/femoco_istate_qubit_count.tex}
\end{subtable}
\hfill
\begin{subtable}[t]{0.48\textwidth}
    \centering
    \caption{Protocol types by lengths.}
    \label{tab:istate_protocol_types}
    \input{graphics/istate_distillation/tables/femoco_istate_protocol_types.tex}
\end{subtable}
\caption{$\keti$-state distillation protocols classified by triangular code distances $d_c$ and lengths. The full distillation protocol lasts for $k+1$ repetition-code cycles, with $k$ being the number of preparatory repetition-code cycles. At given $d_c$, we work with a unique repetition code distance $d_r$, chosen close to saturation in the sense of \zcref{fig:i_distillation_rep_saturating_params}. The total qubit counts (data + ancillas) are then $n_{s}(d_{c}) + (2d_{r}-1)$ in the notations of \eqref{eq: vertice_and_face_number_i_distillation}. 
Any combination of $d_c$ and $k$  is possible, and the $k=1$ value used in the provided Crumble circuits was chosen for illustrative purposes only.  }
\label{tab:istate_distillation_results}
\end{table}

\begin{figure}[!htb]
    \centering
    \begin{subfigure}[b]{0.49\textwidth}
        \centering
        \includegraphics[width=\linewidth]{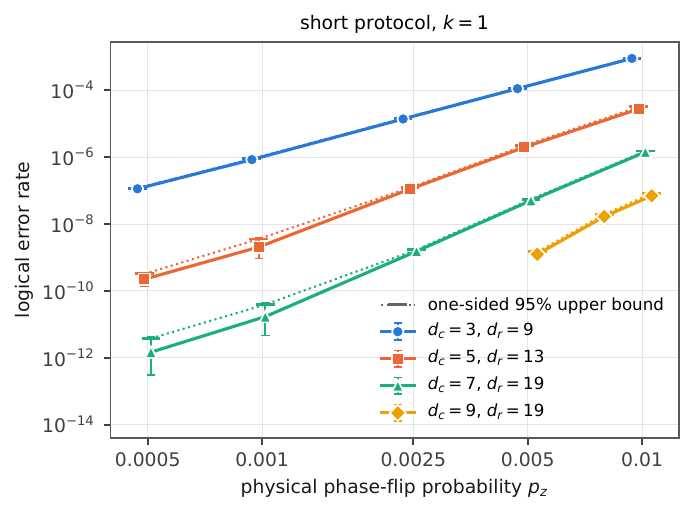}
        \caption{Short protocols.}
        \label{subfig:istate_ler_vs_pz_short}
    \end{subfigure}
    \hfill
    \begin{subfigure}[b]{0.49\textwidth}
        \centering
        \includegraphics[width=\linewidth]{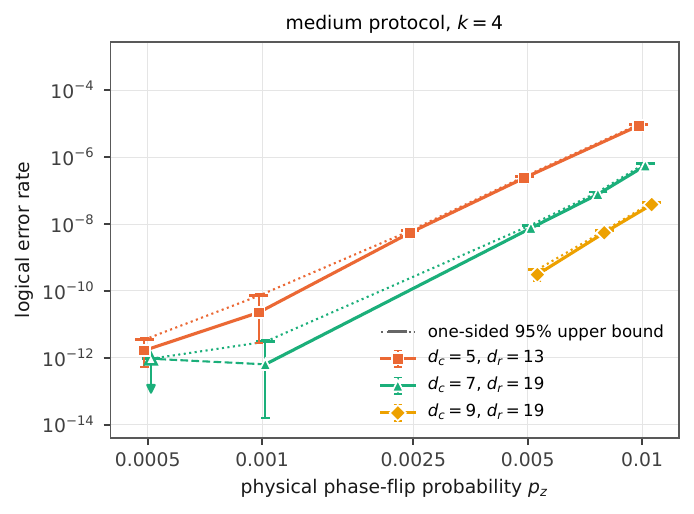}
        \caption{Medium protocols.}
        \label{subfig:istate_ler_vs_pz_medium}
    \end{subfigure}
    \caption{Logical error rate on $\keti$ as a function of $p_z$, for the short and medium protocol types of \zcref{tab:istate_protocol_types}.
    Each point is estimated from $N$ accepted shots, of which $n$ ended in a logical failure.
    Filled markers show the estimate $n/N$.
    Error bars show the exact two-sided 95\% Poisson confidence interval on the true rate, which are asymmetric on a logarithmic scale.
    Hollow markers with downward arrows are points where no failure was observed ($n=0$). There is no
    Poisson estimate in this case, so we use the one-sided 95\% upper bound, $1-0.05^{1/N}$ obtained from the explicit expression for the binomial law. Such a point is set by the number of
    shots, not by the actual protocol's rate, and the true rate may lie well below it. See our code's documentation for more information on the statistical interpretation of our results~\cite{AliceBob2026istatecode}. The curves indicate a power-law behaviour. For saturating values of $d_r$ and $k$, the fits' exponents reach $d_c$, see \zcref{fig:i_distillation_ler_vs_pz_plot}. 
    }
    \label{fig:istate_protocol_plots}
\end{figure}

%% file: graphics/istate_distillation/tables/femoco_istate_qubit_count.tex
\begin{tabular}{cccc}
\toprule
$d_c$ & $d_r$ & Qubits & Circuit \\
\midrule
3 & 9 & 30 & \href{https://algassert.com/crumble\#circuit=Q(0,0)0;Q(4,0)1;Q(12,0)2;Q(6,1)3;Q(10,1)4;Q(4,2)5;Q(6,3)6;Q(0,-2)7;Q(4,-2)8;Q(12,-2)9;Q(16,-2)10;Q(20,-2)11;Q(24,-2)12;Q(28,-2)13;Q(32,-2)14;Q(36,-2)15;Q(9,0)16;Q(3,1)17;Q(9,2)18;Q(7,0)19;Q(1,1)20;Q(7,2)21;Q(2,-2)22;Q(8,-2)23;Q(14,-2)24;Q(18,-2)25;Q(22,-2)26;Q(26,-2)27;Q(30,-2)28;Q(34,-2)29;R_7_8_9_10_11_12_13_14_15;RX_16;R_19;RX_17;R_20;RX_18;R_21;RX_22_23_24_25_26_27_28_29;TICK;RX_0_1_2_3_4_5_6;CX_16_19_17_20_18_21_22_7_23_8_24_9_25_10_26_11_27_12_28_13_29_14;TICK;CX_1_17_0_20_4_18_3_21_22_8_23_9_24_10_25_11_26_12_27_13_28_14_29_15;TICK;CX_2_16_1_19_3_17_5_21;MX_22_23_24_25_26_27_28_29;TICK;CX_0_7_1_8_2_9_3_19_4_16_5_17_6_21;RX_22_23_24_25_26_27_28_29;TICK;SQRT_X_0_1_2_3_4_5_6;CX_16_19_17_20_18_21_22_7_23_8_24_9_25_10_26_11_27_12_28_13_29_14;TICK;M_19;MX_16;M_20;MX_17;M_21;MX_18;CX_22_8_23_9_24_10_25_11_26_12_27_13_28_14_29_15;DT(7,0,6)rec[-5];DT(1,1,6)rec[-3];DT(7,2,6)rec[-1];M_0_1_2_3_4_5_6;TICK;MX_22_23_24_25_26_27_28_29;DT(2,-2,7)rec[-8]_rec[-29];DT(8,-2,7)rec[-7]_rec[-28];DT(14,-2,7)rec[-6]_rec[-27];DT(18,-2,7)rec[-5]_rec[-26];DT(22,-2,7)rec[-4]_rec[-25];DT(26,-2,7)rec[-3]_rec[-24];DT(30,-2,7)rec[-2]_rec[-23];DT(34,-2,7)rec[-1]_rec[-22];TICK;S_7_8_9_10_11_12_13_14_15;TICK;CZ_7_8_7_9_7_10_7_11_7_12_7_13_7_14_7_15_8_9_8_10_8_11_8_12_8_13_8_14_8_15_9_10_9_11_9_12_9_13_9_14_9_15_10_11_10_12_10_13_10_14_10_15_11_12_11_13_11_14_11_15_12_13_12_14_12_15_13_14_13_15_14_15;TICK;MX_7_8_9_10_11_12_13_14_15;TICK;DT(2,-2,10)rec[-9]_rec[-8]_rec[-17];DT(8,-2,10)rec[-8]_rec[-7]_rec[-16];DT(14,-2,10)rec[-7]_rec[-6]_rec[-15];DT(18,-2,10)rec[-6]_rec[-5]_rec[-14];DT(22,-2,10)rec[-5]_rec[-4]_rec[-13];DT(26,-2,10)rec[-4]_rec[-3]_rec[-12];DT(30,-2,10)rec[-3]_rec[-2]_rec[-11];DT(34,-2,10)rec[-2]_rec[-1]_rec[-10];OI(0)rec[-21]_rec[-20]_rec[-22]_rec[-23]_rec[-30]_rec[-37];OI(1)rec[-19]_rec[-21]_rec[-23]_rec[-24]_rec[-28]_rec[-38];OI(2)rec[-18]_rec[-20]_rec[-21]_rec[-19]_rec[-26];OI(3)rec[-24]_rec[-23]_rec[-22]_rec[-9]_rec[-37]_}{Crumble} \\
5 & 13 & 62 & \href{https://algassert.com/crumble\#circuit=Q(0,0)0;Q(4,0)1;Q(12,0)2;Q(16,0)3;Q(24,0)4;Q(6,1)5;Q(10,1)6;Q(18,1)7;Q(22,1)8;Q(4,2)9;Q(12,2)10;Q(16,2)11;Q(6,3)12;Q(10,3)13;Q(18,3)14;Q(12,4)15;Q(16,4)16;Q(10,5)17;Q(12,6)18;Q(0,-2)19;Q(4,-2)20;Q(12,-2)21;Q(16,-2)22;Q(24,-2)23;Q(28,-2)24;Q(32,-2)25;Q(36,-2)26;Q(40,-2)27;Q(44,-2)28;Q(48,-2)29;Q(52,-2)30;Q(56,-2)31;Q(9,0)32;Q(21,0)33;Q(3,1)34;Q(15,1)35;Q(9,2)36;Q(21,2)37;Q(15,3)38;Q(9,4)39;Q(15,5)40;Q(7,0)41;Q(19,0)42;Q(1,1)43;Q(13,1)44;Q(7,2)45;Q(19,2)46;Q(13,3)47;Q(7,4)48;Q(13,5)49;Q(2,-2)50;Q(8,-2)51;Q(14,-2)52;Q(20,-2)53;Q(26,-2)54;Q(30,-2)55;Q(34,-2)56;Q(38,-2)57;Q(42,-2)58;Q(46,-2)59;Q(50,-2)60;Q(54,-2)61;R_19_20_21_22_23_24_25_26_27_28_29_30_31;RX_32;R_41;RX_33;R_42;RX_34;R_43;RX_35;R_44;RX_36;R_45;RX_37;R_46;RX_38;R_47;RX_39;R_48;RX_40;R_49;RX_50_51_52_53_54_55_56_57_58_59_60_61;TICK;RX_0_1_2_3_4_5_6_7_8_9_10_11_12_13_14_15_16_17_18;CX_32_41_33_42_34_43_35_44_36_45_37_46_38_47_39_48_40_49_50_19_51_20_52_21_53_22_54_23_55_24_56_25_57_26_58_27_59_28_60_29_61_30;TICK;CX_1_34_0_43_3_35_2_44_6_36_5_45_8_37_7_46_11_38_10_47_13_39_12_48_16_40_15_49_50_20_51_21_52_22_53_23_54_24_55_25_56_26_57_27_58_28_59_29_60_30_61_31;TICK;CX_2_32_1_41_4_33_3_42_5_34_7_35_6_44_10_36_9_45_11_46_14_38_13_47_15_39_17_49;MX_50_51_52_53_54_55_56_57_58_59_60_61;TICK;CX_0_19_1_20_2_21_3_22_4_23_5_41_6_32_7_42_8_33_9_34_10_44_11_35_12_45_13_36_14_46_15_47_16_38_17_39_18_49;RX_50_51_52_53_54_55_56_57_58_59_60_61;TICK;SQRT_X_0_1_2_3_4_5_6_7_8_9_11_12_13_14_15_16_17_18;SQRT_X_DAG_10;CX_32_41_33_42_34_43_35_44_36_45_37_46_38_47_39_48_40_49_50_19_51_20_52_21_53_22_54_23_55_24_56_25_57_26_58_27_59_28_60_29_61_30;TICK;M_41;MX_32;M_42;MX_33;M_43;MX_34;M_44;MX_35;M_45;MX_36;M_46;MX_37;M_47;MX_38;M_48;MX_39;M_49;MX_40;CX_50_20_51_21_52_22_53_23_54_24_55_25_56_26_57_27_58_28_59_29_60_30_61_31;DT(7,0,6)rec[-17];DT(19,0,6)rec[-15];DT(1,1,6)rec[-13];DT(13,1,6)rec[-11];DT(7,2,6)rec[-9];DT(19,2,6)rec[-7];DT(13,3,6)rec[-5];DT(7,4,6)rec[-3];DT(13,5,6)rec[-1];M_0_1_2_3_4_5_6_7_8_9_10_11_12_13_14_15_16_17_18;TICK;MX_50_51_52_53_54_55_56_57_58_59_60_61;DT(2,-2,7)rec[-12]_rec[-61];DT(8,-2,7)rec[-11]_rec[-60];DT(14,-2,7)rec[-10]_rec[-59];DT(20,-2,7)rec[-9]_rec[-58];DT(26,-2,7)rec[-8]_rec[-57];DT(30,-2,7)rec[-7]_rec[-56];DT(34,-2,7)rec[-6]_rec[-55];DT(38,-2,7)rec[-5]_rec[-54];DT(42,-2,7)rec[-4]_rec[-53];DT(46,-2,7)rec[-3]_rec[-52];DT(50,-2,7)rec[-2]_rec[-51];DT(54,-2,7)rec[-1]_rec[-50];TICK;S_19_20_21_22_23_24_25_26_27_28_29_30_31;TICK;CZ_19_20_19_21_19_22_19_23_19_24_19_25_19_26_19_27_19_28_19_29_19_30_19_31_20_21_20_22_20_23_20_24_20_25_20_26_20_27_20_28_20_29_20_30_20_31_21_22_21_23_21_24_21_25_21_26_21_27_21_28_21_29_21_30_21_31_22_23_22_24_22_25_22_26_22_27_22_28_22_29_22_30_22_31_23_24_23_25_23_26_23_27_23_28_23_29_23_30_23_31_24_25_24_26_24_27_24_28_24_29_24_30_24_31_25_26_25_27_25_28_25_29_25_30_25_31_26_27_26_28_26_29_26_30_26_31_27_28_27_29_27_30_27_31_28_29_28_30_28_31_29_30_29_31_30_31;TICK;MX_19_20_21_22_23_24_25_26_27_28_29_30_31;TICK;DT(2,-2,10)rec[-13]_rec[-12]_rec[-25];DT(8,-2,10)rec[-12]_rec[-11]_rec[-24];DT(14,-2,10)rec[-11]_rec[-10]_rec[-23];DT(20,-2,10)rec[-10]_rec[-9]_rec[-22];DT(26,-2,10)rec[-9]_rec[-8]_rec[-21];DT(30,-2,10)rec[-8]_rec[-7]_rec[-20];DT(34,-2,10)rec[-7]_rec[-6]_rec[-19];DT(38,-2,10)rec[-6]_rec[-5]_rec[-18];DT(42,-2,10)rec[-5]_rec[-4]_rec[-17];DT(46,-2,10)rec[-4]_rec[-3]_rec[-16];DT(50,-2,10)rec[-3]_rec[-2]_rec[-15];DT(54,-2,10)rec[-2]_rec[-1]_rec[-14];OI(0)rec[-39]_rec[-38]_rec[-42]_rec[-43]_rec[-62]_rec[-73];OI(1)rec[-37]_rec[-36]_rec[-40]_rec[-41]_rec[-60]_rec[-71];OI(2)rec[-35]_rec[-39]_rec[-43]_rec[-44]_rec[-58]_rec[-74];OI(3)rec[-34]_rec[-33]_rec[-37]_rec[-41]_rec[-42]_rec[-38]_rec[-56]_rec[-72];OI(4)rec[-32]_rec[-31]_rec[-34]_rec[-38]_rec[-39]_rec[-35]_rec[-54];OI(5)rec[-30]_rec[-36]_rec[-37]_rec[-33]_rec[-52];OI(6)rec[-29]_rec[-28]_rec[-30]_rec[-33]_rec[-34]_rec[-31]_rec[-50];OI(7)rec[-27]_rec[-29]_rec[-31]_rec[-32]_rec[-48];OI(8)rec[-26]_rec[-28]_rec[-29]_rec[-27]_rec[-46];OI(9)rec[-44]_rec[-43]_rec[-42]_rec[-41]_rec[-40]_rec[-13]_rec[-73]_rec[-71]_}{Crumble} \\
7 & 19 & 110 & \href{https://algassert.com/crumble\#circuit=Q(0,0)0;Q(4,0)1;Q(12,0)2;Q(16,0)3;Q(24,0)4;Q(28,0)5;Q(36,0)6;Q(6,1)7;Q(10,1)8;Q(18,1)9;Q(22,1)10;Q(30,1)11;Q(34,1)12;Q(4,2)13;Q(12,2)14;Q(16,2)15;Q(24,2)16;Q(28,2)17;Q(6,3)18;Q(10,3)19;Q(18,3)20;Q(22,3)21;Q(30,3)22;Q(12,4)23;Q(16,4)24;Q(24,4)25;Q(28,4)26;Q(10,5)27;Q(18,5)28;Q(22,5)29;Q(12,6)30;Q(16,6)31;Q(24,6)32;Q(18,7)33;Q(22,7)34;Q(16,8)35;Q(18,9)36;Q(0,-2)37;Q(4,-2)38;Q(12,-2)39;Q(16,-2)40;Q(24,-2)41;Q(28,-2)42;Q(36,-2)43;Q(40,-2)44;Q(44,-2)45;Q(48,-2)46;Q(52,-2)47;Q(56,-2)48;Q(60,-2)49;Q(64,-2)50;Q(68,-2)51;Q(72,-2)52;Q(76,-2)53;Q(80,-2)54;Q(84,-2)55;Q(9,0)56;Q(21,0)57;Q(33,0)58;Q(3,1)59;Q(15,1)60;Q(27,1)61;Q(9,2)62;Q(21,2)63;Q(33,2)64;Q(15,3)65;Q(27,3)66;Q(9,4)67;Q(21,4)68;Q(15,5)69;Q(27,5)70;Q(21,6)71;Q(15,7)72;Q(21,8)73;Q(7,0)74;Q(19,0)75;Q(31,0)76;Q(1,1)77;Q(13,1)78;Q(25,1)79;Q(7,2)80;Q(19,2)81;Q(31,2)82;Q(13,3)83;Q(25,3)84;Q(7,4)85;Q(19,4)86;Q(13,5)87;Q(25,5)88;Q(19,6)89;Q(13,7)90;Q(19,8)91;Q(2,-2)92;Q(8,-2)93;Q(14,-2)94;Q(20,-2)95;Q(26,-2)96;Q(32,-2)97;Q(38,-2)98;Q(42,-2)99;Q(46,-2)100;Q(50,-2)101;Q(54,-2)102;Q(58,-2)103;Q(62,-2)104;Q(66,-2)105;Q(70,-2)106;Q(74,-2)107;Q(78,-2)108;Q(82,-2)109;R_37_38_39_40_41_42_43_44_45_46_47_48_49_50_51_52_53_54_55;RX_56;R_74;RX_57;R_75;RX_58;R_76;RX_59;R_77;RX_60;R_78;RX_61;R_79;RX_62;R_80;RX_63;R_81;RX_64;R_82;RX_65;R_83;RX_66;R_84;RX_67;R_85;RX_68;R_86;RX_69;R_87;RX_70;R_88;RX_71;R_89;RX_72;R_90;RX_73;R_91;RX_92_93_94_95_96_97_98_99_100_101_102_103_104_105_106_107_108_109;TICK;RX_0_1_2_3_4_5_6_7_8_9_10_11_12_13_14_15_16_17_18_19_20_21_22_23_24_25_26_27_28_29_30_31_32_33_34_35_36;CX_56_74_57_75_58_76_59_77_60_78_61_79_62_80_63_81_64_82_65_83_66_84_67_85_68_86_69_87_70_88_71_89_72_90_73_91_92_37_93_38_94_39_95_40_96_41_97_42_98_43_99_44_100_45_101_46_102_47_103_48_104_49_105_50_106_51_107_52_108_53_109_54;TICK;CX_1_59_0_77_3_60_2_78_5_61_4_79_8_62_7_80_10_63_9_81_12_64_11_82_15_65_14_83_17_66_16_84_19_67_18_85_21_68_20_86_24_69_23_87_26_70_25_88_29_71_28_89_31_72_30_90_34_73_33_91_92_38_93_39_94_40_95_41_96_42_97_43_98_44_99_45_100_46_101_47_102_48_103_49_104_50_105_51_106_52_107_53_108_54_109_55;TICK;CX_2_56_1_74_4_57_3_75_6_58_5_76_7_59_9_60_8_78_11_61_10_79_14_62_13_80_16_63_15_81_17_82_20_65_19_83_22_66_21_84_23_67_25_68_24_86_28_69_27_87_29_88_32_71_31_89_33_72_35_91;MX_92_93_94_95_96_97_98_99_100_101_102_103_104_105_106_107_108_109;TICK;CX_0_37_1_38_2_39_3_40_4_41_5_42_6_43_7_74_8_56_9_75_10_57_11_76_12_58_13_59_14_78_15_60_16_79_17_61_18_80_19_62_20_81_21_63_22_82_23_83_24_65_25_84_26_66_27_67_28_86_29_68_30_87_31_69_32_88_33_89_34_71_35_72_36_91;RX_92_93_94_95_96_97_98_99_100_101_102_103_104_105_106_107_108_109;TICK;SQRT_X_0_1_2_3_4_5_6_7_8_9_10_11_12_13_15_17_18_19_20_21_22_23_24_25_26_27_29_30_31_32_33_34_35_36;SQRT_X_DAG_14_16_28;CX_56_74_57_75_58_76_59_77_60_78_61_79_62_80_63_81_64_82_65_83_66_84_67_85_68_86_69_87_70_88_71_89_72_90_73_91_92_37_93_38_94_39_95_40_96_41_97_42_98_43_99_44_100_45_101_46_102_47_103_48_104_49_105_50_106_51_107_52_108_53_109_54;TICK;M_74;MX_56;M_75;MX_57;M_76;MX_58;M_77;MX_59;M_78;MX_60;M_79;MX_61;M_80;MX_62;M_81;MX_63;M_82;MX_64;M_83;MX_65;M_84;MX_66;M_85;MX_67;M_86;MX_68;M_87;MX_69;M_88;MX_70;M_89;MX_71;M_90;MX_72;M_91;MX_73;CX_92_38_93_39_94_40_95_41_96_42_97_43_98_44_99_45_100_46_101_47_102_48_103_49_104_50_105_51_106_52_107_53_108_54_109_55;DT(7,0,6)rec[-35];DT(19,0,6)rec[-33];DT(31,0,6)rec[-31];DT(1,1,6)rec[-29];DT(13,1,6)rec[-27];DT(25,1,6)rec[-25];DT(7,2,6)rec[-23];DT(19,2,6)rec[-21];DT(31,2,6)rec[-19];DT(13,3,6)rec[-17];DT(25,3,6)rec[-15];DT(7,4,6)rec[-13];DT(19,4,6)rec[-11];DT(13,5,6)rec[-9];DT(25,5,6)rec[-7];DT(19,6,6)rec[-5];DT(13,7,6)rec[-3];DT(19,8,6)rec[-1];M_0_1_2_3_4_5_6_7_8_9_10_11_12_13_14_15_16_17_18_19_20_21_22_23_24_25_26_27_28_29_30_31_32_33_34_35_36;TICK;MX_92_93_94_95_96_97_98_99_100_101_102_103_104_105_106_107_108_109;DT(2,-2,7)rec[-18]_rec[-109];DT(8,-2,7)rec[-17]_rec[-108];DT(14,-2,7)rec[-16]_rec[-107];DT(20,-2,7)rec[-15]_rec[-106];DT(26,-2,7)rec[-14]_rec[-105];DT(32,-2,7)rec[-13]_rec[-104];DT(38,-2,7)rec[-12]_rec[-103];DT(42,-2,7)rec[-11]_rec[-102];DT(46,-2,7)rec[-10]_rec[-101];DT(50,-2,7)rec[-9]_rec[-100];DT(54,-2,7)rec[-8]_rec[-99];DT(58,-2,7)rec[-7]_rec[-98];DT(62,-2,7)rec[-6]_rec[-97];DT(66,-2,7)rec[-5]_rec[-96];DT(70,-2,7)rec[-4]_rec[-95];DT(74,-2,7)rec[-3]_rec[-94];DT(78,-2,7)rec[-2]_rec[-93];DT(82,-2,7)rec[-1]_rec[-92];TICK;S_37_38_39_40_41_42_43_44_45_46_47_48_49_50_51_52_53_54_55;TICK;CZ_37_38_37_39_37_40_37_41_37_42_37_43_37_44_37_45_37_46_37_47_37_48_37_49_37_50_37_51_37_52_37_53_37_54_37_55_38_39_38_40_38_41_38_42_38_43_38_44_38_45_38_46_38_47_38_48_38_49_38_50_38_51_38_52_38_53_38_54_38_55_39_40_39_41_39_42_39_43_39_44_39_45_39_46_39_47_39_48_39_49_39_50_39_51_39_52_39_53_39_54_39_55_40_41_40_42_40_43_40_44_40_45_40_46_40_47_40_48_40_49_40_50_40_51_40_52_40_53_40_54_40_55_41_42_41_43_41_44_41_45_41_46_41_47_41_48_41_49_41_50_41_51_41_52_41_53_41_54_41_55_42_43_42_44_42_45_42_46_42_47_42_48_42_49_42_50_42_51_42_52_42_53_42_54_42_55_43_44_43_45_43_46_43_47_43_48_43_49_43_50_43_51_43_52_43_53_43_54_43_55_44_45_44_46_44_47_44_48_44_49_44_50_44_51_44_52_44_53_44_54_44_55_45_46_45_47_45_48_45_49_45_50_45_51_45_52_45_53_45_54_45_55_46_47_46_48_46_49_46_50_46_51_46_52_46_53_46_54_46_55_47_48_47_49_47_50_47_51_47_52_47_53_47_54_47_55_48_49_48_50_48_51_48_52_48_53_48_54_48_55_49_50_49_51_49_52_49_53_49_54_49_55_50_51_50_52_50_53_50_54_50_55_51_52_51_53_51_54_51_55_52_53_52_54_52_55_53_54_53_55_54_55;TICK;MX_37_38_39_40_41_42_43_44_45_46_47_48_49_50_51_52_53_54_55;TICK;DT(2,-2,10)rec[-19]_rec[-18]_rec[-37];DT(8,-2,10)rec[-18]_rec[-17]_rec[-36];DT(14,-2,10)rec[-17]_rec[-16]_rec[-35];DT(20,-2,10)rec[-16]_rec[-15]_rec[-34];DT(26,-2,10)rec[-15]_rec[-14]_rec[-33];DT(32,-2,10)rec[-14]_rec[-13]_rec[-32];DT(38,-2,10)rec[-13]_rec[-12]_rec[-31];DT(42,-2,10)rec[-12]_rec[-11]_rec[-30];DT(46,-2,10)rec[-11]_rec[-10]_rec[-29];DT(50,-2,10)rec[-10]_rec[-9]_rec[-28];DT(54,-2,10)rec[-9]_rec[-8]_rec[-27];DT(58,-2,10)rec[-8]_rec[-7]_rec[-26];DT(62,-2,10)rec[-7]_rec[-6]_rec[-25];DT(66,-2,10)rec[-6]_rec[-5]_rec[-24];DT(70,-2,10)rec[-5]_rec[-4]_rec[-23];DT(74,-2,10)rec[-4]_rec[-3]_rec[-22];DT(78,-2,10)rec[-3]_rec[-2]_rec[-21];DT(82,-2,10)rec[-2]_rec[-1]_rec[-20];OI(0)rec[-67]_rec[-66]_rec[-72]_rec[-73]_rec[-110]_rec[-127];OI(1)rec[-65]_rec[-64]_rec[-70]_rec[-71]_rec[-108]_rec[-125];OI(2)rec[-63]_rec[-62]_rec[-68]_rec[-69]_rec[-106]_rec[-123];OI(3)rec[-61]_rec[-67]_rec[-73]_rec[-74]_rec[-104]_rec[-128];OI(4)rec[-60]_rec[-59]_rec[-65]_rec[-71]_rec[-72]_rec[-66]_rec[-102]_rec[-126];OI(5)rec[-58]_rec[-57]_rec[-63]_rec[-69]_rec[-70]_rec[-64]_rec[-100]_rec[-124];OI(6)rec[-56]_rec[-55]_rec[-60]_rec[-66]_rec[-67]_rec[-61]_rec[-98];OI(7)rec[-54]_rec[-53]_rec[-58]_rec[-64]_rec[-65]_rec[-59]_rec[-96];OI(8)rec[-52]_rec[-62]_rec[-63]_rec[-57]_rec[-94];OI(9)rec[-51]_rec[-50]_rec[-54]_rec[-59]_rec[-60]_rec[-55]_rec[-92];OI(10)rec[-49]_rec[-48]_rec[-52]_rec[-57]_rec[-58]_rec[-53]_rec[-90];OI(11)rec[-47]_rec[-51]_rec[-55]_rec[-56]_rec[-88];OI(12)rec[-46]_rec[-45]_rec[-49]_rec[-53]_rec[-54]_rec[-50]_rec[-86];OI(13)rec[-44]_rec[-43]_rec[-46]_rec[-50]_rec[-51]_rec[-47]_rec[-84];OI(14)rec[-42]_rec[-48]_rec[-49]_rec[-45]_rec[-82];OI(15)rec[-41]_rec[-40]_rec[-42]_rec[-45]_rec[-46]_rec[-43]_rec[-80];OI(16)rec[-39]_rec[-41]_rec[-43]_rec[-44]_rec[-78];OI(17)rec[-38]_rec[-40]_rec[-41]_rec[-39]_rec[-76];OI(18)rec[-74]_rec[-73]_rec[-72]_rec[-71]_rec[-70]_rec[-69]_rec[-68]_rec[-19]_rec[-127]_rec[-125]_rec[-123]_}{Crumble} \\
\bottomrule
\end{tabular}

%% file: graphics/istate_distillation/tables/femoco_istate_protocol_types.tex
\begin{tabular}{cc}
\toprule
type & $k$ \\
\midrule
short & 1 \\
medium & 4 \\
long & 8 \\
\bottomrule
\end{tabular}

%% file: graphics/compilation_graphics/tikz/eq_CSWAP.tex
\usetikzlibrary{arrows,shapes.geometric,decorations.markings,%
                decorations.pathmorphing,backgrounds,fit,shapes}
\pgfdeclarelayer{edgelayer}
\pgfdeclarelayer{nodelayer}
\pgfsetlayers{background,edgelayer,nodelayer,main}
\tikzstyle{every picture}=[scale=0.5]
\tikzstyle{none}=[inner sep=0mm]
\tikzstyle{every loop}=[]
\tikzstyle{Z dot}=[fill={rgb,255: red,221; green,255; blue,221}, draw=black,
                   shape=circle, minimum size=0.2cm, inner sep=0pt]
\tikzstyle{X dot}=[fill={rgb,255: red,255; green,136; blue,136}, draw=black,
                   shape=circle, minimum size=0.2cm, inner sep=0pt]
\tikzstyle{hadamard}=[fill={rgb,255: red,255; green,242; blue,0}, draw=black,
                   shape=rectangle, minimum size=0.17cm, inner sep=0pt]
\tikzstyle{grey dash}=[-, dashed, draw={rgb,255: red,191; green,191; blue,191}]
\tikzstyle{blue_edge}=[-, draw={rgb,255: red,0; green,0; blue,220}]
\tikzstyle{brace edge}=[-, draw={rgb,255: red,0; green,0; blue,220}]
\tikzstyle{red_dash}=[-, dashed, draw={rgb,255: red,220; green,0; blue,0}]

\tikzstyle{Z phase dot}=[fill={rgb,255: red,221; green,255; blue,221}, draw=black,
                   shape=circle, minimum size=0.2cm, inner sep=0.6pt, font=\footnotesize]
\tikzstyle{X phase dot}=[fill={rgb,255: red,255; green,136; blue,136}, draw=black,
                   shape=circle, minimum size=0.2cm, inner sep=0.6pt, font=\footnotesize]
\newcommand{\zxscale}{0.8}
\newsavebox{\zxA}
\newsavebox{\zxB}
\newsavebox{\zxC}
\newcommand{\zxeq}[1]{\;\raisebox{0.5\ht#1}{$~=~$}\;}

\sbox{\zxA}{\scalebox{\zxscale}{\input{graphics/compilation_graphics/tikz/ctrl-SWAP-gate-no-corr-cnot.tikz}}}%
\sbox{\zxB}{\scalebox{\zxscale}{\input{graphics/compilation_graphics/tikz/ctrl-SWAP-gate-no-corr.tikz}}}%
\sbox{\zxC}{\scalebox{\zxscale}{\input{graphics/compilation_graphics/tikz/ctrl_swap_corrections.tikz}}}%
\begin{equation}
\begin{split}
  \usebox{\zxA} &\zxeq{\zxA} \usebox{\zxB} \\[2ex]
                &\zxeq{\zxC} \usebox{\zxC}
\end{split}
\label{eq:cswap-inject}
\end{equation}

%% file: graphics/compilation_graphics/tikz/ctrl-SWAP-gate-no-corr-cnot.tikz
\scalebox{0.8}{
\begin{tikzpicture}
	\begin{pgfonlayer}{nodelayer}
		\node [style=none] (0) at (19, 4) {};
		\node [style=hadamard] (1) at (8, -2) {};
		\node [style=none] (4) at (19, 6) {};
		\node [style=hadamard] (5) at (10, -4) {};
		\node [style=hadamard] (7) at (8, 0) {};
		\node [style=Z dot] (9) at (19, -2) {};
		\node [style=none] (10) at (8, 6) {};
		\node [style=X dot] (11) at (19, -4) {};
		\node [style=X dot] (12) at (16, -4) {};
		\node [style=Z dot] (13) at (19, 0) {};
		\node [style=hadamard] (14) at (8, -4) {};
		\node [style=none] (15) at (19, 2) {};
		\node [style=Z dot] (16) at (12, 0) {};
		\node [style=X dot] (17) at (14, 4) {};
		\node [style=none] (18) at (8, 4) {};
		\node [style=X dot] (19) at (12, 6) {};
		\node [style=hadamard] (20) at (5.75, -2) {};
		\node [style=Z dot] (22) at (16, 2) {};
		\node [style=Z dot] (23) at (14, -2) {};
		\node [style=none] (24) at (8, 2) {};
		\node [style=X dot] (25) at (10, 2) {};
		\node [style=X dot] (26) at (18, 2) {};
		\node [style=Z dot] (27) at (18, 4) {};
		\node [style=Z dot] (28) at (10, 4) {};
		\node [style=none] (29) at (11, 7) {};
		\node [style=none] (30) at (11, 1) {};
		\node [style=none] (31) at (17, 1) {};
		\node [style=none] (32) at (17, 7) {};
		\node [style=none] (33) at (9.5, 4.5) {};
		\node [style=none] (34) at (10.5, 4.5) {};
		\node [style=none] (35) at (10.5, 1.5) {};
		\node [style=none] (36) at (9.5, 1.5) {};
		\node [style=none] (37) at (17.5, 4.5) {};
		\node [style=none] (38) at (18.5, 4.5) {};
		\node [style=none] (39) at (18.5, 1.5) {};
		\node [style=none] (40) at (17.5, 1.5) {};
		\node [style=none] (41) at (10, 5.25) {\textsc{Cnot}};
		\node [style=none] (42) at (14, 7.75) { Injection};
		\node [style=none] (43) at (18, 5.25) {\textsc{Cnot}};
		\node [style=none] (44) at (5.25, 0.5) {};
		\node [style=none] (45) at (8.5, 0.5) {};
		\node [style=none] (46) at (8.5, -4.5) {};
		\node [style=none] (47) at (5.25, -4.5) {};
		\node [style=none] (48) at (6.75, 1.25) {$\ket{C_XC_XX}$};
	\end{pgfonlayer}
	\begin{pgfonlayer}{edgelayer}
		\draw (0.center) to (27);
		\draw (1) to (20);
		\draw (4.center) to (19);
		\draw (5) to (12);
		\draw (7) to (20);
		\draw (9) to (23);
		\draw (10.center) to (19);
		\draw (11) to (12);
		\draw (12) to (22);
		\draw (13) to (16);
		\draw (14) to (20);
		\draw (15.center) to (26);
		\draw (16) to (19);
		\draw (17) to (23);
		\draw (17) to (27);
		\draw (17) to (28);
		\draw (18.center) to (28);
		\draw (22) to (25);
		\draw (22) to (26);
		\draw (24.center) to (25);
		\draw (25) to (28);
		\draw (26) to (27);
		\draw [style=grey dash] (33.center) to (34.center);
		\draw [style=grey dash] (34.center) to (35.center);
		\draw [style=grey dash] (35.center) to (36.center);
		\draw [style=grey dash] (36.center) to (33.center);
		\draw [style=grey dash] (30.center) to (29.center);
		\draw [style=grey dash] (29.center) to (32.center);
		\draw [style=grey dash] (32.center) to (31.center);
		\draw [style=grey dash] (31.center) to (30.center);
		\draw [style=grey dash] (37.center) to (38.center);
		\draw [style=grey dash] (38.center) to (39.center);
		\draw [style=grey dash] (39.center) to (40.center);
		\draw [style=grey dash] (40.center) to (37.center);
		\draw [style=brace edge] (44.center) to (45.center);
		\draw [style=brace edge] (45.center) to (46.center);
		\draw [style=brace edge] (46.center) to (47.center);
		\draw [style=brace edge] (47.center) to (44.center);
		\draw (7) to (16);
		\draw (23) to (1);
		\draw (14) to (5);
	\end{pgfonlayer}
\end{tikzpicture}
}

%% file: graphics/compilation_graphics/tikz/ctrl-SWAP-gate-no-corr.tikz
\scalebox{0.8}{
\begin{tikzpicture}
	\begin{pgfonlayer}{nodelayer}
		\node [style=X dot] (1) at (12, 4) {};
		\node [style=hadamard] (4) at (8, -4) {};
		\node [style=Z dot] (6) at (17, 0) {};
		\node [style=hadamard] (7) at (6, -2) {};
		\node [style=none] (8) at (6, 6) {};
		\node [style=X dot] (9) at (10, 6) {};
		\node [style=X dot] (10) at (17, -4) {};
		\node [style=none] (12) at (17, 6) {};
		\node [style=X dot] (13) at (14, -4) {};
		\node [style=none] (14) at (17, 4) {};
		\node [style=Z dot] (15) at (14, 2) {};
		\node [style=none] (18) at (17, 2) {};
		\node [style=Z dot] (19) at (17, -2) {};
		\node [style=hadamard] (20) at (6, 0) {};
		\node [style=hadamard] (21) at (6, -4) {};
		\node [style=Z dot] (23) at (10, 0) {};
		\node [style=none] (25) at (6, 2) {};
		\node [style=Z dot] (26) at (12, -2) {};
		\node [style=X dot] (28) at (12, 2) {};
		\node [style=hadamard] (29) at (3.75, -2) {};
		\node [style=Z dot] (30) at (14, 4) {};
		\node [style=none] (32) at (6, 4) {};
		\node [style=none] (33) at (3.25, 0.5) {};
		\node [style=none] (34) at (6.5, 0.5) {};
		\node [style=none] (35) at (6.5, -4.5) {};
		\node [style=none] (36) at (3.25, -4.5) {};
		\node [style=none] (37) at (4.75, 1.25) {$\ket{C_XC_XX}$};
	\end{pgfonlayer}
	\begin{pgfonlayer}{edgelayer}
		\draw (1) to (14.center);
		\draw (1) to (30);
		\draw [bend left] (1) to (26);
		\draw (4) to (13);
		\draw (6) to (23);
		\draw (7) to (29);
		\draw (8.center) to (9);
		\draw (9) to (12.center);
		\draw (9) to (23);
		\draw (10) to (13);
		\draw [bend right=15] (13) to (15);
		\draw [bend right] (13) to (30);
		\draw (15) to (25.center);
		\draw (15) to (28);
		\draw (18.center) to (28);
		\draw (19) to (26);
		\draw (20) to (29);
		\draw (21) to (29);
		\draw [bend right=15] (26) to (28);
		\draw (30) to (32.center);
		\draw [style=brace edge] (33.center) to (34.center);
		\draw [style=brace edge] (34.center) to (35.center);
		\draw [style=brace edge] (35.center) to (36.center);
		\draw [style=brace edge] (36.center) to (33.center);
		\draw (20) to (23);
		\draw (26) to (7);
		\draw (21) to (4);
	\end{pgfonlayer}
\end{tikzpicture}
}

%% file: graphics/compilation_graphics/tikz/ctrl_swap_corrections.tikz
\scalebox{0.8}{
\begin{tikzpicture}
	\begin{pgfonlayer}{nodelayer}
		\node [style=hadamard] (11) at (1.75, -4) {};
		\node [style=hadamard] (12) at (1.75, -2) {};
		\node [style=hadamard] (13) at (1.75, 0) {};
		\node [style=hadamard] (14) at (-0.25, -2) {};
		\node [style=none] (22) at (2, 2) {};
		\node [style=none] (23) at (17.75, 2) {};
		\node [style=none] (24) at (17.75, 4) {};
		\node [style=none] (25) at (2, 4) {};
		\node [style=none] (26) at (2, 6) {};
		\node [style=none] (27) at (17.75, 6) {};
		\node [style=X dot] (28) at (9.75, 6) {};
		\node [style=X dot] (29) at (13.75, -4) {};
		\node [style=X dot] (30) at (11.75, 4) {};
		\node [style=X dot] (31) at (11.75, 2) {};
		\node [style=Z dot] (33) at (9.75, 0) {};
		\node [style=Z dot] (34) at (13.75, 4) {};
		\node [style=Z dot] (35) at (13.75, 2) {};
		\node [style=X dot] (36) at (3.75, -2) {};
		\node [style=X dot] (40) at (5.25, 0) {};
		\node [style=Z dot] (41) at (11.75, -2) {};
		\node [style=X dot] (43) at (6.5, 0) {};
		\node [style=X dot] (44) at (6.5, -2) {};
		\node [style=hadamard] (45) at (7, -1) {};
		\node [style=none] (46) at (3.25, -1.5) {};
		\node [style=none] (47) at (4.25, -1.5) {};
		\node [style=none] (48) at (4.25, -4.5) {};
		\node [style=none] (49) at (3.25, -4.5) {};
		\node [style=none] (50) at (4.75, -4.5) {};
		\node [style=none] (51) at (5.75, -4.5) {};
		\node [style=none] (52) at (6.25, -2.5) {};
		\node [style=none] (53) at (7.25, -2.5) {};
		\node [style=none] (54) at (5.75, 0.5) {};
		\node [style=none] (55) at (4.75, 0.5) {};
		\node [style=none] (56) at (6.25, 0.5) {};
		\node [style=none] (57) at (7.25, 0.5) {};
		\node [style=hadamard] (58) at (2.75, -4) {};
		\node [style=Z dot] (59) at (5.25, -4) {};
		\node [style=Z dot] (60) at (3.75, -4) {};
		\node [style=none] (61) at (19, 0) {$m_1$};
		\node [style=none] (62) at (19, -2) {$m_2$};
		\node [style=none] (63) at (19, -4) {$m_3$};
		\node [style=none] (64) at (3.75, -5) {$m_1$};
		\node [style=none] (65) at (5.25, -5) {$m_2$};
		\node [style=none] (66) at (6.75, -3) {$m_3$};
		\node [style=Z dot] (67) at (18, 0) {};
		\node [style=Z dot] (68) at (18, -2) {};
		\node [style=X dot] (70) at (18, -4) {};
		\node [style=none] (71) at (-0.75, 0.5) {};
		\node [style=none] (72) at (2.25, 0.5) {};
		\node [style=none] (73) at (2.25, -4.5) {};
		\node [style=none] (74) at (-0.75, -4.5) {};
		\node [style=none] (75) at (0.75, 1.25) {$\ket{C_XC_XX}$};
	\end{pgfonlayer}
	\begin{pgfonlayer}{edgelayer}
		\draw (11) to (14);
		\draw (12) to (14);
		\draw (13) to (14);
		\draw (22.center) to (35);
		\draw (23.center) to (31);
		\draw (24.center) to (30);
		\draw (25.center) to (34);
		\draw (26.center) to (28);
		\draw (27.center) to (28);
		\draw [bend left=15] (28) to (33);
		\draw [bend right=15] (29) to (35);
		\draw [bend right] (29) to (34);
		\draw (30) to (34);
		\draw [bend left] (30) to (41);
		\draw (31) to (35);
		\draw [bend left=15] (31) to (41);
		\draw (33) to (43);
		\draw (36) to (44);
		\draw (40) to (43);
		\draw (41) to (44);
		\draw (43) to (45);
		\draw (44) to (45);
		\draw [style=grey dash] (47.center) to (48.center);
		\draw [style=grey dash] (48.center) to (49.center);
		\draw [style=grey dash] (49.center) to (46.center);
		\draw [style=grey dash] (46.center) to (47.center);
		\draw [style=grey dash] (50.center) to (55.center);
		\draw [style=grey dash] (55.center) to (54.center);
		\draw [style=grey dash] (54.center) to (51.center);
		\draw [style=grey dash] (51.center) to (50.center);
		\draw [style=grey dash] (52.center) to (53.center);
		\draw [style=grey dash] (53.center) to (57.center);
		\draw [style=grey dash] (57.center) to (56.center);
		\draw [style=grey dash] (56.center) to (52.center);
		\draw (58) to (60);
		\draw (60) to (59);
		\draw (29) to (59);
		\draw [bend right=15] (59) to (40);
		\draw [bend right=15] (60) to (36);
		\draw (68) to (41);
		\draw (67) to (33);
		\draw (29) to (70);
		\draw [style=brace edge] (71.center) to (72.center);
		\draw [style=brace edge] (72.center) to (73.center);
		\draw [style=brace edge] (73.center) to (74.center);
		\draw [style=brace edge] (74.center) to (71.center);
		\draw (13) to (40);
		\draw (36) to (12);
		\draw (11) to (58);
	\end{pgfonlayer}
\end{tikzpicture}
}

%% file: graphics/compilation_graphics/tikz/eq_C_XX.tex
\begin{equation}
  \vcenter{\hbox{\scalebox{\zxscale}{\input{graphics/compilation_graphics/tikz/C_XX_correction.tikz}}}}
  \label{eq:cxx-correction}
\end{equation}

%% file: graphics/compilation_graphics/tikz/C_XX_correction.tikz
\begin{tikzpicture}
	\begin{pgfonlayer}{nodelayer}
		\node [style=none] (0) at (-8, 2) {};
		\node [style=none] (1) at (-8, 0) {};
		\node [style=none] (2) at (-1.75, 2) {};
		\node [style=none] (3) at (-1.75, 0) {};
		\node [style=none] (5) at (-9.25, -2) {$-\pi/2$};
		\node [style=Z dot] (8) at (-6.25, -2) {};
		\node [style=X dot] (9) at (-6.25, 0) {};
		\node [style=X dot] (10) at (-6.25, 2) {};
		\node [style=none] (11) at (-0.25, 0) {$=$};
		\node [style=none] (12) at (1, 2) {};
		\node [style=none] (13) at (1, 0) {};
		\node [style=none] (14) at (6, 2) {};
		\node [style=none] (15) at (6, 0) {};
		\node [style=X dot] (16) at (2.75, 0) {};
		\node [style=X dot] (17) at (2.75, 2) {};
		\node [style=Z dot] (18) at (2.75, 1) {};
		\node [style=none] (19) at (1.75, 1) {$3\pi/2$};
		\node [style=none] (20) at (7.25, 0) {$=$};
		\node [style=none] (21) at (8.5, 2) {};
		\node [style=none] (22) at (8.5, 0) {};
		\node [style=none] (23) at (11.5, 2) {};
		\node [style=none] (24) at (11.75, 0) {};
		\node [style=X dot] (25) at (10.25, 0) {};
		\node [style=X dot] (26) at (10.25, 2) {};
		\node [style=hadamard] (27) at (10.25, 1) {};
		\node [style=X dot] (28) at (-8, -3.25) {};
		\node [style=none] (29) at (-9.25, -3.25) {$\pi/2$};
		\node [style=Z dot] (32) at (-5, -3.25) {};
		\node [style=X dot] (33) at (-8, -4.5) {};
		\node [style=none] (34) at (-9.25, -4.5) {$\pi/2$};
		\node [style=Z dot] (37) at (-3.75, -4.5) {};
		\node [style=X dot] (38) at (-5, 0) {};
		\node [style=X dot] (39) at (-3.75, 2) {};
		\node [style=X dot] (40) at (4, 0) {};
		\node [style=X dot] (41) at (5, 2) {};
		\node [style=none] (42) at (4, -1) {$\pi/2$};
		\node [style=none] (43) at (5, 3) {$\pi/2$};
		\node [style=Z dot] (44) at (-8, -2) {};
		\node [style=Z dot] (45) at (-1.75, -2) {};
		\node [style=Z dot] (46) at (-1.75, -3.25) {};
		\node [style=Z dot] (47) at (-1.75, -4.5) {};
	\end{pgfonlayer}
	\begin{pgfonlayer}{edgelayer}
		\draw (0.center) to (2.center);
		\draw (1.center) to (3.center);
		\draw [bend right=15, looseness=0.75] (8) to (9);
		\draw [bend right=15] (8) to (10);
		\draw (12.center) to (14.center);
		\draw (13.center) to (15.center);
		\draw (17) to (18);
		\draw (18) to (16);
		\draw (21.center) to (23.center);
		\draw (22.center) to (24.center);
		\draw (26) to (27);
		\draw (27) to (25);
		\draw [bend left=15, looseness=0.75] (38) to (32);
		\draw [bend left=15, looseness=0.75] (39) to (37);
		\draw (44) to (8);
		\draw (8) to (45);
		\draw (32) to (46);
		\draw (32) to (28);
		\draw (33) to (37);
		\draw (37) to (47);
	\end{pgfonlayer}
\end{tikzpicture}

%% file: sections/physical_resources.tex
\section{Resource count for the block-encoding of the DFTHC ansatz}\label{sec: Resource Count BE}
In this section, we present a detailed resource estimation for the ground state estimation of FeMoco-54 and FeMoco-76 with the DFTHC ansatz. For the parameters of the ansatz we keep the parameters of \cite[Table XII]{low2026denser} allowing for better comparison. More details can be found at the start of \zcref{sec: rep code resource count}. We assume $d$ cycles per lattice surgery operation and do not explicitly account for initialization and measurement of routing qubits, which are stored in a repetition code in both the repetition-cat architecture as well as the LDPC-cat architecture. We also do not take buffer staging in between routines into consideration. To account for these simplifications, we add 5\% of slack to the duration of the full algorithm.  We do however, account for measurement based uncomputation of lookup table output, as this takes up about 2.7\% of the LDPC routine. 

For subroutines for which we do not precisely describe the volume of the computation, we count the physical resources as follows.
The algorithmic qubits footprint, including ancillae, is $2dn_L$ where $n_L$ is the number of logical qubits.
This count is doubled again to take into account additional routing qubits.
In this simplified estimation, all Toffoli gates are assumed to be performed serially taking $3d$ cycles each.
Since subroutines for which we do not precisely describe the volume of the computation have a small footprint compared to the global computation, this overestimation has a mild impact on the global resource estimate. On the contrary, the Clifford gates are neglected for these subroutines.
For the same reason, this underestimate should have a mild impact on the global resource estimate.
The latter affects controlled quantum-classical addition, quantum-classical comparison, and the reflection.

We make the following choices for each subroutine stated in \zcref{fig: high-level BE circ}: pure state preparation for both \textsc{OuterPrep} and \textsc{InnerPrep}, and lookup table batching with several variants for \textsc{Rprep + Rot}. Toffoli gates are executed serially using a single $\ket{\CXCXX}$-factory, for which we choose the layout with the highest production rate in \zcref{tab:factory_summary}. This still limits the execution to one Toffoli gate every 21 rounds, but introduces the smallest slowdown among the layouts.
The footprint of the algorithmic region is governed by the subroutine utilizing the most spatial resources, in our case the rotation angle loading \textsc{Rprep} together with the application of the Givens rotations \textsc{Rot}. We first discuss a reference configuration, loading the minimal number of rotation angles at a time and, in the repetition-cat architecture, without access corridors for the system qubits. Access corridors are then only present in the compute zone. Within this footprint, the remaining choices, such as whether a \textsc{SelectSwap} subroutine uses clean or dirty ancillae or the address split between \textsc{Select} and \textsc{Swap}, are made to minimize the runtime. Adding access corridors for the system qubits, enlarging the compute zone, or injecting several Toffoli gates in parallel using additional factories would trade footprint for runtime. Second, we investigate the impact of enlarging the compute zone in \zcref{sec:comparison}. A summary of the subroutines used can be found in \zcref{tab:Subroutine summary FeMoCo}.

\begin{table}[t]
    \centering
    \begin{tabularx}{\textwidth}{|X|p{1cm}||X|p{2cm}|} \hline
        High-level operation & Calls & Subroutines &  Calls    \\ \hline 
        \rowcolor{gray!10} &  & Skew-tree \textsc{Qrom} & $n_X$ \\ 
        \rowcolor{gray!10} &  & \textsc{Swap} & $n_X^\textsc{Swap}$ \\
        \rowcolor{gray!10} &  & $b_\coef$-bit \textsc{Adder} & $n_X$ \\
        \rowcolor{gray!10} \multirow{-4}{*}{\textsc{OuterPrep} (\textsc{OuterPrep}$^\dagger$)} & \multirow{-4}{*}{1 (1)} & \textsc{Phaseup} & 1 ($n_X$) \\
        & & Skew-tree \textsc{Qrom}& $n_B$ \\
        & & \textsc{Swap} & $n_B^\textsc{Swap}$\\
        & & $b_\coef$-bit \textsc{Adder} & $n_B$ \\
        \multirow{-4}{*}{\textsc{InnerPrep} (\textsc{InnerPrep}$^\dagger$)} & \multirow{-4}{*}{2 (2)} & \textsc{Phaseup} & 1 ($n_B$)\\
        \rowcolor{gray!10} &  & Quantum-classical \textsc{Comp} & 1 \\ 
        \rowcolor{gray!10} & & $H$ & 2 (1) \\ 
        \rowcolor{gray!10} \multirow{-3}{*}{\textsc{Spin} (\textsc{Spin}$^\dagger$)} & \multirow{-3}{*}{2 (2)} & \textsc{Ctrl-Swap} & $N$ ($N+1$) \\
        & & Quantum classical \textsc{Adder} & 2 \\
        & & $n_B$-sized \textsc{Swap} & 2\\
       &  & Skew-tree \textsc{Qrom} & $J + 1$ \\ 
        &  & $\lambda b_\rot$-sized \textsc{Swap} & $f(\text{variant},J)$\\
         &  & $b_\rot$-bit \textsc{Adder} & $\lambda J $ \\
         \multirow{-6}{*}{\textsc{RPrep} + \textsc{Rot} (\textsc{RPrep} +\textsc{Rot}$^\dagger$)} & \multirow{-6}{*}{2 (2)} & \textsc{Phaseup} & 1 \\
         \rowcolor{gray!10} & & Quantum-classical \textsc{Comp} & 2 \\
         \rowcolor{gray!10} & & $n_B$-bit Toffoli & 2 \\
        \rowcolor{gray!10} \multirow{-3}{*}{Majorana application} & \multirow{-3}{*}{1 (1)} & \textsc{Maj} & 1\\ 
         & & \textsc{InnerRef} & 1 \\
         \multirow{-2}{*}{Reflection} & \multirow{-2}{*}{1} & \textsc{OuterRef} & 1 \\  \hline 
    \end{tabularx}
    \caption{Overview of subroutines used in the compilation for electronic structure simulation. Counts for the adjoint are given in brackets. We omit adjoint subroutine calls that match the non-adjoint version. For \textsc{OuterPrep}, $n_X^\textsc{Swap} \leq n_X$ denotes the number of lookup tables that are costed with clean \textsc{SelectSwap} to minimize cycle time. For small address values, plain skew-tree \textsc{Qrom} is used. For \textsc{InnerPrep}, either dirty or clean \textsc{SelectSwap} is used depending on the architecture and therefore $n_B \leq n_B^\textsc{Swap} \leq 4n_B$. The \textsc{RPrep} and \textsc{Rot} subroutine is costed with lookup table batching. Here, $J = \big\lceil \frac{(N-1)}{\lambda}\big\rceil$ with $\lambda$ being a chosen parameter determined by how many clean ancilla qubits are available. Depending on the layout, lookup table batching is either solely given by \textsc{Qrom} application or uses a \textsc{Swap} network with clean or dirty ancilla. This variability is reflected in $f(\text{variant},J)$. For example, $f(\textsc{Qrom},J) = 0$, while for recycling batching using a dirty \textsc{SelectSwap} with access to only the spin-up system qubits as dirty ancillae, $f(\text{restricted access},J) = J+1$. For the reflection, \textsc{InnerRef} describes the reflection necessary inside the block encoding, whereas \textsc{OuterRef} denotes the reflection performed in between block encoding applications.}\label{tab:Subroutine summary FeMoCo}
\end{table}

\begin{figure}[!htb]
    \centering
    \includegraphics[width=\linewidth]{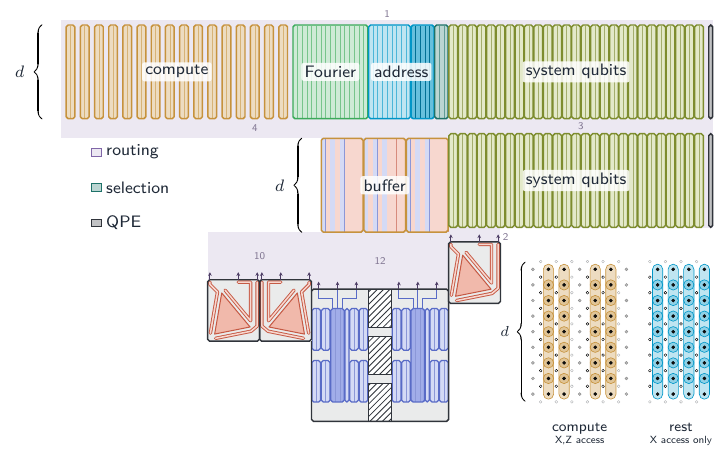}
    \caption{Global layout for the block encoding algorithm for FeMoco-54 in the reference configuration: Light blue and teal denote address qubits, gold denotes the compute zone, olive shows the system qubits, green the Fourier resource state storage and red/blue blocks denote buffer zones.}
    \label{fig:global layout}
\end{figure}

\subsection{Repetition-cat architecture}\label{sec: Rep-cat architecture}

\subsubsection{Global layout}
The global layout is governed by the layout of the most frequently used subroutines \textsc{Qrom}, \textsc{Swap}, and \textsc{Adder}. We imagine the chip as two stacked narrow rectangles each consisting of one row of logical qubits and additional rows of routing qubits on the top and bottom of the logical qubits respectively. We keep the convention from \zcref{sec:compilation} and report counts without explicitly accounting for ancilla qubits. The final physical qubit count is then obtained via simply doubling. A schematic of the global layout for FeMoco-54 in the repetition-cat architecture is given in \zcref{fig:global layout}. The width of the top rectangle can be split into four zones: the compute zone to the left, where \textsc{Qrom} output is stored and phase gradient addition is performed, followed by the Fourier resource state storage zone, the address zone in the middle, where the $X$ and $B$ registers as well as spin selection qubits are stored, and lastly a system storage zone to the right, which gives access to the system qubits and also allocates space for one qubit for quantum phase estimation (QPE). The Fourier resource state is prepared with $b_\rot$ precision and, since it is a product state, can be used for lower precision phase gradient addition as well. The bottom rectangle's width is determined by the largest buffer zone necessary, which in our case is given by \textsc{Adder}, and the other half of the system qubits and an additional QPE qubit.  We account for one row of routing on the top of the chip, spanning the full width of the top rectangle. The two rectangles are separated by either three or four rows of qubits as a routing zone depending on which zones they separate. More precisely, the two halves of system qubits and the QPE qubits are stored on top of each other with 3 rows of routing separating them. In contrast the compute zone, Fourier resource state storage, address and spin selection qubits carry four rows of routing on the bottom. The width of the compute zone is governed by the largest subroutine \textsc{Rprep + Rot}, which we determine in \zcref{sec: Rep-code detailed description of ops}. In full generality, we obtain the following count for the algorithmic core of the chip:
\begin{equation*}
    \textsc{Space}_\text{Algo}^\text{rep} = (w_\text{comp} + b_\rot + n_X + n_{B+1} + 3)(d+5) + (N+2)(d+4) + (N+2)d + 27d, 
\end{equation*}
where $w_\text{comp}$ denotes the width of the compute zone, $b_\rot$ describes the contribution of Fourier resource state storage, $n_X + n_{B+1} + 3$ gives the width of the address registers as well as the two spin selection qubits plus an additional slot for a flag qubit necessary in \textsc{Rprep + Rot}, $N$ accounts for the number of system qubits with the additional +2 accounting for a QPE qubit and an additional access corridor, and lastly $27d$ is the buffer zone necessary for \textsc{Adder}. 

We choose to use one $\ket{\CXCXX}$-factory of minimal space-time volume given in \zcref{fig:factories_layout_3} with a footprint of 1566 qubits as stated in \zcref{tab:factory_summary}. For the $\keti$-factories, we choose factories with $d_C = 7$ using the medium protocol of \zcref{subsec:i state factory} with error rates stated in \zcref{tab:istate_ler}.
The number of $\keti$-factories is determined by the need of the compiled subroutines, with \textsc{Swap} requiring the most $\keti$-states with 4 $\keti$-states at a time.
Thus, we add three $\keti$-state factories with three output repetition codes, leading to a total of $3\cdot 196 = 588$ qubits, where the added qubit count in the $\keti$-factories comes from taking the layout with three outputs into account. The factory placement and the routing area between the buffer and the factories is extended from the routing zone necessary for \textsc{Adder} as described in \cite{Shor2026prep}. More specifically, we place two $\keti$-factories to the left of the $\ket{\CXCXX}$-factory with 10 rows of routing on top, the routing area above the $\ket{\CXCXX}$ factory is additionally enlarged to 12 rows in total. The last $\keti$-factory is placed to the right of the $\ket{\CXCXX}$ factory with 2 rows of routing on top.  Since each $\keti$-factory has width 11 and the $\ket{\CXCXX}$ is 29 qubits wide, we obtain a total footprint of $2(22\cdot 10 + 29\cdot12 + 2\cdot 11) = 2\cdot590 = 1180$.  Overall, we obtain the total physical footprint
\begin{equation}\label{eq:Rep code phys footprint}
     \textsc{Space}_\text{tot}^\text{rep} =  2\textsc{Space}_\text{Algo}^\text{rep} + \underbrace{1180}_\text{factory routing} + \underbrace{2154}_\text{factories} =  2\textsc{Space}_\text{Algo}^\text{rep} +  3334.
\end{equation}

\subsubsection{Detailed description of operations}\label{sec: Rep-code detailed description of ops}
To facilitate access to $r \in [R]$, $c \in [C]$ or $n \in [N]$ in the flattened index $x$, the $r$ component is padded to the next power of two, $2^{n_R}$. Values $(r,c) \in [R]\times[C]$ are stored in the lower values of $x$ and values $n \in [N]$ are stored in the higher values of $x$. This leads to the following flattened index scheme, introduced in \cite{low2026denser},
\begin{equation}
    x(g,n,r,c) = \begin{cases}
        r + 2^{n_R}c, \qquad &g = \SF, \\
        2^{n_R}C + n, \qquad &g \in \{\pos, \nega\}.
    \end{cases}
\end{equation}
With this $r$-padding, there are $X = 2^{n_R} C + N$ address values with $x(\SF, n, r,c) \in [0,2^{n_R}C), x(\pos,n, r,c) \in [2^{n_R}C, 2^{n_R}C + N_+),$ and $x(\nega,n, r,c) \in [2^{n_R}C + N_+, 2^{n_R}C + N)$. In the case of FeMoco-54 for example, $r$-padding leads to an increase from $R=10$ to $2^{n_R} = 16$.

Access to $R$, $C$ or $N$ separately is possible with additional processing. Iteration over $r \in [R]$ or $c \in [C]$ is granted by controlling on a flag qubit that stores $\ket{x < 2^{n_R}C}$. This flag qubit can be computed with a quantum-classical comparison if it has not been computed during a previous iteration. Iteration over $n \in [N]$ is granted by conjugating the unary iteration with the quantum-classical subtraction $\ket{x} \mapsto \ket{x-2^{n_R}C}$.

\paragraph{Outer state preparation.} To save the persistent ancillae that would have been necessary in \textsc{AliasSampling}, the \textsc{OuterPrep} state preparation, which prepares the $\ket{x}$ register with $X = 2^{n_r}C + N$, is implemented using pure state preparation \cite[Appendix E.4]{low2026denser}, \cite{Shende+2005, low2024trading}. This requires data loading of rotation angles and their application using phase gradient addition. The state is prepared sequentially, by performing $n_X$ iterations of a lookup table followed by phase gradient addition. For simplicity, we assume that the number of address values of the lookup table at each level is given by $2^i$, where $i = 0, \ldots, n_X-1$ even though the last stage technically only needs to load $X-2^{n_X-1}$ data values. We cost both the lookup table as well as phase gradient addition using compilation specific numbers determined previously. 

For the lookup table, we choose a \textsc{SelectSwap} circuit with clean ancilla with target register size determined by the precision of the rotation angles given by $b_\coef$. The \textsc{Select} part is implemented using our compilation of the skew-tree \textsc{Qrom}. The size of the \textsc{Swap} network is determined by the clean ancilla budget set by the output register of \textsc{RPrep}. We do not use dirty ancillae here, since the system qubits only have X access in this layout. For each \textsc{SelectSwap}, we choose the optimal number of \textsc{Swap}s to reduce the number of cycles subject to this constraint.  Observe that in the global layout in \zcref{fig:global layout}, the address register sits to the right of the compute zone. Although in the description in \zcref{sec: QROM comp rep cat-code} and \zcref{sec: Swap rep-code comp}, the address sits to the left of the compute zone, the same sequence of operations and cost still applies, since we can simply mirror the layout outlined in the previous sections.

Once the output register is loaded, we uncompute the additional \textsc{Swap} registers freeing ancillae for the following phase gradient addition. Furthermore, we need to interleave the Fourier resource state with the output register and additional ancillae in between, since this is the starting layout of our compiled \textsc{Adder} operation. Note that we only need to move $b_\coef$ qubits of the $b_\rot$ qubits of the Fourier resource state.  The correct starting layout for the addition is obtained by moving one Fourier resource state qubit and output qubit at a time utilizing the three data qubit routing zone (in grey in \zcref{fig:global layout}), which amounts to $ b_\coef d$ cycles. This could be further optimized by using simultaneous lattice surgery and smarter scheduling similar to the upcoming discussion for \textsc{Rprep} and \textsc{Rot}, but since the preparation time in \textsc{OuterPrep} and \textsc{InnerPrep} amounts to less than $0.5\%$ of the full block-encoding time, we deem its benefit negligible. Once the layout is prepared, we perform phase gradient addition using the compilation of \textsc{Adder}. Next, the Fourier resource state must be moved back into its storage zone, adding an additional $\lceil b_\coef/3 \rceil d$ cycles. Lastly, we need to account for phases incurred by MBU of the register storing the rotation angles.  In the forward implementation, all phases can be combined to a single \textsc{Phaseup} at the end, which can potentially be pushed further and combined with another \textsc{Phaseup} later in the algorithm. For the implementation of the adjoint, each \textsc{Phaseup} must be performed on the spot. For costing the \textsc{Phaseup}, see the paragraph on uncomputation below.

\paragraph{Inner state preparation.} This is again done using pure state preparation. The key difference here is that for each index pair $(r,c)$, a state $\ket{\psi_{r,c}}$ is prepared. This leads to $n_{B+1}$ lookup tables over address ranges of respective sizes $2^{n_R} C 2^i$, where $i = 0, \ldots, n_{B+1}-1$. Note that to iterate over $(r,c) \in [R]\times[C]$ from $\ket{x}$, we can use a \textsc{Qrom} over $X$ ending at the appropriate address value. For costing, we use the same reasoning as in \textsc{OuterPrep} and take into account the additional cost of iterating over $2^{n_R}C$ times more addresses. 

\paragraph{Spin considerations.} Before applying the rotations through \textsc{Rot}, we need to take into account the spin in the 2-electron SOS factors. This is implemented using a controlled swap between the system qubits controlled on the spin qubit $\ket{\sigma}$. Before we use $\ket{\sigma}$ as a control, we first apply a Hadamard gate.  Although the system qubits to be swapped are located on top of each other in the global layout, the compilation of a controlled-\textsc{Swap} gate injection, as described in \zcref{sec: Swap rep-code comp}, still applies. More precisely,  the red routing ancilla in \zcref{fig:compiled_layout_CSWAP_No_correction} for the parity-\textsc{Cnot} between the two qubits to be swapped is now laid out vertically. This might lead to a crossing between the dark green routing ancilla from the magic state to the swapped qubits and the green-yellow routing ancilla from the purple $\ket{i}$-state to the red ancilla, which can be implemented at the same time using simultaneous lattice surgery \cite{Simultaneous2026prep}. Thus, the same cost considerations still apply. Regarding the corrections, since all \textsc{Ctrl-Swap} gates are controlled by the same qubit, we can combine any necessary $\sqrt{X}$ corrections into one correction to be performed at the very end of this subroutine. Furthermore, as each \textsc{Ctrl-Swap} gate only concerns each spin-up and spin-down qubit exactly once, the swap correction can be implemented by relabeling qubits in the compiler without slowing down the computation. The parity-\textsc{Cnot} correction can either be implemented by considering the larger auto-corrected magic state including an additional Bell pair, or with the fork code approach utilizing the unused compute zone as storage. Neither approach incurs additional spatial cost since there are enough ancillae available at that time of the computation.

Observe that for this subroutine we need $Z$ access to the system qubits. However, since we are only swapping two system qubits at a time, it suffices to have $Z$ access to the system qubits in question. To do so, we add one additional ancilla corridor at the end of the system qubit storage (seen in grey to the very right of the olive system qubits in \zcref{fig:global layout}). This gives $Z$ access to the last system qubits, and therefore, we can perform a \textsc{Ctrl-Swap} controlled by the spin qubit $\ket{\sigma}$ swapping the last two qubits. Afterwards, we move the previously swapped system qubits into the corresponding ancilla corridor. This can be done by preparing the qubits in the ancilla corridor in $\ket{0}$ and applying a transversal \textsc{Cnot} between each data qubit, which contributes approximately one cycle.
We have now freed up $Z$ access for the second-last pair of system qubits and can repeat the procedure in an accordion like fashion until all system qubits are swapped. In total, we incur the time necessary to perform $N$ \textsc{Ctrl-Swap} gates, as well as $N$ extra cycles for the moves.

Additionally, we use this step in the algorithm to prepare the comparison flag qubit $\ket{x \geq 2^{n_R}C}$. Although this is not explicitly used during the forward application of \textsc{Spin}, it will be used in \textsc{RPrep} to prepare the flattened register, as well as for the application of the Majorana operator \textsc{Maj}. For the latter, we use the qubit efficient logical circuit described in \cite[Appendix E.2]{low2026denser}. In their circuit, the flag qubit is still used after uncomputing the \textsc{Ctrl-Swap}s of \textsc{Spin} in \textsc{Spin}$^\dagger$ to perform a \textsc{Ctrl-Swap} between $\ket{\zeta}$ and $\ket{\sigma}$. Once this has been performed, the flag qubit can be uncomputed. Thus, we account for it as part of the \textsc{Spin} subroutine. Since \textsc{Spin} does not use the compute zone or the full buffer zone, many ancillae are available during this computation. Hence, we use a Toffoli efficient quantum-classical comparator that requires $n-1$ Toffoli gates and $n-1$ clean ancillae to compare $x$ with an $n$-bit number \cite[Lemma 9]{low2026denser}.  We also use this step to account for another Hadamard gate applied to $ \ket{\zeta}$ in \textsc{Spin}$^\dagger$, which is necessary for Majorana selection.

\begin{table}[!hbtp]
    \centering
    \begin{subtable}{\textwidth}
    \centering
    \begin{tabular}{|c|c|c|c|}\hline
        Method & Parameters & Toffoli cost & Clean ancilla \\ \hline
        \rowcolor{gray!10} & $K=2$, $\lambda = 2$ & 8372   & 32 (48) \\
       \rowcolor{gray!10} \multirow{-2}{*}{dirty \textsc{SelectSwap}, restricted access} & $K = 2$, $\lambda = 3$ & 6289 & 48 (72) \\
        & $K = 4$, $\lambda = 2$ & 9824 & 32 (48)\\
        & $K = 2$, $\lambda = 4$ & 5808  & 64 (96) \\
       \multirow{-2}{*}{dirty \textsc{SelectSwap}, full access} & $K = 2$, $\lambda = 5$ & 5530 & 80 (120) \\
       \rowcolor{gray!10} & $K=2$, $\lambda = 1$ & 13303 & 32 (48) \\
      \rowcolor{gray!10}  \multirow{-2}{*}{clean \textsc{SelectSwap}} & $K=2$, $\lambda = 2$ & 7209 & 64 (96) \\
       & $\lambda = 2$ & 12879 & 32 (48) \\
       & $\lambda = 3$ & 8586 & 48 (72) \\
        & $\lambda = 4$ & 6678 & 64 (96) \\
       \multirow{-4}{*}{\textsc{Qrom}} & $\lambda = 5$ & 5247 & 80 (120) \\ \hline
    \end{tabular}
    \caption{FeMoco-54: $N=54$, $b_\rot = 16$, addresses $N + 2^{n_R}B = 486$}
    \end{subtable}

    \vspace{0.5em}
    
    \begin{subtable}{\textwidth}
    \centering
    \begin{tabular}{|c|c|c|c|}\hline
        Method & Parameters & Toffoli cost & Clean ancilla \\ \hline
       \rowcolor{gray!10}  & $K=2$, $\lambda =2$ & 21255 & 30 (45) \\
        \rowcolor{gray!10} & $K=2$, $\lambda = 3$ &  14950 &  45 (69) \\
        \rowcolor{gray!10} & $K=2$, $\lambda = 4$ & 12100 & 60 (90) \\
       \rowcolor{gray!10} \multirow{-4}{*}{dirty \textsc{SelectSwap}, restricted access} & $K=2$, $\lambda = 5$ & 10160 & 75 (114) \\
        & $K=4$, $\lambda = 2$ & 16760  & 30 (45) \\
        & $K=4$, $\lambda = 3$ & 15270  & 45 (69) \\
       \multirow{-3}{*}{dirty \textsc{SelectSwap}, full access} & $K=2$, $\lambda = 6$ & 9975& 90 (135) \\
       \rowcolor{gray!10} & $K=2$, $\lambda = 1$ & 37500 & 30 (45) \\
       \rowcolor{gray!10} & $K=2$, $\lambda = 2$ & 19570 &  60 (90) \\
      \rowcolor{gray!10}  \multirow{-3}{*}{clean \textsc{SelectSwap}} & $K=2$, $\lambda = 3$  & 13250 &  90 (135) \\
        & $\lambda = 2$ & 37164 & 30(45) \\
        & $\lambda = 3$ & 24450 &  45 (69)\\
        & $\lambda = 4$ & 18582 & 60 (90) \\
        & $\lambda = 5$ & 14670 & 75 (114) \\
       \multirow{-5}{*}{\textsc{Qrom}}  & $\lambda = 6$ & 12714 & 90 (135) \\ \hline
    \end{tabular}
     \caption{FeMoco-76: $N=76$, $b_\rot = 15$, addresses $N + 2^{n_R}B = 988$}
    \end{subtable}
    \caption{Logical cost break down for different lookup table batching versions for \textsc{Rprep}. Each method requires a serial application of Givens rotations that is not included in the cost. We ignore the finite size effect and assume that all lookup tables load $\lambda b_\rot$  angle bits. Dirty \textsc{SelectSwap} with restricted access uses up to $N+3$ ancilla qubits and is costed using recycling batching described in \zcref{sec: Lookup table batching}. Dirty \textsc{SelectSwap} with full access can use the full slate of system qubits, but needs to perform several iterations of recycling batching to free up system qubits before applying rotations. Here $K$ denotes the number of addresses swapped in the \textsc{Swap} network and $\lambda$ the number of  rotation angles loaded per batch. We restrict ourselves to $\lambda \geq 2$ for both dirty \textsc{SelectSwap} variants and the \textsc{Qrom} as a consequence of the inequalities in \eqref{eq: output condition I} and \eqref{eq: output condition II}. The clean ancilla cost consists of the size of the output register $\lambda b_\rot$ and in case of the clean \textsc{SelectSwap} additional $(K-1)\lambda b_\rot$ qubits. The number in brackets includes the necessary routing corridors to guarantee $Z$ access for the target register. Access corridors necessary for the system qubits for dirty \textsc{SelectSwap} are not included.}
    \label{tab:RPrep Rot lookup table batching options}
\end{table}

\paragraph{Loading of rotation angles and application of Givens rotations.} Since we are loading the rotation angles in batches, we combine the application of the rotations using phase gradient addition with the loading of the rotation angles. This has the benefit of reducing the number of logical qubits necessary but leads to a stark increase in Toffoli gates as the rotation angles need to be reloaded for the uncomputation of Givens rotations. As mentioned above, this subroutine is the bottleneck of the algorithm, contributing both the largest spatial and the largest temporal footprints. Before loading the rotation angles, we compute a composite flattened register $\ket{y}$ that holds $Y = N + 2^{n_R}B$  addresses following the procedure detailed in \cite[Appendix E.3]{low2026denser}, which we review in the computational basis now for convenience. Since the ancilla holding $\ket{x \geq 2^{n_R}C}$ is already available from the previous \textsc{Spin} application, we start by performing a controlled quantum-classical addition adding $2^{n_R}(B-C)$ to $\ket{x}$ if $\ket{x \geq 2^{n_R}C} = \ket{1}$. This essentially forces the $N$ contribution to start at $2^{n_R}B$ instead of $2^{n_R}C$ for $\ket{x \geq 2^{n_R}C} = \ket{1}$. Next, observe that if $\ket{x \geq 2^{n_R}C} = \ket{0}$, i.e., if $x < 2^{n_R}C$, the $\ket{x}$ register splits into low and high bits corresponding to $\ket{r}$ and $\ket{c}$ respectively. Thus, performing a controlled swap of the $\ket{c}$ and $\ket{b}$ register if $\ket{x \geq 2^{n_R}C} = \ket{0}$ produces the desired flattened register, since it holds all values from $[0, 2^{n_R}B)$ when $\ket{x \geq 2^{n_R}C} = \ket{0}$ and $[2^{n_R}B, 2^{n_R}B + N-1]$ when $\ket{x \geq 2^{n_R}C} = \ket{1}$. Note that to facilitate $Z$ access for the \textsc{Swap}, we can again utilize the accordion solution of \textsc{Spin}. More precisely, moving in the access corridor from the last \textsc{Spin} application and freeing another access corridor utilizing the compute zone, the $\ket{c}$ and $\ket{b}$ register can be swapped one qubit at a time with the access corridors scanning through the qubits. We omit the time it takes to set up and return the access corridors appropriately as it is negligible and accounted for in the slack. In sum, we can now directly access $Y$ by iterating over the $\ket{x}$ register. After the rotation angle loading and Givens rotations are successfully applied, we perform a necessary \textsc{Phaseup} and uncompute the flattened register by reversing the above operations, leaving the flag ancilla $\ket{x \geq 2^{n_R}C}$ intact, as it is still used for the application of the Majorana operator.

To reduce the spatial footprint, recall that the buffer zone for a \textsc{Qrom} takes up significantly less space than the buffer necessary for \textsc{Adder}. More concretely, the buffer for \textsc{Adder} requires 32 logical qubits whereas \textsc{Qrom} requires only 12, giving access to 20 potential ancilla qubits. Instead of adding the ancilla for the \textsc{Qrom} into the top row of logical qubits and subsequently increasing the width of the largest subroutine, we use the available space in the buffer zone. In the mirrored framework the ancillae are placed to the left of the buffer and the compilation steps outlined in \zcref{fig: compiled_and} can still be performed simultaneously with the qubit displacement now routing to the left as opposed to upwards. This in turn means that we would only require 2 routing corridors between address/target register and the buffer/ancilla area. However, as we need the larger routing zone for other subroutines, we gain no additional spatial benefit. Similarly, the parity-\textsc{Cnot} is compiled analogously to \zcref{fig: compiled_parityNOT}. 

Additionally, we implement the loading of rotation angles and their application in batches using lookup table batching as described in \zcref{sec: Lookup table batching} to save space.  In principle, we have the option of performing batching with skew-tree \textsc{Qrom}, or \textsc{SelectSwap} with dirty or clean ancilla. The size of the output register for lookup table batching with dirty \textsc{SelectSwap} is governed by the number of dirty ancillae available. The bulk of the dirty ancillae is given by the system qubits plus three additional qubits from spin considerations. Note that since the rotations are applied to the spin-down system qubits, not all of them can be used as dirty ancillae at the same time. This can be overcome by grouping the dirty ancillae affected by specific rotation angles, excluding one of these groups from the \textsc{SelectSwap}, and then moving on to another group. However, this in particular requires the dirty ancilla to be reset to their original state in between these grouped \textsc{SelectSwaps}, leading to multiple applications of recycling lookup table batching. In contrast, if we restrict the dirty ancilla budget to only the spin-up system qubits, which are not affected by the rotation application, one application of recycling lookup table batching suffices.  In this case, the number of rotation angles that we are able to load in one go, denoted $\lambda$, is given by $(K-1)\lambda b_\rot \approx N$, where $K$ denotes the swap parameter. 
We compare the different lookup table options including lookup table batching with clean \textsc{SelectSwap} and skew-tree \textsc{Qrom} only for a similar size output register for both FeMoco-54 and FeMoco-76 in \zcref{tab:RPrep Rot lookup table batching options}.  We see that for a small output register the dirty \textsc{SelectSwap} variants are favorable in terms of Toffoli count, but as soon as more qubits are available, other options such as the plain skew-tree \textsc{Qrom} start to dominate. However, in order to use the system qubits as dirty ancilla in the repetition-cat architecture, they need to be equipped with access corridors to provide $Z$ access. While this could be done by placing select shifting access corridors in between swaps similar to what is done in \textsc{Spin}, the complexity of shifting and its temporal impact increase for a true \textsc{Swap} network. Thus, we cost dirty \textsc{SelectSwap} as requiring all access corridors and subsequently refrain from using such subroutines in the reference-configuration compilation for the repetition-cat architecture.

\begin{figure}[ht]
\centering
\begin{subfigure}[b]{0.45\textwidth}
\centering
  \includegraphics[width=\linewidth]{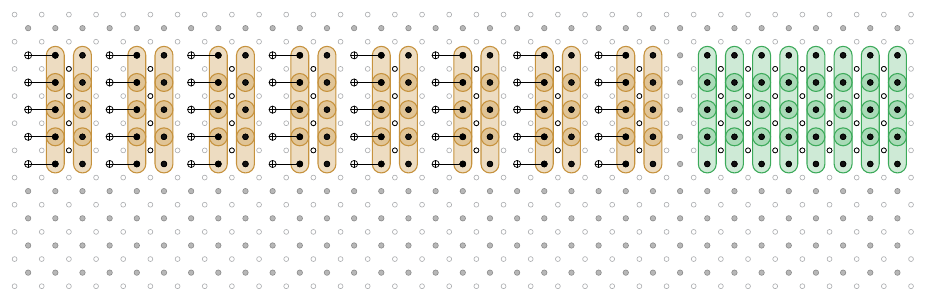}
    \caption{}
\end{subfigure}
\begin{subfigure}[b]{0.45\textwidth}
\centering
  \includegraphics[width=\linewidth]{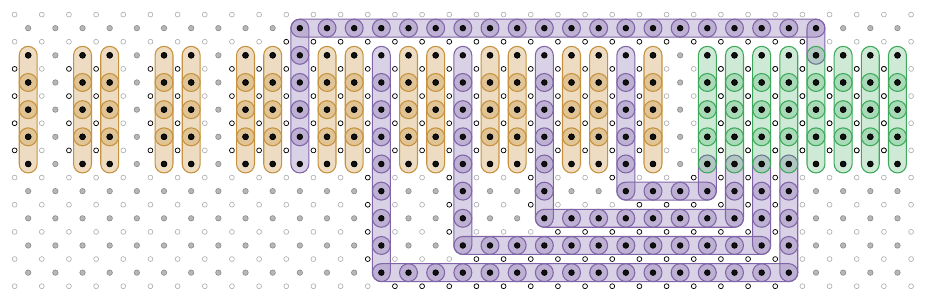}
    \caption{}
\end{subfigure}

\medskip

\begin{subfigure}[b]{0.45\textwidth}
\centering
  \includegraphics[width=\linewidth]{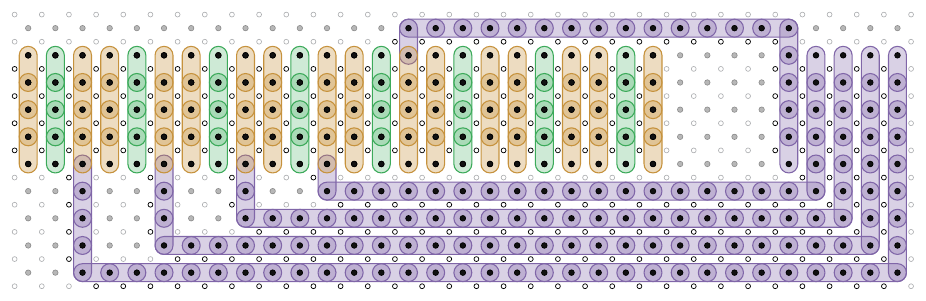}
    \caption{}
\end{subfigure}
\begin{subfigure}[b]{0.45\textwidth}
\centering
  \includegraphics[width=\linewidth]{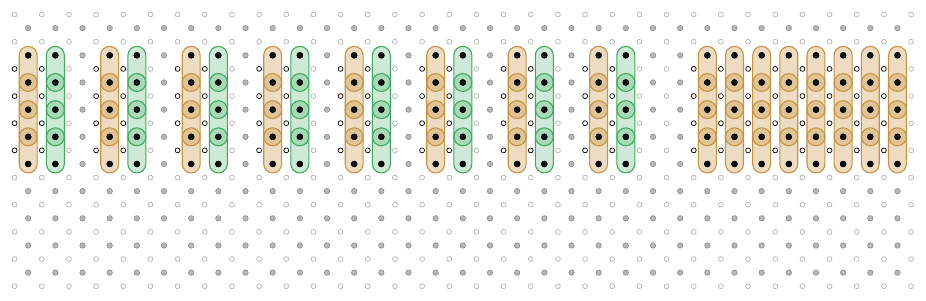}
    \caption{}
\end{subfigure}

\caption{Illustration of qubit movement to prepare the layout necessary for \textsc{Adder}. Panel (a) shows two rotation angles depicted in yellow and the Fourier resource state in green to the right. The rotation angles are loaded with a lookup table and are thus stored with $Z$ access. The first pair holds the least significant qubit for both rotation angles, the second pair the second-least for both and so forth. We begin by shifting the qubits of the first rotation angle to the left using a layer of transversal \textsc{Cnot} gates. Next, the Fourier resource state is moved into the gaps in the rotation angle register utilizing the routing corridor as depicted in panel (b). Note that we assume that the least significant qubit of the Fourier resource state is the rightmost qubit. Panel (c) depicts the displacement of the second rotation angle into the storage area freed up by the Fourier resource state. Finally, panel (d) shows the layout ready for the addition and also the parity-\textsc{Cnot} gate between the Fourier resource state and a system qubit.} \label{fig: qubit movement}
\end{figure}

After a batch of rotation angles has been loaded, we apply the Givens rotations to the system qubits via phase gradient addition. Similarly to \textsc{OuterPrep} and \textsc{InnerPrep}, this requires the preparation of the compute area to fit the initial adder layout by interleaving $b_\rot$ rotation angle qubits, the full Fourier resource state and $b_\rot$ empty columns for clean ancilla use. In contrast to the initial state preparation subroutines however, we cannot uncompute rotation angles and thus have a tighter budget to maneuver. This turns out to be unproblematic though, as long as enough rotation angle batches are loaded, since the lookup table output is stored facilitating Z access, i.e., as pairs of two output qubits with an access corridor in between. Thus, for $\lambda b_\rot$ output qubits, we have access to $\lceil \lambda b_\rot/2 \rceil$ ancillae. In order to use the desired adder compilation, we need $2b_\rot$ free logical qubit slots. To ensure that we can free up these slots in the compute area, as illustrated in \zcref{fig: qubit movement}, we need to choose $\lambda$ such that 
\begin{equation}\label{eq: output condition I}
    \lceil \lambda b_\rot/2 \rceil + (\lambda-1)b_\rot \geq 2 b_\rot.
\end{equation} 
This is never true for $\lambda = 1$, but always holds for $\lambda \geq 2$. It remains to ensure that we have enough space to move rotation angles that are blocking the adder layout out of the compute area, i.e., $\lambda$ needs to be chosen such that
\begin{equation}\label{eq: output condition II}
\lceil \lambda b_\rot/2 \rceil \geq b_\rot.
\end{equation}
We see that $\lambda \geq 2$ is necessary and sufficient for both inequalities. However, we still need to make sure that the adder compilation requirement that the adder buffer zone sits slightly to the right of the adder compute zone is satisfied. More precisely, we require that only one buffer overlaps with the compute zone and that only 8 qubits overlap. Assuming that the buffer sits directly aligned under the right edge of the compute zone, this is always satisfied for $\lambda \geq 3$ as $b_\rot$ is sufficiently large. For $\lambda =2$, we would need to shift the compute zone over. We avoid this in our layout by storing the address register and the Fourier resource state to the right of the compute zone and allocating the buffer aligned with the right edge of the address and spin area (see \zcref{fig:global layout}). Observe that for clean \textsc{SelectSwap}, $\lambda = 1$ is possible since in this case there are at least two copies of the output, which implies that there is enough space in the compute area.

After having ensured sufficient access to ancillae, we cost the reshuffling time $t_\mathrm{shuffle}$. We start by considering the first batch of two rotation angles. Note that we load the first two rotation angles as pairs of qubits from each rotation angle, i.e., the leftmost pair holds $u_{b_1, 1}$ and $u_{b_2, 1}$ followed by $u_{b_1, 2}$ and $u_{b_2, 2}$ and continued until $u_{b_1, b_\rot}$ and $u_{b_2, b_\rot}$. First, we shift the rotation angle into the access corridor to the left using transversal \textsc{Cnot} gates. This takes less than 1 cycle. Next, we move the Fourier resource state from the storage area to the right of the compute zone into the first $b_\rot$ ancilla routing corridors in the output register. Taking into account that the routing's zone height is four in this section of the chip and that we have access to a routing row on top, this can be done in $\lceil b_\rot/5 \rceil d$ cycles assuming the least-significant qubit of the Fourier resource state is stored right-most. Lastly, we move the second batch of rotation angles into the Fourier resource state storage area contributing another $\lceil b_\rot/5 \rceil d$ cycles. A graphical depiction of this maneuver can be found in \zcref{fig: qubit movement}. Since the Fourier resource state is a catalytic state, it remains in the addition layout as the output of the addition. However, the output is shifted to the right by one qubit, so the Fourier resource state needs to be shifted over one after each addition using transversal \textsc{Cnot} gates again contributing less than 1 cycle. Thus, after uncomputing the other adder input, we only need to move in the new rotation angles to perform our next addition. The first move can again be performed in $\lceil b_\rot/5 \rceil d$ cycles by moving the rotation angles out of the storage zone. Similarly, any subsequent move can also be performed in $\lceil b_\rot/5 \rceil d$ cycles assuming that the other rotation angles are loaded in descending order, i.e., with the least significant qubit to the right.  Once all rotation angles have been applied, the Fourier state must be moved back to storage for another $\lceil b_\rot/5 \rceil d$. In total, we obtain $t_\mathrm{shuffle} = (4 + \lambda-2)\lceil b_\rot/5 \rceil d +\lambda  = (2 + \lambda)\lceil b_\rot/5 \rceil d +\lambda$.

Now that compatibility of the layout has been taken into account it remains to cost the \textsc{Cnot} conjugation for the phase gradient addition as well as consider any \textsc{Phaseup}s necessary from uncomputing the rotation angles. The \textsc{Cnot} conjugations needs to be applied in between the Fourier resource state and the target system qubit. In our color-changed compilation, the Fourier resource state acts as the control and the system qubits are the target. Thus, we only require $X$ access to the system qubits for this step in the computation. The adder layout with interleaved ancillae already provides the necessary $Z$ access for the Fourier resource state. For each phase gradient addition we need to perform two parity-\textsc{Cnot} gates leading to a total time contribution of $2\lambda d$ per lookup table batch. In terms of uncomputation, all phase corrections can be combined into a single \textsc{Phaseup}, which is to be performed before the flattened $\ket{y}$ register is uncomputed. For the costing of this procedure see the \textsc{Phaseup} section below. 

\paragraph{Majorana operator.} After the application of the Givens rotations, the application of the Majorana operator can be performed on a single qubit. As is common in the literature, Majorana operators are transformed into Pauli strings using the Jordan-Wigner transform. The main component of this subroutine is then to ensure that the correct Majorana operator is applied depending on the value of $g$. This is achieved using comparison flag qubits and multi-controlled Toffoli gates. We follow the description of \textsc{GComp} introduced in \cite[Table XXI, Figure 70]{low2026denser}. For convenience and since we are only doing a coarse cost accounting of the Majorana operator, we review this subroutine with respect to the computational basis avoiding any color-change discussion. First, observe that the already available flag qubit $\ket{x \geq 2^{n_R}C}$ allows us to distinguish between one- and two-electron terms. To distinguish between the positive and negative one-electron terms, we need to compute another flag qubit, namely $\ket{x \geq 2^{n_R}C + N_+}$. As the compute zone and part of the buffer area are unused, we do so again by performing an $n$-bit quantum-classical comparison using $n-1$ ancillae and  $n-1$ Toffoli gates and storing the flag qubit in the compute zone. All Pauli operations are applied to the zeroth spin-down qubit $\ket{\psi_\downarrow}_0$. We start by applying a \textsc{Cnot} controlled on $\ket{x\geq 2^{n_R}C + N_+} = \ket{0}$ followed by a $\CZ$ between $\ket{\zeta}$ and $\ket{\psi_\downarrow}_0$ if $g \neq \SF$, i.e., controlled on $\ket{x \geq 2^{n_R}C} = \ket{1}$. Afterwards, a $\CZ$ between the flag $\ket{x\geq 2^{n_R}C}$ and $\ket{\psi_\downarrow}_0$ is applied. Lastly, another \textsc{Cnot}, this time controlled on $\ket{x\geq 2^{n_R}C + N_+} = \ket{1}$, is performed. Now, the flag $\ket{x\geq 2^{n_R}C + N_+}$ can be uncomputed. It remains to ensure that, in the block-encoding of the two-electron term, no Pauli operator is applied for $b=B$ as this corresponds to the identity in \eqref{eq:2-elec O}. Thus, we apply a multi-controlled Toffoli gate subject to the condition $b=B$. This can be decomposed into  $\lceil \log (B+1) \rceil$ many Toffoli gates using as many ancillae. We omit any cost from uncomputation of \textsc{And} gate ladders, as this subroutines amounts to less than $0.1\%$ of the full block encoding cost and is therefore taken into account through the addition of $5\%$ slack when reporting the final duration. 

\paragraph{Reflection.} The block encoding algorithm requires two reflections: one reflection for the inner block encoding  and another for the full block encoding. Reflections are implemented as multi-controlled $\CXX$ gates, which can be performed using a ladder of \textsc{And} gates. More precisely, a reflection over $n$ qubits is costed as $n-2$ \textsc{And} gates using $n-2$ ancilla. For the inner reflection, we reflect over the $B$ register as well as $\ket{\zeta}$ incurring a Toffoli cost of $n_{B+1} - 1 $ \textsc{And} gates. The outer reflection over the $\ket{x}$ and $\ket{b}$ registers and $\ket{\sigma}$ and $\ket{\zeta}$ qubits is also controlled on two additional qubits necessary for phase estimation, as outlined in \cite{Low+FastQuantumSA2025, low2026denser}. Thus, we count $n_X + n_{B+1} +2$ \textsc{And} gates and ancilla. As above, we omit any cost from uncomputation of \textsc{And} gate ladders, as this subroutines amounts to less than $0.1\%$ of the full block encoding cost. 

\paragraph{Uncomputation using \textsc{Phaseup}.}
Phase corrections on the address registers arise for example as a byproduct of \textsc{Qrom} applications either through the injection and uncomputation of \textsc{And} gates or through uncomputation of lookup tables in general. As long as the affected address register is only used as a control, we can combine phase corrections into a single \textsc{Phaseup}. Thus we perform \textsc{Phaseup} at the end of \textsc{OuterPrep}, \textsc{InnerPrep}, before uncomputing the flattened index in \textsc{RPrep + Rot}, and for each lookup uncomputation in \textsc{OuterPrep}$^\dagger$, \textsc{InnerPrep}$^\dagger$, since the address register shrinks in the process. As described in \zcref{sec: phase fixup}, we have the option of implementing the \textsc{Phaseup} using either the space-space split or the space-time split. If there are enough ancilla available, we opt for whichever option is faster. For the space-space split we use the compilation outlined in \zcref{sec: PhaseFixup comp}. If not enough ancilla are available, we revert to the space-time split. In particular, we restrict the size of the unary-one-hot encoding to fit the ancilla budget available. Once the ancilla budget is set, we pick the most efficient option over all admissible space-time splits in terms of number of rounds using the compilation of both the \textsc{Swap} network and also the skew-tree \textsc{Qrom}. To account for phase corrections arising from the skew-tree \textsc{Qrom} we continue applying smaller \textsc{Phaseups} in the space-time split until all phase corrections are resolved. It remains to account for the conjugation by a layer of Hadamard gates. Since the Hadamards act on the target qubits of the \textsc{Swap} network as well as the target of the \textsc{Qrom}, they are stored with $Z$ access. Thus, each Hadamard application takes $d$ cycles utilizing a free access corridor for qubit displacement as described in \zcref{sec: Multi-target C_XX and H}.

\subsubsection{Resource counts for FeMoco-54 and FeMoco-76}\label{sec: rep code resource count}
We now provide a detailed resource estimate of the phase estimation algorithm for FeMoco-54 and FeMoco-76 with respect to the reference configuration using the compilation of \textsc{Qrom} and \textsc{Swap} in \zcref{sec:compilation} as well as the \textsc{Adder} compilation of \cite[Section C.5]{Shor2026prep}. To choose the appropriate parameters $(R,B,C)$ for the DFTHC ansatz in \eqref{eq: DFTHC SOS}, we use the same parameter sets as in \cite[Table XII]{low2026denser} to ensure comparability. We also use the same precision for state preparation coefficients, $b_\coef$, as well as rotation angles, $b_\rot$. Specifically, we set $(N, R, B, C) = (54,10,27,27)$ with $(b_\coef, b_\rot) = (9,16)$ for FeMoco-54 and $(N, R, B, C) = (76,15,57,19)$ with $(b_\coef, b_\rot) = (9,15)$ for FeMoco-76. We load the minimal number of rotation angles at a time without the need of adding additional clean ancilla space for the addition, i.e., $\lambda = 2$. This gives a compute zone of 48 qubits for FeMoco-54 and 45 qubits for FeMoCo-76 taking into account $Z$ access corridors.

We quantify the duration, $t_\mathrm{S}$, of a single run of the algorithm in syndrome extraction cycles, the probability $\varepsilon_\mathrm{alg}$ that at least one logical error occurs in the algorithmic region, and the probabilities $\varepsilon_{\CXCXX}$ and $\varepsilon_{\keti}$ that at least one consumed resource state is faulty. Since the factory couples to the algorithm only through the delivered states, any undetected logical error there manifests as a faulty resource state and is accounted for in $\varepsilon_{\CXCXX}$ and $\varepsilon_{\keti}$. Thus, we obtain $\varepsilon_\mathrm{alg}$ from the duration in syndrome extraction cycles and the physical footprint of the algorithmic region, comprising the data and ancilla qubits of each subroutine outlined in the previous section, together with the volume for the factory routing area. For the latter, we deliberately overestimate the contribution by counting all 1180 physical qubits over the full duration of the computation. The volume coming from the algorithmic region is made up of two components: qubits sitting idle during the computation as well as qubits actively involved in the subroutine. Since idle and active qubits are subject to the same logical error rate, this distinction serves only as bookkeeping. Note that idle assignments often change during a subroutine and are to be set for each subroutine constituting the larger subroutines of \textsc{OuterPrep}, \textsc{InnerPrep}, \textsc{Spin}, and \textsc{RPrep + Rot}. For example, while the Fourier resource state is idle during the data loading via \textsc{Qrom} in \textsc{OuterPrep}, $b_\coef$ qubits of the Fourier resource state are involved in the shuffling procedure and \textsc{Adder} for phase gradient addition. The system qubits sit idle during all of \textsc{OuterPrep}. Regardless the volume is then computed by multiplying the corresponding duration and physical footprint for both active and idle components and summing all contributions. 

Considering all subroutines described in the previous sections and taking their volume into account gives the physical volume $V_\text{BE}$ for one application of the block encoding. To obtain the contribution of the algorithmic region to the total physical volume, $V_\text{BE}$ is multiplied by the number of calls to the block encoding for the DFTHC ansatz, i.e., $\#\text{Calls} = \lceil \pi\lambda_\text{eff}/(2\sigma_\text{PEA})\rceil$, where $\lambda_\text{eff}$ denotes the effective block encoding constant for the DFTHC ansatz \cite{low2026denser, Low+FastQuantumSA2025}. We again use the parameters from \cite[Table XII]{low2026denser} giving $\sigma_\text{PEA} = \qty{1}{\milli\hartree}$, $\lambda_\text{eff} = \qty{21.4}{\hartree}$ for FeMoco-54, and $\lambda_\text{eff} = \qty{43.7}{\hartree} $ for FeMoco-76.  We omit the cost of initial state preparation for the system qubits and any additional overhead of phase estimation beyond the block encoding calls, as is standard in the literature. Additionally, we do not account for Fourier resource state preparation, since it can be prepared at the beginning of the computation with negligible impact to the overall resource count.  In sum, $V_\text{phys,tot} = (\#\text{Calls})V_\text{BE} + 1180 t_\mathrm{S}$, where the last summand is the factory routing zone contribution. Recall here, that $t_S$ is the duration of the algorithm in syndrome extractions cycles. 

Once the total physical volume $V_\text{phys,tot}$ is obtained, we can compute the effective logical volume $V_\text{log,eff}$, i.e.,  the number of logical-qubit cycles, by dividing by the physical footprint $2d-1$ of a repetition-code logical qubit and by the $d$ syndrome extraction cycles per logical cycle. In symbols, $V_\text{log,eff} = V_\text{phys,tot}/(d[2d-1])$. Using this quantity we obtain the probability that at least one logical error occurs via
\begin{equation*}
    \varepsilon_\mathrm{alg} = 1 - (1-p_Z^\mathrm{rep})^{V_\text{log,eff}},
\end{equation*}
where $p_Z^\mathrm{rep}$ denotes the logical phase-flip error probability per logical qubit after $d$ error correction cycles. The latter is given using the same ansatz as \eqref{eq: LDPC ansatz} with the repetition code parameters given in \cite[Table III]{Ruiz_2025}. To determine the error contribution given by the resource states, we simply count the number of resource states consumed and multiply them by their corresponding output error rate, i.e., $\varepsilon_{\CXCXX} = (\#\text{\ket{\CXCXX}})p_{\CXCXX}$ and $\varepsilon_{\keti} = (\#\keti\text{-states})p_{\keti}$. The number of $\ket{\CXCXX}$ corresponds exactly to the Toffoli count of the algorithm. The number of $\keti$ states is determined using the compilation outlined in previously from the corresponding buffer states. Since buffer layouts change between subroutines, we initialize each buffer catalysis state fresh before the start of the subroutine. More precisely, \textsc{Qrom} uses $2\ket{i}$ states per \textsc{And} gate with an additional 2 $\keti$ catalysts; \textsc{Swap} uses 4 $\keti$ states per buffer state with $3\keti$ catalysts and \textsc{Adder} uses 2 $\ket{i}$ states per Toffoli gate with $3\keti$ catalysts and an additional $2\keti$ per phase correction. Additionally, we account $2\keti$ states per Hadamard application and $4\keti$ states for Toffoli gates that have not been explicitly compiled as we are assuming an autocorrected state and utilize catalysis as described in \zcref{app: Clifford}. 

Since the probability that at least one of the error sources causes a failure is at most the sum of their individual probabilities, we take the failure probability to be $p_\mathrm{fail} = \varepsilon_\mathrm{alg} + \varepsilon_{\CXCXX} + \varepsilon_{\keti}$, which gives an expected runtime $\langle t \rangle = t/(1-p_\mathrm{fail})$, where $t = t_S \cdot \qty{1}{\micro\second}$. Here, we again follow \cite{low2026denser}, this time in accounting for failed runs by dividing the runtime by the success probability. We then determine the appropriate code distance $d$ of the algorithmic region by choosing the one with the shortest expected duration. The total spatial footprint is obtained using the global layout formula \eqref{eq:Rep code phys footprint}. The results can be found in \zcref{tab: Rep code error accounting}. For a phase-flip error of $p_Z = 5\times 10^{-4}$, a repetition code distance of $d = 15$ suffices for FeMoCo-54 leading to $11\,192$ physical qubits and an expected runtime of 23 hours accounting for $5\%$ temporal slack as described at the beginning of the section. The failure probability for both FeMoco-54 as well as FeMoco-76 stays below $5\%$ for $p_Z = 5 \times 10^{-4}$, the error from $\ket{\CXCXX}$ states starts to dominate in the $p_Z = 10^{-3}$ regime, contributing about 60\% of the failure probability. This is most apparent for FeMoco-76, where for a distance $d\geq 19$ the failure probability remains essentially constant just under $30\%$.

\begin{table}[hbpt]
    \centering
    \begin{tabular}{c|c|c|c|c|c|c|c|c}\hline
       Molecule & $p_Z$ & $d$ & $\varepsilon_\mathrm{alg}$ & $\varepsilon_\mathrm{\CXCXX}$ & $\varepsilon_\mathrm{\keti}$ & Physical qubits & One-shot & $\langle t\rangle$ \\ \hline
       \rowcolor{gray!10}  & $5\times 10^{-4}$ & 15 & 0.0011 & 0.00058 & 0.0059 & $11\,192$ & 0.94 days & 0.95 days \\
       \rowcolor{gray!10} \multirow{-2}{*}{FeMoco-54} & $10^{-3}$ & 17 & 0.0065 & 0.043 & 0.019 & $12\,072$ & 1.01 days & 1.09 days \\
        & $5\times 10^{-4}$ & 15 & 0.0058 & 0.0027 & 0.027 & $12\,568$ & 4.33 days & 4.49 days \\
       \multirow{-2}{*}{FeMoco-76} & $10^{-3}$ & 17 & 0.035 & 0.20 & 0.089 & $13\,612$ & 4.66 days & 6.90 days \\ \hline
    \end{tabular}
    \caption{Runtime and total physical qubit count for FeMoco-54 and FeMoco-76 for different phase-flip errors in the repetition-cat architecture. The physical qubit count includes ancilla and contributions from factories and factory routing. The duration accounts for $5\%$ additional slack for initialization, measurements and buffer staging. The expected duration is computed by normalizing with $1-p_\mathrm{fail}$, where $p_\mathrm{fail} = \varepsilon_\mathrm{alg} + \varepsilon_\mathrm{\CXCXX} + \varepsilon_\mathrm{\keti}$.}
    \label{tab: Rep code error accounting}
\end{table}

We end this section by providing a breakdown of costs across subroutines and operations for one block-encoding of FeMoco-54 at $d=15$ summarized in \zcref{tab:be_femoco54_rep}. As described in \zcref{sec: Rep-code detailed description of ops}, \textsc{RPrep + Rot} is the subroutine with the largest physical footprint. From the description it is also not surprising that it is the largest subroutine in terms of duration. This is evident in \zcref{tab:be_femoco54_rep_counts}, detailing that \textsc{RPrep + Rot} takes up  $65.1\%$ all syndrome extraction cycles of the full block-encoding run. \textsc{InnerPrep} and its adjoint are the next largest routine taking up $33.3\%$ of the total time. Thus, these two subroutines essentially give the full algorithm, with the remaining $1.5\%$ being dominated by \textsc{OuterPrep}. The breakdown in the same table by primitive also makes it clear that skew tree  \textsc{Qrom} constitutes the main part of the algorithm with 87.0\% round share, 91.5\% of Toffolis and 84.3\% $\keti$-states. This is mainly due to the fact that dirty \textsc{SelectSwap} is a non-feasible subroutine in a repetition-cat architecture focusing on a small spatial footprint. \zcref{tab:be_femoco54_rep_volume} gives a breakdown of the effective logical volume for one block-encoding run including idle qubits, without taking factories and factory routing into account. Idle system qubits constitute nearly half the effective logical volume with \textsc{Qrom} taking up slightly more than a third. In general, idle qubits account for 59.6\% of the logical volume.

\begin{table}[hbtp]
    \centering

    \begin{subtable}[t]{\textwidth}
    \centering
    \begin{tabular}{c|c|c|c|c|c|c}\hline
        Subroutine & Rounds & Share & Toffolis & Share & $\keti$-states & Share \\ \hline
        \rowcolor{gray!10} \textsc{Rprep + Rot} ($\times 4$) & $1\,503\,384$ & 65.1\% & $55\,224$ & 64.3\% & $118\,768$ & 63.6\% \\
        \textsc{InnerPrep}$^\dagger$ ($\times 2$) & $401\,590$ & 17.4\% & $15\,354$ & 17.9\% & $34\,542$ & 18.5\% \\
        \rowcolor{gray!10} \textsc{InnerPrep} ($\times 2$) & $367\,682$ & 15.9\% & $14\,230$ & 16.6\% & $29\,846$ & 16.0\% \\
        \textsc{OuterPrep}$^\dagger$ ($\times 1$) & $13\,534$ & 0.6\% & $408$ & 0.5\% & $1\,267$ & 0.7\% \\
        \rowcolor{gray!10} \textsc{OuterPrep} ($\times 1$) & $12\,304$ & 0.5\% & $377$ & 0.4\% & $1\,103$ & 0.6\% \\
        \textsc{Spin} ($\times 4$) & $6\,136$ & 0.3\% & $248$ & 0.3\% & $1\,016$ & 0.5\% \\
        \rowcolor{gray!10} \textsc{Maj} ($\times 2$) & $2\,670$ & 0.1\% & $56$ & 0.1\% & $236$ & 0.1\% \\
        \textsc{OuterRef} ($\times 1$) & $720$ & $<0.1\%$ & $16$ & $<0.1\%$ & $66$ & $<0.1\%$ \\
        \rowcolor{gray!10} \textsc{InnerRef} ($\times 1$) & $180$ & $<0.1\%$ & $4$ & $<0.1\%$ & $18$ & $<0.1\%$ \\
        \hline
        Primitive & Rounds & Share & Toffolis & Share & $\keti$-states & Share \\ \hline
        \rowcolor{gray!10} \textsc{Qrom} & $2\,007\,090$ & 87.0\% & $78\,576$ & 91.5\% & $157\,440$ & 84.3\% \\
        \textsc{Adder} & $159\,516$ & 6.9\% & $3\,798$ & 4.4\% & $15\,954$ & 8.5\% \\
        \rowcolor{gray!10} Phaseup & $83\,978$ & 3.6\% & $2\,863$ & 3.3\% & $10\,548$ & 5.6\% \\
        Shuffle & $29\,556$ & 1.3\% & -- & -- & -- & -- \\
        \rowcolor{gray!10} \textsc{Swap} & $12\,268$ & 0.5\% & $508$ & 0.6\% & $2\,192$ & 1.2\% \\
        Rest & $15\,792$ & 0.7\% & $172$ & 0.2\% & $728$ & 0.4\% \\
        \hline
        Total & $2\,308\,200$ & 100\% & $85\,917$ & 100\% & $186\,862$ & 100\% \\ \hline
    \end{tabular}
    \caption{Number of syndrome extraction rounds, Toffoli gates and $\keti$-states by subroutine and by primitive. The subroutine numbers take the number of calls (in brackets) into account, so that they make up the whole block encoding. Both subroutines and primitives break down the same totals. Here, shuffle denotes the procedure to prepare the compute zone for the \textsc{Adder} layout. Rest combines comparator$^*$, quantum-classical-addition$^*$, rotation, \textsc{Ref}$^*$, remaining routines for Majorana application$^*$ and MBU, i.e., the primitives with less than 0.1\% each of the effective volume in \zcref{tab:be_femoco54_rep_volume}, where rotation refers to the parity-\textsc{Cnot} gates necessary for phase gradient addition and MBU counts explicit MBU of lookup table outputs.}\label{tab:be_femoco54_rep_counts}
    \end{subtable}
    \par\bigskip
    \begin{subtable}[t]{\textwidth}
    \centering
    \begin{tabular}{c|c|c}\hline
        Primitive & Logical qubit-cycles & Share \\ \hline
        \rowcolor{gray!10} System (idle) & $1.66\times 10^{7}$ & 48.4\% \\
        \textsc{Qrom} & $1.22\times 10^{7}$ & 35.6\% \\
        \rowcolor{gray!10} Fourier state (idle) & $2.29\times 10^{6}$ & 6.7\% \\
        Address and spin qubits (idle) & $1.04\times 10^{6}$ & 3.0\% \\
        \rowcolor{gray!10} \textsc{Adder} & $1.04\times 10^{6}$ & 3.0\% \\
        Phaseup & $4.31\times 10^{5}$ & 1.3\% \\
        \rowcolor{gray!10} QPE qubits (idle) & $3.08\times 10^{5}$ & 0.9\% \\
        Output (idle) & $1.19\times 10^{5}$ & 0.3\% \\
        \rowcolor{gray!10} Flag (idle) & $1.00\times 10^{5}$ & 0.3\% \\
        \textsc{Swap} & $7.71\times 10^{4}$ & 0.2\% \\
        \rowcolor{gray!10} Shuffle & $2.99\times 10^{4}$ & 0.1\% \\
        Rest & $3.94\times 10^{4}$ & 0.1\% \\
        \hline
        Total & $3.43\times 10^{7}$ & 100\% \\ \hline
    \end{tabular}
    \caption{Effective logical volume $V_\mathrm{log,eff}$ by primitive, i.e., physical qubit-rounds divided by $d(2d-1)$. Idle output qubits refer to already loaded rotation angles, idle during other operations. Idle qubits account for 59.6\% of the algorithmic effective volume. Rest combines comparator$^*$, quantum-classical-addition$^*$, rotation, \textsc{Ref}$^*$, remaining routines for Majorana application$^*$ and MBU, each below 0.1\%.}\label{tab:be_femoco54_rep_volume}
    \end{subtable}
    \caption{Cost breakdown of one block encoding for FeMoco-54 in the repetition-cat architecture at $d = 15$, with \textsc{Rprep + Rot} implemented by lookup table batching using skew-tree \textsc{Qrom} loading $\lambda = 2$ rotation angles at a time. $^*$ denotes routines whose duration is estimated at $3d$ per Toffoli. Such non-compiled routines total 0.3\% of the rounds, 0.1\% of the logical qubit-cycles, 0.2\% of the Toffolis.}
    \label{tab:be_femoco54_rep}
\end{table}

\subsection{LDPC-cat architecture}\label{sec: LDPC-cat architecture}

\subsubsection{Global layout}
\begin{figure}[ht]
    \centering
    \includegraphics[width=\linewidth]{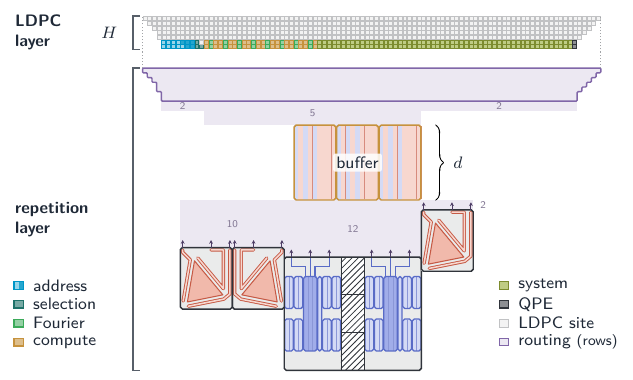}
    \caption{Global layout for the LDPC cat architecture. The top shows the LDPC layer with the $X$ logical operators of the corresponding logical qubits marked by colored squares. In the LDPC layer, shades of blue correspond to address qubits, teal denotes spin selection qubits $\ket{\sigma}, \ket{\zeta}$ as well as space for a flag qubit for selection over $g$, green are Fourier resource state qubits, yellow corresponds to space for lookup table output as well as addition operations, olive are system qubits and slate denotes two qubits reserved for quantum phase estimation (QPE). The repetition code layer includes routing underneath the LDPC layer, as well as a buffer zone, extra routing rows and $\ket{\CXCXX}$ (blue) and $\ket{i}$ (red) state factories with associated factory routing. }
    \label{fig:LDPC global layout}
\end{figure}

For the global layout in the LDPC-cat architecture the chip is split into an LDPC layer holding the algorithmic qubits and a repetition code layer for routing, buffer storage and resource state production. \zcref{fig:LDPC global layout} shows a schematic of the layout. The LDPC layer is given by an area of width $L$ and height of $H$ rows with an additional offset on each side to give enough space for $Z$ logical operators totaling $\Delta_H$. For more details on this type of construction, we refer the reader to \cite{Ruiz_2025}. Algorithmic qubits are distinguished by their unique logical $X$ operator located in the bottom two rows of the LDPC layer. The two address registers, $X$ and $B$, are stored to the very left of the LDPC layer, taking up a width $\ell_\text{address} = \lceil (n_X + n_{B+1})/2 \rceil$, followed by three qubits for spin selection and necessary flags during the computation. To the right, space is allocated for lookup table output as well as phase gradient addition. More precisely, two Fourier resource state qubits are stored vertically followed by a two-by-two square of spare qubits until the all Fourier resource state qubits are allocated. Any additional compute qubits are added to the right. A minimal width of $\lceil 3b_\rot/2 \rceil$ guarantees that the addition layout can be easily prepared at no extra spatial cost. While the area now also acts as Fourier resource state storage, we still refer to it as compute zone and denote its width by $\ell_\textsc{comp}$. To the right of the compute zone, the system qubits are stored, taking up a width of $N$. Finally, at the right edge of the chip the two QPE qubits are stored on top of each other. In total the LDPC layer can hold up to $2L$ logical qubits, where $L = \ell_\text{address} + \ell_\text{comp} + N + 3$ denotes the width of the chip. In total, the LDPC layer contains $HL+ \Delta_H$ data qubits and $HL+ \Delta_H - 2L = (H-2)L + \Delta_H$ ancilla qubits.  

The LDPC layer is part of a flip chip with a repetition code routing layer underneath. In addition to this routing layer we add two rows of routing at the bottom spanning the full width $L$. We also place an additional three rows of routing underneath the compute zone until the right edge of the buffer zone. The buffer zone sits below this routing right aligned with the fifth qubit column from the right of the compute zone. Overall we obtain the following physical qubit count not accounting for ancilla
\begin{align*}
    \text{Space}^\text{LDPC}_\text{alg} &= \underbrace{HL + \Delta_H}_{\text{LDPC layer}} + \underbrace{HL + \Delta_H + 2L + 3(\ell_\text{comp} + 22) + 27d_\mathrm{rep}}_{\text{repetition layer}}.
\end{align*}
Note that the 22 qubit offset is the offset necessary for the reference configuration when loading $\lambda = 2$ rotation angles and it is possible to reduce this when loading more angles at the same time. However, as the qubit count is just an estimate, we leave it as a deliberate overcount.
The remainder of the chip is given by factory routing as well as a factory zone analogous to the repetition code layout resulting in the total physical footprint
\begin{equation}
    \text{Space}^\text{LDPC}_\text{tot} = 2\Big( (H-1)L + \Delta_H\Big) + 2\Big(HL + \Delta_H + 2L + 3(\ell_\text{comp} + 22) + 27d_\mathrm{rep}\Big) + 3334.
\end{equation}

\subsubsection{Operations overview} 
The operations for the LDPC-cat architecture follow the descriptions for the repetition-cat architecture as outlined in \zcref{sec: Rep-code detailed description of ops} very closely. The main difference is that now all algorithmic qubits do provide $Z$ access through the routing qubits underneath the LDPC layer. Thus, options such as dirty \textsc{SelectSwap} are now feasible. This can reduce the number of resource states necessary as well as speed up the operation. In contrast, other operations such as qubit movements, or measurements can take longer as outlined in \zcref{sec:compilation}. This particularly affects measurement based uncomputation of lookup table output as is used in clean \textsc{SelectSwap} in \textsc{OuterPrep} and also during \textsc{RPrep} + \textsc{Rot}. Furthermore, application of a Hadamard gate to a logical qubit in the LDPC layer described in \zcref{sec: Multi-target C_XX and H} takes thrice as long as the corresponding operation in the repetition-cat architecture, since the qubit needs to be moved in and out of the LDPC layer. This affects applications of dirty \textsc{SelectSwap} as the conjugation by a Hadamards now requires $3d$ cycles per Hadamard gate. In the remainder of this section, we discuss any changes from the operations described in \zcref{sec: Rep-code detailed description of ops}. Operations that stay the same on the high-level and where only the cost accounting changes are omitted.

\begin{figure}
    \centering
    \includegraphics[width=\linewidth]{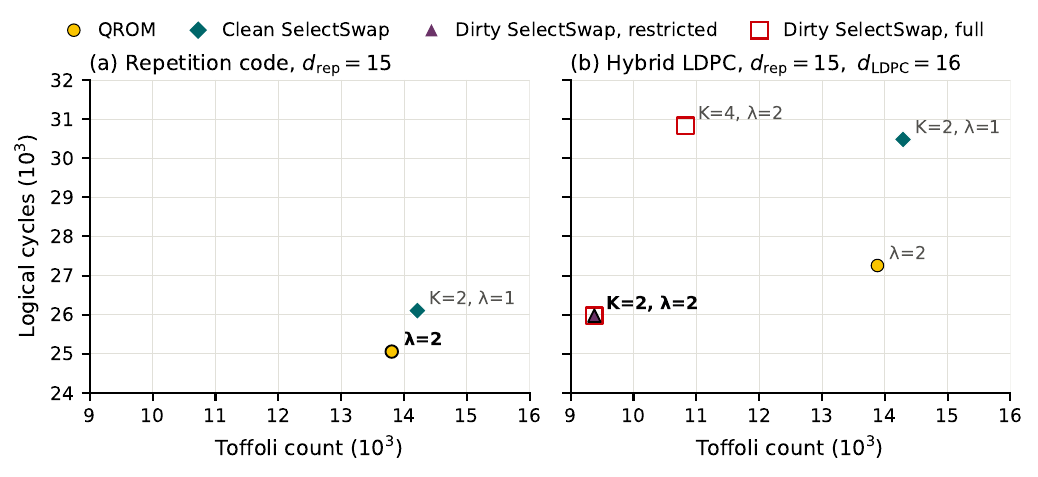}
    \caption{A visualization of admissible parameters for loading two rotation angles at a time both in the repetition-cat architecture and LDPC-cat architecture for FeMoco-54. The left plot shows all admissible options for the repetition-cat architecture, while the right plot shows all options for the LDPC-cat architecture. Both plots share a $y$-axis, depicting the number of logical cycles for the recycling batching of loading $N-1$ rotation angles computed using the compilation outlined in \zcref{sec:compilation}, and each $x$-axis shows the number of Toffolis. Dirty \textsc{SelectSwap}, restricted refers to only using the spin-up qubits and dirty \textsc{SelectSwap} full uses all system qubits. The chosen option is depicted in bold in both cases. Since system qubits are stored without $Z$ access, they are not admissible as dirty qubits for a \textsc{Swap} network in that regime. In contrast, storing system qubits in the LDPC layer allows for $Z$ access and thus dirty \textsc{SelectSwap} is a viable routine.}
    \label{fig:RPrep + Rot TradeOff}
\end{figure}

\paragraph{Usage of dirty \textsc{SelectSwap} in \textsc{InnerPrep} and \textsc{RPrep}.} The main difference in subroutines using lookup table operations, is the ability to use system qubits as dirty qubits. For the cost accounting, we use the compilation of the skew-tree \textsc{Qrom} with ancilla qubits stored in repetition code in the buffer zone as given in \eqref{eq: QROM LDPC time non-slow} and \eqref{eq: QROM LDPC space buffer anc} matching the duration in the repetition cat-architecture and the \textsc{Swap} network compilation in \zcref{sec: Swap LDPC comp} taking roughly twice as long as the corresponding repetition code version.  In principle, the possibility to use dirty qubits affects \textsc{OuterPrep}, \textsc{InnerPrep} and \textsc{RPrep}. However, since dirty \textsc{SelectSwap} requires more \textsc{Swap} applications and the \textsc{Swap} compilation in LDPC takes roughly twice as long per Toffoli gate than the skew-tree \textsc{Qrom}, it is not efficient for small address values. Thus, \textsc{OuterPrep} still uses only plain \textsc{Qrom} and clean \textsc{SelectSwap} analogously to the repetition-cat architecture. For \textsc{InnerPrep}, clean \textsc{SelectSwap} is fully replaced by dirty \textsc{SelectSwap} reducing the Toffoli count and duration.  For \textsc{RPrep}, the options of dirty \textsc{SelectSwap} using half the system qubits or all system qubits as dirty ancillae as described but ultimately not used in \zcref{sec: Rep-code detailed description of ops}, are now available since we \textsc{Qrom} output is stored in the LDPC layer with $Z$ access. We refer again to \zcref{tab:RPrep Rot lookup table batching options} for an overview of options for both FeMoco-54 and FeMoco-76. Note that analogously to the repetition-cat architecture, in order to ensure enough space for the layout for \textsc{Adder}, we need to load at least two rotation angles at a time, i.e., we only consider $\lambda = 2$ (or $\lambda = 1$ for clean \textsc{SelectSwap}). While the table only compares Toffoli count, \zcref{fig:RPrep + Rot TradeOff} also compares the duration of the lookup table batching procedure in logical cycles for both architectures using comparable distances for FeMoco-54 when loading $\lambda =2$ rotation angles at a time. As discussed the plain skew-tree \textsc{Qrom} is the most qubit and time efficient for the repetition-cat architecture, while the dirty \textsc{SelectSwap} using only the spin-up qubits is most efficient across both metrics for the LDPC-cat architecture. The plot for the LDPC-cat architecture also highlights the role of the slower \textsc{Swap} network. Even though the dirty \textsc{SelectSwap} using all system qubits as dirty ancilla with a swap parameter of $K=4$ uses $22\%$ less Toffoli gates than the plain skew-tree \textsc{Qrom} version, it is $13\%$ slower. 

\paragraph{Preparation of the compute zone for \textsc{Adder}.} To avoid qubit movement and since we do not need to allocate access corridors in the LDPC layer, we store the Fourier resource state in the compute layer already ready for \textsc{Adder}. More specifically, we allocate one operand qubit slot followed by a stored Fourier resource state qubit and another operand qubit slot. This is visualized in \zcref{fig:LDPC global layout} using the yellow-green-yellow sequence. We start this pattern with the least significant Fourier resource state qubit in the bottom row to the left of the spin selection qubits and populate the bottom row in ascending order with qubits $k \le \lfloor b_\rot/2 \rfloor$. The top row is filled analogously with qubits $\lfloor b_\rot/2\rfloor + 1, \dots, 2\lfloor b_\rot/2\rfloor$, starting on top of the least significant one. If $b_\rot$ is even this pattern occupies $3 b_\rot/2$ columns densely. If $b_\rot$ is odd, the last application of the pattern is arranged in a triangle shape inside a $2 \times 2$ square, i.e., the most significant Fourier resource state qubit is stored in the bottom right corner of the square with one operand slot to the left and another on top. Observe that since $Z$ access is given for the Fourier resource state, the parity-\textsc{Cnot} required before and after \textsc{Adder} to finish the phase gradient addition does not require any extra thought. We account for this with $d$ cycles each. However, the shuffling of the summands requires more care, which we discuss in the following two paragraphs. 

For \textsc{Outer} and \textsc{InnerPrep}, $b_\coef$ coefficients are loaded and only the first $b_\coef$ qubits of the Fourier resource state take part in the addition. The output of the lookup tables is loaded directly to the left of the appropriate Fourier resource state qubits, ready to be added. Thus, no movement is necessary to prepare the addition layout for this subroutine. One thing to note however is that in case the data was loaded using clean \textsc{SelectSwap}, extra lookup table output needs to be uncomputed to free up the clean ancilla slots for the addition. As mentioned previously, this requires access to the $Z$ logical operator for one logical cycle. To avoid routing congestion we cost this serially as an upper bound. Next the newly freed ancilla are prepared in $\ket{+}_L$ to be ready for the addition. This also has to be done using a QND measurement, but now only access to the $X$ logical operator on the tip is required. Hence, we can perform the measurements at the same time totaling $d$ cycles. Note that after the addition has been successfully performed, the Fourier resource state has moved one slot to the right. Observe that after the application of \textsc{Adder}, the summand in the second slot is uncomputed using an $M_Z$ measurement. Thus, up to Pauli frame updates, the pattern in the compute zone consists of the first rotation angle in slot one, ancilla in $\ket{0}_L$ in slot two and the shifted over Fourier resource state in slot three. Thus, we can move the Fourier resource state over using \textsc{Cnot} gates followed by an $M_X$ measurement of the third slot. We assume each \textsc{Cnot} is performed serially due to routing congestion, contributing a total of $b_\coef d$ cycles. We perform the $M_X$ measurements again simultaneously with the same reasoning as in the discussion above contributing $d$ cycles in total.

For \textsc{RPrep}, we load the rotation angles as follows: the first rotation angle into slot one to the left of the Fourier resource state and the second rotation angle into slot three to the right of the Fourier resource state. Any additional rotation angles would be stored thereafter. Thus, to prepare for the first phase gradient addition, we need to move the second set of rotation angles out of the LDPC layer. We store the $b_\rot$ rotation angle bits in the repetition layer underneath the system qubits and if necessary also underneath the address and spin selection qubits. Thus, we need to ensure enough space in the routing layer underneath. More specifically, the layout needs to satisfy $b_\rot d_\rep \leq (N+1 + \ell_\text{address} + 2 -4)H $, where the $-4$ accounts for the fact that the compilation of \textsc{Adder} requires extra routing to the right of the compute zone. The qubits are moved out using a \textsc{Cnot} gate which routes into the storage area followed by an $M_X$ measurement of the LDPC layer. This also guarantees that the ancilla are correctly prepared for \textsc{Adder}. To avoid routing conflicts, this is again costed serially, costing $2d$ cycles per move for a total of $2b_\rot d$.  After the first iteration of \textsc{Adder}, we need to move the Fourier resource state back to its original placement for a total of $(b_\rot +1)d$ cycles using the same reasoning as in the previous paragraph. Next, the first rotation angles need to be uncomputed using another $M_Z$ measurement. As this again requires access to their $Z$ logical operator, the operation is performed serially for a total of $b_\rot d$ cycles. The first slot is now in $\ket{0}_L$ up to Pauli frame updates and the stored rotation angles can be moved out using an $M_{XX}$. Since the qubits are packed very tightly in the routing layer underneath the system qubits, we cost this operation also serially to avoid congestion and access constrictions, leading to a contribution of $b_\rot d$.  The considerations for any additional rotation angles in a given batch are very similar with the same accounting for the Fourier resource state movement. The only difference being that the move now happens from one LDPC location to another and can be performed using \textsc{Cnot} gates followed by an $M_X$ measurement. To ensure sufficient routing access, we again cost this serially with $(b_\rot +1)d$ total time contribution. In total, we obtain $t_\text{shuffle} = 3d b_\rot + (\lambda -2)(b_\rot + 1)d + \lambda b_\rot d + \lambda(b_\rot + 1)d$, where the last summand describes the contribution of moving the Fourier resource state back after each addition.

\subsubsection{Resource counts}\label{sec: LDPC resource counts}
To compute the resource count for the LDPC-cat architecture, we follow the same procedure as outlined for the repetition-cat architecture \zcref{sec: rep code resource count}. While the considerations for the error contribution of resource states are analogous to the repetition-cat architecture, the error stemming from the algorithmic qubits now splits into errors occurring in the LDPC layer and errors occurring in the repetition layer. For both layers, we compute the probability $\varepsilon_\text{alg}^\text{layer}$ for $\text{layer} \in \{\mathrm{LDPC}, \mathrm{rep}\}$ that at least one logical error occurs in the layer. For each layer, we consider the effective logical volume for the layer, $V_\mathrm{log,eff}^\mathrm{rep}$ and $V_\mathrm{log,eff}^\mathrm{LDPC}$, where $V_\mathrm{log,eff}^\mathrm{rep}$ is obtained from the physical volume in qubitcycles from the repetition layer including factory routing normalized by $d_\mathrm{rep}(2d_\mathrm{rep}-1)$. $V_\mathrm{log,eff}^\mathrm{LDPC}$ is computed from the number of logical qubits times syndrome extraction cycles in the LDPC layer, normalized by $d_\mathrm{LDPC}$. We obtain
\begin{equation*}
    \varepsilon_\text{alg}^\text{rep} = 1-(1-p_Z^\text{rep})^{V_\mathrm{log,eff}^\mathrm{rep}}, \qquad \varepsilon_\text{alg}^\text{LDPC} = 1-(1-p_Z^\text{LDPC})^{V_\mathrm{log,eff}^\mathrm{LDPC}}, 
\end{equation*}
where $p_Z^\mathrm{rep}$ and $p_Z^\mathrm{LDPC}$ denote the logical phase-flip probability per logical qubit after $d_\mathrm{rep}$ and $d_\mathrm{LDPC}$ syndrome extraction cycles, respectively. Analogously to the repetition-cat architecture, $p_Z^\mathrm{rep}$ is obtained using the ansatz in \eqref{eq: LDPC ansatz} with the parameters from \cite[Table III]{Ruiz_2025} and $p_Z^\mathrm{LDPC}$ is given by the same ansatz with the parameters from \zcref{tab: LDPC ABC}. We then set $\varepsilon_\mathrm{alg} = \varepsilon_\text{alg}^\text{rep} + \varepsilon_\text{alg}^\text{LDPC}$ using the same argument as used for $p_\mathrm{fail}$ in the repetition-cat architecture leading to $p_\mathrm{fail} = \varepsilon_\text{alg}^\text{rep} + \varepsilon_\text{alg}^\text{LDPC} + \varepsilon_{\CXCXX} + \varepsilon_{\keti}$. The appropriate code distances are then determined by choosing the shortest expected duration analogously to \zcref{sec: rep code resource count}. A summary of results is given in \zcref{tab: LDPC code error accounting}. 
The contribution of the LDPC layer $\varepsilon_\text{alg}^\text{LDPC}$ dominates the algorithmic error, accounting for around 90\% of $\varepsilon_\mathrm{alg}$ for both molecules at $p_Z = 5 \times 10^{-4}$ and 96\% for FeMoco-76 at $p_Z = 10^{-3}$. Only for FeMoco-54 at $p_Z = 10^{-3}$ does $\varepsilon_\text{alg}^\text{rep}$ dominate, constituting roughly two thirds of $\varepsilon_\mathrm{alg}$.
In the case of FeMoco-54 with $p_Z = 5 \times 10^{-4}$, the full algorithm runs on $7\,164$ physical qubits  with an expected runtime of 0.98 days. This is a 36\% reduction in physical qubits at the expense of a slightly longer run time, more specifically a 3\% increase in comparison with the repetition-cat architecture (\zcref{tab: Rep code error accounting}). For FeMoco-76 the comparison is positive in both regards, since the ability to use dirty \textsc{SelectSwap} reduces the overall runtime by 26\% in addition to a 38\% reduction in physical qubits. This illustrates the trade-off between some operations being slower compiled on the LDPC-cat architecture, but the compact $Z$ access giving flexibility that can be exploited in algorithm design. Furthermore, the algorithm for the LDPC-cat architecture is more Toffoli efficient, which reduces the error contribution coming from $\CXCXX$ states. This is particularly impactful for FeMoco-76 at $p_Z = 10^{-3}$ which has a failure probability of less than 20\% for the LDPC-cat architecture but 32\% for the repetition-cat architecture.

\begin{table}[hbpt]
    \centering
    \begin{tabular}{c|c|c|c|c|c|c|c|c|c}\hline
    FeMoco & $p_Z$ & $d_\mathrm{LDPC}$ & $d_\mathrm{rep}$ & $\varepsilon_\mathrm{alg}$ & $\varepsilon_\mathrm{\CXCXX}$ & $\varepsilon_\mathrm{\keti}$ & Physical & One-shot & $\langle t\rangle$ \\
     &  &  &  &  &  &  & qubits & (days) & (days) \\ \hline
    \rowcolor{gray!10}  & $5\times 10^{-4}$ & 16 & 15 & 0.0035 & 0.00041 & 0.0055 & $7\,164$ & 0.97 & 0.98 \\
    \rowcolor{gray!10} \multirow{-2}{*}{54} & $10^{-3}$ & 22 & 17 & 0.0039 & 0.030 & 0.018 & $7\,636$ & 1.26 & 1.33 \\
     & $5\times 10^{-4}$ & 16 & 15 & 0.014 & 0.0014 & 0.020 & $7\,818$ & 3.20 & 3.31 \\
    \multirow{-2}{*}{76} & $10^{-3}$ & 22 & 19 & 0.0058 & 0.10 & 0.067 & $8\,486$ & 4.17 & 5.06 \\ \hline
\end{tabular}
\caption{Runtime and total physical qubit count for FeMoco-54 and FeMoco-76 for different phase-flip errors in the LDPC-cat architecture. The physical qubit count includes ancilla and contributions from factories and factory routing. The duration accounts for $5\%$ additional slack for initialization, measurements and buffer staging. The expected duration is computed by normalizing with $1-p_\mathrm{fail}$, where $p_\mathrm{fail} = \varepsilon_\mathrm{alg} + \varepsilon_\mathrm{\CXCXX} + \varepsilon_\mathrm{\keti}$. The algorithmic error contribution is obtained additively from the contribution of the repetition layer and the LDPC layer.}
    \label{tab: LDPC code error accounting}
\end{table}

\zcref{tab:be_femoco54_ldpc} gives a detailed account for one round of the block encoding in terms of round share and resource state share for FeMoco-54. While the split between \textsc{RPrep + Rot} and \textsc{InnerPrep} stays essentially the same as in \zcref{tab:be_femoco54_rep}, the ability to use lookup table batching with dirty \textsc{SelectSwap} in \textsc{RPrep + Rot} and dirty \textsc{SelectSwap} in \textsc{InnerPrep} reduces the round share of \textsc{Qrom} to 45\% with the share of \textsc{Swap} increasing to 31.9\%. Notably, the preparation of the \textsc{Adder} layout as well as uncomputation of lookup table output now takes 6.2\% and 2.7\% respectively, whereas for the repetition-cat architecture these procedures were basically negligible taking less than 1.4\%. Another interesting difference is the rounds spent applying Hadamard gates. Comparing the explicit application of Hadamard gates required for dirty \textsc{Swap} and \textsc{Phaseup}, we see that this is essentially negligible at 0.3\% for the repetition-cat architecture with the only source being the \textsc{Phaseup}, but takes up 17.2\% of the block encoding rounds in the LDPC-cat architecture. This is due to both an increase in Hadamard count because of dirty \textsc{SelectSwap} and also the fact that a Hadamard application in the LDPC-cat architecture takes thrice as many rounds than in the repetition-cat architecture. To isolate the effect of the different architectures, we compare the architectures using the same algorithm in the next section.

\begin{table}[hbtp]
    \centering
    \begin{subtable}[t]{\textwidth}
    \centering
    \begin{tabular}{c|c|c|c|c|c|c}\hline
        Subroutine & Rounds & Share & Toffolis & Share & $\keti$-states & Share \\ \hline
        \rowcolor{gray!10} \textsc{Rprep + Rot} ($\times 4$) & $1\,662\,316$ & 69.8\% & $37\,504$ & 62.7\% & $113\,680$ & 65.1\% \\
        \textsc{InnerPrep}$^\dagger$ ($\times 2$) & $370\,096$ & 15.5\% & $11\,300$ & 18.9\% & $31\,310$ & 17.9\% \\
        \rowcolor{gray!10} \textsc{InnerPrep} ($\times 2$) & $302\,048$ & 12.7\% & $9\,852$ & 16.5\% & $25\,966$ & 14.9\% \\
        \textsc{OuterPrep}$^\dagger$ ($\times 1$) & $17\,935$ & 0.8\% & $414$ & 0.7\% & $1\,257$ & 0.7\% \\
        \rowcolor{gray!10} \textsc{OuterPrep} ($\times 1$) & $15\,986$ & 0.7\% & $383$ & 0.6\% & $1\,093$ & 0.6\% \\
        \textsc{Spin} ($\times 4$) & $8\,616$ & 0.4\% & $248$ & 0.4\% & $1\,016$ & 0.6\% \\
        \rowcolor{gray!10} \textsc{Maj} ($\times 2$) & $2\,848$ & 0.1\% & $56$ & 0.1\% & $236$ & 0.1\% \\
        \textsc{OuterRef} ($\times 1$) & $768$ & $<0.1\%$ & $16$ & $<0.1\%$ & $66$ & $<0.1\%$ \\
        \rowcolor{gray!10} \textsc{InnerRef} ($\times 1$) & $192$ & $<0.1\%$ & $4$ & $<0.1\%$ & $18$ & $<0.1\%$ \\
        \hline
        Primitive & Rounds & Share & Toffolis & Share & $\keti$-states & Share \\ \hline
        \rowcolor{gray!10} \textsc{Qrom} & $1\,074\,116$ & 45.1\% & $40\,322$ & 67.5\% & $80\,980$ & 46.4\% \\
        \textsc{Swap} & $759\,316$ & 31.9\% & $11\,942$ & 20.0\% & $64\,520$ & 36.9\% \\
        \rowcolor{gray!10} \textsc{Adder} & $159\,516$ & 6.7\% & $3\,798$ & 6.4\% & $15\,954$ & 9.1\% \\
        Phaseup & $159\,249$ & 6.7\% & $3\,543$ & 5.9\% & $12\,460$ & 7.1\% \\
        \rowcolor{gray!10} Shuffle & $148\,384$ & 6.2\% & -- & -- & -- & -- \\
        MBU & $63\,680$ & 2.7\% & -- & -- & -- & -- \\
        \rowcolor{gray!10} Rest & $16\,544$ & 0.7\% & $172$ & 0.3\% & $728$ & 0.4\% \\
        \hline
        Total & $2\,380\,805$ & 100\% & $59\,777$ & 100\% & $174\,642$ & 100\% \\ \hline
    \end{tabular}
    \caption{Number of syndrome extraction rounds, Toffoli gates and $\keti$-states by subroutine and by primitive. The subroutine numbers take the number of calls (in brackets) into account, so that they make up the whole block encoding. Both subroutines and primitives break down the same totals. Here, shuffle denotes the procedure to prepare the compute zone for the \textsc{Adder} layout and MBU refers to explicit uncomputation of lookup table output. Rest combines rotation, comparator$^*$, quantum-classical-addition$^*$, \textsc{Ref}$^*$ and remaining routines for Majorana application$^*$, i.e., the primitives with less than 0.1\% each of the effective volume in \zcref{tab:be_femoco54_ldpc_volume}, where rotation refers to the parity-\textsc{Cnot} gates necessary for phase gradient addition.}\label{tab:be_femoco54_ldpc_counts}
    \end{subtable}
    \par\bigskip
    \begin{subtable}[t]{\textwidth}
    \centering
    \begin{tabular}{c|c|c|c|c}\hline
        Primitive & Repetition layer & LDPC layer & Total & Share \\ \hline
        \rowcolor{gray!10} System (idle) & -- & $1.15\times 10^{7}$ & $1.15\times 10^{7}$ & 37.0\% \\
        \textsc{Qrom} & $3.28\times 10^{6}$ & $4.95\times 10^{6}$ & $8.23\times 10^{6}$ & 26.5\% \\
        \rowcolor{gray!10} \textsc{Swap} & $2.47\times 10^{6}$ & $2.77\times 10^{6}$ & $5.24\times 10^{6}$ & 16.9\% \\
        Fourier state (idle) & -- & $2.22\times 10^{6}$ & $2.22\times 10^{6}$ & 7.1\% \\
        \rowcolor{gray!10} Address and spin qubits (idle) & -- & $1.14\times 10^{6}$ & $1.14\times 10^{6}$ & 3.7\% \\
        \textsc{Adder} & $6.18\times 10^{5}$ & $4.60\times 10^{5}$ & $1.08\times 10^{6}$ & 3.5\% \\
        \rowcolor{gray!10} Phaseup & $3.74\times 10^{5}$ & $2.39\times 10^{5}$ & $6.13\times 10^{5}$ & 2.0\% \\
        QPE qubits (idle) & -- & $2.98\times 10^{5}$ & $2.98\times 10^{5}$ & 1.0\% \\
        \rowcolor{gray!10} Output (idle) & $2.62\times 10^{5}$ & $1.24\times 10^{4}$ & $2.74\times 10^{5}$ & 0.9\% \\
        Shuffle & $9.26\times 10^{4}$ & $1.45\times 10^{5}$ & $2.38\times 10^{5}$ & 0.8\% \\
        \rowcolor{gray!10} Flag (idle) & -- & $1.04\times 10^{5}$ & $1.04\times 10^{5}$ & 0.3\% \\
        MBU & $3.90\times 10^{4}$ & $6.09\times 10^{4}$ & $9.99\times 10^{4}$ & 0.3\% \\
        \rowcolor{gray!10} Rest & $3.46\times 10^{4}$ & $1.35\times 10^{4}$ & $4.81\times 10^{4}$ & 0.2\% \\
        \hline
        Total & $7.16\times 10^{6}$ & $2.39\times 10^{7}$ & $3.11\times 10^{7}$ & 100\% \\ \hline
    \end{tabular}
    \caption{Effective logical volume $V_\mathrm{log,eff}$ by primitive and layer, i.e., the repetition layer's physical qubit-rounds divided by $d_\mathrm{rep}(2d_\mathrm{rep}-1)$ and the LDPC layer's logical qubit-rounds divided by $d_\mathrm{LDPC}$. The share is of the total. Idle output qubits refer to already loaded rotation angles, idle during other operations. Idle qubits account for 50.0\% of the algorithmic effective volume. Rest combines rotation, comparator$^*$, quantum-classical-addition$^*$, \textsc{Ref}$^*$ and remaining routines for Majorana application$^*$, each below 0.1\%.}\label{tab:be_femoco54_ldpc_volume}
    \end{subtable}
    \caption{Cost breakdown of one block encoding for FeMoco-54 in the LDPC-cat architecture at $d_\mathrm{rep} = 15$, $d_\mathrm{LDPC} = 16$, with \textsc{Rprep + Rot} implemented by lookup table batching using restricted dirty \textsc{SelectSwap} ($K = 2$) loading $\lambda = 2$ rotation angles at a time, and \textsc{InnerPrep} using dirty \textsc{SelectSwap}. $^*$ denotes routines whose duration is estimated at $3d$ per Toffoli. Such non-compiled routines total 0.4\% of the rounds, 0.1\% of the logical qubit-cycles, 0.3\% of the Toffolis.}
    \label{tab:be_femoco54_ldpc}
\end{table}

\begin{table}[hbtp]
    \centering
    \begin{tabular}{c|c|c|c|c|c|c|c|c|c}\hline
    FeMoco & Variant & $d_\mathrm{rep}$ & $d_\mathrm{LDPC}$ & $\varepsilon_\mathrm{alg}$ & $\varepsilon_\mathrm{\CXCXX}$ & $\varepsilon_\mathrm{\keti}$ & Physical & One-shot & $\langle t\rangle$ \\
     &  &  &  &  &  &  & qubits & (days) & (days) \\ \hline
    \rowcolor{gray!10}  & Rep & 15 & -- & 0.0011 & 0.00058 & 0.0059 & $11\,192$ & 0.94 & 0.95 \\
    \rowcolor{gray!10}  & Rep, dirty & 15 & -- & 0.00097 & 0.00040 & 0.0055 & $13\,028$ & 0.67 & 0.68 \\
    \rowcolor{gray!10}  & LDPC & 15 & 16 & 0.0035 & 0.00041 & 0.0055 & $7\,164$ & 0.97 & 0.98 \\
    \rowcolor{gray!10} \multirow{-4}{*}{54} & LDPC, rep & 15 & 16 & 0.0040 & 0.00059 & 0.0060 & $7\,164$ & 1.08 & 1.10 \\
    \hline
     & Rep & 15 & -- & 0.0058 & 0.0027 & 0.027 & $12\,568$ & 4.33 & 4.49 \\
     & Rep, dirty & 15 & -- & 0.0040 & 0.0014 & 0.020 & $15\,152$ & 2.14 & 2.20 \\
     & LDPC & 15 & 16 & 0.014 & 0.0014 & 0.020 & $7\,818$ & 3.20 & 3.31 \\
    \multirow{-4}{*}{76} & LDPC, rep & 15 & 16 & 0.022 & 0.0028 & 0.027 & $7\,818$ & 4.83 & 5.09 \\ \hline
\end{tabular}
\caption{Error accounting with the lookup options of either architecture in both, at $p_Z = 5\times 10^{-4}$ for all variants. The rep code with dirty \textsc{SelectSwap} needs $Z$ access to the system qubits it borrows, which the repetition-cat layout provides only with access corridors in between them, one per two system qubits ($N$ in total): they are included in the physical qubits count and taken as present for the whole run in its volume to compute $\varepsilon_\mathrm{alg}$. The physical qubit count includes ancilla, factories and factory routing. The durations include $5\%$ slack and the expected duration is the one-shot duration divided by $1-p_\mathrm{fail}$, with $p_\mathrm{fail} = \varepsilon_\mathrm{alg} + \varepsilon_\mathrm{\CXCXX} + \varepsilon_\mathrm{\keti}$. Rep and LDPC refer to each architecture with its own chosen lookup options. Rep, dirty is the repetition-cat architecture with the LDPC-cat architecture's lookup options for the same molecule: dirty \textsc{SelectSwap} in \textsc{InnerPrep} and in \textsc{Rprep + Rot}, the latter with $K = 2$, $\lambda = 2$ on half the system qubits for FeMoco-54 and with $K = 4$, $\lambda = 2$ on all system qubits for FeMoco-76. Lastly, LDPC, rep is the LDPC architecture with the repetition code's option of \textsc{Qrom} lookup table batching and clean \textsc{SelectSwap} in \textsc{InnerPrep}.}
    \label{tab:cross_error_accounting}
\end{table}

\subsection{Trading footprint for runtime}\label{sec:comparison}
\begin{figure}
    \centering
    \includegraphics[width=\linewidth]{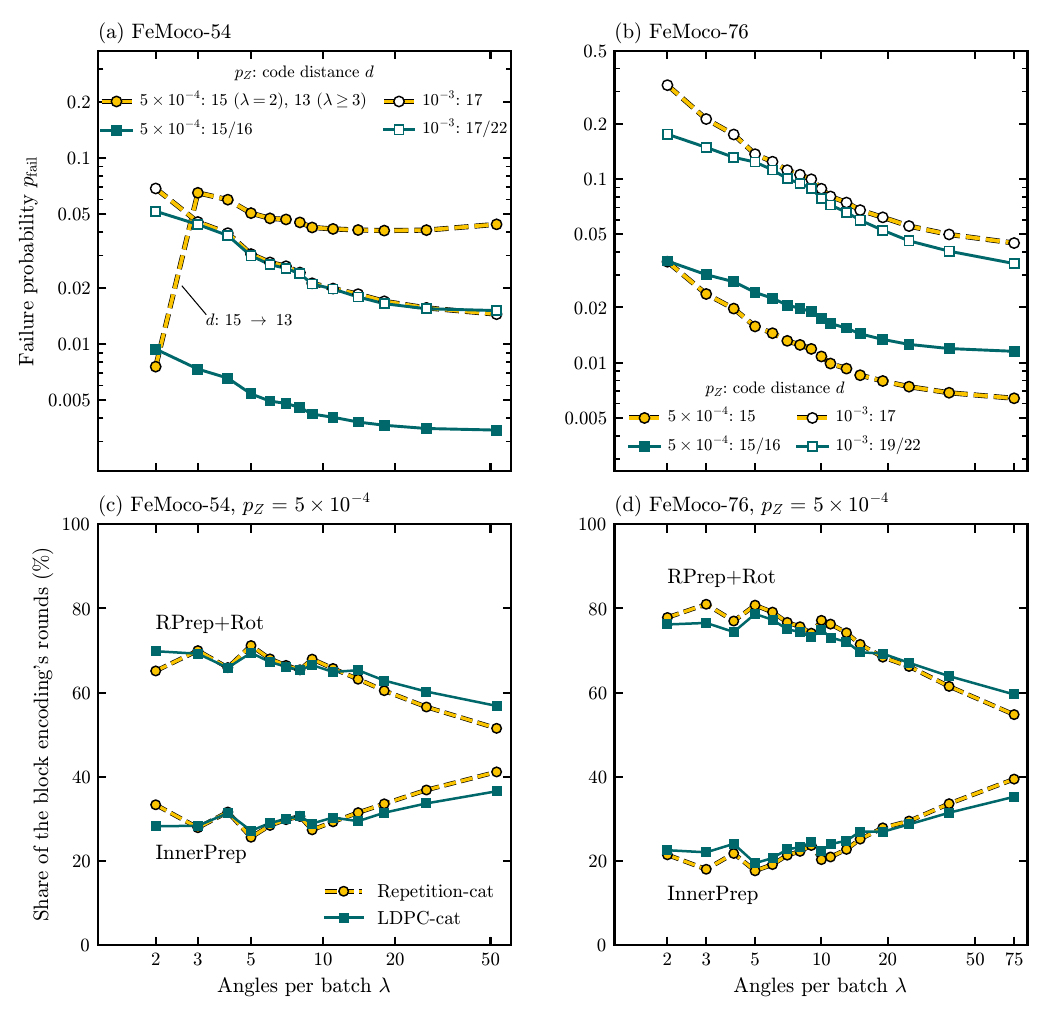}
    \caption{Panels (a) and (b) show the failure probability against the number of angles per batch $\lambda$. Each point is the configuration with the shortest expected runtime for its compute-zone size, with the code distance chosen by the same criterion. For FeMoco-54 on the repetition-cat architecture this lowers the distance from $d = 15$ to $13$ from $\lambda = 3$ on, at the price of a higher failure probability. Panels (c) and (d) show the share of the block encoding's rounds spent in \textsc{RPrep + Rot} and \textsc{InnerPrep} again as a function of $\lambda$. At $p_Z = 10^{-3}$ the shares differ by at most two percentage points and are not shown. The remaining rounds are spent in \textsc{OuterPrep}, \textsc{PhaseFixup} and the other subroutines.}
    \label{fig:Enlarge compute zone: Failure prob}
\end{figure}

\zcref{sec: rep code resource count} and \zcref{sec: LDPC resource counts} give the results for the fastest option for the reference configuration ($\lambda =2$) for each architecture. In this section, we increase the compute zone budget as a function of the rotation angles $\lambda$ loaded, to further highlight the strengths and weaknesses of each architecture and to take advantage of our closed from compilation. The additional clean qubits also enlarge the \textsc{SelectSwap} networks of \textsc{OuterPrep} and \textsc{InnerPrep}, which draw on the same compute zone. All other choices stay fixed: a single $\CXCXX$-factory, \textsc{InnerPrep} by pure state preparation, and serial phase-gradient additions. For each $\lambda$ from 2 to $N-1$, we select the lookup options and the code distance with the shortest expected runtime, as for the reference configuration. \zcref{fig:tradeoff lambda sweep intro} shows the resulting expected runtime against the number of physical qubits, and \zcref{fig:Enlarge compute zone: Failure prob} the failure probability and the share of the block encoding's rounds spent in \textsc{RPrep + Rot} and \textsc{InnerPrep}. Adding any of the fixed options can only lower these curves, so they give an upper bound for the space--time trade-off achievable with our compiled primitives.

The size of the compute zone is determined by $\lambda b_\rot$ in both architectures. More precisely, for the repetition-cat architecture, the compute zone contributes $\lceil 3 \lambda b_\rot/2 \rceil (d+4)$, thus increasing $\lambda \to \lambda + 1$ increases the qubit count by $24(d+5)$ for FeMoco-54, which equals 960 qubits including ancilla for $d=15$ and 864 qubits at $d=13$. For LDPC, an increase in $\lambda$ by one costs $2(H-1)\lceil b_\rot/2 \rceil$ qubits in the LDPC layer and roughly $2(H+5)$ qubits in the repetition layer both counts including ancilla, i.e., 288 qubits for $d=16$. Note that this LDPC number is a slight overcount, as the 22 qubit offset to ensure that the buffer zone sits enough to the right of the \textsc{Adder} memory zone is not necessary anymore with large enough compute zone. The LDPC-cat architecture thus pays at least 3 times fewer qubits per step.

\zcref{fig:tradeoff lambda sweep intro} shows that in both architectures the runtime first changes rapidly then flattens. For FeMoco-54 in the repetition-cat architecture, the first step costs no qubits at all: the shorter run at a larger compute zone allows for $d = 13$ instead of $d = 15$, which offsets the qubit increase from the wider compute zone ($11\,176$ against $11\,192$ physical qubits). The expected runtime drops from 0.95 to 0.62 days, at the price of a failure probability of 6.5\% instead of 0.8\% as seen in panel (a) of \zcref{fig:Enlarge compute zone: Failure prob}. The flattening  of \zcref{fig:tradeoff lambda sweep intro} follows from the round shares in panels (c) and (d) of \zcref{fig:Enlarge compute zone: Failure prob}. \textsc{RPrep + Rot} shrinks with the number of sweeps, but \textsc{InnerPrep} profits from the larger compute zone only until its \textsc{SelectSwap} network saturates. Its share grows from 21–33\% at $\lambda = 2$ to 35–41\% when all rotation angles are loaded at once. The space--time volume is minimal at $\lambda = 9$ for FeMoco-54, with 0.109 and 0.088 megaqubit-hours for the repetition-cat and LDPC-cat architecture respectively. The minimum for FeMoco-76 is reached at $\lambda = 15$ with 0.50 megaqubit-hours for the repetition-cat and $\lambda = 19$ with 0.33 for the LDPC-cat architectures.

The cheaper steps of the LDPC-cat architecture decide which architecture is faster at a given footprint. up to about $16\,000$ (FeMoco-54) and $24\,000$ (FeMoco-76) physical qubits, the LDPC-cat architecture reaches the shorter runtime, because it can afford a larger compute zone. For FeMoco-76, the lead then changes hands twice: the repetition-cat architecture is 1\% faster from $24\,300$ physical qubits, and the LDPC-cat architecture 10\% faster from $27\,500$, where it loads all rotation angles at once. Beyond $16\,000$ (FeMoco-54) and $28\,000$ (FeMoco-76) physical qubits, the repetition-cat architecture is faster, because at equal $\lambda$ its operations requiring $Z$ access are cheaper, as we discuss for the reference configuration in the remainder of this subsection. 

\paragraph{Architectural comparison.} In addition to enlarging the compute zone per architecture, we compare the numbers for the same algorithm run both on the repetition-cat architecture as well as the LDPC-cat architecture. \zcref{tab:cross_error_accounting} compares the time and error contribution for the same algorithm for the reference configuration run on the repetition-cat architecture and the LDPC-cat architecture for FeMoco-54 and FeMoco-76 for $p_Z = 5 \times 10^{-4}$. If we only use \textsc{Qrom} lookup table batching and clean \textsc{SelectSwap} with the LDPC-cat architecture, we observe a 15\% increase relative to the repetition-cat architecture. We note that dirty-ancilla data loading reduces the runtime by 10\% for the LDPC architecture in comparison to solely using \textsc{Qrom} and clean \textsc{SelectSwap}. In contrast, adding access corridors to the system qubits allowing for dirty \textsc{SelectSwap} in both \textsc{Rot + RPrep} as well as \textsc{InnerPrep}, reduces the runtime by 29\% to an expected runtime of 0.68 days. The additional 54 access corridors lead to a physical qubit increase of 1836. Thus, the same reduction in Toffoli gates leads to only a 10\% runtime reduction for the LDPC-cat architecture but a 29\% for the repetition-cat architecture. However, the repetition-cat speed-up comes at a 16.4\% increase in spatial footprint while the LDPC-cat architecture's footprint does not change.

The reason for the LDPC slowdown for FeMoco-54 is two-fold: on the one hand distance discretization, i.e., only having distances $d=12,16,22$ available leads to a slowdown of $\approx 7\%$ compared to distance $d=15$, and on the other hand \textsc{Swap}, Hadamards and MBU are slower. More precisely, MBU of lookup table output takes up 2.4\% in the case of LDPC with the \textsc{Qrom} lookup table batching compared to less than 0.1\% in the repetition-cat architecture. Since the dirty-ancilla data loading reduces the runtime on the LDPC-cat architecture by only 10\%, it does not compensate the 15\% slowdown, and the LDPC-cat architecture remains slower than the repetition-cat architecture. For FeMoco-76, changing the algorithm reduces the runtime by 35\%. This far outweighs the 13\% slowdown due to slower operations in the LDPC-cat architecture compared to the repetition-cat architecture. Thus, the algorithm optimized for the LDPC-cat architecture is faster than the one optimized for the repetition-cat architecture for the reference configuration. 

\zcref{tab:cross_femoco54_primitives} shows the breakdown across primitives for both the dirty \textsc{SelectSwap} option in the repetition-cat architecture (\zcref{tab:cross_femoco54_rep_dirty_options}) as well as for the LDPC-cat architecture (\zcref{tab:cross_femoco54_LDPC_rep_options}) for FeMoco-54 at distance $d_\mathrm{rep} = 15, d_\mathrm{LDPC} = 16$. Comparing \zcref{tab:cross_femoco54_rep_dirty_options} with its LDPC counterpart \zcref{tab:be_femoco54_ldpc_counts} we observe that in the repetition-cat architecture \textsc{Qrom} still constitutes 63\% of all rounds in total with \textsc{Swap} constituting less than a quarter of all rounds. This is in contrast with the LDPC-cat architecture where \textsc{Qrom} rounds were less than half of all rounds and \textsc{Swap} constituted nearly a third. Analogously, looking at the breakdown of the \textsc{Qrom} only rotation loading for the LDPC-cat architecture in \zcref{tab:cross_femoco54_LDPC_rep_options}, we see that while the number of rounds for \textsc{Qrom} is nearly the same as in the repetition-code architecture, it takes up 8 percentage points less share. The discrepancy in the number of rounds (LDPC takes $80\,474$ more rounds) comes mainly from the difference in distance and the fact that for \textsc{OuterPrep}, one lookup table chooses the \textsc{Qrom} only version for the LDPC-cat architecture but \textsc{SelectSwap} in the repetition-cat architecture. This is due to the fact that we pick the \textsc{SelectSwap} split using duration not Toffoli count, leading to the \textsc{SelectSwap} being slightly disadvantaged in the LDPC-cat architecture. The difference in \textsc{Qrom} share is mostly taken up by more expensive preparation of the \textsc{Adder} layout and MBU.

\begin{table}[hbtp]
    \centering

    \begin{subtable}[t]{\textwidth}
    \centering
    \begin{tabular}{c|c|c|c|c|c|c}\hline
        Primitive & Rounds & Share & Toffolis & Share & $\keti$-states & Share \\ \hline
        \rowcolor{gray!10} \textsc{Qrom} & $1\,031\,898$ & 62.8\% & $40\,292$ & 68.2\% & $80\,920$ & 46.8\% \\
        \textsc{Swap} & $322\,740$ & 19.6\% & $11\,960$ & 20.2\% & $64\,864$ & 37.5\% \\
        \rowcolor{gray!10} \textsc{Adder} & $159\,516$ & 9.7\% & $3\,798$ & 6.4\% & $15\,954$ & 9.2\% \\
        Phaseup & $83\,978$ & 5.1\% & $2\,863$ & 4.8\% & $10\,548$ & 6.1\% \\
        \rowcolor{gray!10} Shuffle & $29\,556$ & 1.8\% & -- & -- & -- & -- \\
        Rest & $15\,780$ & 1.0\% & $172$ & 0.3\% & $728$ & 0.4\% \\
        \hline
        Total & $1\,643\,468$ & 100\% & $59\,085$ & 100\% & $173\,014$ & 100\% \\ \hline
    \end{tabular}
    \caption{Rep code, dirty \textsc{SelectSwap}, $d = 15$. Its access corridors cost no rounds. Rest combines comparator$^*$, quantum-classical-addition$^*$, rotation, \textsc{Ref}$^*$, remaining routines for Majorana application$^*$ and MBU, each below 0.1\% of the effective volume.}\label{tab:cross_femoco54_rep_dirty_options}
    \end{subtable}
    \par\bigskip
    \begin{subtable}[t]{\textwidth}
    \centering
    \begin{tabular}{c|c|c|c|c|c|c}\hline
        Primitive & Rounds & Share & Toffolis & Share & $\keti$-states & Share \\ \hline
        \rowcolor{gray!10} \textsc{Qrom} & $2\,087\,564$ & 78.7\% & $78\,606$ & 90.8\% & $157\,500$ & 83.4\% \\
        \textsc{Adder} & $159\,516$ & 6.0\% & $3\,798$ & 4.4\% & $15\,954$ & 8.5\% \\
        \rowcolor{gray!10} Phaseup & $159\,249$ & 6.0\% & $3\,543$ & 4.1\% & $12\,460$ & 6.6\% \\
        Shuffle & $148\,384$ & 5.6\% & -- & -- & -- & -- \\
        \rowcolor{gray!10} MBU & $64\,512$ & 2.4\% & -- & -- & -- & -- \\
        \textsc{Swap} & $17\,276$ & 0.7\% & $490$ & 0.6\% & $2\,112$ & 1.1\% \\
        \rowcolor{gray!10} Rest & $16\,544$ & 0.6\% & $172$ & 0.2\% & $728$ & 0.4\% \\
        \hline
        Total & $2\,653\,045$ & 100\% & $86\,609$ & 100\% & $188\,754$ & 100\% \\ \hline
    \end{tabular}
    \caption{LDPC, with \textsc{Qrom} only lookup table batching, $d_\mathrm{rep} = 15$, $d_\mathrm{LDPC} = 16$. Rest combines rotation, comparator$^*$, quantum-classical-addition$^*$, \textsc{Ref}$^*$ and remaining routines for Majorana application$^*$, each below 0.1\% of the effective volume.}\label{tab:cross_femoco54_LDPC_rep_options}
    \end{subtable}
        \caption{Rounds, Toffoli gates and $\keti$-states of one block encoding for FeMoco-54 by primitive, with the lookup options of either architecture in both. The rep code with dirty \textsc{SelectSwap} needs Z access to the system qubits it borrows, which the repetition-cat layout provides only with access corridors under them, one per two system qubits ($N$ in all). $^*$ denotes routines whose duration is estimated at $3d$ per Toffoli.}
    \label{tab:cross_femoco54_primitives}
\end{table}

\subsection{Additional layer of bit-flip protection}\label{sec:elevator-codes}
Our previous calculations have assumed no physical $X$ errors occur. In this section we analyze the use of an additional layer of active $X$ error correction through asymmetric distance quantum codes. The methodology in this section closely follows that of \cite{Shor2026prep} and details can be found there. We estimate that the physical $X$ error rate needed to execute the algorithm using only classical error correcting codes as $p_X\approx10^{-16}$ using Equation (C76) from \cite{Shor2026prep}. In this section we will calculate the resources required to execute the algorithm for FeMoco-54 if we assume a physical $X$ error rate of only $p_X=10^{-11}$. The estimation method and Elevator code \cite{shanahan2026elevator} LER models used are the same as \cite{Shor2026prep}, using the logical volume and effective number of logical qubits to estimate the resources needed to run the algorithm with $[15,9,d_X=3]$ Elevator codes. Under an assumption of a physical $Z$ error rate of $p_Z=5\times10^{-4}$ the logical volume of the algorithm is $\text{Vol}_\text{log}=1.75\times10^{12}$, the effective number of logical qubits is $n_\text{logical}^{(\text{eff})}=338$ and the inner repetition codes used are $d_Z=17$. This gives per shot logical failure rates of $\epsilon^Z_\text{logical}$ = $4.2\times 10^{-3}$ and $\epsilon^X_\text{logical}=1.6\times10^{-3}$ leading to an overall per shot probability of no logical failures above 99\%. Using $n^{(\text{eff})}_\text{logical}=338$ and the overhead of the $[15,9,d_X=3]$ Elevator code, $n_\text{Over}=\frac{16}{9}(2d_Z-1)$, we find a physical qubit resource requirement of $19\,830$.

The process can be repeated under the assumption of a physical error rate of $p_Z=10^{-3}$. In this case, the logical volume is $\text{Vol}_\text{logical}=1.58\times10^{12}$ and the effective number of logical qubits is $n_\text{logical}^{(\text{eff})}=321$. The repetition codes used are $d_z=21$ and the number of physical qubits needed to achieve a per shot probability of no logical error above 99\% is $23\,397$.

Following \cite{Shor2026prep} we will estimate the resources needed to implement an asymmetrically concatenated LDPC code \cite{ConcatenatedLDPC2026prep} by finding the proportional overhead required to go from a repetition code to an Elevator code and applying the same proportional overhead to the LDPC code. If $p_Z=5\times10^{-4}$ this is a 77\% increase, meaning the physical qubits required for an LDPC architecture would go from $7\,164$ to $12\,694$. If $p_Z=10^{-3}$ this is a 94\% increase meaning the physical qubits required would go from $7\,636$ to $14\,800$. 

%% file: sections/ldpc_circuit_noise.tex
\section{Logical error rate of optimized LDPC codes under generic circuit-level noise}
\label{sec:ldpc_circuit_noise}
Here we provide the scaling of the logical error rates of the optimized LDPC codes. We refer to the codes by the usual $[n,k,d]$ notation as $[85,34,9]$, $[102,34,12]$, $[119,34,16]$ and $[136,34,22]$ (Table~II shapes of \cite{Ruiz_2025}). The codes live on a lattice of size $H\times L$, the codes are quasi-cyclic by lateral translation, i.e. the same stabilizer shape is repeated in the horizontal direction. We work with the same ansatz as~\cite{Ruiz_2025} for the logical error rate after $d$ rounds of syndrome measurements
\begin{equation}\label{eq: LDPC ansatz}
    p_Z^{\mathrm{tot}}(p) = A\,d\,(B\,p)^{C\,t}, \qquad t = \lfloor (d+1)/2\rfloor,
\end{equation}
which is a conversion of the block failure rate $P$ via $p_Z^{\mathrm{tot}} = 1-(1-P)^{1/k}$, where $P$ is the probability that at least one logical qubit fails.
\subsection{Noise model}
We use a variant of generic circuit-level noise restricted to phase-flips, where every preparation, idling or readout is followed (preceded) by a $Z$ error with probability $p$ and every \textsc{Cnot} gate is followed by any of the three configurations of $Z$ errors ($IZ,ZI,ZZ$) on control and target with probability $p/3$.
\subsection{Decoders and direct Monte Carlo sampling}
We employ BP+OSD (min-sum with adaptive scaling or with fixed scaling factor $0.625$, OSD-CS order 60) and a most-likely error (MLE) decoder as an integer linear program with the HiGHS optimization solver. We let the final readout round be noiseless to ensure we end up in the codespace. We then perform direct Monte Carlo sampling, i.e., we sample fault configurations at a given $p$ and count how often the decoder decision leads to a logical error.
\subsection{Results}
\begin{table}[!ht]
\centering\footnotesize\setlength{\tabcolsep}{4pt}
\begin{tabular}{lllllccc}
\hline
code & $A$ & $B$ & $C$ & anchor (MLE fails/shots @ $p$) & $P$ per $d$ cycles & $p_Z^{\mathrm{tot}}$ & $p_Z^{\mathrm{tot}}/d$\\
\hline
$[102,34,12]$ & 0.1 & 25.8 & 1 & 8/3M @ 0.25\,\% & $1.9\,(1.2\text{--}2.8)\times10^{-10}$ & $5.5\times10^{-12}$ & $4.6\times10^{-13}$\\
$[119,34,16]$ & 0.1 & 29.8 & 1 & 2/360k @ 0.45\,\% & $1.3\,(0.47\text{--}3)\times10^{-13}$ & $3.9\times10^{-15}$ & $2.4\times10^{-16}$\\
$[136,34,22]$ & 0.1 & $\le$42.5 & 1 & 0/2M @ 0.45\,\% (limit) & $<3.0\times10^{-17}$ & $<8.8\times10^{-19}$ & $<4.0\times10^{-20}$\\
\hline
\end{tabular}
\caption{Ansatz parameters of \eqref{eq: LDPC ansatz} for the MLE decoder with $C=1$ and $A=0.1$ fixed and $B$ set by the lowest sampled point ($t=6,8,11$ for $d=12,16,22$), and the resulting values at
$p_Z=5\times10^{-4}$ (block failure $P$ with its 68\,\% Poisson interval from the anchor count, mantissas in parentheses; per-logical $p_Z^{\mathrm{tot}}$; per logical qubit and error-correction cycle, $p_Z^{\mathrm{tot}}/d$). A zero count (``limit'') is replaced by its one-sided 68\,\% upper limit $1.83/N$.}
\label{tab: LDPC ABC}
\end{table}
When decoding with both variants of BP+OSD, we find values of $C$ not close to unity and in particular can become larger than one. Together with our finding that BP+OSD fails on fault configurations of weight less than $t$ (weight 6 at $d=22$ and $d=16$ for the adaptive setting, weight 9 and 6 for the $0.625$ setting, verified by decoding the syndromes with the circuit-level decoder at $p=0.45\,\%$) this had us switch to a most-likely error decoder, where we found no cases of decoder failure below $t$. However, also for the MLE decoder, we find that across the sampled interval, the logical error rate is dominated by failure configurations of weight higher than the minimal weight: the local slope of $\ln P$ vs $\ln p$ between the two lowest points is 8, 11 and 16 for $t=6,8,11$. We are thus still in a crossover regime, where the ansatz is not strictly warranted and a $C>1$ would lead to unwarranted extrapolations. We therefore opt for the conservative strategy to use the lowest sampled $p$ data point as an anchor to fit to an ansatz with fixed $C=1$. The resulting parameters are listed in \zcref{tab: LDPC ABC}.
\begin{figure}[!ht]
    \centering
    \includegraphics[width=0.8\linewidth]{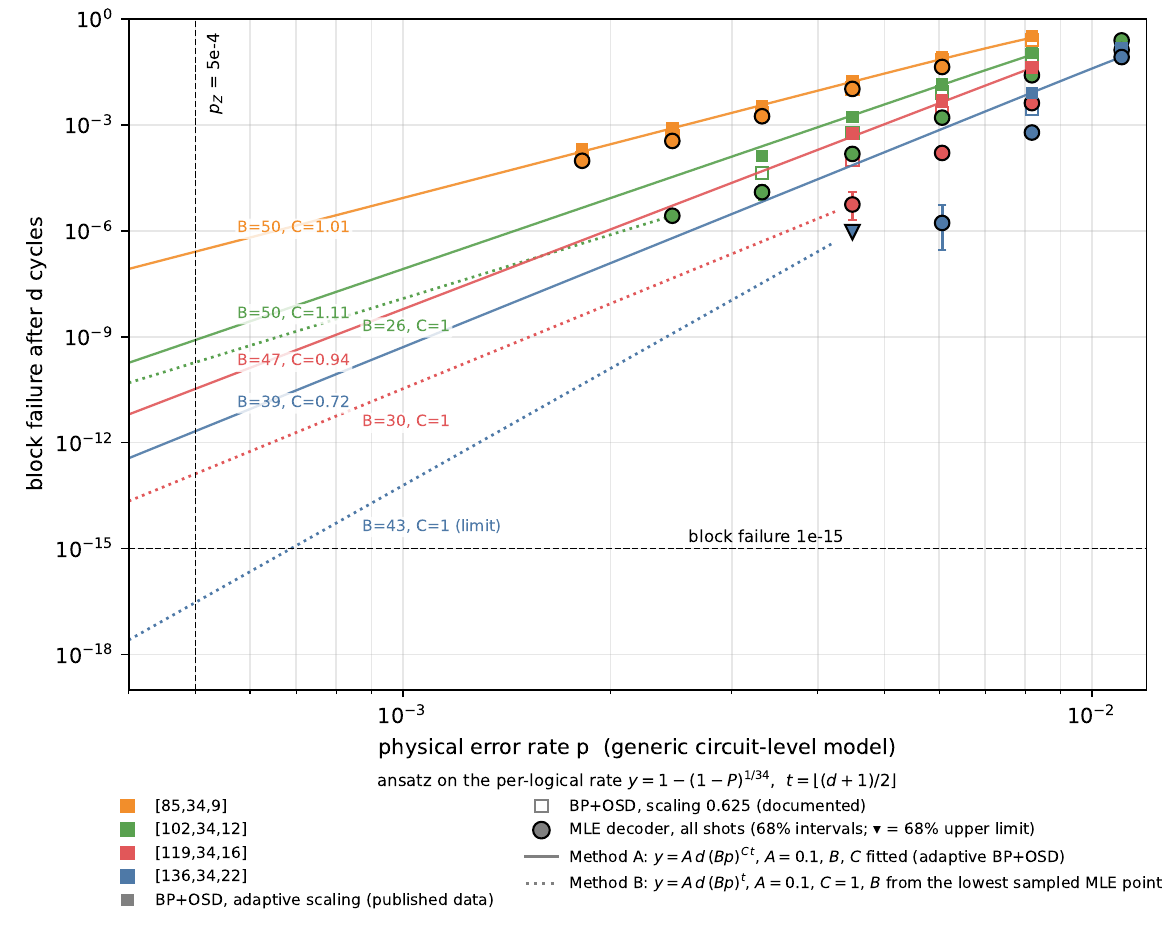}
    \caption{Block failure rate after $d$ cycles for the optimized LDPC codes of~\cite{Ruiz_2025} under generic circuit-level noise. BP+OSD (filled squares: adaptive scaling, open squares: scaling factor 0.625) and MLE decoder (circles, 68\% intervals, triangle: 68\% upper limit), identical shots per point. Solid lines: fits of \eqref{eq: LDPC ansatz} to the adaptive BP+OSD data with $A=0.1$ and $B$, $C$ free. Dotted lines: the $C=1$ ansatz of \zcref{tab: LDPC ABC} from the lowest sampled MLE point. The $Y$ axis is $P$. the $B$, $C$ labels refer to the per-logical ansatz $p_Z^{tot}$.}
    \label{fig:ldpc_cln_results}
\end{figure}

Note that the optimized LDPC codes can be grown in the horizontal direction by choosing the lateral dimension $L$ and extending the same stabilizer shapes. Since changing $L$ does not change the shape of logical operators but just the number of them, the per-logical results of \zcref{tab: LDPC ABC} extend to different choices for $L$.

%% file: sections/acknowledgments.tex
\section*{Acknowledgments and contributions}
\addcontentsline{toc}{section}{Acknowledgments}
We thank Jérémie Guillaud for creating a nice research environment, Dahmun Gourdazi for always having conversations to lean on, and all members of the Firmware department at Alice\&Bob for stimulating and insightful discussions.
PS acknowledges EPSRC CDT grant EP/Y035097/1.

The block encoding overview (\zcref{sec:alg description}), the review of subroutines (\zcref{sec:subroutines}) and the resource counts for the DFTHC block encoding (\zcref{sec: Rep-cat architecture, sec: LDPC-cat architecture, sec:comparison}) were written by LK. The compilation (\zcref{sec:compilation}) and the introduction (\zcref{sec:intro}) were written by HJ and LK, except the \zcref{subsec:i state factory} on $\ket{i}$ factories which was written by ABM. VH and AP contributed to \zcref{sec:compilation}. \zcref{sec:ldpc_circuit_noise} on LDPC code performance was written by MR and PS wrote \zcref{sec:elevator-codes} on adding an additional layer of bit-flip protection. \zcref{app:factorization} on Hamiltonian factorization was written by LK, \zcref{app:istate_distillation} on $\ket{i}$ factories was written by ABM and \zcref{app: Clifford} by HJ and VH. EG supervised the work on distillation factories and CV supervised \zcref{subsec:i state factory}. VL supervised the project. Claude (Anthropic) assisted in editing and writing of the paper as well as coding assistance for the resource estimation. The LDPC simulations of \zcref{sec:ldpc_circuit_noise} were also carried out with the help of Claude Code.

%% file: sections/appendix.tex
\section{Details on Hamiltonian Factorizations}
\label{app:factorization}

\subsection{Positivity of the Coulomb operator}\label{app: Positivity Coulomb op}
For the double factorization ansatz it is necessary to verify that the Coulomb operator is indeed positive semi-definite. Recall the definition of the Coulomb operator for indices $i,j,k,\ell$
\begin{equation}\label{eq: appendix definition Coulomb op}
    h_{ijk\ell}^{(2)} = \int_{\R^3} \int_{\R^3} \frac{\phi_i(\bfr_1) \phi_j(\bfr_1) \phi_k(\bfr_2) \phi_\ell(\bfr_2)}{\Vert \bfr_1 - \bfr_2 \Vert} \mathrm{d}\bfr_1 \mathrm{d} \bfr_2 = \int_{\R^3} \int_{\R^3} \frac{\rho_{ij}(\bfr_1)\rho_{k\ell}(\bfr_2)}{\Vert \bfr_1 - \bfr_2 \Vert} \mathrm{d}\bfr_1 \mathrm{d} \bfr_2,
\end{equation}
where we introduced the orbital pair density function $\rho_{ij}(\bfr) = \phi_i (\bfr) \phi_j(\bfr)$ consisting of products of the real molecular orbitals. Now consider the potential $V_{ij}(\bfr)$ associated to $\rho_{ij}(\bfr)$, defined as the solution to the Poisson equation, i.e., $\Delta V_{ij}(\bfr) = - \rho_{ij}(\bfr)$. Since we are operating in three-dimensional Euclidean space, the solution to the Poisson equation is given by $V_{ij}(\bfr) = \frac{1}{4\pi} \int_{\R^3} \frac{\rho_{ij}(\bfr^\prime)}{\Vert \bfr - \bfr^\prime \Vert} \mathrm{d}\bfr^\prime$. Thus, we can rewrite \eqref{eq: appendix definition Coulomb op} as
\begin{align}\label{eq: appendix int by parts}
    h_{ijk\ell}^{(2)} = 4\pi \int_{\R^3} \rho_{ij}(\bfr) V_{k\ell}(\bfr) \mathrm{d}\bfr = - 4\pi \int_{\R^3} \Delta V_{ij}(\bfr) V_{k\ell}(\bfr) \mathrm{d}\bfr = 4\pi \int_{\R^3} \nabla V_{ij}(\bfr) \cdot \nabla V_{k\ell} (\bfr) \mathrm{d}\bfr,
\end{align}
where the last equality is due to integration by parts. With the description in \eqref{eq: appendix int by parts} we readily observe that the Coulomb operator is positive semi-definite when viewed as a real symmetric matrix $h^{(2)}_{(ij)(k\ell)}$. To see this let $x \in \R^{N^2}$ and consider
\begin{align*}
    x^\transpose h^{(2)} x &= \sum_{(ij), (k\ell)} x_{ij} h^{(2)}_{(ij),(k\ell)} x_{k\ell} = 4\pi \int_{\R^3} \Big( \sum_{(ij)} x_{ij} \nabla V_{ij}(\bfr) \Big) \cdot \Big( \sum_{(k\ell)} x_{k\ell} \nabla V_{k\ell}(\bfr) \Big) \mathrm{d}\bfr  \\
    &= 4\pi \int_{\R^3} \Big\Vert \nabla \big( \sum_{(ij)} x_{ij} V_{ij}(\bfr)\big)\Big\Vert_2^2 \mathrm{d}\bfr \geq 0
\end{align*}
showing the desired.

\subsection{Sum-of-square description of the DFTHC Ansatz}\label{app: DFTHC Details}
To obtain an SOS expression for $H_\DFTHC$, first note that $\Tilde{h}^{(1)}_{ij}$ is symmetric and thus, we can consider its eigendecomposition $\Tilde{h}^{(1)}_{ij} = \sum_{s\in [N]} \mu_s v_{si} v_{sj} = \sum_{s}( \mu_s^+ - \mu_s^-) v_{si} v_{sj}$, where $\mu_s^+ = \max(\mu_s,0)$ and $\mu_s^- = \max(-\mu_s,0)$. Next, observe that we can write $a_{\bfu_b^{(r)},\sigma} = \sum_{i\in [N]} u_{bi}^{(r)} a_{i\sigma}$ and still obtain $\{a_{\bfu_b^{(r)},\sigma}, a_{\bfv_b^{(r)},\tau}^\dagger \} = 2 \delta_{\bfu \bfv} \delta_{\sigma \tau}$.  Then, using the anti-commutation relations and grouping eigenvectors with non-negative and negative eigenvalues respectively, we can write 
\begin{align*}
    \sum_{i,j \in [N]}\Tilde{h}^{(1)}_{ij} a_{i\sigma}^\dagger a_{j\sigma} = \sum_{\mu_s\geq 0} \mu_s^+ a_{\bfv_s^+, \sigma}^\dagger a_{\bfv_s^+, \sigma}  + \sum_{\mu_s< 0} \mu_s^{-} a_{\bfv_s^-, \sigma} a_{\bfv_s^-, \sigma}^\dagger - 2\sum_{\mu_s< 0} \mu_s^-\I.
\end{align*}
This shows that setting $\beta_{i\sigma}^\ell = \mu_\ell^+$ and $\delta_{i\sigma}^\ell = \mu_\ell^-$ in \eqref{eq: General SOS summand} recovers the one-body terms in the SOS expression. Anticipating that the block encoding is expressed in terms of Majorana operators, we express the annihilation and creation operators as follows:
\begin{equation*}
    a_{i\sigma} = \frac{\gamma_{i\sigma, 0}+ i \gamma_{i\sigma, 1}}{2}, \qquad a_{i\sigma}^\dagger = \frac{\gamma_{i\sigma, 0}- i \gamma_{i\sigma, 1}}{2}, \qquad \{\gamma_{i\sigma, x }, \gamma_{j \tau, y}\}= 2 \delta_{ij} \delta_{\sigma \tau} \delta_{xy},
\end{equation*}
where $\{A, B\} = AB + BA$ denotes the usual anti-commutator. While the higher-order terms in \eqref{eq: HDFTC full} are already given as a square, the switch to Majorana operators preserves this expression as a square and additionally allows us to express the higher order terms how they appear in the block encoding. Again we can write $\gamma_{\bfu_b^{(r)},\sigma, x} = \sum_{i\in [N]} u_{bi}^{(r)} \gamma_{i\sigma, x}$ and still obtain $\{\gamma_{\bfu_b^{(r)},\sigma, x}, \gamma_{\bfu_{b^\prime}^{(r)},\tau, y} \} = 2 \delta_{\bfu_b \bfu_{b^\prime}} \delta_{\sigma \tau} \delta_{xy}$ since $\bfu_b$ and $\bfu_{b^\prime}$ are orthogonal. Then, $a_{\bfu, \sigma}^\dagger a_{\bfu, \sigma} = \tfrac12 ( \I + \tfrac{i}{2} \gamma_{\bfu, \sigma, 0} \gamma_{\bfu, \sigma, 1})$ and we obtain 
\begin{equation*}
    W^{(rc)} \I + \sum_{b\in [B]} w_b^{(rc)} \sum_{i,j} u_{bi}^{(r)} u_{bj}^{(r)} E_{ij} = \Big(W^{(rc)} + \sum_{b\in [B]} w_b^{(rc)}\Big) \I + \frac{i}{2} \sum_{b\in [B]} \sum_{\sigma \in \{0,1\}} w_b^{(rc)} \gamma_{\bfu_b^{(r)}, \sigma, 0 } \gamma_{\bfu_b^{(r)}, \sigma, 1 },
\end{equation*}
which exactly gives $O_{\mathrm{SF}, rc}$.

%% file: sections/app_i_state_dist.tex
\section{\texorpdfstring{$\keti$}{i}-state distillation}
\label{app:istate_distillation}

We describe here the low-level details of the protocol in \zcref{subsec:i state factory}.

\begin{figure}[tbp]
    \centering
    \includegraphics[width=0.75\textwidth]{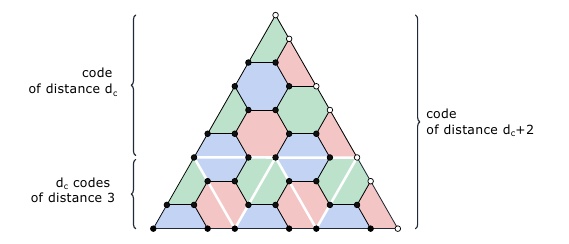}
    \caption{Recursive structure of the triangular code, shown for $d_c=5$. The code of distance $d_c+2$ is obtained by merging the code of distance $d_c$ with a strip of $d_c$ distance-$3$ codes, alternately upright and inverted. The white qubits are a possible choice of support for a pair or logical operators $X_L$ and $Z_L$, which indeed anticommute as $d_c$ is odd. Plenty of other equivalent representatives exist: for instance, multiplying $X_L$  with all blue $X$ stabilizers yields the operator $\overline{X}$ acting on every qubits.}
    \label{fig:triangular_code_induction}
\end{figure}

\paragraph{Review on triangular codes}
The stabilizer structure of the usual [[7,1,3]] Steane code can be represented by a triangular diagram. 
It is the distance-$3$ representative of a family of CSS topological error correction codes kown as triangular codes. They are the simplest color codes and can be defined by induction on the distance  $d_c$, see \zcref{fig:triangular_code_induction}.

The triangular code has as many qubits as the figure from \zcref{fig:triangular_code_induction} has of vertices, and as many $X$-$Z$ stabilizer dual pairs as the figure has of faces, and it can be proved that $d_c$ is its distance, i.e.,\ the minimum number of phase flips and bit flips that is undetected by noiseless stabilizer measurements.
Denote by $n_v(d_c)$ and $n_f(d_c)$ the respective numbers of vertices and of faces of the triangular code of odd distance $d_c$. It is easy to establish from the inductive structure represented in \zcref{fig:triangular_code_induction} that $n_{s}(d_c+2)=n_{s}(d_c)+3 (d_c+1) $ and $n_{f}(d_c+2)=n_{f}(d_c)+3 \frac{d_c+1}{2}$, hence 
\begin{equation} \label{eq: vertice_and_face_number_i_distillation}
    n_{s}(d_c)= 1 + 3 \frac{d_c^{2}-1}{4} \, , \qquad n_{f}(d_c) = 3 \frac{d_c^{2}- 1}{8} \, .
\end{equation}

For any odd $d_c$, the associated triangular code thus encodes exactly $n_s(d_c) - 2 n_f(d_c)=1$ logical qubit. The operators $\overline{Z}, \overline{X}$ 
acting transversally on every qubit of the code are a natural choice of logical operators, see \cite[\textsection C]{kubica_2015}. Another equivalent choice is to fix one edge of the triangle and act transversally on its $d_c$ qubits only, see \zcref{fig:triangular_code_induction}. 
By a standard result, any one-qubit Clifford gate on the triangular code can be realized transversally, although all qubits are not necessarily acted on by the same gate. For instance, the $X^{1/2}$ logical gate \eqref{eq: X14 X12 and S defs} can be implemented by acting jointly with $X^{1/2}$ or $(X^{1/2})^{-1}$ on each qubit, following a certain explicit bipartition, see \cite[\textsection C]{kubica_2015} for details in the analogous case of $S$.

We will adapt to our purposes the ``superdense'' version of the triangular code syndrome extraction circuit which was first proposed in \cite{gidney2023new}, see also \cite{lacroix2025scaling}. It has the specificity of using two entangled ancillae for each face in order to simultaneously measure the associated $X$ and $Z$ stabilizers. Although the triangular code has both connectivity-$4$ and connecticity-$6$ stabilizers, the superdense syndrome extraction circuit only requires each ancilla to be connected via \textsc{Cnot}s to a maximum of three data qubits.

\paragraph{Noise and logical error rate}

The goal of our distillation protocol is to prepare a repetition code in a logical $\keti$-state, the difficulty being to do it in a way that does not fail in the presence of physical noise. 

We will note 
\begin{equation}
    \ket{\pm \ii} \coloneqq \frac{\ket{0} + \ii \ket{1}}{\sqrt{2}}\, ,
\end{equation}
so that $\keti = \ket{+\ii}$.
In the context of quantum error correction, in which the output resource state will be consumed after the distillation, errors are as always discretized, and we will thus define a logical error on an output $\keti$ state in terms of logical bit flips and phase flips. 
Because $X \keti \propto Z\keti = \ket{-\ii}$, it turns out bit flips and phase flips play similar roles on an $\keti$ state. We accordingly define the logical error rate of our protocol to be the probability of its output state to be projected onto $\ket{-\ii}$ 
when it is ideally measured in the basis $(\keti, \ket{-\ii})$, which amounts to performing an ideal $Y$ measurement.

We will exclusively work with cat qubits as physical qubits and assume their noise bias to be infinite in the following sense: idle noise consists of phase flips only and the usual bias-preserving physical operations from \cite[\textsection IV]{Guillaud_2019}) do not cause any bit flip either. For our numerical tests, we more specifically worked with the same circuit-level noise model as for unfolded distillation simulations \cite{Ruiz_2026}. This first means that we assumed every physical operation to take the same amount of time, decomposing our circuit into timesteps where operations can be run in parallel because they are performed on disjoint sets of qubits. This also implies that both idle qubits and gates undergo noise at each timestep. We take a model where this noise is parametrized by a single value $p_z$ representing the probability of idle phase-flip errors, see details in \zcref{table:istate_distillation_noise_tab}.  
The objective of our simulations is to evaluate the ability of our distillation protocol to reach low logical error rates for
\begin{equation} \label{eq: pz values}
    p_z \in \{10^{-3}, \, 5.10^{-4}\} \, .
\end{equation}
We will define the circuit-level distance of a circuit to be the minimum number of faults producing an undetected logical error under our circuit-level noise model described in \zcref{table:istate_distillation_noise_tab}. It is necessarily lower or equal to its distance, as the latter notion does not account for errors arising from gate applications. In the case of the phase-flip protected repetition code, both notions of distances coincide since the standard stabilizer extraction circuit is fault-tolerant: no isolated fault is propagated to multiple errors on the data qubits.

\begin{table*}[ht]
    \centering
    \begin{tabular}{ |c|c| } 
    \hline
    Circuit element at a given timestep & Noise (\texttt{stim} notation)\\ 
    \hline
    Physical $\ket{+}$ preparation &  \texttt{Z\_ERROR} \\ 
    Physical $\ket{0}$ preparation &  Perfect \\ 
    $X$ basis measurement & \texttt{Z\_ERROR}  \\
    $Z$ basis measurement & Perfect \\
    Physical \textsc{Cnot} & \texttt{PAULI\_CHANNEL\_2} (biased) \\
    $X^{1/2}$ or $(X^{1/2})^{-1}$ & \texttt{DEPOLARIZE1} (unbiased) \\
    $S$ & \texttt{Z\_ERROR} \\
    Idle & \texttt{Z\_ERROR} \\
    \hline
    \end{tabular}  
    \caption{Circuit-level noise in distillation protocol simulations. We need no more gates for our circuit than what is listed in this table, see examples in \zcref{tab:istate_distillation_results} (but note that these examples of stim circuits also include a noiseless final logical $Y$ measurement that is not part of the protocol). All noise channels are applied with fixed probability $p_z$. In the case of the \textsc{Cnot}, which is the most frequent gate in our circuit, this $p_z$ is evenly distributed between $Z_1$, $Z_2$, $Z_1Z_2$. The noise is assumed to be independently distributed among operations and time steps.}
    \label{table:istate_distillation_noise_tab}
\end{table*}

It is usually expected that the logical error rate of a circuit with circuit-level distance $d$ and protected against circuit-level noise by quantum error correction satisfies $\propto p_z^{(d+1)/2}$ at first order. This can be properly proved under some mild assumptions on the spatial dispersion of errors and was  for instance numerically checked for the phase-flip repetition code \cite[III eq (2)]{Ruiz_2025}. Nonetheless, the unfolded distillation protocol \cite{Ruiz_2026} that we briefly introduce in \zcref{subsec:i state factory} used the noise model from \zcref{table:istate_distillation_noise_tab} and still managed to reach regimes where the simulated logical error rates became proportional to $p_z^{d}$. This is because this protocol takes advantage of the full detecting power of the distillation code, and not only of its corrective capabilities. More precisely, it includes a selection procedure deciding that some errors must lead to the complete reboot of the protocol upon detection, discarding the resource state produced as being invalid. While this comes at the cost of a longer resource state generation time as the protocol needs to be repeated until a valid result is returned, the main benefit is a lower output logical rate because detecting errors is easier than correcting them. 
Going into the details of unfolded distillation, selection is carried out on the basis of distillation code stabilizer only, while error correction is at the same time performed on the repetition code side. The authors of \cite{Ruiz_2026} chose the repetition code distance to satisfy $d_r \geq 2d'-1$, where $d'$ is the distillation code's distance, so its expected contribution to the logical error rate is $\propto p_z^{(d_r+1)/2} = O (p_z^d)$. Meanwhile, their distillation code's circuit-level distance coincides with its usual distance $d'$ as its stabilizer extraction circuits are fault-tolerant: this is due to repeated $X$ stabilizers being unable to cause hook errors because of the infinite noise bias. Because of selection, its contribution should then be $\propto p_z^d$, which is consistent with their numerical values.

\begin{figure}[tbp]
    \centering
    \includegraphics[width=0.75\textwidth]{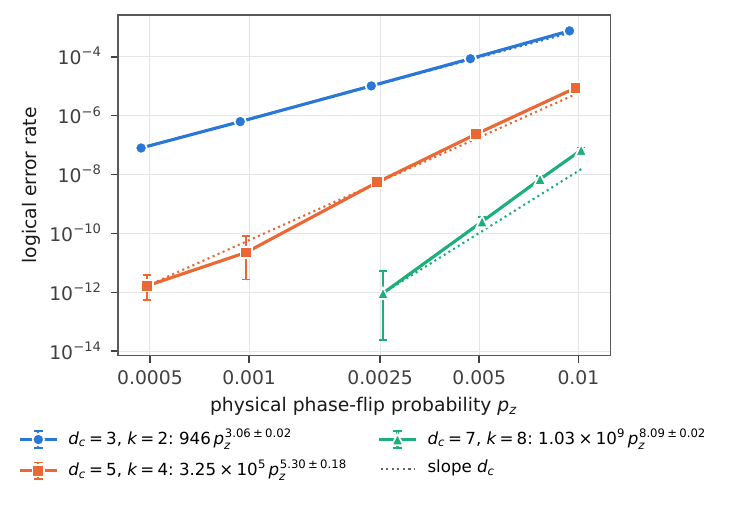}
    \caption{$\keti$-state distillation logical error rate as a function of $p_z$ in logarithmic scales for saturating values of repetition-code parameters $k$ and $d_r$, see \zcref{fig:i_distillation_rep_saturating_params}. The curves confirm a power-law behaviour, although our Monte Carlo simulations with a limited number of shots were not able to capture low values of $p_z$ properly. We interpret the curve being even steeper than the expected $d_c$ slope as higher-order terms playing a role when $p_z$ is high. }
    \label{fig:i_distillation_ler_vs_pz_plot}
\end{figure}

\begin{figure}[!tbp]
    \centering
    \begin{subfigure}[b]{0.55\textwidth}
        \centering
        \includegraphics[width=\linewidth]{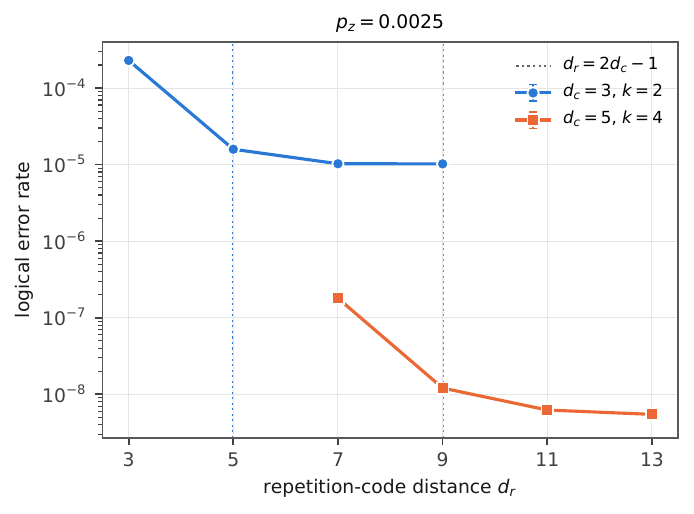}
        \caption{Logical error rate as a function of the repetition-code distance $d_r$ at fixed $p_z=2.5.10^{-4}$.}
        \label{subfig:i_distillation_ler_vs_dr}
    \end{subfigure}

    \vspace{1em}

    \begin{subfigure}[b]{0.85\textwidth}
        \centering
        \includegraphics[width=\linewidth]{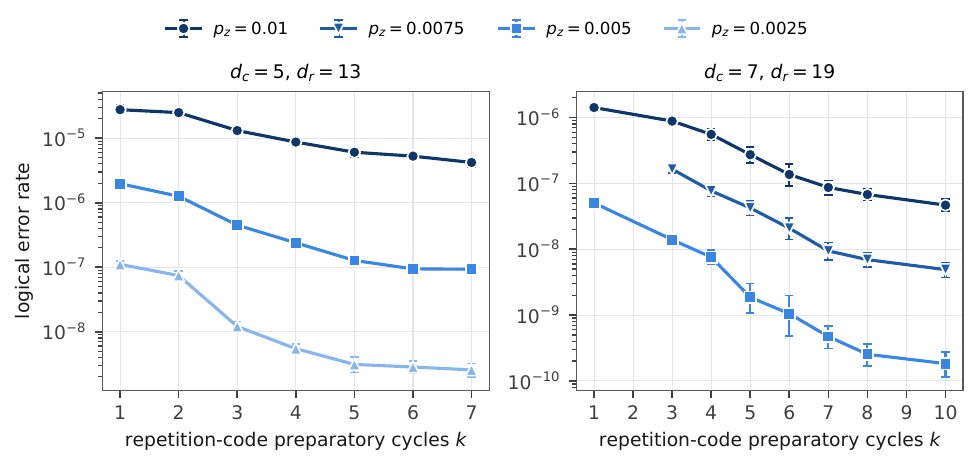}
        \caption{Logical error rate as a function of the number of preparatory cycles.}
        \label{subfig:i_distillation_ler_vs_n_prep}
    \end{subfigure}
    \caption{$\keti$-state distillation logical error rate as a function of (a) the repetition-code distance $d_r$ and (b) the repetition-code cycles number $k$ before the \textsc{Cnot}. The curves seem to point at a saturation regime reached around $d_r \geq 2d_c-1$ at fixed $k$, which was expected as the repetition code cycle's contribution to the error rate is $\propto p_z^{\lfloor \frac{d_r-1}{2} \rfloor}$. At fixed $d_r$, it appears that the protocol has to be lengthened to a slightly larger number $k$ of repetition-code preparatory cycles than $d_c$ in order to saturate the logical error rate. }
    \label{fig:i_distillation_rep_saturating_params}
\end{figure}

We here report similar results on our $\keti$-state distillation circuit: for saturating values of repetition-code parameters, see \zcref{fig:i_distillation_rep_saturating_params}, our protocol seems to reach a regime where its circuit-level distance reaches $d_c$, see \zcref{fig:i_distillation_ler_vs_pz_plot}. Our selection strategy, detailed in the next paragraph, has led to fairly high acceptance rates for our range \eqref{eq: pz values} of $p_z$, see \zcref{fig:i_distillation_acceptance_plot}. We provide detailed simulation data for the different protocol parameters proposed in \zcref{subsec:i state factory}  in \zcref{tab:istate_ler}.

\paragraph{Compiled distillation circuit}

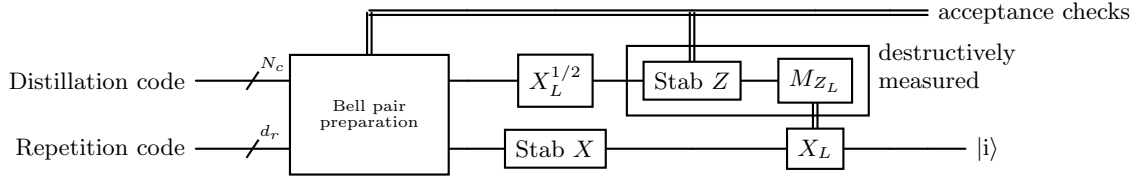
\begin{figure*}[tbp]
    \centering
    \begin{quantikz}[wire types={n,q,q}, row sep={0.9cm,between origins}, column sep=0.5cm]
        & [1cm] & \setwiretype{c} & & & & & \rstick{acceptance checks} \\
        \lstick{Distillation code} & [1cm] \gate[wires=2,style={minimum width=2.1cm}]{\substack{\text{Bell pair}\\ \text{preparation}}} \qwbundle{N_c} \wire[u][1]{c} & & \gate{X_L^{1/2}} & \gate{\text{Stab }Z} \wire[u][1]{c} \gategroup[1,steps=2,style={inner xsep=3pt,inner ysep=0pt},label style={label position=right,anchor=west,align=left}]{destructively\\measured} & \gate{M_{Z_L}} \wire[d][1]{c} & \setwiretype{n} & & \\
        \lstick{Repetition code} & [1cm] \qwbundle{d_r} & & \gate{\text{Stab }X} & & \gate{X_L} & & & \rstick{$\keti$}
    \end{quantikz}
    \caption{The $\keti$-state distillation protocol. Only the $N_c$ distillation code and $d_r$ repetition code data qubits are represented. Ancillas are only used in the Bell pair preparation and Stab $X$ boxes because the final $Z$ stabiliser measurement $\mathrm{Stab} Z$ and logical $Z$ value $M_{Z_L}$ are reconstructed from the destructive measurement of all distillation code data qubits in the $Z$ basis. Our implementation of the Bell pair preparation is represented in \zcref{fig:istate_distillation_bell_pair_prep}.}
    \label{fig:istate_distillation_high_lvl}
\end{figure*}

We now elaborate on the protocol that we described at a higher level in \zcref{subsec:i state factory} and represented in \zcref{fig:istate_distillation_high_lvl}. As mentioned there, the quantum circuit starts by entangling a repetition code and a triangular code into a Bell pair logical state. To do so, we first prepare the triangular code in a logical $\ket{+}_L$ state and the repetition code in a logical $\ket{0}_L$ state. We proceed as follows: each triangular code data qubit is prepared in a $\ket{+}$ state, each repetition code data qubit in a $\ket{0}$ state, and we then measure all triangular code $Z$ stabilizers and repetition code $X$ stabilizers. Since both codes are CSS, this indeed yields logical $\ket{+}_L$ and $\ket{0}_L$ states if the syndrome is trivial. Yet,  even in the case of a nontrivial syndrome, which is almost inevitable since the measurement outcome is here nondeterministic even in the absence of noise, we simply keep it in memory by classical software throughout the circuit as a `by-product' to account for in subsequent observable measurements. More precisely, we thus prepare the state $\ket{+}_L \otimes \ket{0}_L$ only up to a Pauli operation that we keep under control.

On the triangular code side, we implement the $Z$ stabilizer measurement routines following the superdense model, see \zcref{fig:stabZ_extraction_circuit}. This implies using two ancillae per face even though $X$ stabilizers are not measured. In addition to making the protocol implementable on a square lattice, this extra ancilla overhead on $Z$ stabilizers has the benefit of mitigating hook errors, which are already an issue at the step of the $\ket{+}_L$ preparation. 
$Z$ stabilizer measurements have indeed the exact opposite behavior regarding phase flips compared to their $X$ counterparts: they yield very reliable results but any phase flip occurring on ancillae during the stabilizer measurement routine creates hook errors on data qubits, see \zcref{fig:stabZ_extraction_circuit}. As a result, they eventually directly contribute to the logical error rate by reducing the code's circuit-level distance.
We have therefore decided to use only one round of $Z$ stabilizer measurements, see \zcref{fig:istate_distillation_bell_pair_prep}. This is the opposite of $\ketXmag$-state unfolded distillation  \cite{Ruiz_2026}, which made the number of distillation code rounds a tunable parameter since the $X$ stabilizers they exclusively used benefited from successive measurements: their results are indeed unreliable due to the noise bias. 
Furthermore, we have protected distillation code ancillae against phase flips via a set of ``flags'', inspired by \cite{chamberland2018flag, ChaoReichardt2018}. They are additional distance-2 $X$ stabilizers at the level of ancillae that can be measured for free with the superdense layout, see again \zcref{fig:stabZ_extraction_circuit}. Because we have numerically confirmed that hook errors cut the circuit-level distance by two, we take a hard line here and do selection based on these flags: if a hook error was detected, the protocol's output is simply discarded. In practice, this addition has greatly improved numerical logical error rates.

\input{graphics/istate_distillation/distillation_figure}

Once the system has been prepared up to a byproduct into a $\ket{+}_L \otimes \ket{0}_L$ state, we apply a logical \textsc{Cnot} to entangle it into a Bell pair. To do so, we use a transversal, hence fault tolerant, implementation of the two-qubit logical gate, see \zcref{fig:istate_logical_cnot}. Note that the physical \textsc{Cnot}s involved in this logical operation commute with the ones involved in the $Z$ stabilizer measurement routines, which allows us to reorder them in a way that reduces physical qubit idleness as much as possible. At this step, any phase flip in the support of the logical \textsc{Cnot} on the repetition code side propagates to the distillation code. 
Numerical checks confirm that feeding the decoder with a sufficiently large number of additional repetition code $X$ stabilizer measurements performed before the logical \textsc{Cnot} is enough to mitigate this phenomenon. We keep the number of preparatory repetition code cycles as a major tunable trade-off parameter for our simulations.

The whole point of introducing the triangular code is to benefit from its transversal $X^{1/2}$ logical gate, which is a simpler and more robust implementation of this logical operation than what is possible on the phase-flip-protected repetition code. Following a standard textbook teleportation protocol, see \zcref{fig:istate_distillation_high_lvl}, we apply a logical $X^{1/2}$ on the distillation code side of the newly prepared logical Bell pair before proceeding with a classically controlled logical $Z$ measurement, putting the repetition code into the desired logical $\keti$ state 
in the absence of noise. 
The classically controlled gate being the last operation involving the distillation code in the circuit, it is efficiently implemented by destructively measuring all distillation code data qubits in the $Z$ basis, see also \zcref{fig:istate_distillation_high_lvl}. Even if this measurement is lethal for the quantum state on the distillation code side, it indeed provides the required information for the controlled gate to be applied, or more precisely for the system's Pauli frame to be properly updated. 
Note that, despite the infinite noise bias, the logical $Z$ measurement can be disturbed by physical errors because of the $(X^{1/2})^{\pm 1}$ gates applied on the distillation code data qubits: not only do they cause bit flips by \zcref{table:istate_distillation_noise_tab}, they also propagate existing $Z$ errors into $Y$ errors. 
To address this source of noise, we here capitalize on the destructive $Z$ measurements at the end of circuit and reconstruct the distillation code $Z$ stabilizer measurements for free. 
Borrowing a key idea from unfolded distillation \cite{Ruiz_2026}, we select on the basis of these reconstructed distillation $Z$ syndromes with the objective of reaching a regime  where the logical error rate is proportional to $p_z^{d_c}$, with $d_c$ the distance of the triangular code chosen.

As already suggested in \cite[V]{Ruiz_2026}, we note that a theoretically equivalent teleportation circuit can be obtained by replacing the $X^{1/2}$ operation by an $S$, and the eventual $Z$ measurement by an $X$ measurement. While the change is small, the $X^{1/2}$ version of the circuit we use appears at first sight to be more advantageous in view of our noise model \zcref{table:istate_distillation_noise_tab}, hence our choice.

We simulated our protocol on stim \cite{gidney2021stim} via sinter to perform high-performance simulations. All our simulation results were performed with the Tesseract decoder \cite{beni2025tesseractsearchbaseddecoderquantum}. In the example circuits from \zcref{tab:istate_qubit_count}, the $\keti$ state left in the repetition code is checked by an ideal logical $Y$ measurement for the purpose of the simulation. It is built from an $S$ gate, a chain of $CZ$ gates and an $X$-basis measurement, all of them noiseless. This readout is a measuring device of the simulation and is not part of the protocol.

\begin{table}[!htb]
\centering
\begin{subtable}[t]{\textwidth}
\centering
\caption{Logical error rate (LER) on $\keti$ state distillation at $p_z = 10^{-3}$.}
\label{tab:istate_ler_1e-3}
    \input{graphics/istate_distillation/tables/femoco_istate_ler_1e-3.tex}
\end{subtable}

\vspace{1em}

\begin{subtable}[t]{\textwidth}
\centering
\caption{Logical error rate (LER) on $\keti$ state distillation at $p_z = 5\times10^{-4}$.}
\label{tab:istate_ler_5e-4}
    \input{graphics/istate_distillation/tables/femoco_istate_ler_5e-4.tex}
\end{subtable}
\caption{$\keti$ state distillation results for the protocols of \zcref{tab:istate_distillation_results}. The ``short$^*$''-type line was especially added to cover the case $d_c=3$, $k=2$ that we use in the main text. By ``time'', we mean the total number $t_i= k+1$ of repetition code cycles performed over the distillation circuit.  Computing the logical error rate via Monte Carlo simulations on stim consists of performing a Bernouilli sampling. By 95\% upper bound, we mean that if the LER was a uniform random variable taking values between 0 and 1, then the probability, conditioned to the observed counts, that it takes value below this upper bound is 95 \%. This claim is based on a standard Poisson approximation, see our code's documentation~\cite{AliceBob2026istatecode}. The fitted values claimed for the long protocol were obtained on the basis of the fit given in \zcref{fig:i_distillation_ler_vs_pz_plot}.}
\label{tab:istate_ler}
\end{table}

\input{graphics/istate_distillation/distillation_plots}

%% file: graphics/istate_distillation/distillation_figure.tex
\begin{figure*}[tbp]
    \centering
    \begin{subfigure}[t]{\textwidth}
        \centering
        \begin{quantikz}[wire types={n,q,q}, column sep=8mm, row sep=6mm]
            & & & & & \\
            \lstick{$\ket{+}^{\otimes N_c}$} & \qwbundle{} & \gate{\text{Stab }Z} \wire[u][1]["\textstyle\text{flag}"{above,pos=1}]{c} & \ctrl{1} & \\
            \lstick{$\ket{0}^{\otimes d_r}$} & \qwbundle{} & \gate{\text{Stab }X}\gategroup[1,steps=1,style={dashed,rounded corners,inner xsep=3pt,inner ysep=2pt}]{$\times k$} & \targ{} &
        \end{quantikz}
    \end{subfigure}

    \caption{Logical Bell pair preparation circuit. The $Z$ stabiliser measurement is described in Figure~\ref{fig:stabZ_extraction_circuit} while the logical \textsc{Cnot} is represented in Figure~\ref{fig:istate_logical_cnot}. The unique $Z$ stabiliser and the first $X$ stabiliser initialise the state in a logical $\ket{+}\otimes \ket {0}$ state. The number $(k - 1) \geq 0$ counts additional preparatory repetition code stabiliser cycles.}
    \label{fig:istate_distillation_bell_pair_prep}
\end{figure*}
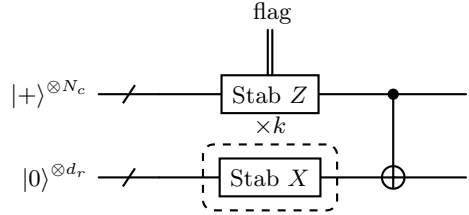

\begin{figure*}
    \begin{subfigure}[b]{0.38\textwidth}
        \centering
        \includegraphics[width=0.7\linewidth]{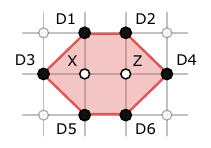}
        \caption{Superdense layout of one $Z$ stabiliser face}
        \label{subfig:istate_superdense_hexagon}
    \end{subfigure}
    \hfill
    \begin{subfigure}[b]{0.58\textwidth}
        \centering
        \begin{quantikz}[row sep={7.5mm,between origins}, column sep=3mm]
            \lstick{D6} & & \ctrl{3}\gategroup[8,steps=3,style={dashed,rounded corners,inner xsep=2pt},background]{$Z$ stabiliser} & & & & \\
            \lstick{D4} & & & \ctrl{2} & & & \\
            \lstick{D2} & & & & \ctrl{1} & & \\
            \lstick{Z\quad$\ket{0}$} & \targ{} & \targ{} & \targ{} & \targ{} & \targ{} & \meterD{M_Z} \\
            \lstick{X\!\!\quad$\ket{+}$} & \ctrl{-1} & \targ{} & \targ{} & \targ{} & \ctrl{-1} & \meterD{M_X} \\
            \lstick{D5} & & & & \ctrl{-1} & & \\
            \lstick{D3} & & & \ctrl{-2} & & & \\
            \lstick{D1} & & \ctrl{-3} & & & &
        \end{quantikz}
        \caption{$Z$ stabiliser extraction circuit.}
        \label{subfig:istate_custom_superdense}
    \end{subfigure}
    \caption{Triangular code $Z$ stabiliser circuit. Measuring a face's $Z$ stabiliser and its flag can be performed simultaneously via a version of the superdense circuit of \cite{gidney2023new} where the original $X$ stabiliser \textsc{Cnot}s have been removed.  Quadrilateral faces are trimmed hexagons and are covered by the same circuit deprived from the corresponding two qubits.  A single phase-flip error on either ancilla before or during the sequence of three pairs of \textsc{Cnot} gates is propagated onto the data qubits whose gates are yet to be applied. The resulting hook error was produced by a single fault but may have weight up to half the weight of the stabiliser, so that the circuit-level distance of the triangular code may fall to about half its distance $d_c$. We follow an idea of \cite{chamberland2018flag, ChaoReichardt2018} and counter this phenomenon via a $X$ measurement flag counting modulo 2 the phase flips occuring on either ancilla over the extraction circuit.}
    \label{fig:stabZ_extraction_circuit}
\end{figure*}

\begin{figure*}[tbp]
    \centering
    \includegraphics[width=0.5\textwidth]{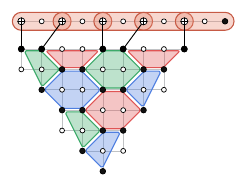}
\caption{Logical \textsc{Cnot} from the triangular code to the repetition code. It is a logical operator since it normalises the stabiliser group of the joint code: it does commute with all repetition code stabilisers and triangular code $Z$ stabilisers, but conjugates a triangular code $X$ stabiliser into itself only if the latter has no qubit involved in the physical \textsc{Cnot}s, otherwise there is an additional factor corresponding to the opposite repetition code stabiliser. We interpret it as a logical \textsc{Cnot} due to its commutation relations with logical operators. Note that although the triangular code layout was specifically chosen for not requiring more that nearest-neighbour connectivity on a square grid, for the purpose of our simulations, we allow diagonal physical \textsc{Cnot} here.}
    \label{fig:istate_logical_cnot}
\end{figure*}

%% file: graphics/istate_distillation/tables/femoco_istate_ler_1e-3.tex
\begin{tabular}{ccccc}
\toprule
$d_c$ & type & LER, 95\% upper bound (failures / accepted) & Acceptance ($\alpha$) & Time $\div$ acceptance ($t_i/\alpha$) \\
\midrule
3 & short & $8.8\times10^{-7}$ \; ($1202 / 1.43\times10^{9}$) & 95.2\% & 2.1 \\
3 & short$^*$ & $6.6\times10^{-7}$ \; ($897 / 1.43\times10^{9}$) & 95.2\% & 3.1 \\
5 & short & $3.6\times10^{-9}$ \; ($9 / 4.39\times10^{9}$) & 87.7\% & 2.3 \\
5 & medium & $7.2\times10^{-11}$ \; ($2 / 8.78\times10^{10}$) & 87.8\% & 5.7 \\
7 & short & $3.9\times10^{-11}$ \; ($4 / 2.33\times10^{11}$) & 77.8\% & 2.6 \\
7 & medium & $3.0\times10^{-12}$ \; ($1 / 1.56\times10^{12}$) & 77.8\% & 6.4 \\
7 & long & n/a \; (fit: ${\sim}6\times10^{-16}$) & 77.8\% & 11.6 \\
\bottomrule
\end{tabular}

%% file: graphics/istate_distillation/tables/femoco_istate_ler_5e-4.tex
\begin{tabular}{ccccc}
\toprule
$d_c$ & type & LER, 95\% upper bound (failures / accepted) & Acceptance ($\alpha$) & Time $\div$ acceptance ($t_i/\alpha$) \\
\midrule
3 & short & $1.2\times10^{-7}$ \; ($1105 / 9.76\times10^{9}$) & 97.6\% & 2.0 \\
3 & short$^*$ & $8.5\times10^{-8}$ \; ($779 / 9.76\times10^{9}$) & 97.6\% & 3.1 \\
5 & short & $3.2\times10^{-10}$ \; ($21 / 9.37\times10^{10}$) & 93.7\% & 2.1 \\
5 & medium & $3.5\times10^{-12}$ \; ($5 / 3.00\times10^{12}$) & 93.7\% & 5.3 \\
7 & short & $3.8\times10^{-12}$ \; ($3 / 2.03\times10^{12}$) & 88.2\% & 2.3 \\
7 & medium & $9.4\times10^{-13}$ \; ($0 / 3.18\times10^{12}$) & 88.2\% & 5.7 \\
7 & long & n/a \; (fit: ${\sim}2\times10^{-18}$) & 88.2\% & 10.2 \\
\bottomrule
\end{tabular}

%% file: graphics/istate_distillation/distillation_plots.tex
\begin{figure}[tbp]
    \centering
    \begin{subfigure}[b]{0.6\textwidth}
        \centering
        \includegraphics[width=\linewidth]{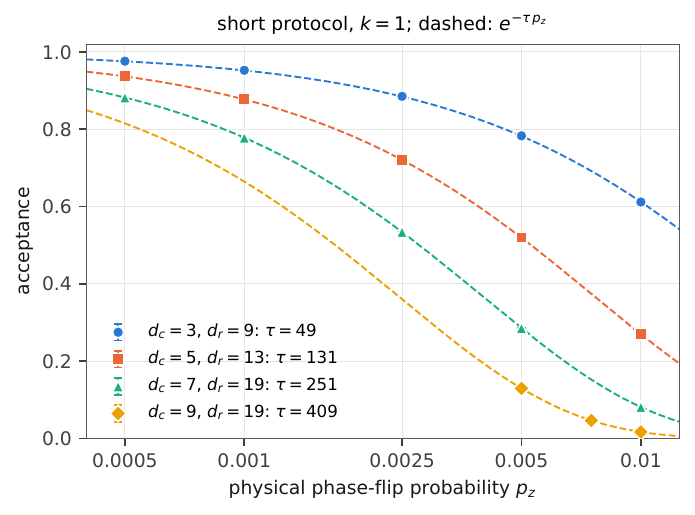}
        \caption{Acceptance of the short protocols vs $p_z$, with fits $e^{-\tau p_z}$.}
        \label{subfig:i_distillation_acceptance_vs_pz}
    \end{subfigure}
    \hfill
    \begin{minipage}[b]{0.37\textwidth}
        \centering
        \begin{subfigure}{\linewidth}
            \centering
            \small
            \begin{tabular}{rrrr}
            \toprule
            $d_c$ & $n_s(d_c)$ & $\tau$ & $\tau/n_s(d_c)$ \\
            \midrule
            3 & 7 & 49 & 7.01 \\
            5 & 19 & 131 & 6.90 \\
            7 & 37 & 251 & 6.78 \\
            9 & 61 & 409 & 6.70 \\
            \bottomrule
            \end{tabular}
            \caption{Fitted constant $\tau$ vs qubit counts.}
            \label{subfig:i_distillation_acceptance_table}
        \end{subfigure}

        \begin{subfigure}{\linewidth}
            \centering
            \includegraphics[width=0.85\linewidth]{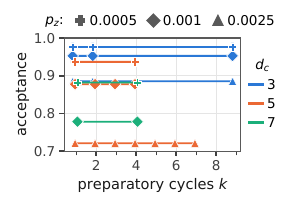}
            \caption{Acceptance vs preparatory cycles.}
            \label{subfig:i_distillation_acceptance_vs_n_prep}
        \end{subfigure}
    \end{minipage}
    \caption{Acceptance of the $\keti$-distillation protocol. A protocol output is accepted if{f} no acceptance check fires, see  \zcref{fig:istate_distillation_high_lvl}. Reasoning at the level of the circuit's detector error model, acceptance is determined by which of the mutually independent error mechanism fires, and it follows from our error model \zcref{table:istate_distillation_noise_tab} that each mechanism fires with probability $p_z$ multiplied by a specific combinatorial factor. By a standard aproximation of Poisson-binomial distributions which is valid when $p_z$ is small, acceptance has an exponential fit, see details in our code's documentation documentation~\cite{AliceBob2026istatecode}. This is here confirmed by numerics in \zcref{subfig:i_distillation_acceptance_vs_pz}, and \zcref{subfig:i_distillation_acceptance_table} seems to indicate that the constant $\tau$ is proportional to the number of data qubits in the triangular code. We additionally observe in \zcref{subfig:i_distillation_acceptance_vs_n_prep} that acceptance is independent of the number $k$ of repetition-code preparatory cycles, legitimating the limited number of curves displayed on \zcref{subfig:i_distillation_acceptance_vs_pz}.}
    \label{fig:i_distillation_acceptance_plot}
\end{figure}

%% file: sections/app_Clifford_gates.tex
\section{Logical Clifford gates on cat qubit architecture}\label{app: Clifford}

In this appendix, we present the different circuits used to implement Clifford gates on phase-flip repetition code and phase-flip LDPC architectures.
Indeed, the concatenation of cat qubit bosonic encoding with a classical code does add some constraints on the accessible gate set~\cite{ Guillaud_2019, barizien2025accessiblequantumgatesclassical}.
We focus here on two types of logical Clifford gates, those akin to \textsc{Cnot}s gates and those that require the injection of $\ket{i}$ states, whose distillation protocols are described in \zcref{subsec:i state factory} and \zcref{app:istate_distillation}.

\subsection{\textsc{Cnot} and variants}
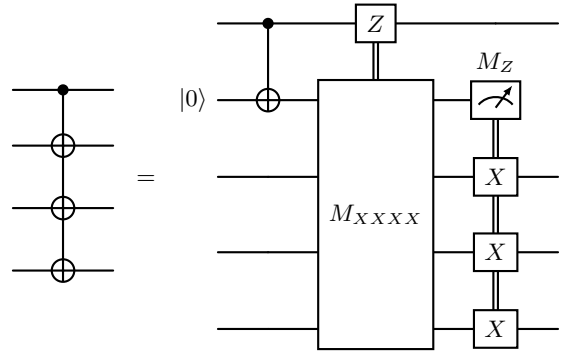
\begin{figure}[!htb]
    \centering
    \begin{quantikz}[align equals at=2.5]
  & \ctrl{3} & \\
  & \targ{}  & \\
  & \targ{}  & \\
  & \targ{}  &
\end{quantikz}
\;=\;
\begin{quantikz}[align equals at=3]
  & \ctrl{1} &  \gate{Z} && \\
  \lstick{$\ket{0}$} & \targ{} \setwiretype{q}
    & \gate[4]{M_{XXXX}}\vcw{-1}
    & \meter{M_Z}\vcw{3}  &\setwiretype{n} \\
  & & & \gate{X} & \\
  & & & \gate{X} & \\
  & & & \gate{X} &
\end{quantikz}
    \caption{Long range multi-target  \textsc{Cnot} circuit identity of~\cite{Gouzien_2023}.
    The  \textsc{Cnot} on the right hand-side is performed in a local transversal manner.
    The $M_{XXXX}$ measurement is performed via lattice surgery.}
    \label{fig:multi-target-cnot}
\end{figure}

For completeness, we start by reminding in~\zcref{fig:multi-target-cnot} the circuit identity used in~\cite{Gouzien_2023} to perform long-range \textsc{Cnot} and multi-target \textsc{Cnot} gates. 
In this circuit, the \textsc{Cnot} between the control qubit and the logical auxiliary qubit initialized in $\ket{0}$ is a local transversal \textsc{Cnot}.
In repetition code architecture, this requires a '$Z$-access' to the control logical qubit, namely a corridor of routing qubits next to its repetition code in order to initialize another repetition code with which transversal \textsc{Cnot}s can be performed.
Note that this auxiliary repetition code can be longer than the size of the control qubit repetition code. What matters is that every qubit in the support of the control $Z_L$ is involved in a \textsc{Cnot}. Since the $X_L$ of the second repetition code is equivalently supported on any data qubits, there is no need to target every data qubit of the auxiliary qubit.
In particular, this feature is useful for performing the $M_{XXXX}$ right after the transversal $\ket{0}$ initialization. 

\begin{figure}[!htb]
  \centering
  \begin{equation*}
    \vcenter{\hbox{%
      \begin{quantikz}[column sep=0.55cm, row sep=0.7cm]
        \lstick{} & \ctrl{2}& & \qw \\
        \lstick{} && \ctrl{1} & \qw \\
        \lstick{} & \targ{} & \targ{} & \qw 
      \end{quantikz}}}
      \;=\;
    \vcenter{\hbox{%
      \begin{quantikz}[column sep=0.55cm, row sep=0.7cm]
        \lstick{}          & \ctrl{2} && \gate{Z}\vcw{1}  & \qw & \qw \\
        \lstick{}          &       &\ctrl{1} & \gate{Z}\vcw{1}  & \qw & \qw \\
        \lstick{$\ket{0}$} & \targ{}  & \targ{}  & \gate[2]{M_{XX}} & \meter{M_Z}\vcw{1} &\setwiretype{n} \\
        \lstick{}          & \qw      &&                  & \gate{X} & \qw
      \end{quantikz}}}
  \end{equation*}
  \caption{Circuit identity of parity-controlled \textsc{not} gadget. }
  \label{fig:paritynot-gadget}
\end{figure}
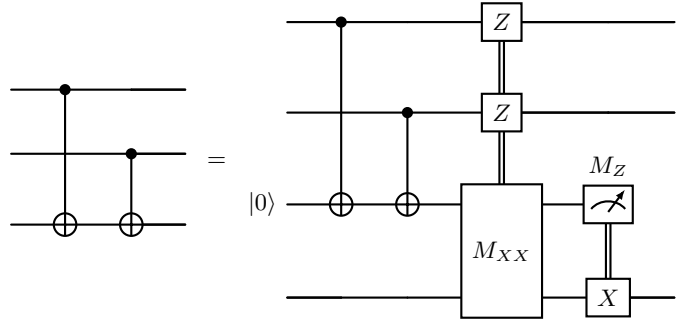

In a very similar fashion, the circuit identity of \zcref{fig:paritynot-gadget} allows one to perform a {parity-controlled} \textsc{Not} gate, i.e.\@ a sequence of \textsc{Cnot} all sharing the same target.

Both circuit identities can be verified easily using $ZX$-calculus and obviously, they can be combined by doing a $M_{X\ldots X}$ instead of a $M_{XX}$ measurement in \zcref{fig:paritynot-gadget} to perform a {parity-controlled} multi-target \textsc{not} gate.
\begin{figure}[!htb]
    \centering
    \input{graphics/compilation_graphics/tikz/parity_control_multi_target_proof_short}
    \caption{$ZX$-calculus proof of a parity-controlled multi-target \textsc{not} circuit for every possible random measurement outcomes $a,b\in\mathbb{F}_2$. 
    Phases are in units of $\pi$ and scalars are omitted.
    The outcome $a$ is $\pi$-copied onto the targets, where it cancels with the $X$ corrections, the phase $b$ is $\pi$-copied through the ancilla onto the controls, where it cancels the $Z$ corrections.
    A final bialgebra yields the target circuit. 
    The $ZX$ rules used are: ($f$) spider fusion, ($\pi$) $\pi$-copy, ($b$) bialgebra.}
    \label{fig:placeholder}
\end{figure}
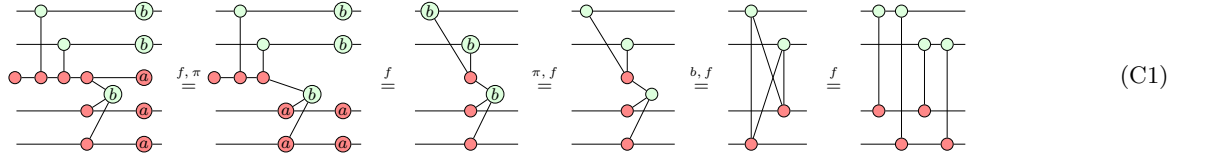

The $X$-logical operator measurement is performed by lattice surgery adapted to phaseflip classical codes using routing qubits, as originally introduced in~\cite{Gouzien_2023} and extended in~\cite{Ruiz_2025, Simultaneous2026prep}.
As a consequence, an arbitrary sequence of \textsc{Cnot}s with disjoint set of control and target qubits can be performed in $d+O(1)$ rounds of cycle extraction on a repetition code cat qubits architecture.
The $d$ rounds come from the duration of the $M_{X\ldots X}$ measurement, while the $O(1)$ gathers the transversal initialization of the auxiliary repetition code and the local transversal \textsc{Cnot}s with each control logical qubit. 
Those circuit identities are compiled in the main text onto the repetition code and the LDPC layout.
We round this duration to simply $d$ rounds.

\subsection{Hadamard and S gates}
\label{app: hadamard}
\begin{figure}[!htb]
\centering
\input{graphics/compilation_graphics/tikz/catalytic_resource_state_proof_v2.tikz}
\caption{ZX-calculus proof that the \textsc{Cnot} network acting on the three $\ket{i}$ states (qubits 2--4 of \zcref{fig:H_w_cnot}), prepares $\ket{i}$ times the two-qubit $\ket{CZ}$ state on the first and third qubits.
Scalars are omitted.
Rules: ($f$) spider fusion, ($eu$) Euler decomposition $Z_{\pi/2}=X_{-\pi/2}HX_{-\pi/2}$, ($h$) color change, ($lc$) local complementation~\cite{duncan2009graph,Duncan_2020}, ($\pi$) $\pi$-copy.}\label{fig: catalytic state}
\end{figure}
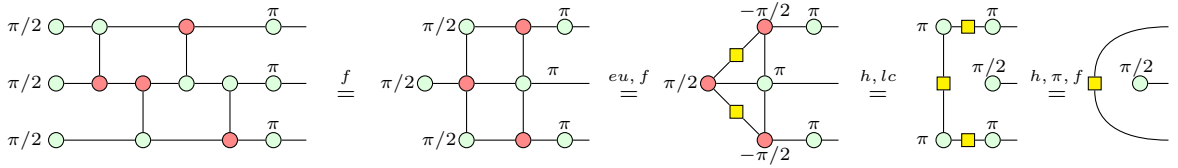

Hadamard and $S$ gates are achieved by $\ket{i}$ state injection methods. 
It was known that the Hadamard gates could be performed by consuming three $\ket{i}$ states, see Appendix J.2 of~\cite{Chamberland_2022}.
Here we rather use a catalysis injection scheme that requires $\ket{i}$ as well but consumes only 2 of them.
The third one can be used again in the next Hadamard gate.
More precisely, the three input $\ket{i}$ undergo a sequence of \textsc{cnot}s transforming this $\ket{i}^{\otimes3}$ product state into a $\ket{\CZ}\ket{i}$ state.
The $H$ gate is then injected using the two qubit resource state $\ket{\CZ}$ while the state $\ket{i}$ can be reused.
This transformation of the resource state is proved using $ZX$-calculus in \zcref{fig: catalytic state}.

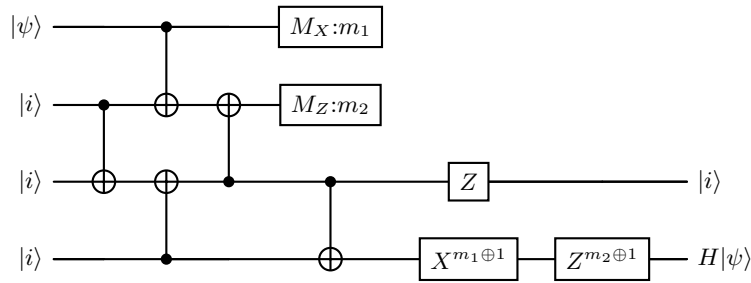
\begin{figure}[!htb]
    \centering
    \input{graphics/compilation_graphics/App_clifford/Hadamard_injection}
    \caption{Out-of-place catalyzed Hadamard injection circuit with $Z$-access (\textsc{Cnot}) to $\ket{\psi}$.}
    \label{fig:H_w_cnot}
\end{figure}

If a '$Z$-access' of the logical qubit $\ket{\psi}$ on which we want to perform the $H$ gate is available, as is the case in resource state buffers, then the interaction with $\ket{\psi}$ is made through a local transversal \textsc{Cnot} gate, see \zcref{fig:H_w_cnot}.
In a ZX point of view, this is nothing more than connecting the input wire of our qubit $\ket{\psi}$ to the top wire of the resource state depicted in \zcref{fig: catalytic state} returning $H\ket{\psi}$ on the bottom wire of the resource states. 

\zcref{fig:H_w_cnot} is a circuit identity that performs Hadamard gates with only '$X$-access' to the logical qubit through a $M_{XX}$ interaction instead of \textsc{Cnot}.
 It is easy to show that both circuit identities are equivalent using standard $ZX$-calculus identities.
This is useful in the case where repetition codes are stored densely, see system qubits storage in~\zcref{fig:global layout}.

\begin{figure}[!htb]
    \centering
    \input{graphics/compilation_graphics/App_clifford/Hadamard_injection_MXX}
    \caption{Out-of-place catalyzed Hadamard injection circuit with $X$-access ($M_{XX}$) to $\ket{\psi}$.}
    \label{fig:H_w_Mxx}
\end{figure}
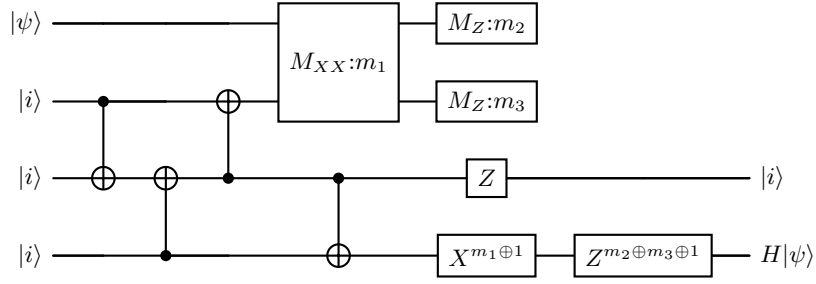

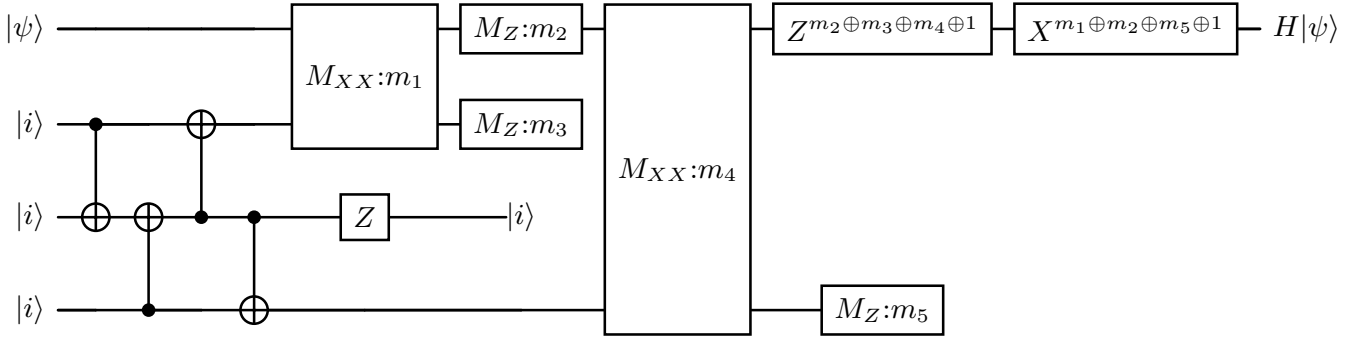
\begin{figure}[!htb]
    \centering
    \input{graphics/compilation_graphics/App_clifford/Hadamard_injection_LDPC}
    \caption{In-place catalyzed Hadamard injection circuit on a logical qubit $\ket{\psi}$.
    If $\ket{\psi}$ is stored in a LDPC code, the $M_Z$ measurement of $\ket{\psi}$ (outcome $m_2$) is non-destructive~\cite{Ruiz_2025}.
    Otherwise for a repetition code architecture it can be replaced by a transversal measurement and a $\ket{0}$ reset.}
    \label{fig:H_inplace}
\end{figure}

Both circuits are out-of-place, leaving the output $H\ket{\psi}$ state in a distinct repetition code.
When $\ket{\psi}$ is stored in the LDPC layer, we apply an in-place Hadamard using the circuit of \zcref{fig:H_inplace}.
It first teleports $H\ket{\psi}$ into a repetition code as in \zcref{fig:H_w_Mxx}, with the $M_Z$ measurement of $\ket{\psi}$ performed non destructively as originally proposed in~\cite{Ruiz_2025}, and then teleports it back into the same LDPC logical qubit with a subsequent $M_{XX}$ measurement.

\subsection{\texorpdfstring{$\CXX$ gate and variants}{CxX gate and variants}}\label{App: CxX}

The $\CXX$ and its parity-multi-target variant represent most of the cost of the \textsc{Phase-fixup} subroutine described in \zcref{sec: phase fixup} and compiled in \zcref{sec: PhaseFixup comp}.
This gate consumes $\ket{i}$ state resources pre-processed in the form of a catalytic state, see~\zcref{fig: catalytic state}.

\begin{figure}[!htb]
    \begin{center}
  \begin{quantikz}
    \lstick{$q_1$}     & \qw      & \qw       & \qw       & \gate[2]{M_{XX}\outcome{m_{1}}} & \qw                        & \gate{X^{m_{2}\xor m_{3}\xor 1}} & \qw                              \rstick{$q_1$}    \\
    \lstick{$\ket{i}$} & \ctrl{1} & \qw       & \targ{}   &                                 & \gate{M_Z\outcome{m_{3}}}  & \setwiretype{n}                  &                                                     \\
    \lstick{$\ket{i}$} & \targ{}  & \targ{}   & \ctrl{-1} & \ctrl{1}                        & \gate{Z}                   & \qw                              &                               \rstick{$\ket{i}$} \\
    \lstick{$\ket{i}$} & \qw      & \ctrl{-1} & \qw       & \targ{}                         & \gate[2]{M_{XX}\outcome{m_{2}}} & \gate{M_Z\outcome{m_{4}}} & \setwiretype{n}                                         \\
    \lstick{$q_2$}     & \qw      & \qw       & \qw       & \qw                             &                                                         & \gate{X^{m_{1}\xor m_{4}\xor 1}} & \qw \rstick{$q_2$}
  \end{quantikz}%
\end{center}
    \caption{In-place catalyzed $\CXX$ gate injection circuit.}
    \label{fig:cxX}
\end{figure}
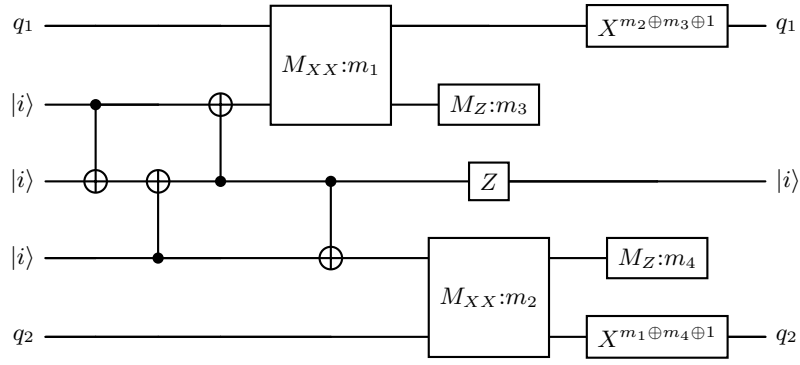

\begin{figure}[!htb]
    \begin{center}
  \begin{quantikz}
    \lstick{$c_1$}     & \qw      & \qw       & \qw       & \gate[3]{M_{XXX}\outcome{m_{1}}} & \qw                             & \gate{X^{m_{2}\xor m_{3}\xor 1}} & \qw \rstick{$c_1$}     \\
    \lstick{$c_2$}     & \qw      & \qw       & \qw       &                                  & \qw                             & \gate{X^{m_{2}\xor m_{3}\xor 1}} & \qw \rstick{$c_2$}     \\
    \lstick{$\ket{i}$} & \ctrl{1} & \qw       & \targ{}   &                                  & \gate{M_Z\outcome{m_{3}}}       & \setwiretype{n}                  &                                  &                        \\
    \lstick{$\ket{i}$} & \targ{}  & \targ{}   & \ctrl{-1} & \ctrl{1}                         & \gate{Z}                        & \qw                              & \qw  \rstick{$\ket{i}$} \\
    \lstick{$\ket{i}$} & \qw      & \ctrl{-1} & \qw       & \targ{}                          & \gate[3]{M_{XXX}\outcome{m_{2}}} & \gate{M_Z\outcome{m_{4}}}       & \setwiretype{n}                  &                        \\
    \lstick{$t_1$}     & \qw      & \qw       & \qw       & \qw                              &                                 & \gate{X^{m_{1}\xor m_{4}\xor 1}} & \qw \rstick{$t_1$}     \\
    \lstick{$t_2$}     & \qw      & \qw       & \qw       & \qw                              &                                 & \gate{X^{m_{1}\xor m_{4}\xor 1}} & \qw \rstick{$t_2$}
  \end{quantikz}%
\end{center}
    \caption{In-place catalyzed $C_{XX}XX$ injection circuit.
    An $X_{t_1}X_{t_2}$ gate is applied if the parity of $X_{c_1}X_{c_2}$ is $-1$ or equivalently the other way around. }
    \label{fig: CxxXX}
\end{figure}
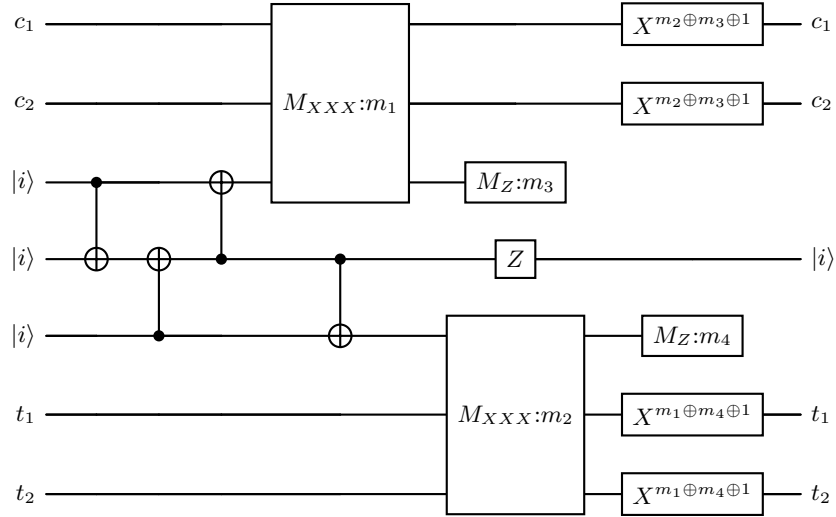

Note that, as opposed to the \textsc{cnot} gate, the $\CXX$ is symmetric regarding control and target. 
As a consequence, it doesn't make sense to differentiate parity-controlled $\CXX$ from multi-target $\CXX$ and we directly exhibit a circuit identity that injects a parity control multi target $\CXX$ gate in \zcref{fig: CxxXX}.

\begin{figure}[!htb]
\centering
\input{graphics/compilation_graphics/App_clifford/cxx_multi_proof.tikz}
\caption{ZX-calculus proof that the in-place injection of~\zcref{fig: CxxXX} implements the four gates $C_XX$ between every control and every target, shown for $2$ control and $2$ targets. 
Scalars are omitted and multiple of $\pi$ phases are written as boolean variables.
Each $M_{X\cdots X}$,$M_{Z}$ with outcome $m_k$ induces corrections that are $x_C=m_2\oplus m_3\oplus1$ and $x_T=m_1\oplus m_4\oplus1$. After fusion and Pauli pushing (third diagram), a bialgebra turns this into a complete bipartite set of Hadamard edges, which unfuses into the 4 desired gates. Rules: ($f$) spider fusion, ($h$) color change, ($\pi$) $\pi$-copy, ($b$) bialgebra.}
\label{fig:cxx-proof}
\end{figure}
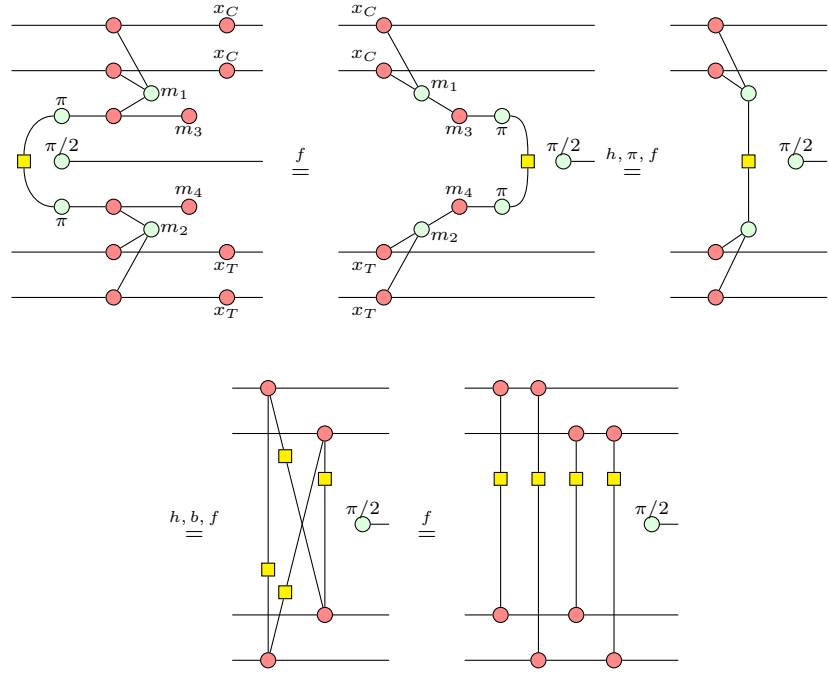

The advantage of this circuit compared to naive sequential application of \zcref{fig:cxX} is that it consumes a single catalyzed state. 
The equivalence between the circuit of \zcref{fig: CxxXX} and a sequence of four $\CXX$ gates is proven using ZX-calculus in \zcref{fig:cxx-proof}.

\clearpage

%% file: graphics/compilation_graphics/tikz/parity_control_multi_target_proof_short.tex
\begin{equation}
  \vcenter{\hbox{\scalebox{\zxscale}{\input{graphics/compilation_graphics/tikz/parity_control_multi_target_proof_short.tikz}}}}
  \label{eq:parity_multi_target_zx_proof}
\end{equation}

%% file: graphics/compilation_graphics/tikz/parity_control_multi_target_proof_short.tikz
\begin{tikzpicture}
    \begin{pgfonlayer}{nodelayer}
        \node [style=none] (0) at (0.00, 4.40) {};
        \node [style=none] (1) at (4.91, 4.40) {};
        \node [style=none] (2) at (0.00, 3.30) {};
        \node [style=none] (3) at (4.91, 3.30) {};
        \node [style=none] (4) at (0.00, 1.10) {};
        \node [style=none] (5) at (4.91, 1.10) {};
        \node [style=none] (6) at (0.00, 0.00) {};
        \node [style=none] (7) at (4.91, 0.00) {};
        \node [style=Z dot] (8) at (0.86, 4.40) {};
        \node [style=Z dot] (9) at (1.62, 3.30) {};
        \node [style=X dot] (10) at (0.00, 2.20) {};
        \node [style=X dot] (11) at (0.86, 2.20) {};
        \node [style=X dot] (12) at (1.62, 2.20) {};
        \node [style=X dot] (13) at (2.38, 2.20) {};
        \node [style=X dot] (14) at (2.38, 1.10) {};
        \node [style=X dot] (15) at (2.38, 0.00) {};
        \node [style=Z phase dot] (16) at (3.24, 1.65) {$b$};
        \node [style=X phase dot] (17) at (4.27, 2.20) {$a$};
        \node [style=Z phase dot] (18) at (4.27, 4.40) {$b$};
        \node [style=Z phase dot] (19) at (4.27, 3.30) {$b$};
        \node [style=X phase dot] (20) at (4.27, 1.10) {$a$};
        \node [style=X phase dot] (21) at (4.27, 0.00) {$a$};
        \node [style=none] (22) at (5.75, 2.20) {$\overset{\raisebox{0.5ex}{$\scriptstyle f,\,\pi$}}{=}$};
        \node [style=none] (23) at (6.59, 4.40) {};
        \node [style=none] (24) at (11.50, 4.40) {};
        \node [style=none] (25) at (6.59, 3.30) {};
        \node [style=none] (26) at (11.50, 3.30) {};
        \node [style=none] (27) at (6.59, 1.10) {};
        \node [style=none] (28) at (11.50, 1.10) {};
        \node [style=none] (29) at (6.59, 0.00) {};
        \node [style=none] (30) at (11.50, 0.00) {};
        \node [style=Z dot] (31) at (7.45, 4.40) {};
        \node [style=Z dot] (32) at (8.21, 3.30) {};
        \node [style=X dot] (33) at (6.59, 2.20) {};
        \node [style=X dot] (34) at (7.45, 2.20) {};
        \node [style=X dot] (35) at (8.21, 2.20) {};
        \node [style=X phase dot] (36) at (8.96, 1.10) {$a$};
        \node [style=X phase dot] (37) at (8.96, 0.00) {$a$};
        \node [style=Z phase dot] (38) at (9.83, 1.65) {$b$};
        \node [style=Z phase dot] (39) at (10.85, 4.40) {$b$};
        \node [style=Z phase dot] (40) at (10.85, 3.30) {$b$};
        \node [style=X phase dot] (41) at (10.85, 1.10) {$a$};
        \node [style=X phase dot] (42) at (10.85, 0.00) {$a$};
        \node [style=none] (43) at (12.34, 2.20) {$\overset{\raisebox{0.5ex}{$\scriptstyle f$}}{=}$};
        \node [style=none] (44) at (13.18, 4.40) {};
        \node [style=none] (45) at (16.69, 4.40) {};
        \node [style=none] (46) at (13.18, 3.30) {};
        \node [style=none] (47) at (16.69, 3.30) {};
        \node [style=none] (48) at (13.18, 1.10) {};
        \node [style=none] (49) at (16.69, 1.10) {};
        \node [style=none] (50) at (13.18, 0.00) {};
        \node [style=none] (51) at (16.69, 0.00) {};
        \node [style=Z phase dot] (52) at (13.72, 4.40) {$b$};
        \node [style=Z phase dot] (53) at (15.07, 3.30) {$b$};
        \node [style=X dot] (54) at (15.07, 2.20) {};
        \node [style=X dot] (55) at (15.07, 1.10) {};
        \node [style=X dot] (56) at (15.07, 0.00) {};
        \node [style=Z phase dot] (57) at (15.88, 1.65) {$b$};
        \node [style=none] (58) at (17.52, 2.20) {$\overset{\raisebox{0.5ex}{$\scriptstyle \pi,\,f$}}{=}$};
        \node [style=none] (59) at (18.36, 4.40) {};
        \node [style=none] (60) at (21.87, 4.40) {};
        \node [style=none] (61) at (18.36, 3.30) {};
        \node [style=none] (62) at (21.87, 3.30) {};
        \node [style=none] (63) at (18.36, 1.10) {};
        \node [style=none] (64) at (21.87, 1.10) {};
        \node [style=none] (65) at (18.36, 0.00) {};
        \node [style=none] (66) at (21.87, 0.00) {};
        \node [style=Z dot] (67) at (18.90, 4.40) {};
        \node [style=Z dot] (68) at (20.25, 3.30) {};
        \node [style=X dot] (69) at (20.25, 2.20) {};
        \node [style=X dot] (70) at (20.25, 1.10) {};
        \node [style=X dot] (71) at (20.25, 0.00) {};
        \node [style=Z dot] (72) at (21.06, 1.65) {};
        \node [style=none] (73) at (22.71, 2.20) {$\overset{\raisebox{0.5ex}{$\scriptstyle b,\,f$}}{=}$};
        \node [style=none] (74) at (23.54, 4.40) {};
        \node [style=none] (75) at (26.24, 4.40) {};
        \node [style=none] (76) at (23.54, 3.30) {};
        \node [style=none] (77) at (26.24, 3.30) {};
        \node [style=none] (78) at (23.54, 1.10) {};
        \node [style=none] (79) at (26.24, 1.10) {};
        \node [style=none] (80) at (23.54, 0.00) {};
        \node [style=none] (81) at (26.24, 0.00) {};
        \node [style=Z dot] (82) at (24.35, 4.40) {};
        \node [style=Z dot] (83) at (25.43, 3.30) {};
        \node [style=X dot] (84) at (25.43, 1.10) {};
        \node [style=X dot] (85) at (24.35, 0.00) {};
        \node [style=none] (86) at (27.08, 2.20) {$\overset{\raisebox{0.5ex}{$\scriptstyle f$}}{=}$};
        \node [style=none] (87) at (27.92, 4.40) {};
        \node [style=none] (88) at (31.48, 4.40) {};
        \node [style=none] (89) at (27.92, 3.30) {};
        \node [style=none] (90) at (31.48, 3.30) {};
        \node [style=none] (91) at (27.92, 1.10) {};
        \node [style=none] (92) at (31.48, 1.10) {};
        \node [style=none] (93) at (27.92, 0.00) {};
        \node [style=none] (94) at (31.48, 0.00) {};
        \node [style=Z dot] (95) at (28.57, 4.40) {};
        \node [style=Z dot] (96) at (29.32, 4.40) {};
        \node [style=Z dot] (97) at (30.08, 3.30) {};
        \node [style=Z dot] (98) at (30.83, 3.30) {};
        \node [style=X dot] (99) at (28.57, 1.10) {};
        \node [style=X dot] (100) at (30.08, 1.10) {};
        \node [style=X dot] (101) at (29.32, 0.00) {};
        \node [style=X dot] (102) at (30.83, 0.00) {};
    \end{pgfonlayer}
    \begin{pgfonlayer}{edgelayer}
        \draw (0) to (8);
        \draw (8) to (18);
        \draw (18) to (1);
        \draw (2) to (9);
        \draw (9) to (19);
        \draw (19) to (3);
        \draw (10) to (11);
        \draw (11) to (12);
        \draw (12) to (13);
        \draw (13) to (17);
        \draw (8) to (11);
        \draw (9) to (12);
        \draw (13) to (16);
        \draw (16) to (14);
        \draw (16) to (15);
        \draw (4) to (14);
        \draw (14) to (20);
        \draw (20) to (5);
        \draw (6) to (15);
        \draw (15) to (21);
        \draw (21) to (7);
        \draw (23) to (31);
        \draw (31) to (39);
        \draw (39) to (24);
        \draw (25) to (32);
        \draw (32) to (40);
        \draw (40) to (26);
        \draw (33) to (34);
        \draw (34) to (35);
        \draw (31) to (34);
        \draw (32) to (35);
        \draw (35) to (38);
        \draw (38) to (36);
        \draw (38) to (37);
        \draw (27) to (36);
        \draw (36) to (41);
        \draw (41) to (28);
        \draw (29) to (37);
        \draw (37) to (42);
        \draw (42) to (30);
        \draw (44) to (52);
        \draw (52) to (45);
        \draw (46) to (53);
        \draw (53) to (47);
        \draw (52) to (54);
        \draw (53) to (54);
        \draw (54) to (57);
        \draw (57) to (55);
        \draw (57) to (56);
        \draw (48) to (55);
        \draw (55) to (49);
        \draw (50) to (56);
        \draw (56) to (51);
        \draw (59) to (67);
        \draw (67) to (60);
        \draw (61) to (68);
        \draw (68) to (62);
        \draw (67) to (69);
        \draw (68) to (69);
        \draw (69) to (72);
        \draw (72) to (70);
        \draw (72) to (71);
        \draw (63) to (70);
        \draw (70) to (64);
        \draw (65) to (71);
        \draw (71) to (66);
        \draw (74) to (82);
        \draw (82) to (75);
        \draw (76) to (83);
        \draw (83) to (77);
        \draw (78) to (84);
        \draw (84) to (79);
        \draw (80) to (85);
        \draw (85) to (81);
        \draw (82) to (84);
        \draw (82) to (85);
        \draw (83) to (84);
        \draw (83) to (85);
        \draw (87) to (95);
        \draw (95) to (96);
        \draw (96) to (88);
        \draw (89) to (97);
        \draw (97) to (98);
        \draw (98) to (90);
        \draw (91) to (99);
        \draw (99) to (100);
        \draw (100) to (92);
        \draw (93) to (101);
        \draw (101) to (102);
        \draw (102) to (94);
        \draw (95) to (99);
        \draw (96) to (101);
        \draw (97) to (100);
        \draw (98) to (102);
    \end{pgfonlayer}
\end{tikzpicture}

%% file: graphics/compilation_graphics/tikz/catalytic_resource_state_proof_v2.tikz
\begin{tikzpicture}
    \begin{pgfonlayer}{nodelayer}
        \node [style=Z dot] (0) at (1.25, 3.00) {};
        \node [style=none] (1) at (0.41, 3.00) {\scriptsize$\pi/2$};
        \node [style=Z dot] (2) at (2.40, 3.00) {};
        \node [style=X dot] (3) at (4.70, 3.00) {};
        \node [style=Z dot] (4) at (7.00, 3.00) {};
        \node [style=none] (5) at (7.00, 3.40) {\scriptsize$\pi$};
        \node [style=none] (6) at (7.92, 3.00) {};
        \node [style=Z dot] (7) at (1.25, 1.50) {};
        \node [style=none] (8) at (0.41, 1.50) {\scriptsize$\pi/2$};
        \node [style=X dot] (9) at (2.40, 1.50) {};
        \node [style=X dot] (10) at (3.55, 1.50) {};
        \node [style=Z dot] (11) at (4.70, 1.50) {};
        \node [style=Z dot] (12) at (5.85, 1.50) {};
        \node [style=Z dot] (13) at (7.00, 1.50) {};
        \node [style=none] (14) at (7.00, 1.90) {\scriptsize$\pi$};
        \node [style=none] (15) at (7.92, 1.50) {};
        \node [style=Z dot] (16) at (1.25, 0.00) {};
        \node [style=none] (17) at (0.41, 0.00) {\scriptsize$\pi/2$};
        \node [style=Z dot] (18) at (3.55, 0.00) {};
        \node [style=X dot] (19) at (5.85, 0.00) {};
        \node [style=Z dot] (20) at (7.00, 0.00) {};
        \node [style=none] (21) at (7.00, 0.40) {\scriptsize$\pi$};
        \node [style=none] (22) at (7.92, 0.00) {};
        \node [style=none] (23) at (8.92, 1.50) {$\overset{\raisebox{0.5ex}{$\scriptstyle f$}}{=}$};
        \node [style=Z dot] (24) at (12.11, 3.00) {};
        \node [style=none] (25) at (11.38, 3.00) {\scriptsize$\pi/2$};
        \node [style=X dot] (26) at (13.61, 3.00) {};
        \node [style=Z dot] (27) at (14.71, 3.00) {};
        \node [style=none] (28) at (14.71, 3.40) {\scriptsize$\pi$};
        \node [style=none] (29) at (15.41, 3.00) {};
        \node [style=Z dot] (30) at (11.01, 1.50) {};
        \node [style=none] (31) at (10.28, 1.50) {\scriptsize$\pi/2$};
        \node [style=X dot] (32) at (12.11, 1.50) {};
        \node [style=Z dot] (33) at (13.61, 1.50) {};
        \node [style=none] (34) at (14.41, 1.83) {\scriptsize$\pi$};
        \node [style=none] (35) at (15.41, 1.50) {};
        \node [style=Z dot] (36) at (12.11, 0.00) {};
        \node [style=none] (37) at (11.38, 0.00) {\scriptsize$\pi/2$};
        \node [style=X dot] (38) at (13.61, 0.00) {};
        \node [style=Z dot] (39) at (14.71, 0.00) {};
        \node [style=none] (40) at (14.71, 0.40) {\scriptsize$\pi$};
        \node [style=none] (41) at (15.41, 0.00) {};
        \node [style=none] (42) at (16.41, 1.50) {$\overset{\raisebox{0.5ex}{$\scriptstyle eu,\,f$}}{=}$};
        \node [style=X dot] (43) at (18.50, 1.50) {};
        \node [style=none] (44) at (17.77, 1.50) {\scriptsize$\pi/2$};
        \node [style=X dot] (45) at (20.00, 3.00) {};
        \node [style=none] (46) at (20.00, 3.40) {\scriptsize$-\pi/2$};
        \node [style=X dot] (47) at (20.00, 0.00) {};
        \node [style=none] (48) at (20.00, -0.40) {\scriptsize$-\pi/2$};
        \node [style=Z dot] (49) at (20.00, 1.50) {};
        \node [style=none] (50) at (20.40, 1.83) {\scriptsize$\pi$};
        \node [style=Z dot] (51) at (21.30, 3.00) {};
        \node [style=none] (52) at (21.30, 3.40) {\scriptsize$\pi$};
        \node [style=Z dot] (53) at (21.30, 0.00) {};
        \node [style=none] (54) at (21.30, 0.40) {\scriptsize$\pi$};
        \node [style=none] (55) at (22.00, 3.00) {};
        \node [style=none] (56) at (22.00, 1.50) {};
        \node [style=none] (57) at (22.00, 0.00) {};
        \node [style=hadamard] (58) at (19.25, 2.25) {};
        \node [style=hadamard] (59) at (19.25, 0.75) {};
        \node [style=none] (60) at (23.00, 1.50) {$\overset{\raisebox{0.5ex}{$\scriptstyle h,\,lc$}}{=}$};
        \node [style=Z dot] (61) at (24.73, 3.00) {};
        \node [style=none] (62) at (24.18, 3.00) {\scriptsize$\pi$};
        \node [style=Z dot] (63) at (24.73, 0.00) {};
        \node [style=none] (64) at (24.18, 0.00) {\scriptsize$\pi$};
        \node [style=Z dot] (65) at (26.03, 3.00) {};
        \node [style=none] (66) at (26.03, 3.40) {\scriptsize$\pi$};
        \node [style=Z dot] (67) at (26.03, 0.00) {};
        \node [style=none] (68) at (26.03, 0.40) {\scriptsize$\pi$};
        \node [style=Z dot] (69) at (26.03, 1.50) {};
        \node [style=none] (70) at (25.93, 1.92) {\scriptsize$\pi/2$};
        \node [style=none] (71) at (26.73, 3.00) {};
        \node [style=none] (72) at (26.73, 1.50) {};
        \node [style=none] (73) at (26.73, 0.00) {};
        \node [style=hadamard] (74) at (24.73, 1.50) {};
        \node [style=hadamard] (75) at (25.38, 3.00) {};
        \node [style=hadamard] (76) at (25.38, 0.00) {};
        \node [style=none] (77) at (27.73, 1.50) {$\overset{\raisebox{0.5ex}{$\scriptstyle h,\,\pi,\,f$}}{=}$};
        \node [style=hadamard] (78) at (28.73, 1.50) {};
        \node [style=Z dot] (79) at (29.93, 1.50) {};
        \node [style=none] (80) at (29.83, 1.92) {\scriptsize$\pi/2$};
        \node [style=none] (81) at (30.73, 3.00) {};
        \node [style=none] (82) at (30.73, 1.50) {};
        \node [style=none] (83) at (30.73, 0.00) {};
    \end{pgfonlayer}
    \begin{pgfonlayer}{edgelayer}
        \draw (0) to (2);
        \draw (2) to (3);
        \draw (3) to (4);
        \draw (4) to (6);
        \draw (7) to (9);
        \draw (9) to (10);
        \draw (10) to (11);
        \draw (11) to (12);
        \draw (12) to (13);
        \draw (13) to (15);
        \draw (16) to (18);
        \draw (18) to (19);
        \draw (19) to (20);
        \draw (20) to (22);
        \draw (2) to (9);
        \draw (18) to (10);
        \draw (11) to (3);
        \draw (12) to (19);
        \draw (30) to (32);
        \draw (24) to (32);
        \draw (36) to (32);
        \draw (32) to (33);
        \draw (24) to (26);
        \draw (33) to (26);
        \draw (36) to (38);
        \draw (33) to (38);
        \draw (26) to (27);
        \draw (38) to (39);
        \draw (27) to (29);
        \draw (33) to (35);
        \draw (39) to (41);
        \draw (43) to (49);
        \draw (45) to (49);
        \draw (47) to (49);
        \draw (45) to (51);
        \draw (47) to (53);
        \draw (51) to (55);
        \draw (49) to (56);
        \draw (53) to (57);
        \draw (43) to (58);
        \draw (58) to (45);
        \draw (43) to (59);
        \draw (59) to (47);
        \draw (65) to (71);
        \draw (69) to (72);
        \draw (67) to (73);
        \draw (61) to (74);
        \draw (74) to (63);
        \draw (61) to (75);
        \draw (75) to (65);
        \draw (63) to (76);
        \draw (76) to (67);
        \draw [in=180, out=90, looseness=1.1] (78) to (81);
        \draw [in=180, out=-90, looseness=1.1] (78) to (83);
        \draw (79) to (82);
    \end{pgfonlayer}
\end{tikzpicture}

%% file: graphics/compilation_graphics/App_clifford/Hadamard_injection.tex
\begin{center}
  \begin{quantikz}
    \lstick{\ket{\psi}} & \qw      & \ctrl{1}  & \qw       & \gate{M_X\outcome{m_{1}}} & \setwiretype{n}                           &                          &                                      \\
    \lstick{\ket{i}}    & \ctrl{1} & \targ{}   & \targ{}   & \gate{M_Z\outcome{m_{2}}} &  \setwiretype{n}                        &                          &                                      \\
    \lstick{\ket{i}}    & \targ{}  & \targ{}   & \ctrl{-1} & \ctrl{1}                  & \gate{Z}                 & \qw                      & \qw \rstick{\ket{i}}                 \\
    \lstick{\ket{i}}    & \qw      & \ctrl{-1} & \qw       & \targ{}                   & \gate{X^{m_{1}\xor 1}}   & \gate{Z^{m_{2}\xor 1}}   & \qw \rstick{$H\mathord{\ket{\psi}}$}
  \end{quantikz}%
\end{center}

%% file: graphics/compilation_graphics/App_clifford/Hadamard_injection_MXX.tex
\begin{center}
  \begin{quantikz}
    \lstick{$\ket{\psi}$} & \qw      & \qw       & \qw       & \gate[2]{M_{XX}\outcome{m_{1}}} & \gate{M_Z\outcome{m_{2}}}  & \setwiretype{n}                  &                                                     \\
    \lstick{$\ket{i}$}     & \ctrl{1} & \qw       & \targ{}   &                                 & \gate{M_Z\outcome{m_{3}}}  & \setwiretype{n}                  &                                                     \\
    \lstick{$\ket{i}$}     & \targ{}  & \targ{}   & \ctrl{-1} & \ctrl{1}                        & \gate{Z}                   & \qw                              & \qw \rstick{$\ket{i}$}               \\
    \lstick{$\ket{i}$}     & \qw      & \ctrl{-1} & \qw       & \targ{}                         & \gate{X^{m_{1}\xor 1}}     & \gate{Z^{m_{2}\xor m_{3}\xor 1}} & \qw \rstick{$H{\ket{\psi}}$}
  \end{quantikz}%
\end{center}

%% file: graphics/compilation_graphics/App_clifford/Hadamard_injection_LDPC.tex
\providecommand{\outcome}[1]{\mathord{:}#1}

\begin{center}
  \resizebox{\linewidth}{!}{%
  \begin{quantikz}[column sep=0.25cm]
    \lstick{$\ket{\psi}$} & \qw      & \qw       & \qw       & \qw      & \gate[2]{M_{XX}\outcome{m_{1}}} & \gate{M_Z\outcome{m_{2}}} & \gate[4]{M_{XX}\outcome{m_{4}}} & \gate{Z^{m_{2}\xor m_{3}\xor m_{4}\xor 1}} & \gate{X^{m_{1}\xor m_{2}\xor m_{5}\xor 1}} & \qw \rstick{$H\mathord{\ket{\psi}}$} \\
    \lstick{$\ket{i}$}     & \ctrl{1} & \qw       & \targ{}   & \qw      &                                 & \gate{M_Z\outcome{m_{3}}} & \setwiretype{n}                 &                                            &                                            & \\
    \lstick{$\ket{i}$}     & \targ{}  & \targ{}   & \ctrl{-1} & \ctrl{1} & \gate{Z}                        &  \push{\ket{i}}       & \setwiretype{n}                &                                            &                                            & \\
    \lstick{$\ket{i}$}     & \qw      & \ctrl{-1} & \qw       & \targ{}  & \qw                             & \qw                       &                                 & \gate{M_Z\outcome{m_{5}}}                  & \setwiretype{n}                            &
  \end{quantikz}}%
\end{center}

%% file: graphics/compilation_graphics/App_clifford/cxx_multi_proof.tikz
\begin{tikzpicture}
    \begin{pgfonlayer}{nodelayer}
        \node [style=none] (0) at (0.00, 7.20) {};
        \node [style=none] (1) at (0.00, 6.00) {};
        \node [style=none] (2) at (0.00, 1.20) {};
        \node [style=none] (3) at (0.00, 0.00) {};
        \node [style=hadamard] (4) at (0.38, 3.60) {};
        \node [style=Z dot] (5) at (1.38, 4.80) {};
        \node [style=none] (6) at (1.38, 5.20) {\scriptsize$\pi$};
        \node [style=Z dot] (7) at (1.38, 2.40) {};
        \node [style=none] (8) at (1.38, 2.00) {\scriptsize$\pi$};
        \node [style=Z dot] (9) at (1.38, 3.60) {};
        \node [style=none] (10) at (1.38, 4.00) {\scriptsize$\pi/2$};
        \node [style=X dot] (11) at (2.75, 7.20) {};
        \node [style=X dot] (12) at (2.75, 6.00) {};
        \node [style=X dot] (13) at (2.75, 4.80) {};
        \node [style=Z dot] (14) at (3.75, 5.40) {};
        \node [style=none] (15) at (4.37, 5.40) {\scriptsize$m_1$};
        \node [style=X dot] (16) at (2.75, 2.40) {};
        \node [style=X dot] (17) at (2.75, 1.20) {};
        \node [style=X dot] (18) at (2.75, 0.00) {};
        \node [style=Z dot] (19) at (3.75, 1.80) {};
        \node [style=none] (20) at (4.37, 1.80) {\scriptsize$m_2$};
        \node [style=X dot] (21) at (4.75, 4.80) {};
        \node [style=none] (22) at (4.75, 4.40) {\scriptsize$m_3$};
        \node [style=X dot] (23) at (4.75, 2.40) {};
        \node [style=none] (24) at (4.75, 2.80) {\scriptsize$m_4$};
        \node [style=X dot] (25) at (5.75, 7.20) {};
        \node [style=none] (26) at (5.75, 7.60) {\scriptsize$x_C$};
        \node [style=X dot] (27) at (5.75, 6.00) {};
        \node [style=none] (28) at (5.75, 6.40) {\scriptsize$x_C$};
        \node [style=X dot] (29) at (5.75, 1.20) {};
        \node [style=none] (30) at (5.75, 0.80) {\scriptsize$x_T$};
        \node [style=X dot] (31) at (5.75, 0.00) {};
        \node [style=none] (32) at (5.75, -0.40) {\scriptsize$x_T$};
        \node [style=none] (33) at (6.75, 7.20) {};
        \node [style=none] (34) at (6.75, 6.00) {};
        \node [style=none] (35) at (6.75, 3.60) {};
        \node [style=none] (36) at (6.75, 1.20) {};
        \node [style=none] (37) at (6.75, 0.00) {};
        \node [style=none] (38) at (7.70, 3.60) {$\overset{\raisebox{0.5ex}{$\scriptstyle f$}}{=}$};
        \node [style=none] (39) at (8.65, 7.20) {};
        \node [style=none] (40) at (8.65, 6.00) {};
        \node [style=none] (41) at (8.65, 1.20) {};
        \node [style=none] (42) at (8.65, 0.00) {};
        \node [style=X dot] (43) at (9.90, 7.20) {};
        \node [style=none] (44) at (9.40, 7.60) {\scriptsize$x_C$};
        \node [style=X dot] (45) at (9.90, 6.00) {};
        \node [style=none] (46) at (9.40, 6.40) {\scriptsize$x_C$};
        \node [style=Z dot] (47) at (10.90, 5.40) {};
        \node [style=none] (48) at (11.52, 5.60) {\scriptsize$m_1$};
        \node [style=X dot] (49) at (9.90, 1.20) {};
        \node [style=none] (50) at (9.40, 0.80) {\scriptsize$x_T$};
        \node [style=X dot] (51) at (9.90, 0.00) {};
        \node [style=none] (52) at (9.40, -0.40) {\scriptsize$x_T$};
        \node [style=Z dot] (53) at (10.90, 1.80) {};
        \node [style=none] (54) at (11.52, 1.60) {\scriptsize$m_2$};
        \node [style=X dot] (55) at (11.90, 4.80) {};
        \node [style=none] (56) at (11.90, 4.40) {\scriptsize$m_3$};
        \node [style=Z dot] (57) at (13.03, 4.80) {};
        \node [style=none] (58) at (13.03, 4.40) {\scriptsize$\pi$};
        \node [style=hadamard] (59) at (13.71, 3.60) {};
        \node [style=Z dot] (60) at (13.03, 2.40) {};
        \node [style=none] (61) at (13.03, 2.80) {\scriptsize$\pi$};
        \node [style=X dot] (62) at (11.90, 2.40) {};
        \node [style=none] (63) at (11.90, 2.80) {\scriptsize$m_4$};
        \node [style=Z dot] (64) at (14.65, 3.60) {};
        \node [style=none] (65) at (14.65, 4.00) {\scriptsize$\pi/2$};
        \node [style=none] (66) at (15.53, 7.20) {};
        \node [style=none] (67) at (15.53, 6.00) {};
        \node [style=none] (68) at (15.53, 3.60) {};
        \node [style=none] (69) at (15.53, 1.20) {};
        \node [style=none] (70) at (15.53, 0.00) {};
        \node [style=none] (71) at (16.48, 3.60) {$\overset{\raisebox{0.5ex}{$\scriptstyle h,\,\pi,\,f$}}{=}$};
        \node [style=none] (72) at (17.43, 7.20) {};
        \node [style=none] (73) at (17.43, 6.00) {};
        \node [style=none] (74) at (17.43, 1.20) {};
        \node [style=none] (75) at (17.43, 0.00) {};
        \node [style=X dot] (76) at (18.68, 7.20) {};
        \node [style=X dot] (77) at (18.68, 6.00) {};
        \node [style=Z dot] (78) at (19.55, 5.40) {};
        \node [style=X dot] (79) at (18.68, 1.20) {};
        \node [style=X dot] (80) at (18.68, 0.00) {};
        \node [style=Z dot] (81) at (19.55, 1.80) {};
        \node [style=hadamard] (82) at (19.55, 3.60) {};
        \node [style=Z dot] (83) at (20.80, 3.60) {};
        \node [style=none] (84) at (20.80, 4.00) {\scriptsize$\pi/2$};
        \node [style=none] (85) at (21.68, 7.20) {};
        \node [style=none] (86) at (21.68, 6.00) {};
        \node [style=none] (87) at (21.68, 3.60) {};
        \node [style=none] (88) at (21.68, 1.20) {};
        \node [style=none] (89) at (21.68, 0.00) {};
        \node [style=none] (90) at (4.89, -6.00) {$\overset{\raisebox{0.5ex}{$\scriptstyle h,\,b,\,f$}}{=}$};
        \node [style=none] (91) at (5.84, -2.40) {};
        \node [style=none] (92) at (5.84, -3.60) {};
        \node [style=none] (93) at (5.84, -8.40) {};
        \node [style=none] (94) at (5.84, -9.60) {};
        \node [style=X dot] (95) at (6.84, -2.40) {};
        \node [style=X dot] (96) at (8.34, -3.60) {};
        \node [style=X dot] (97) at (8.34, -8.40) {};
        \node [style=X dot] (98) at (6.84, -9.60) {};
        \node [style=Z dot] (99) at (9.34, -6.00) {};
        \node [style=none] (100) at (9.34, -5.60) {\scriptsize$\pi/2$};
        \node [style=none] (101) at (10.09, -2.40) {};
        \node [style=none] (102) at (10.09, -3.60) {};
        \node [style=none] (103) at (10.09, -6.00) {};
        \node [style=none] (104) at (10.09, -8.40) {};
        \node [style=none] (105) at (10.09, -9.60) {};
        \node [style=hadamard] (106) at (6.84, -7.20) {};
        \node [style=hadamard] (107) at (8.34, -4.80) {};
        \node [style=hadamard] (108) at (7.29, -4.20) {};
        \node [style=hadamard] (109) at (7.29, -7.80) {};
        \node [style=none] (110) at (11.04, -6.00) {$\overset{\raisebox{0.5ex}{$\scriptstyle f$}}{=}$};
        \node [style=none] (111) at (11.99, -2.40) {};
        \node [style=none] (112) at (11.99, -3.60) {};
        \node [style=none] (113) at (11.99, -8.40) {};
        \node [style=none] (114) at (11.99, -9.60) {};
        \node [style=X dot] (115) at (12.99, -2.40) {};
        \node [style=X dot] (116) at (12.99, -8.40) {};
        \node [style=hadamard] (117) at (12.99, -4.80) {};
        \node [style=X dot] (118) at (13.99, -2.40) {};
        \node [style=X dot] (119) at (13.99, -9.60) {};
        \node [style=hadamard] (120) at (13.99, -4.80) {};
        \node [style=X dot] (121) at (14.99, -3.60) {};
        \node [style=X dot] (122) at (14.99, -8.40) {};
        \node [style=hadamard] (123) at (14.99, -4.80) {};
        \node [style=X dot] (124) at (15.99, -3.60) {};
        \node [style=X dot] (125) at (15.99, -9.60) {};
        \node [style=hadamard] (126) at (15.99, -4.80) {};
        \node [style=Z dot] (127) at (16.99, -6.00) {};
        \node [style=none] (128) at (16.99, -5.60) {\scriptsize$\pi/2$};
        \node [style=none] (129) at (17.74, -2.40) {};
        \node [style=none] (130) at (17.74, -3.60) {};
        \node [style=none] (131) at (17.74, -6.00) {};
        \node [style=none] (132) at (17.74, -8.40) {};
        \node [style=none] (133) at (17.74, -9.60) {};
    \end{pgfonlayer}
    \begin{pgfonlayer}{edgelayer}
        \draw (0) to (11);
        \draw (11) to (25);
        \draw (25) to (33);
        \draw (1) to (12);
        \draw (12) to (27);
        \draw (27) to (34);
        \draw (2) to (17);
        \draw (17) to (29);
        \draw (29) to (36);
        \draw (3) to (18);
        \draw (18) to (31);
        \draw (31) to (37);
        \draw [in=180, out=90, looseness=1.0] (4) to (5);
        \draw [in=180, out=-90, looseness=1.0] (4) to (7);
        \draw (5) to (13);
        \draw (13) to (21);
        \draw (7) to (16);
        \draw (16) to (23);
        \draw (9) to (35);
        \draw (14) to (11);
        \draw (14) to (12);
        \draw (14) to (13);
        \draw (19) to (16);
        \draw (19) to (17);
        \draw (19) to (18);
        \draw (39) to (43);
        \draw (43) to (66);
        \draw (40) to (45);
        \draw (45) to (67);
        \draw (41) to (49);
        \draw (49) to (69);
        \draw (42) to (51);
        \draw (51) to (70);
        \draw (47) to (43);
        \draw (47) to (45);
        \draw (53) to (49);
        \draw (53) to (51);
        \draw (47) to (55);
        \draw (55) to (57);
        \draw [in=90, out=0, looseness=1.0] (57) to (59);
        \draw [in=0, out=-90, looseness=1.0] (59) to (60);
        \draw (60) to (62);
        \draw (62) to (53);
        \draw (64) to (68);
        \draw (72) to (76);
        \draw (76) to (85);
        \draw (73) to (77);
        \draw (77) to (86);
        \draw (74) to (79);
        \draw (79) to (88);
        \draw (75) to (80);
        \draw (80) to (89);
        \draw (78) to (76);
        \draw (78) to (77);
        \draw (81) to (79);
        \draw (81) to (80);
        \draw (78) to (82);
        \draw (82) to (81);
        \draw (83) to (87);
        \draw (91) to (95);
        \draw (95) to (101);
        \draw (92) to (96);
        \draw (96) to (102);
        \draw (93) to (97);
        \draw (97) to (104);
        \draw (94) to (98);
        \draw (98) to (105);
        \draw (99) to (103);
        \draw (95) to (106);
        \draw (106) to (98);
        \draw (96) to (107);
        \draw (107) to (97);
        \draw (95) to (108);
        \draw (108) to (97);
        \draw (96) to (109);
        \draw (109) to (98);
        \draw (111) to (115);
        \draw (113) to (116);
        \draw (115) to (117);
        \draw (117) to (116);
        \draw (115) to (118);
        \draw (114) to (119);
        \draw (118) to (120);
        \draw (120) to (119);
        \draw (112) to (121);
        \draw (116) to (122);
        \draw (121) to (123);
        \draw (123) to (122);
        \draw (121) to (124);
        \draw (119) to (125);
        \draw (124) to (126);
        \draw (126) to (125);
        \draw (118) to (129);
        \draw (124) to (130);
        \draw (122) to (132);
        \draw (125) to (133);
        \draw (127) to (131);
    \end{pgfonlayer}
\end{tikzpicture}